\documentclass[usenatbib]{mn2e}
\usepackage{apjfonts,amsfonts,amsmath,amssymb,bm,ctable,verbatim}

\newcommand{\Alf}{{Alfv\'en}}
\newcommand{\bhat}{\hat{\bf b}}

\newcommand{\gizmourl}{\href{http://www.tapir.caltech.edu/~phopkins/Site/GIZMO.html}{\url{http://www.tapir.caltech.edu/~phopkins/Site/GIZMO.html}}}

\newcommand{\paperone}{Paper {\small I}}
\newcommand{\papertwo}{Paper {\small II}}

\newcommand{\etal}{et al.}

\newcommand{\acknowledgments}[1]{\begin{small}\section*{Acknowledgments}\end{small}{\noindent #1}\vspace{5pt}}
\newcommand{\datastatement}[1]{\begin{small}\section*{Data Availability Statement}\end{small}{\noindent #1}\vspace{5pt}}

\title[CR Spectra in Galaxies]{First Predicted Cosmic Ray Spectra, Primary-to-Secondary Ratios, and Ionization Rates from MHD Galaxy Formation Simulations}

\author[Hopkins \etal]{
\parbox[t]{\textwidth}{
Philip F.~Hopkins$^1$, Iryna S.\ Butsky$^{1,2}$,  
Georgia V.~Panopoulou$^1$, Suoqing Ji$^1$, \\
Eliot Quataert$^3$, Claude-Andr{\'e} Faucher-Gigu{\`e}re$^4$, Du\v{s}an Kere\v{s}$^5$
}\vspace*{4pt} \\
$^1$ TAPIR, Mailcode 350-17, California Institute of Technology, Pasadena, CA 91125, USA. E-mail:phopkins@caltech.edu \\
$^2$ Astronomy Department, University of Washington, Seattle, WA 98195, USA \\
$^3$ Department of Astrophysical Sciences, Princeton University, Peyton Hall, Princeton, NJ 08544, USA \\
$^4$ Department of Physics and Astronomy and CIERA, Northwestern University, 2145 Sheridan Road, Evanston, IL 60208, USA \\ 
$^5$ Department of Physics, Center for Astrophysics and Space Science, University of California at San Diego, 9500 Gilman Drive, La Jolla, CA 92093, USA \\ 
}

\date{}
\begin{document}
\maketitle

\begin{abstract}
We present the first simulations evolving resolved spectra of cosmic rays (CRs) from MeV-TeV energies (including electrons, positrons, (anti)protons, and heavier nuclei), in live kinetic-MHD galaxy simulations with star formation and feedback. We utilize new numerical methods including terms often neglected in historical models, comparing Milky Way analogues with phenomenological scattering coefficients $\nu$ to Solar-neighborhood (LISM) observations (spectra, B/C, $e^{+}/e^{-}$, $\bar{p}/p$, $^{10}$Be/$^{9}$Be, ionization, $\gamma$-rays). We show it is possible to reproduce observations with simple single-power-law injection and scattering coefficients (scaling with rigidity $R$), similar to previous (non-dynamical) calculations. We also find: (1) The circum-galactic medium in realistic galaxies necessarily imposes a $\sim10\,$kpc CR scattering halo, influencing the required $\nu(R)$. (2) Increasing the normalization of $\nu(R)$ re-normalizes CR secondary spectra but also changes primary spectral slopes, owing to source distribution and loss effects. (3) Diffusive/turbulent reacceleration is unimportant and generally sub-dominant to gyroresonant/streaming losses, which are sub-dominant to adiabatic/convective terms dominated by $\sim0.1-1\,$kpc turbulent/fountain motions. (4) CR spectra vary considerably across galaxies; certain features can arise from local structure rather than transport physics. (5) Systematic variation in CR ionization rates between LISM and molecular clouds (or Galactic position) arises naturally without invoking alternative sources. (6) Abundances of CNO nuclei require most CR acceleration occurs around when reverse shocks form in SNe, not in OB wind bubbles or later Sedov-Taylor stages of SNe remnants.
\end{abstract}


\section{Introduction}
\label{sec:intro}

The propagation and dynamics of cosmic rays (CRs) in the interstellar medium (ISM) and circum/inter-galactic medium (CGM/IGM) is an unsolved problem of fundamental importance for space plasma physics as well as star and galaxy formation and evolution \citep[see reviews in][]{Zwei13,zweibel:cr.feedback.review,2018AdSpR..62.2731A,2019PrPNP.10903710K}. For decades, the state-of-the-art modeling of Galactic (Milky Way; MW) CR propagation has largely been dominated by idealized analytic models, where a population of CRs is propagated through a time-static MW model, with simple or freely-fit assumptions about the ``halo'' or thick disk around the galaxy and no appreciable circum-galactic medium (CGM)\footnote{The term ``halo'' is used differently in CR and galaxy literature. In most CR literature, the ``halo'' is generally taken to have a size $\sim1-10$\,kpc, corresponding to the ``thick disk'' or ``disk-halo interface'' region in galaxy formation/structure terminology. In the galaxy community, the gaseous ``halo'' usually refers to the circum-galactic medium (CGM), with scale-lengths $\sim 20-50\,$kpc and extent $\sim 200-500\,$kpc \citep{tumlinson:2017.cgm.review}.} with ``escape'' (as a leaky box or flat halo-diffusion type model) outside of some radius \citep{blasi:cr.propagation.constraints,strong:2001.galprop,vladimirov:cr.highegy.diff,gaggero:2015.cr.diffusion.coefficient,2016ApJ...819...54G,2016ApJ...824...16J,cummings:2016.voyager.1.cr.spectra,2016PhRvD..94l3019K,evoli:dragon2.cr.prop}. 

These calculations generally ignore phase structure or inhomogeneity in the ISM/CGM, magnetic field structure (anisotropic CR transport), streaming, complicated inflow/outflow/fountain and turbulent motions within the galaxy, and time-variability of galactic structure and ISM phases \citep[although see e.g.][]{2012JCAP...01..011B,2016ApJ...824...16J,2018ApJ...869..176L,2018JCAP...07..051G}, even though, for example, secondary production rates depend on the local gas density which varies by {\em several orders of magnitude} in both space and time (even at a given galacto-centric radius) as CRs propagate through the ISM. Likewise, the injection itself being proportional to e.g.\ SNe rates is strongly clustered in both space and time and specifically related to certain ISM phases \citep[see][]{evans:1999.sf.gmc.review,vazquez-semadeni:2003.turb.reg.sfr,mac-low:2004.turb.sf.review,2015MNRAS.454..238W,2018MNRAS.481.3325F}, and other key loss terms depend on e.g.\ local ionized vs.\ neutral fractions, magnetic and radiation energy densities -- quantities that can vary by ten orders of magnitude within the MW \citep{wolfire:1995.neutral.ism.phases,evans:1999.sf.gmc.review,draine:ism.book}. And these static models cannot, by construction, capture non-linear effects of CRs actually modifying the galaxy/ISM structure through which they propagate. This in turn means that most inferred physical quantities such as CR diffusivities. residence times, re-acceleration efficiencies, and ``convective'' speeds (let alone their dependence on CR energy or ISM properties) are potentially subject to order-of-magnitude systematic uncertainties. That is not to say these static-Galaxy models are simple, however: their complexity focuses on evolving an enormous range of CR energies from $\lesssim\,$MeV to $\gtrsim$\,PeV, including a huge number of different species, and incorporating state-of-the-art nuclear networks for detailed spallation, annihilation, and other reaction rates \citep[recently, see][]{2018ApJ...869..176L,2018AdSpR..62.2731A}. 

Meanwhile, simulations of galaxy structure, dynamics, evolution, and formation have made tremendous progress incorporating and reproducing detailed observations of the 
time-dependent, multi-phase complexity of the ISM and CGM \citep{hopkins:fb.ism.prop,kim:tigress.ism.model.sims,grudic:sfe.gmcs.vs.obs,benincasa:2020.gmc.lifetimes.fire,keating:co.h2.conversion.mw.sims,gurvich:2020.fire.vertical.support.balance}, 
galaxy inflows/outflows/fountains 
\citep{narayanan:co.outflows,hayward.2015:stellar.feedback.analytic.model.winds,muratov:2016.fire.metal.outflow.loading,angles.alcazar:particle.tracking.fire.baryon.cycle.intergalactic.transfer,hafen:2018.cgm.fire.origins,2019arXiv191001123H,hopkins:2020.cr.outflows.to.mpc.scales,ji:fire.cr.cgm},
and turbulent motions \citep{hopkins:frag.theory,hopkins:2012.intermittent.turb.density.pdfs,guszejnov:imf.var.mw,escala:turbulent.metal.diffusion.fire,guszejnov:universal.scalings,rennehan:turb.diff.implementation.fancy}, 
magnetic field structure and amplification
\citep{su:2016.weak.mhd.cond.visc.turbdiff.fx,su:fire.feedback.alters.magnetic.amplification.morphology,su:2018.stellar.fb.fails.to.solve.cooling.flow,hopkins:2019.mhd.rdi.periodic.box.sims,guszejnov:2020.mhd.turb.isothermal.imf.cannot.solve,2018MNRAS.479.3343M}, 
dynamics of mergers and spiral arms and other gravitational phenomena 
\citep{hopkins:clumpy.disk.evol,hopkins:stellar.fb.mergers,hopkins:2013.merger.sb.fb.winds,fitts:mergers.in.dwarf.form,ma:2016.disk.structure,garrisonkimmel:fire.morphologies.vs.dm,moreno:2019.fire.merger.suite},
star formation \citep{grudic:sfe.cluster.form.surface.density,orr:ks.law,orr:non.eqm.sf.model,grudic:sfe.gmcs.vs.obs,grudic:2019.imf.sampling.fx.on.gmc.destruction,garrison.kimmel:2019.sfh.local.group.fire.dwarfs,wheeler:ultra.highres.dwarfs,ma:2020.globular.form.highz.sims,grudic:starforge.methods}, 
and stellar ``feedback'' from supernovae 
\citep{martizzi:sne.momentum.sims,gentry:clustered.sne.momentum.enhancement,rosdahl:2016.sne.method.isolated.gal.sims,hopkins:sne.methods,2018MNRAS.478..302S,kawakatu:2020.obscuration.torus.from.stellar.fb.in.torus},
stellar mass-loss
\citep{wiersma:2009.enrichment,conroy:2014.agb.heating.quenching,2018A&ARv..26....1H},
radiation \citep{hopkins:rad.pressure.sf.fb,hopkins:2019.grudic.photon.momentum.rad.pressure.coupling,hopkins:radiation.methods,wise:2012.rad.pressure.effects,rosdahl:m1.method.ramses,2018ApJ...859...68K,emerick:rad.fb.important.stromgren.ok}, 
and jets \citep{burzle:2011.protostellar.outflows,offner:2011.rad.protostellar.outflows,hansen:2012.lowmass.sf.radsims,guszejnov:protostellar.feedback.stellar.clustering.multiplicity},
resolving the dynamics of those feedback mechanisms interacting with the ISM.
However, these calculations (including our own) treat the high-energy astro-particle physics in an incredibly simple fashion. Most ignore it entirely. Even though there has been a surge of work in recent years arguing that CRs could have major dynamical effects on both the phase (temperature-density) structure and dynamics (inflow/outflow rates, strength of turbulence, bulk star formation rates) of galaxies \citep[see][]{jubelgas:2008.cosmic.ray.outflows,uhlig:2012.cosmic.ray.streaming.winds,Boot13,Wien13,Hana13,Sale14,Sale14cos,Chen16,Simp16,Giri16,Pakm16,Sale16,wiener:2017.cr.streaming.winds,Rusz17,Buts18,farber:decoupled.crs.in.neutral.gas,Jaco18,Giri18}, essentially all of these studies have treated CRs with a ``single bin'' approximation. evolving a single fluid representing ``the CRs'' 
\citep[recently, see][]{Sale16,chan:2018.cosmicray.fire.gammaray,Buts18,su:turb.crs.quench,hopkins:cr.mhd.fire2,ji:fire.cr.cgm,ji:20.virial.shocks.suppressed.cr.dominated.halos,bustard:2020.crs.multiphase.ism.accel.confinement,thomas:2021.cr.prop.single.bin}. 
Even if one is only interested in the dynamical effects of CRs on the gas itself, so assumes the CR pressure is strongly dominated by $\sim\,$GeV protons, this could be inaccurate in many circumstances. For example, certain terms which should ``shift'' CRs in their individual energies or Lorentz factors and therefore change their emission/loss/transport properties instead simply rescale ``up'' or ``down'' the CR energy density in single-bin models, effectively akin to ``creating'' new CRs. 

More importantly, even if ``single-bin'' models allow for a reasonable approximate estimation of bulk CR pressure effects on gas, a ``single-bin'' CR model  precludes comparing to the vast majority of observational constraints. Essentially, it restricts comparison to a handful of galaxy-integrated $\sim$\,GeV $\gamma$-ray detections in nearby star-forming galaxies \citep{lacki:2011.cosmic.ray.sub.calorimetric,tang:2014.ngc.2146.proton.calorimeter,griffin:2016.arp220.detection.gammarays,fu:2017.m33.revised.cr.upper.limit,wjac:2017.4945.gamma.rays,wang:2018.starbursts.are.proton.calorimeters,lopez:2018.smc.below.calorimetric.crs}, which in turn means that theoretical CR transport models are fundamentally under-constrained (see \citealt{hopkins:cr.transport.constraints.from.galaxies}). Because the $\gamma$-rays constrain a galactic-ISM-integrated quantity over a narrow range of CR energies, different physically-motivated models which reproduce the same $\gamma$-ray luminosity can predict qualitatively different CR transport in the ISM/CGM/IGM (depending on how they scale with properties as noted above), as well as totally different effects of CRs on outflows, accretion, and galaxy formation \citep{hopkins:2020.cr.transport.model.fx.galform}. This also precludes comparing to the  enormous wealth of detailed Solar system CR data covering a huge array of species, as well as the tremendous amount of spatially-resolved synchrotron data from large numbers of galaxies spanning the densest regions of the ISM through the diffuse CGM, and all galaxy types. While there have been important preliminary efforts to model these more detailed datasets with variations of post-processing or tracer-species calculations \citep[see][]{pinzke:2017.analytic.cr.halo.electron.model,gaches:2018.protostellar.cr.acceleration,offner:2019.cr.feedback.accretion.chem.disks,winner:2019.cr.electron.post-processing.tracers,vazza:2021.cr.electrons.icm.tracer.particles,werhahn:2021.electron.spectra.models,wehahn:2021.gamma.rays,werhahn:2021.cr.calorimetry.simulated.galaxies}, these necessarily neglect the dynamics above, and are more akin to the ``time static'' analytic models in some ways.

In this manuscript, we therefore generalize our previous explicit CR transport models from previous studies to a resolved CR spectrum of electrons, positrons, protons, anti-protons, and heavier nuclei spanning energies $\sim$MeV to $\sim$TeV. This makes it possible to explicitly forward-model from cosmological initial conditions quantities including the CR electron and proton spectra, B/C and radioactive isotope ratios, and detailed observables including synchrotron spectra, alongside Galactic magnetic field and halo and ISM structure. We show that for plausible injection assumptions the simulations can reproduce the observed Solar neighborhood values. We explicitly account for and explore the roles of a wide range of processes including: anisotropic diffusion and streaming, gyro-resonant plasma instability losses, ``adiabatic'' CR acceleration, diffusive/turbulent re-acceleration, Coulomb and ionization losses, catastrophic/hadronic losses (and $\gamma$-ray emission), Bremstrahhlung, inverse Compton (accounting for time-and-space-varying radiation fields), and synchrotron terms. In \S~\ref{sec:methods}, we outline the numerical methods and treatment of spectrally-resolved CR populations, and describe our simulation initial conditions. In \S~\ref{sec:results} we summarize the qualitative results, and explore the effects of each of the different pieces of physics in turn. We also compare with observational constraints and attempt to present some simplified analytic models that explain the relevant scalings. We conclude in \S~\ref{sec:conclusions}. The Appendices contain various additional details, showing typical CR drift velocities and loss timescales (\S~\ref{sec:vdrift.tloss}), predicted CR spectra for additional parameter choices (\S~\ref{sec:appendix:additional}), detailed numerical methods and validation tests (\S~\ref{sec:numerics}), and mock observational diagnostics of the simulation magnetic fields (\S~\ref{sec:bfields}).

\begin{figure*}
	\includegraphics[width=0.33\textwidth]{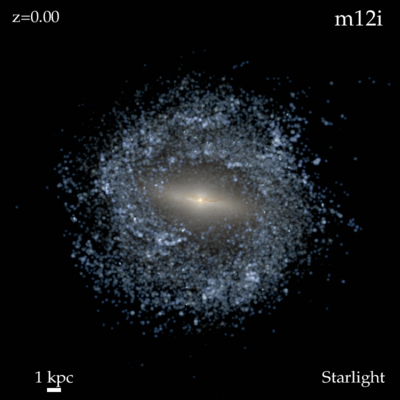}
	\includegraphics[width=0.33\textwidth]{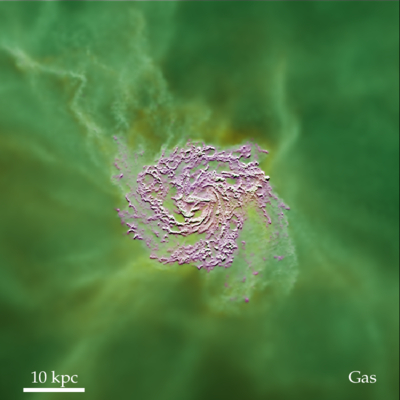}
	\includegraphics[width=0.33\textwidth]{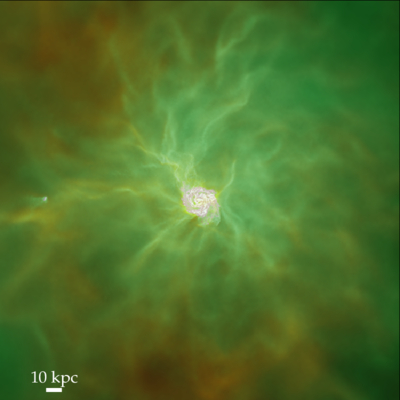}
	\\
	\includegraphics[width=0.33\textwidth]{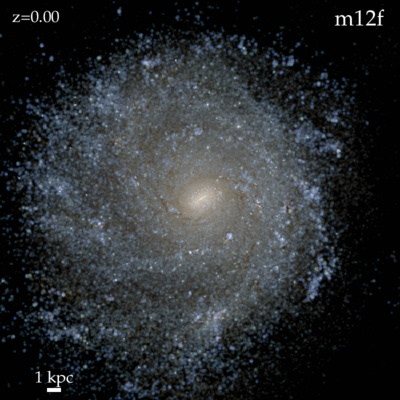}
	\includegraphics[width=0.33\textwidth]{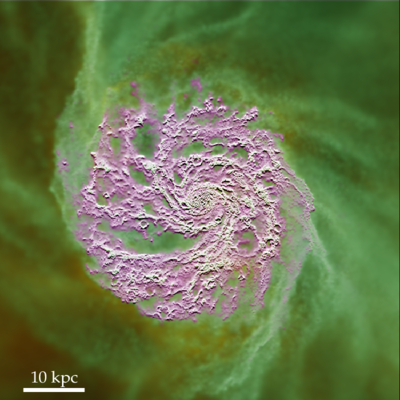}
	\includegraphics[width=0.33\textwidth]{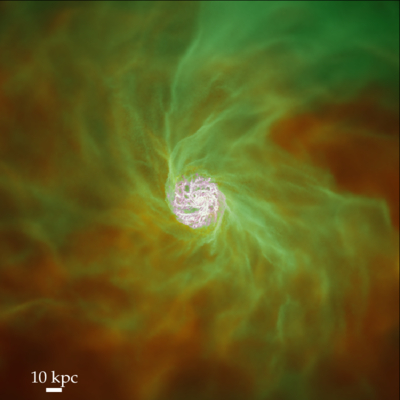}
	\\
	\includegraphics[width=0.33\textwidth]{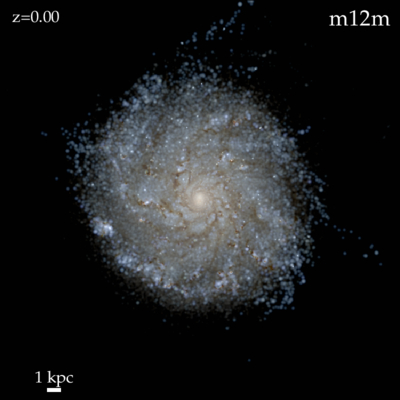}
	\includegraphics[width=0.33\textwidth]{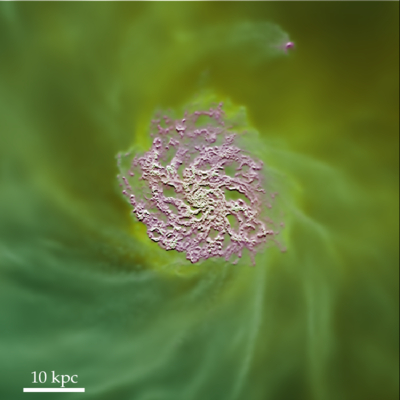}
	\includegraphics[width=0.33\textwidth]{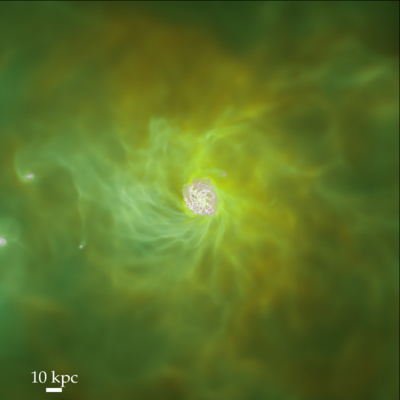}
	\vspace{-0.1cm}
	\caption{Mock images of the simulations studied here, selected as Milky Way (MW)-like galaxies near $z\approx 0$ from the FIRE cosmological simulation project. The three galaxies are {\bf m12i} (our ``fiducial'' galaxy, {\em top}), {\bf m12f} ({\em middle}), and {\bf m12m} ({\em bottom}), all broadly MW-like but different in detail with e.g.\ different extended outer gas/stellar disks, different bar/spiral arm strengths, and different detailed spatial distribution of gas \&\ star formation within the disk. 
	{\em Left:} Hubble Space Telescope-style $ugr$ composite image ray-tracing starlight (attenuated by dust in the simulations as \citealt{hopkins:dust}) with a log-stretch ($\sim4\,$dex surface-brightness range).
	{\em Middle:} Gas portrayed with a 3-band volume render showing ``hot'' ($T\gg 10^{5}\,$K, {\em red}), ``warm/cool'' ($T\sim 10^{4}-10^{5}\,$K, {\em green}), and ``cold (neutral)'' ($T\ll 10^{4}\,$K, {\em magenta}) phases.
	{\em Right:} Gas again, but on larger scales, more clearly showing the continuing gas distribution well into the circum-galactic medium and ``halo'' up to $\gtrsim100\,$kpc beyond the galactic disk.
	\label{fig:images}}
\end{figure*}

\section{Methods}
\label{sec:methods}

\subsection{Non-CR Physics}
\label{sec:methods:overview}

The simulations here extend those in several previous works including \citet{chan:2018.cosmicray.fire.gammaray}, \citet{hopkins:cr.mhd.fire2} (\paperone), and \citet{hopkins:cr.transport.constraints.from.galaxies} (\papertwo), where additional numerical details are described. We only briefly summarize these and the non-CR physics here. The simulations are run with {\small GIZMO}\footnote{A public version of {\small GIZMO} is available at \gizmourl} \citep{hopkins:gizmo}, in its meshless finite-mass MFM mode (a mesh-free finite-volume Lagrangian Godunov method). All simulations include magneto-hydrodynamics (MHD), solved as described in \citep{hopkins:mhd.gizmo,hopkins:cg.mhd.gizmo} with fully-anisotropic Spitzer-Braginskii conduction and viscosity \citep[implemented as in \papertwo; see also][]{hopkins:gizmo.diffusion,su:2016.weak.mhd.cond.visc.turbdiff.fx}. Gravity is solved with adaptive Lagrangian force softening (matching hydrodynamic and force resolution). We treat cooling, star formation, and stellar feedback following the FIRE-2 implementation of the Feedback In Realistic Environments (FIRE) physics \citep[all details in][]{hopkins:fire2.methods}; as noted in \S~\ref{sec:physics} our conclusions are robust to variations in detailed numerical implementation of FIRE. We explicitly follow the enrichment, chemistry, and dynamics of 11 abundances \citep[H, He, Z, C, N, O, Ne, Mg, Si, S, Ca, Fe;][]{colbrook:passive.scalar.scalings,escala:turbulent.metal.diffusion.fire}; gas cooling chemistry from $\sim 10-10^{10}\,$K accounting for a range of processes including metal-line, molecular, fine-structure, photo-electric, and photo-ionization, including local sources and the \citet{faucher-giguere:2009.ion.background} meta-galactic background (with self-shielding) and tracking detailed ionization states; and star formation in gas which is dense ($>1000\,{\rm cm^{-3}}$), self-shielding, thermally Jeans-unstable, and locally self-gravitating \citep{hopkins:virial.sf,grudic:sfe.cluster.form.surface.density}. Once formed, stars evolve according to standard stellar evolution models accounting explicitly for the mass, metal, momentum, and energy injection via individual SNe (Ia \&\ II) and O/B or AGB-star mass-loss \citep[for details see][]{hopkins:sne.methods}, and radiation (including photo-electric and photo-ionization heating and radiation pressure with a five-band radiation-hydrodynamic scheme; \citealt{hopkins:radiation.methods}). Our initial conditions (see Fig.~\ref{fig:images}) are fully-cosmological ``zoom-in'' simulations, evolving a large box from redshifts $z\gtrsim 100$, with resolution concentrated in a $\sim 1-10\,$Mpc co-moving volume centered on a ``target'' halo of interest. While there are many smaller galaxies in that volume, for the sake of clarity we focus just on the properties of the ``primary'' (i.e.\ best-resolved) galaxies in each volume.

\subsection{CR Physics \&\ Methods}
\label{sec:methods:crs}

\subsubsection{Overview \&\ Equations Solved}

Our CR physics implementation essentially follows the combination of \papertwo\ \&\ \citet{hopkins:m1.cr.closure} with \citet{girichidis:cr.spectral.scheme}. We explicitly evolve the CR distribution function (DF): $f = f({\bf x},\,{\bf p},\,t,\,s,\,...)$, as a function of position ${\bf x}$, CR momentum ${\bf p}$, time $t$, and CR species $s$. An extensive summary of the numerical details and some additional validation tests are presented in  Appendix~\ref{sec:numerics},  but we summarize the salient physics here.

We assume a gyrotropic DF for the phase angle $\phi$ and evolve the first two pitch-angle ($\mu \equiv \hat{\bf p} \cdot \bhat$) moments of the focused CR transport equation \citep{1997JGR...102.4719I,2001GeoRL..28.3831L}, to leading order in $\mathcal{O}(u/c)$ (where ${\bf u}$ is the fluid velocity) on macroscopic scales much larger than CR gyro-radii,\footnote{Of course, certain kinetic processes and plasma instabilities on gyro scales can only be resolved and properly treated in particle-in-cell (PIC) or MHD-PIC simulations of the sort in e.g.\ \citet{bai:2015.mhd.pic,bai:2019.cr.pic.streaming,mignone:2018.mhd.pic,holcolmb.spitkovsky:saturation.gri.sims,ji:2021.mhd.pic.rsol,ji:2021.cr.mhd.pic.dust.sims}. But recall that CR gyro radii are $\sim 0.1\,{\rm au}\,(R/{\rm GV})\,(|{\bf B}|/\mu{\rm G})^{-1}$, vastly smaller than our resolution at all rigidities we consider.} for an arbitrary $f=f(p,\,\mu, ...)$. From \citet{hopkins:m1.cr.closure}, this gives the equations solved:
\begin{align}
\label{eqn:f0} 
D_{t} \bar{f}_{0} +  \nabla \cdot  (v\,\bhat\,\bar{f}_{1}) &=  {j_{0}} + \mathbb{D}:\nabla{\bf u}\,\left[ 3\,\bar{f}_{0} + p\,\frac{\partial \bar{f}_{0}}{\partial p} \right] \\
\nonumber &\frac{1}{p^{2}}\frac{\partial }{\partial p}\left[ p^{2}\,\left\{ S_{\ell}\,\bar{f}_{0}  
+ \tilde{D}_{p \mu}\,\bar{f}_{1}
+ \tilde{D}_{p p}\, \frac{\partial \bar{f}_{0}}{\partial p} 
\right\}\right]  \\
\label{eqn:f1} 
D_{t} \bar{f}_{1} +  
v\,\Delta(\bar{f}_{0}) &= - \left[ \tilde{D}_{\mu\mu}\,\bar{f}_{1} + \tilde{D}_{\mu p}\,\frac{\partial \bar{f}_{0}}{\partial p} \right]
+ {j_{1}} \\
\nonumber
\tilde{D}_{p p}  = \chi\,\frac{p^{2}\,v_{A}^{2}}{v^{2}}\,\bar{\nu} 
\  , \  \
& \tilde{D}_{p \mu} = \frac{p\,\bar{v}_{A}}{v}\,\bar{\nu} 
\  , \ \
\tilde{D}_{\mu\mu} = \bar{\nu} 
\  , \  \
\tilde{D}_{\mu p} = \chi\,\frac{p\,\bar{v}_{A}}{v}\,\bar{\nu}
\end{align}
where $\bar{f}_{n} \equiv \langle \mu^{n}\, f \rangle_{\mu}$ is the $n$'th pitch-angle moment (so e.g.\ $\bar{f}_{0}$ is the isotropic part of the DF, and $\bar{f}_{1} = \langle \mu \rangle\,\bar{f}_{0}$), $D_{t} X \equiv \partial_{t}  X + \nabla \cdot({\bf u}\,X) \equiv \rho\,{\rm d}_{t}(X/\rho)$ is the conservative co-moving derivative, $v=\beta\,c$ is the CR velocity, $p=\gamma\,\beta\,m_{s}\,c$ the CR momentum, $\bhat \equiv {\bf B}/|{\bf B}|$ the unit magnetic field vector, $j_{n}$ represent injection \&\ catastrophic losses, $S_{\ell}$ represents continuous loss processes described below, $v_{A}$ is \Alf\ speed, the coefficients $\bar{D}$ are defined in terms of the scattering rate $\bar{\nu} \equiv \bar{\nu}_{+} + \bar{\nu}_{-}$, the signed $\bar{v}_{A} \equiv v_{A}\,(\bar{\nu}_{+} - \bar{\nu}_{-})/(\bar{\nu}_{+} + \bar{\nu}_{-})$, and the operator $\Delta(q) \equiv \bhat \cdot \nabla \left( \chi\,q \right) + \nabla \cdot [ (1-3\,\chi)\,q\,\bhat ]$ and Eddington tensor $\mathbb{D} \equiv \chi\,\mathbb{I} + ( 1-3\,\chi )\,\bhat\bhat$ are defined in terms of $\chi$:
\begin{align}
\chi &\equiv \frac{1-\langle \mu^{2} \rangle}{2} = \frac{1}{2}\,\left[ 1 - \frac{\bar{f}_{2}}{\bar{f}_{0}} \right]
\end{align}
where $\langle \mu \rangle \equiv \bar{f}_{1}/\bar{f}_{0}$. The moments hierarchy for $\bar{f}_{2}$ is closed by the assumed M1-like relation $\langle \mu^{2} \rangle \approx (3+4\,\langle \mu\rangle^{2}) / (5 + 2\,\sqrt{4-3\,\langle \mu\rangle^{2}})$, which is exact for both a near-isotropic DF (the case of greatest practical relevance, as argued in e.g.\ \citealt{thomas:2021.compare.cr.closures.from.prev.papers}) or a maximally-anisotropic DF ($\langle \mu \rangle \rightarrow \pm 1$), or any DF which can  be made approximately isotropic via some Lorentz transformation \citep{hopkins:m1.cr.closure}. All of the variables above should be understood to be functions of ${\bf x}$ and $t$, etc. The CRs act on the gas+radiation field as well: the appropriate collisional/radiative terms are either thermalized or added to the total radiation or magnetic energy, and the CRs exert forces on the gas in the form of the Lorentz force (proportional to the perpendicular CR pressure gradient) and parallel force from scattering, as detailed in \papertwo\ and \citet{hopkins:m1.cr.closure}. As defined therein the CR pressure tensor $\mathbb{P} = \int d^{3}{\bf p}\,( {\bf p}\,{\bf v})\,f$ is anisotropic following $\mathbb{D}$.

Note that if the ``flux'' equation Eq.~\ref{eqn:f1} reaches local steady-state with $|D_{t} \bar{f}_{1}| \ll |\bar{\nu}\,\bar{f}_{1}|$, which occurs on a scattering time $\sim \bar{\nu}^{-1}$ (generally short compared to other timescales of interest in our simulations, so this is often a reasonable approximation), then we have $\chi\rightarrow 1/3$, $\mathbb{D} \rightarrow \mathbb{I}/3$, $v\,\bar{f}_{1} \rightarrow -\bar{v}_{A}\,p\,\partial_{p}\,\bar{f}_{0} - (v^{2}/3\,\bar{\nu})\,\bhat\cdot\nabla \bar{f}_{0}$. In this case Eq.~\ref{eqn:f0} for $D_{t} \bar{f}_{0}$ reduces to the familiar Fokker-Planck equation with a streaming speed $\propto \bar{v}_{A}$ and anisotropic/parallel diffusivity $\kappa_{\|} = v^{2}/3\bar{\nu}$.

The spatial discretization follows the gas mesh: each gas cell $j$ represents some finite-volume domain $V_{j}$, which carries a cell-averaged $f_{j}({\bf p},\,t,\,s,\,...\,|\,{\bf x}_{j})$. Each species $s$ is then treated with its own explicitly-evolved spectrum ${\bf p}$, discretized into a number of intervals or bins $n$, defined by a range of momenta $p^{-}_{n,\,s} < p < p^{+}_{n,\,s}$ ($p\equiv |{\bf p}|$) within each cell $j$.
To ensure manifest conservation we evolve the conserved variables of CR number $N_{j,\,n,\,s}(t)$ and kinetic energy $E^{\rm kin}_{j,\,n,\,s}(t)$ integrated over each interval in space and momentum: 
\begin{align}
N_{j,\,n,\,s}(t) &\equiv \int _{V_{j}} n_{j,\,n,\,s}\,d^{3}{\bf x} \equiv \int_{V_{j}}\int_{p_{n,\,s}^{-}}^{p_{n,\,s}^{+}} f_{j,\,n,\,s}(...) \, d^{3}{\bf x} \,d^{3}{\bf p}  \\ 
E^{\rm kin}_{j,\,n,\,s}(t) &\equiv \int_{V_{j}} \epsilon_{j,\,n,\,s}\,d^{3}{\bf x} \equiv \int_{V_{j}} \int_{p_{n,\,s}^{-}}^{p_{n,\,s}^{+}} T_{s}(p)\,f_{j,\,n,\,s}(...) \, d^{3}{\bf x} \,d^{3}{\bf p}
\end{align}
where $d^{3}{\bf p} \equiv p^{2}\,dp\,d\Omega = p^{2}\,dp\,d\phi\,d\mu$ and $T_{s}(p) \equiv (p^{2}\,c^{2} + m_{s}^{2}\,c^{4})^{1/2} - m_{s}\,c^{2}$ is the CR kinetic energy. Note we could equivalently evolve the total CR energy as by definition $E^{\rm tot}_{j,\,n,\,s} \equiv E^{\rm kin}_{j,\,n,\,s} + N_{j,\,n,\,s}\,m_{s}\,c^{2}$ (or $e_{j,\,n,\,s} = \epsilon_{j,\,n,\,s} + n_{j,\,n,\,s}\,m_{s}\,c^{2}$). 

\subsubsection{Spatial Evolution \&\ Coupling to Gas}
\label{sec:methods.spatial.evol}

Operator-splitting (1) spatial evolution, (2) momentum-space operations, and (3) injection, the spatial part of Eqs.~\ref{eqn:f0}-\ref{eqn:f1} can be written as a normal hyperbolic/conservation law for $\bar{f}_{0}$: $D_{t}\bar{f}_{0} = -\nabla \cdot (v\,\bhat\,\bar{f}_{1})$, and Eq.~\ref{eqn:f1} for the flux $\bar{f}_{1}$. That is discretized and integrated on the spatial mesh defined by the gas cells identically in structure to our two-moment formulation for the CR number density or energy and their fluxes from e.g.\ \papertwo\ and \citet{chan:2018.cosmicray.fire.gammaray,hopkins:m1.cr.closure}, and solved with the same finite-volume method. Because the detailed form of the scattering rates $\bar{\nu}$ are orders-of-magnitude uncertain (see review in \papertwo), we neglect details such as bin-boundary flux terms and differences in diffusion coefficients for number and energy across the finite width of a momentum bin (i.e.\ use the ``bin centered'' $\bar{\nu}$).\footnote{\label{footnote:bin.centering}This ``bin-centered'' approximation (along with simple finite-sampling effects owing to our finite-size bins) leads to a well-understood numerical artifact (shown in \citealt{girichidis:cr.spectral.scheme}, \citealt{ogrodnik:2021.spectral.cr.electron.code}, and our \S~\ref{sec:numerical.tests}) wherein small ``step'' features appear between the edges of different spectral ``bins'' (i.e.\ the slopes do not join continuously, because the variation of the effective spatial diffusivity continuously across the bin is neglected, so it  changes discretely bin-to-bin). This is evident in e.g.\ our Fig.~\ref{fig:demo.cr.spectra.fiducial} and essentially all our CR spectra, but we show the effect is small compared to $\sim 1\sigma$ variations in the spectra and much smaller than physical variations from different scattering-rate assumptions or Galactic environments.} At this level, the spatial equations for $(\bar{f}_{0},\,\bar{f}_{1}$) are exactly equivalent to the two-moment equations for e.g.\ $(n,\,{\bf F}_{n})$ or $(\epsilon,\,{\bf F}_{\epsilon})$ (where ${\bf F}_{q}$ is the flux of $q$) in \citet{hopkins:m1.cr.closure}, integrated separately for each $j,\,n,\,s$.

Per \papertwo\ and \citet{hopkins:m1.cr.closure}, it is convenient to write the CR forces on the gas in terms of ``bin integrated'' variables, which can then be integrated into the Reimann solver or hydrodynamic source terms. 
Performing the relevant integrals within each bin $n$ for species $s$, to obtain the total energy $e_{n,s} = \int_{n} 4\pi\,p^{2}\,dp\,E(p)$, total energy flux $F_{e,\,n,s}=\int_{n} 4\pi\,p^{2}\,dp\,E(p)\,v$, and scalar isotropic-equivalent pressure $P_{0,\,n,s} = \int_{n} 4\pi\,p^{2}\,dp\,(p\,v/3)$, the  force on the gas can then be represented as a sum over all bins: 
\begin{align}
\label{eqn:gas.momentum} D_{t}(\rho\,{\bf u}) + ... =  \sum_{s} & \sum_{n} {\Bigl[}  -( \mathbb{I} - \bhat\bhat ) \cdot \left[ \nabla \cdot \left( \mathbb{D}_{n,\,s}\,P_{0,\,n,s} \right) \right] \\ 
\nonumber & + \frac{\bar{\nu}_{n,s}}{c^{2}}\,\left[ F_{e,\,n,s} - 3\,\chi_{n,s}\,\bar{v}_{A}\,(e_{n,s} + P_{0,\,n,s}) \right] \,\bhat\ {\Bigr]}
\end{align}

\subsubsection{Momentum-Space Evolution}
\label{sec:methods:momentum.maintext}

Within each cell and bin $j,\,n,\,s$ at each time $t$, we operator-split the momentum-space terms in Eq.~\ref{eqn:f0} (those inside $p^{-2}\,\partial_{p}\,[p^{2}\{ ... \}]$) and integrate these following the method in \citet{girichidis:cr.spectral.scheme}, to which we refer for details and only briefly summarize here. We evolve the DF as an {\em independent} power law in momentum in each interval, with slope $\psi$, as $\bar{f}_{0,\,j,\,n,\,s}({\bf x},\,t) = \bar{f}_{0,\,j,\,n,\,s}[p_{n,\,s}^{0}]\, (p / p_{n,\,s}^{0})^{\psi_{j,\,n,\,s}}$, where $p_{n,\,s}^{0}$ is the bin-centered momentum. 
Note there is a strict one-to-one relationship between e.g.\ the pair $(n_{j,\,n,\,s},\,\epsilon_{j,\,n,\,s})$ and ($ \bar{f}_{0,\,j,\,n,\,s}[p_{n,\,s}^{0}], \, \psi_{j,\,n,\,s})$, so we work with whichever is convenient.

In a timestep $\Delta t$, processes which modify the momentum $p$ of CRs (the $S_{\ell}$ term in Eq.~\ref{eqn:f0}) give rise to some $\dot{p} = -S_{\ell} = F(p,\,s,\,{\bf U}({\bf x},\,t),\,...)$ which is some function of the local plasma state ${\bf U}({\bf x},\,t)$ (gas/magnetic/radiation field properties) and CR species and rigidity. If we operator-split these terms (so ${\bf U}$ is constant over $\Delta t$ within cell $j$), and begin from a power-law DF $f_{j,\,n,\,s}$ as specified above, then we can solve {exactly} for the final momentum $p_{f}=p(t+\Delta t)$ (and therefore rigidity or energy) of each CR with some initial $p_{i}=p(t)$ obeying $\dot{p}$ above. Even if the integrals cannot be analytically solved, they can be numerically integrated to arbitrary precision. This allows us to exactly calculate the final CR energy $E^{\rm kin,\,tot}_{j,\,n,\,s}(t + \Delta t) = \int_{\Omega_{j}}\,d^{3}\,{\bf x}\,\int_{p_{i}}\, T(p_{f}[p_{i}])\,f^{i}_{j,\,n,\,s}(p_{i})\,dp_{i}$ and likewise $N^{\rm tot}_{j,\,n,\,s}(t + \Delta t)$ for the CR population which ``began'' in the bin, as well as the final energy and number which remain ``in the bin'' (i.e.\ with momenta in the interval $p_{n,\,s}^{-} < p_{s}^{f} < p_{n,\,s}^{+}$), e.g.\ $E^{\rm kin,\,bin}_{j,\,n,\,s}(t + \Delta t) = \int_{\Omega_{j}}\,d^{3}\,{\bf x}\,\int_{p_{n,\,s}^{-} < p_{f}}^{p_{f} < p_{n,\,s}^{+}}\, T(p_{f}[p_{i}])\,f^{i}_{j,\,n,\,s}(p_{i})\,dp_{i}$. The difference (e.g.\ $E^{\rm kin,\,tot}_{j,\,n,\,s} - E^{\rm kin,\,bin}_{j,\,n,\,s}$) gives the flux of energy or number which goes to the next bin (representing CRs ``moving down'' or up a bin as they lose or gain energy). After each update to $E$ and $N$, we re-solve for the corresponding DF slope $\psi$ and normalization. Because this is purely local, it can be sub-cycled and parallelized efficiently. For a given $\dot{p}=...$, we calculate the time $\delta t_{j,\,n,\,s}$ which would be required for a CR to move from one ``edge'' of the bin to another (e.g. to cool from $p^{+}_{n,\,s}$ to $p^{-}_{n,\,s}$), for each bin, and for the lowest-energy bin to cool to zero. To integrate stably we require the subcycle timestep $\Delta t_{j} \le C_{\rm cour}\,{\rm MIN}(\delta t_{j,\,n,\,s})$, with $C_{\rm cour}$ the usual Courant factor and the minimum over all bins and species. 

Catastrophic losses (e.g.\ fragmentation and decay) eliminate CRs entirely so appear directly in e.g.\ $j$ as $\partial_{t} f = -...$ reducing $f$, $n$, $\epsilon$ together. We can therefore simply integrate these within each bin similar to the procedure above, but remove the losses rather than transferring them to the neighboring bin.

In this paper we consider spectra of protons, nuclei, electrons, and positrons, with 11 intervals/bins for each leptonic species and 8 intervals for each hadronic. For leptons these intervals span rigidities ($10^{-3}$ - $5.62\times10^{-3}$, $5.62\times10^{-3}$ - $1.78\times10^{-2}$, $1.78\times10^{-2}$ - $5.62\times10^{-2}$, $5.62\times10^{-2}$ - $0.178$, $0.178$ - $0.562$, $0.562$ - $1.78$, $1.78$ - $5.62$, $5.62$ - $17.8$, $17.8$ - $56.2$, $56.2$ - $178$, $178$ - $1000$)\,GV. For hadronic species the ranges are identical but we do not explicitly evolve the three lowest-$R$ intervals because these contain negligible energy and are highly non-relativistic. This corresponds to evolving CRs with kinetic energies over a nearly identical range for nuclei and leptons from $< 1\,$MeV to $>1$\,TeV. This is summarized in Table~\ref{tbl:intervals}, which gives the upper and lower rigidity boundaries between each of our bins for both leptons and hadrons, with representative values of $T$, $\gamma$, and $\beta$ for species like $e^{-}$, $e^{+}$, $p$, $\bar{p}$.

\subsubsection{Injection \&\ First-Order Acceleration}
\label{sec:methods:injection}

By definition our treatment of the CRs averages over gyro orbits (assuming gyro radii are smaller than resolved scales), so first-order Fermi acceleration cannot be resolved but is instead treated as an injection term $j$. Algorithmically, injection is straightforward and treated as in \papertwo, generalized to the spectrally-resolved method here: sources (e.g.\ SNe) inject some CR energy and number into neighbor gas cells alongside radiation, mechanical energy, metals, etc. We simply assume an injection spectrum (and ratio of leptons-to-hadrons injected), and use it to calculate {exactly} the $\Delta E^{\rm kin}_{j,\,n,\,s}$ and $\Delta N_{j,\,n,\,s}$ injected in a cell given the desired total injected CR energy $\Delta E^{\rm kin}_{{\rm tot},\,j} = \sum_{s}\sum_{n} \Delta E^{\rm kin}_{j,\,n,\,s}$. 

The relative normalization of the injection spectra for heavier species $s$ (relative to $p$ or $e^{-}$) is set by assuming the test-particle limit, given the  abundance of that species $N_{s}$ within the injection shock, e.g.\   $dN_{s}/d\beta = (N_{s,\,j}/N_{{\rm H},\,j})\,d N_{{\rm H}}/d\beta$. This is only important for CNO, as the primary injection of other species we follow (beyond $p$ and $e^{-}$) is negligible.\footnote{At Solar abundances, $N_{\rm B}/N_{\rm H} \sim 3\times10^{-10}$, $N_{\rm Be}/N_{\rm H} \sim 10^{-11}$, $N_{e^{+}}/N_{e^{-}} \ll 10^{-12}$, and $N_{\bar{p}}/N_{\rm H} \ll 10^{-12}$, all many orders-of-magnitude lower than the ratios observed in CRs.}
Because the acceleration is un-resolved, to calculate the ratio of heavy-element to $p$ nuclei ($N_{s}/N_{\rm H}$), we need to make some assumption about where/when most of the acceleration occurs: for example, for pure core-collapse SNe ejecta (averaging over the IMF), $N_{\rm O,\,ej}/N_{\rm H,\,ej} \sim 0.015$ \citep[e.g.][and references therein]{nomoto:2013.yield.update.review,nugrid:yields,limongi.chieffi:2018.rotating.star.yields}, while for the ISM at Solar abundances $N_{\rm O,\,ISM}/N_{\rm H,\,ISM} \sim 0.0005$ \citep{lodders:updated.solar.abundances.review}. 
For initial ejecta mass $M_{\rm ej}$, if we assume most of the acceleration occurs at some time when the swept-up ISM mass passing through the shock (which increases rapidly in time) is $\sim M_{\rm swept}$, then $N_{s}/N_{\rm H} \approx (N^{\prime}_{\rm s,\,ej}\,M_{\rm ej} + N^{\prime}_{\rm s,\,ISM}\,M_{\rm swept}) / (N^{\prime}_{\rm H,\,ej}\,M_{\rm ej} + N^{\prime}_{\rm H,\,ISM}\,M_{\rm swept})$ (where $N^{\prime}_{\rm s} \equiv dN_{\rm s} / dM$ is the number of species $s$ in the ejecta or ISM, per unit mass). Equivalently we could write this in terms of the shock velocity relative to its initial value, assuming we are somewhere in the energy-conserving Sedov-Taylor phase.  
In either case, $N^{\prime}_{\rm s,\,ej}$ and $N^{\prime}_{\rm s,\,ISM}$ are given by the abundances of the stellar ejecta and the ISM gas cell into which the CRs are being injected, which follow the detailed FIRE stellar evolution and yield models and reproduce extensive metallicity studies of galactic stars and the ISM \citep{ma:2015.fire.mass.metallicity,ma:2016.disk.structure,ma:radial.gradients,muratov:2016.fire.metal.outflow.loading,escala:turbulent.metal.diffusion.fire,bonaca:gaia.structure.vs.fire,vandevoort:deuterium.fire,wheeler:ultra.highres.dwarfs}.

\subsection{CR Loss/Gain Terms Included}
\label{sec:loss.terms.included}

Our simulations self-consistently include adiabatic/turbulent/convective terms, diffusive re-acceleration, ``streaming'' or gyro-resonant losses, Coulomb, ionization, catastrophic/hadronic/fragmentation/pionic and other collisional, radioactive decay, annihilation, Bremstrahhlung, inverse Compton, and synchrotron terms, with the scalings below.

\subsubsection{Catastrophic \&\ Continuous Losses}
\label{sec:collisional.losses}

For protons and nuclei, we include Coulomb and ionization losses, catastrophic/collisional/fragmentation/ionization losses, and radioactive decay. 
Coulomb and ionization losses scale essentially identically with momentum as $\dot{p} \equiv \dot{T}_{\rm C}\,({\rm d}p/{\rm d}T)$ with $\dot{T}_{\rm C} \approx  -3.1 \times10^{-7}\,{\rm eV\,s^{-1}\,cm^{3}}\, Z_{\rm cr}^{2}\,\beta^{-1}\,[ n_{e} + 0.57\,n_{\rm neutral} ]$ and ${\rm d}p/{\rm d}T = 1/v = 1/\beta\,c$ (the difference being whether they operate primarily in ionized or neutral gas; \citealt{1972Phy....60..145G}), where $n_{e}$ is the free (thermal) electron density (the Coulomb term), and $n_{\rm neutral}$ the neutral number density (ionization term). 

For protons, we take the total pion/catastrophic loss rate to be 
$\dot{f}_{\rm cr}(p) = -n_{\rm n}\,\beta\,c\,\sigma_{{\rm eff},p}\,f_{\rm cr}(p)$ with $\sigma_{{\rm eff},p}\approx 21.3\,\beta^{-1}\,\Theta(T[p] - 0.28\,{\rm GeV})$\,mb 
%$\dot{f}(p) = -6.37\times10^{-16}\,{\rm s^{-1}\,cm^{3}}\,n_{\rm n}\,\Theta(T[p] - 0.28\,{\rm GeV})\,f_{\rm cr}(p)$ 
\citep{Mann94,guo.oh:cosmic.rays}, 
where $n_{\rm n}$ is the nucleon number density ($\approx \rho/m_{p}$), $\Theta(x) = 0$ for $x<0$, $=1$ for $x \ge 1$. For heavier nuclei, we take the total fragmentation/catastrophic loss rate to be $\dot{f}_{\rm cr}(p)=-n_{\rm n}\,\beta\,c\,\sigma_{s}\,f_{\rm cr}(p)$ with $\sigma_{s}=45\,A^{0.7}\,(1+0.016\,\sin{[1.3-2.63\ln{A}]})$\,mb (with $A$ the atomic mass number) at $\ge 2\,$GeV and $\sigma_{s} = \sigma_{s}(>2\,{\rm GeV})\,[1 - 0.62\,\exp{(-T/{0.2\,{\rm GeV}})}\,\sin{(1.57553\,[T/{\rm GeV}]^{0.28})}]$ at $<2\,$GeV from \citet{Mann94}, with the cross-sections for secondary production of various relevant species described below. 
For antimatter ($\bar{p}$), we include annihilation with $\dot{f}_{\rm cr}(p)=-n_{\rm H}\,\beta\,c\,\sigma_{p\bar{p}}\,f_{\rm cr}(p)$ where $n_{\rm H}$ is the number density of hydrogen nuclei ($\approx X_{\rm H}\,\rho/m_{p}$) and $\sigma_{p \bar{p}}\approx 1.5\,{\rm mb}\,(-107.9+29.43\,x-1.655\,x^{2}+189.9\,e^{-x/3})$ (with $x\equiv \ln(R/{\rm GV})$; \citealt{evoli:dragon2.cr.prop}). 
For radioactive species ($^{10}$Be), the loss rate scales as $\dot{f}_{\rm cr}(p)=-f_{\rm cr}(p)/(\gamma\,t_{1/2,\,s}/\ln{2})$, where $t_{1/2,\,s}$ is the rest-frame half-life of the species ($t_{1/2}=1.51\,$Myr for $^{10}$Be).

For electrons and positrons, we include Bremstrahhlung, ionization, Coulomb, inverse Compton, and synchrotron losses, plus annihilation. 
At our energies of interest we always assume electrons/positrons are relativistic for the calculation of loss rates. 
For Bremstrahhlung we take $\dot{p} = -(3/2\pi)\,\alpha_{\rm fs}\,\sigma_{\rm T}\,c\,[\sum_{Z}\,n_{\rm Z}\,Z\,(Z+1)]\,(\ln{[2\gamma]}-1/3)\,p$, where $\sigma_{\rm T}$ is the Thompson cross-section, $\alpha_{\rm fs}$ the fine-structure constant, and $n_{\rm Z}$ the number density of ions (determined self-consistently using the ionization fractions computed in our radiation-chemistry solver) with charge $Z$ \citep{1970RvMP...42..237B}. 
For ionization we adopt $\dot{p} = -(3/4)\,m_{e}\,c^{2}\,\sigma_{\rm T}\,n_{\rm neutral}\,\ln{(2\,\gamma^{3}/\alpha_{\rm fs}^{4})}$ \citep{1965AnAp...28..171G},\footnote{For lepton ionization, using the more extended Bethe-Bloch formula appropriately corrected for light (electron/positron) species from \citet{ginzburg:1979.book}, $\dot{p} = (3/4)\,m_{e}\,c^{2}\,\sigma_{\rm T}\,\beta^{-2}\,\sum_{s_{\rm gas}}\,n_{{\rm neutral},s_{\rm gas}}\,Z_{s_{\rm gas}}\,\Phi_{s_{\rm gas}}$ with $\Phi_{s_{\rm gas}} \equiv \ln{\{  ([\gamma-1]\,\beta^{2}\,\gamma^{2}\,m_{e}^{2}\,c^{4}) / (2\,I_{s_{\rm gas}}^{2}) \} } - (2/\gamma-1/\gamma^{2})\,\ln{2} + 1/\gamma^{2} + (1-1/\gamma)^{2}/8 $;  $s_{\rm gas}=$\,H, He at Solar abundances with $(I_{\rm H},\,I_{\rm He})=(13.6,\,24.6)\,$eV gives a result which differs by $\lesssim 4\%$ from the simpler \citet{1965AnAp...28..171G} expression at all energies we consider.} while for Coulomb we have $\dot{p}=-(3/2)\,m_{e}\,c^{2}\,\sigma_{\rm T}\,n_{e}\,\beta^{-2}\,\{ \ln[m_{e}\,c^{2}\,\beta\,\sqrt{\gamma-1}/\hbar\,\omega_{\rm pl}] + \ln[2]\,(\beta^{2}/2+1/\gamma) + 1/2 + (\gamma-1)^{2}/16\,\gamma^{2} \}$ with the plasma frequency $\omega_{\rm pl}^{2}\equiv 4\pi\,e^{2}\,n_{e}/m_{e}$ \citep{1972Phy....60..145G}. 
Ignoring Klein-Nishina corrections (unimportant at the energies of interest), for inverse Compton and synchrotron we have $\dot{p} = -(4/3)\,\sigma_{\rm T}\,\gamma^{2}\,(u_{\rm rad} + u_{\rm B})$ \citep[e.g.][]{rybicki.lightman:1986.radiative.processes.book}, where $u_{\rm rad}$ and $u_{\rm B}$ are the local radiation energy density and magnetic field energy density (given self-consistently from summing all five [ionizing, FUV, NUV, optical/NIR, IR] bands followed in our radiation-hydrodynamics approximation in-code, plus the un-attenuated CMB, and from our explicitly-evolved magnetic fields). 

Positron annihilation is treated as other catastrophic terms with $\dot{f}_{\rm cr}(p)=-n_{\rm e}\,\beta\,c\,\sigma_{(e^{+}e^{-})}\,f_{\rm cr}(p)$ with the Dirac $\sigma_{(e^{+}e^{-})} \approx \pi\,r_{0,\,e}^{2}\,(\gamma_{\ast}^{2}+1)^{-1}\,[(\gamma_{\ast}^{2}+4\,\gamma_{\ast}+1)\,(\gamma_{\ast}^{2}-1)^{-1}\,\ln{(\gamma_{\ast}+\sqrt{\gamma_{\ast}^{2}-1})} - (\gamma_{\ast}+3)\,(\gamma_{\ast}^{2}-1)^{-1/2}]$ where $\gamma_{\ast}$ is the positron Lorentz factor in the electron frame and $r_{0,\,e}$ is the classical electron radius. 

Following \papertwo\ and \citet{guo.oh:cosmic.rays}, the Coulomb losses and a fraction $=1/6$ of the hadronic losses (from thermalized portions of the cascade) are thermalized (added to the gas internal energy), while a portion of the ionization losses are thermalized corresponding to the energy less ionization potential. Other radiative and collisional losses are assumed to go primarily into escaping radiation.

\subsubsection{Secondary Products: Fragmentation \&\ Decay}

Our method allows for an arbitrary set of primary species, each of which can produce an arbitrary set of secondary species (which can themselves also produce secondaries, in principle): energy and particle number are transferred bin-to-bin in secondary-producing reactions akin to the bin-to-bin fluxes within a given species described above. For computational reasons, however, it is impractical to integrate a detailed extended species network like those in codes such as {\small GALPROP} or {\small DRAGON} on-the-fly. We therefore adopt an intentionally highly-simplified network, intended to capture some of the most important secondary processes: we evolve spectra for $e^{-}$, $e^{+}$, $\bar{p}$, and nuclei for H (protons), B, CNO, stable Be ($^{7}$Be + $^{9}$Be) and unstable $^{10}$Be. 

For collisional secondary production from some ``primary'' species $s$ with momentum $p=p_{s}$ (or $T=T_{s}(p)$), which produces a species $s^{\prime}$ with momentum $p^{\prime}=p^{\prime}_{s^{\prime}}$  ($T^{\prime}=T_{s^{\prime}}(p^{\prime})$) with an effective production cross-section $\sigma_{s\rightarrow s^{\prime}}$, we generically have 
$\dot{f}_{{\rm cr},\,s^{\prime}}(p^{\prime})\,d^{3}{\bf p}^{\prime} = n_{\rm n}\,\beta_{s}(p)\,c\,\sigma_{s\rightarrow s^{\prime}}(p\rightarrow p^{\prime})\,f_{{\rm cr},\,s}(p)\,d^{3}{\bf p}$.

We consider secondary $e^{-}$ and $e^{+}$ produced by protons via pion production, with standard branching ratios ($\sim 1/3$ to each) and because our spectral bins are relatively coarse-grained assume the energy distribution of the injected leptons from a given proton energy $T$ is simply given by the expected mean factor $T_{e^{\pm}}=\alpha_{pe^{\pm}}\,T_{p}$ \citep[with the weighted mean $\alpha_{p e^{\pm}}\approx 0.12$ given by integrating over the spectra of secondary energies at the scales of interest; see][]{galprop:1997.positron.electron,dibernardo:2013.cr.positron.analysis,reinert.winkler:2018.antiproton.positron.wimp.cosmic.ray.comparison}, so 
$\sigma_{p\rightarrow e^{\pm}}(T_{e^{\pm},{\rm final}} = \alpha_{p e^{\pm}}\,T_{p,{\rm initial}}) \approx (1/3)\,\sigma_{{\rm eff},p}$. We similarly treat the production rate for $\bar{p}$ from $p$ (which overwhelmingly dominates production) with the effective integrated  cross-section $\sigma_{p \rightarrow \bar{p}} \approx 1.4\,{\rm mb}\,\sqrt{\tilde{s}_{p}}^{0.6}\,\exp{[-(17/\sqrt{\tilde{s}_{p}})^{1.4}]}$ with $\sqrt{\tilde{s}_{p}} =  1.87654\,\sqrt{1 + T_{\rm GeV}/1.87654}$ (which includes production of e.g.\ $\bar{n}$ which rapidly decay to $\bar{p}$) 
with again a weighted-mean energy $T^{\prime}\approx 0.1\,T$ of the primary \citep[][and references therein]{dimauro:2014.antiproton.cx,winkler:2017.cr.antiprotons,korsmeier:2018.antiproton.cx,evoli:2018.dragon2.cr.cross.sections}. 

The vast majority of B and Be stem from fragmentation of C, N, and O. Rather than follow C, N, and O separately, since their primary spectra and dynamics are quite similar, we simply follow a ``CNO'' bin, which is the sum of C, N, and O individually (so for processes like fragmentation we simply sum the weighted cross-sections of each) assuming Solar-like C-to-N-to-O ratios within each bin and cell. We have also experimented with following C and O separately, and find our approximation produces negligible $\sim 10\%$-level errors, much smaller than other physical uncertainties in our models. We then calculate production cross-sections for B, stable Be ($^{7}$Be + $^{9}$Be), and $^{10}$Be appropriately integrated over species and isotopes, from the fits tabulated in \citet{moskalenko:2003.galprop.cx,tommassetti:2015.cr.frag.cx.uncertainties,korsmeier:2018.antiproton.cx,evoli:2018.dragon2.cr.cross.sections}%\footnote{We use a look-up table for this, but the results can be approximated for the effective CNO bin by the following fitting functions. Take $\sigma_{\rm CNO\rightarrow Q} \approx \sigma_{0,\,Q}(T)$ for $T_{\rm min}\le T\le T_{\rm max}$, with $\sigma_{\rm CNO\rightarrow Q}=\sigma_{0,\,Q}(T=T_{\rm min})$ for $T<T_{\rm min}$ and $\sigma_{\rm CNO\rightarrow Q}=\sigma_{0,\,Q}(T=T_{\rm max})$ for $T>T_{\rm max}$, and $\sigma_{0,\,Q}/{\rm mb} \equiv b_{-1} + 10^{b_{0} + b_{1}\,x + b_{2}\,x^{2} + b_{3}\,x^{3}+b_{4}\,x^{4} + b_{5}\,x^{5} + b_{6}\,x^{6} + b_{7}\,x^{7} + b_{e,1}\,\exp{\{-b_{e,2}\,(x - b_{e,3})^{2}  \}}}$ where $x\equiv \log_{10}{(T/{\rm GeV})}$, $(T_{\rm min},\,T_{\rm max})=(0.01,\,100)\,$GeV and $(b_{-1},\,b_{0},\,b_{1},\,b_{2},\,b_{3},\,b_{4},\,b_{5},\,b_{6},\,b_{7},\,b_{e,1},\,b_{e,2},\,b_{e,3})$ is given by (0,\,1.885,\,-0.05649,\,-0.1311,\,0.1134,\,0.08120,\,-0.06574,\,-0.01160,\,0.009620,\,0.2340,\,10.81,\,-1.247) for Q\,=\,B, (0,\,1.183,\,0.1163,\,0.01653,\,-0.1132,\,-0.03376,\,0.05772,\,0.006850,\,-0.008764,\,0.4059,\,17.13,\,-1.285) for Q\,=\,$^{7,9}$Be, and (0.1076,\,0.5341,\,0.3848,\,-0.5158,\,-0.2261,\,0.5101,\,0.04493,\,-0.2383,\,0.06890,\,0,\,0,\,0) for Q\,=\,$^{10}$Be.}
. Here $p^{\prime}$ is calculated assuming constant energy-per-nucleon in fragmentation (i.e.\  $T(p^{\prime},\,s^{\prime})= (N^{\rm nuc}_{s^{\prime}}/N^{\rm nuc}_{s})\,T(p,\,s)$ with $N^{\rm nuc}$ the nucleon number). For completeness we also follow ${\rm B} \rightarrow {\rm Be}$ with $\sigma_{\rm B\rightarrow^{7,9}Be}\approx12\,T_{\rm GeV}^{-0.022}$\,mb and $\sigma_{\rm B\rightarrow^{10}Be}\approx12.5\,T_{\rm GeV}^{0.018}$\,mb (again assuming constant energy-per-nucleon).

For radioactive decay, we consider $^{10}{\rm Be}\rightarrow ^{10}$B with $\dot{f}_{{\rm cr},{\rm B}}^{\rm decay} = -\dot{f}_{{\rm cr}, ^{10}{\rm Be}}^{\rm decay}$, i.e.\ each primary produces one secondary, with negligible energy loss ($T^{\prime}\approx T$), but this is negligible as a source of secondary B production.

\subsubsection{Adiabatic and Streaming/Gyro-Resonant/Re-Acceleration Terms}
\label{sec:adiabatic.streaming.diffreaccel.terms}

From the focused-transport equation and quasi-linear scattering theory, there are three ``re-acceleration'' and/or second-order Fermi (Fermi-II) terms, all of which we include: the ``adiabatic'' or ``convective'' term $\mathbb{D}:\nabla{\bf u}$, the ``gyro-resonant'' or ``streaming'' loss term $\propto D_{p\mu}$ and the ``diffusive'' or ``micro-turbulent'' reacceleration term $\propto D_{pp}$. 
These immediately follow from the usual focused CR transport equation  plus linear scattering terms, and can be written as a mean evolution in momentum space (see \S~\ref{sec:numerics:momentum}) as:
%As shown in \citet{hopkins:m1.cr.closure}, these can be re-written in momentum space after gyro and pitch-angle averaging: 
\begin{align}
\label{eqn:reaccel} \frac{\dot{p}}{p} &= -\mathbb{D}:\nabla{\bf u} - \langle\mu\rangle\,\frac{\bar{D}_{p \mu}}{p} + \frac{\bar{D}_{p p}}{p^{2}}\,\frac{p}{\bar{f}_{0}}\,\frac{\partial \bar{f}_{0}}{\partial p} \\ 
\nonumber &= -\mathbb{D}:\nabla{\bf u} - \bar{\nu}\,\left[ \frac{\bar{f}_{1}}{\bar{f}_{0}}\,\frac{\bar{v}_{A}}{v} + \chi\,\psi\,\frac{v_{A}^{2}}{v^{2}} \right] 
%\label{eqn:reaccel} \frac{\dot{p}}{p} &= -\mathbb{D}:\nabla{\bf u} - \langle\mu\rangle\,\frac{\bar{D}_{p \mu}}{p} + 2\,(1+\beta^{2})\,\frac{\bar{D}_{p p}}{p^{2}} \\ 
%\nonumber &= -\mathbb{D}:\nabla{\bf u} - \bar{\nu}\,\left[ \frac{\bar{f}_{1}}{\bar{f}_{0}}\,\frac{\bar{v}_{A}}{v} - 2\,\chi\,(1+\beta^{2})\,\frac{v_{A}^{2}}{v^{2}} \right] 
\end{align}
where $\psi=\psi_{j,n,s}$ is the local power-law slope of the three-dimensional CR DF (defined as $\bar{f}_{0} \propto p^{\psi}$; see \S~\ref{sec:methods:momentum.maintext}), so for all energies and Galactic conditions we consider $\psi \sim -4$ is $<0$.\footnote{Note that \citet{hopkins:m1.cr.closure} wrote a similar expression to our Eq.~\ref{eqn:reaccel}, but with $\psi$ replaced by $-2\,(1+\beta^{2})$ in the $\bar{D}_{pp}$ term. Their expression came from considering the behavior of the mean momentum of a ``packet'' of CRs with a $\delta$-function-like DF, as opposed to the simpler behavior here where we consider a piecewise power-law so $(p/\bar{f}_{0})\,(\partial \bar{f}_{0}/\partial p) = \psi$ by definition. Nevertheless, it is striking that over the energy range MeV-TeV, these give quite-similar prefactors (both $\sim-4$) despite reflecting wildly different DFs.} The physical nature and importance of these is discussed below and in detail in \citet{hopkins:m1.cr.closure}, but briefly, this includes all ``re-acceleration'' terms to leading order in $\mathcal{O}(u/c)$, and generalize the expressions commonly seen for these. The ``adiabatic'' (non-intertial frame) term reduces to the familiar $-(\nabla \cdot {\bf u})/3$ as the DF becomes isotropic ($\chi\rightarrow 1/3$, $\mathbb{D}\rightarrow\mathbb{I}/3$), but extends to anisotropic DFs and is valid even in the zero-scattering limit. The $D_{pp}$ term produces a positive-definite momentum/energy gain $\dot{p}/p = |\psi|\,\chi\,v_{A}^{2}/v^{2}$ (since $\psi<0$ for any physical DF of interest  here); for the commonly adopted assumptions that give rise to the isotropic strong-scattering Fokker-Planck equation for CR transport  we would have $\bar{D}_{pp} \rightarrow p^{2}\,v_{A}^{2}/9\,D_{xx}$ and recover the usual ``diffusive re-acceleration'' expressions, but again the term here is more general, accounting for finite $\beta$ and weak-scattering/anisotropic-$f(\mu)$ effects. The $D_{p\mu}$ term is often ignored in historical MW CR transport models (which implicitly assume $\bar{\nu}_{+}=\bar{\nu}_{-}$) but this gives rise to the ``gyro-resonant'' or ``streaming'' losses \citep{Wien13,wiener:cr.supersonic.streaming.deriv,Rusz17,thomas.pfrommer.18:alfven.reg.cr.transport}. Specifically, since gyro-resonant instabilities/perturbations are excited by the CR flux in one direction (and damped in the other), if these contribute non-negligibly to the scattering rates then generically $\bar{\nu}_{+} \gg \bar{\nu}_{-}$ or $\bar{\nu}_{+} \ll \bar{\nu}_{-}$, so $\bar{v}_{A} \approx \pm v_{A}$, in which case the $\bar{D}_{\mu p}$ term is almost always dominant over the $\bar{D}_{p p}$ term dimensionally. In this regime (i.e.\ if $\bar{\nu}_{+} \ne \bar{\nu}_{-}$), then in flux-steady-state ($D_{t}\bar{f}_{1} \rightarrow 0$) the combined $\bar{D}_{p\mu}$ and $D_{pp}$ term in Eq.~\ref{eqn:reaccel} becomes negative-definite with $\dot{p}/p \sim -v_{A}/3\,\ell_{\rm grad}$ where $\ell_{\rm grad} \equiv P_{0} / |\bhat \cdot \nabla P_{0}|$. 

Given the CR energies of interest, in our default simulations we will assume self-confinement contributes non-negligibly (or other effects prevent exact $\bar{\nu}_{+},\,\bar{\nu}_{-}$ cancellation; see \S~\ref{sec:reaccel}) so $\bar{v}_{A}\,\bar{f}_{1} \approx v_{A}\,|\bar{f}_{1}|$, self-consistently including all terms in Eq.~\ref{eqn:reaccel}. We run and discuss alternate tests with $\bar{v}_{A} \rightarrow 0$ and different expressions for $D_{pp}$ or the ``diffusive reacceleration'' terms but generically find none of these change our conclusions.

\begin{figure*}
	\includegraphics[width=0.48\textwidth]{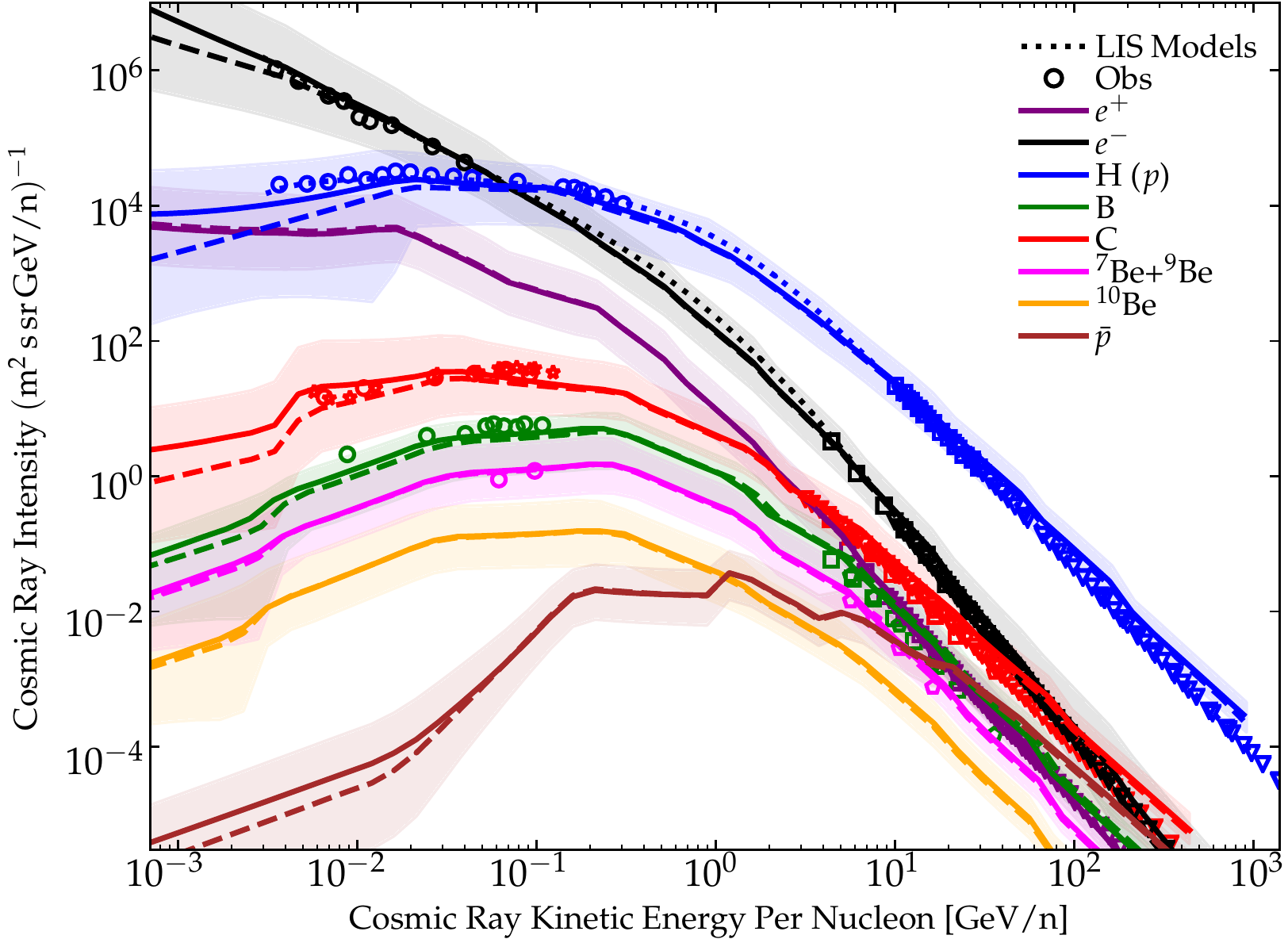}
	\includegraphics[width=0.48\textwidth]{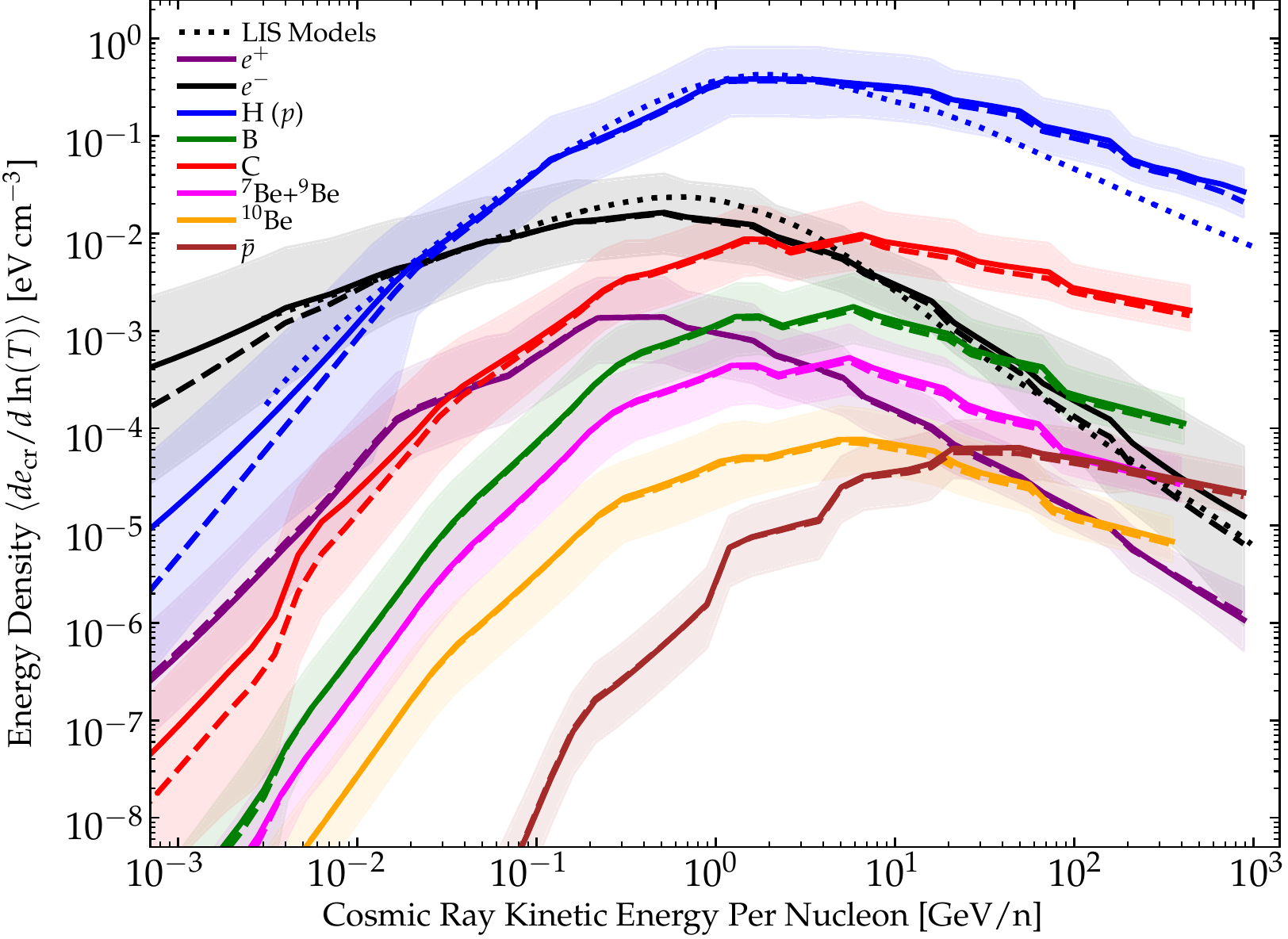} \\
	\hspace{0.2cm}\includegraphics[width=0.46\textwidth]{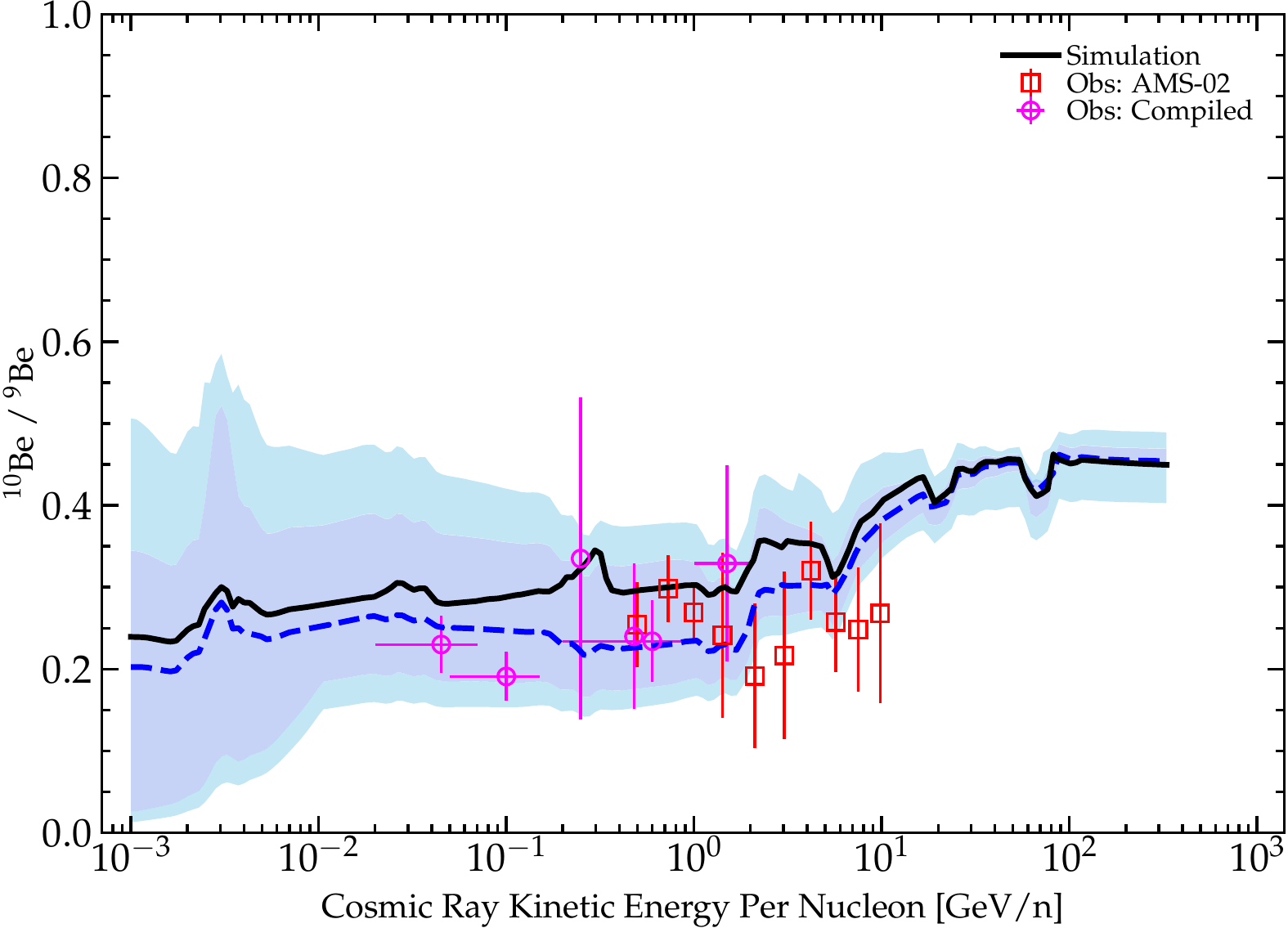}
	\hspace{0.3cm}\includegraphics[width=0.46\textwidth]{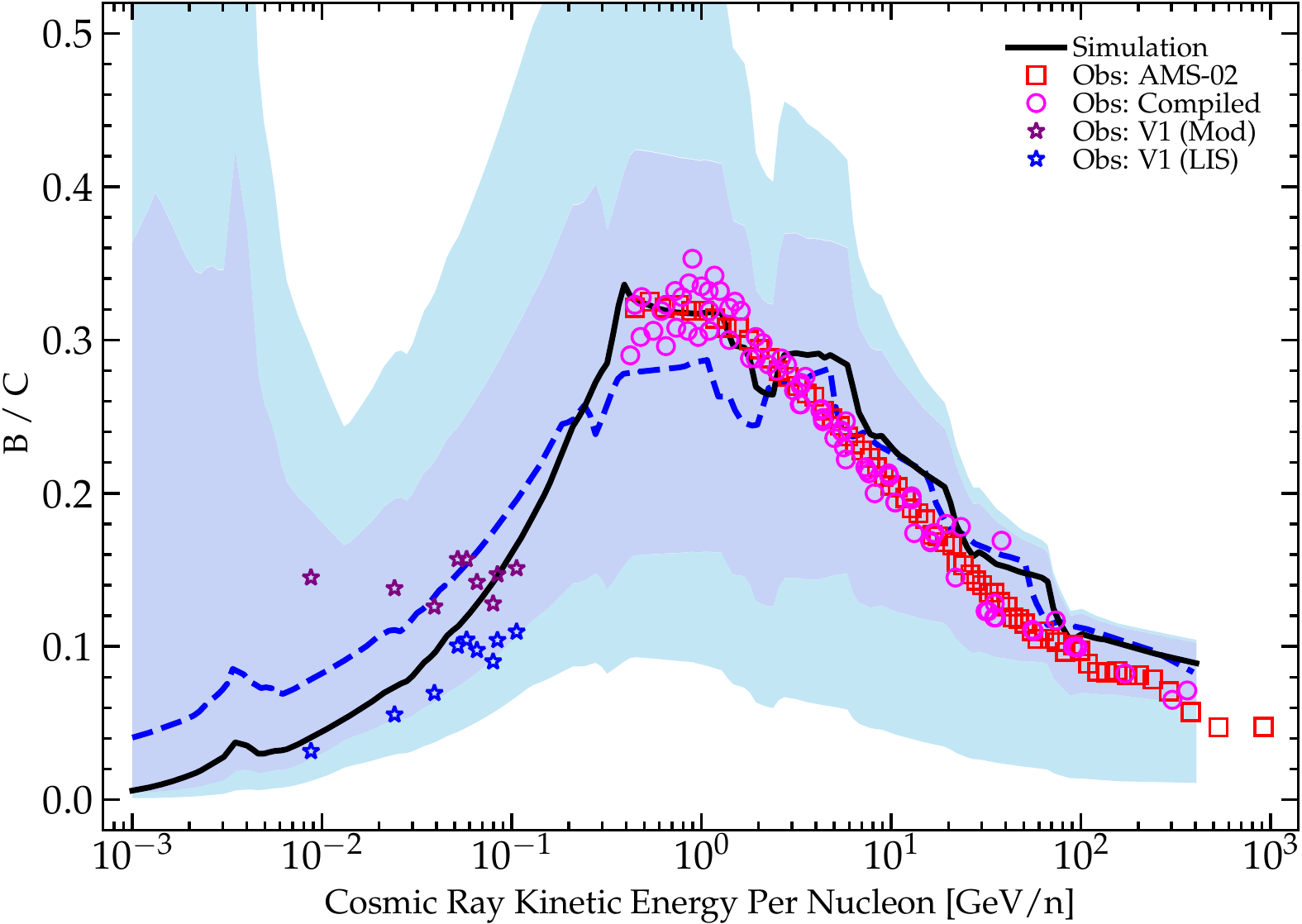} \\
	\includegraphics[width=0.47\textwidth]{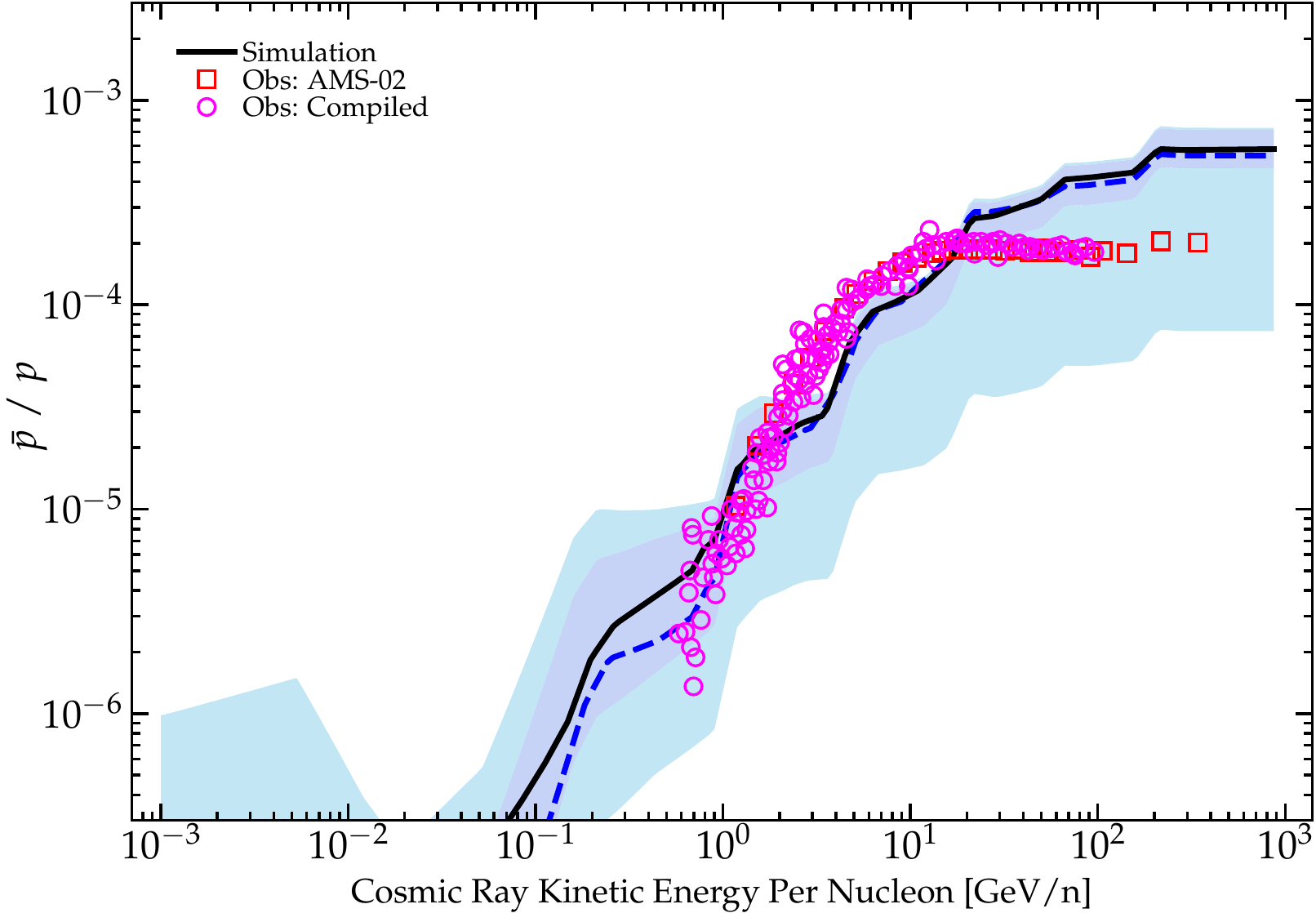}
	\hspace{0.05cm}\includegraphics[width=0.47\textwidth]{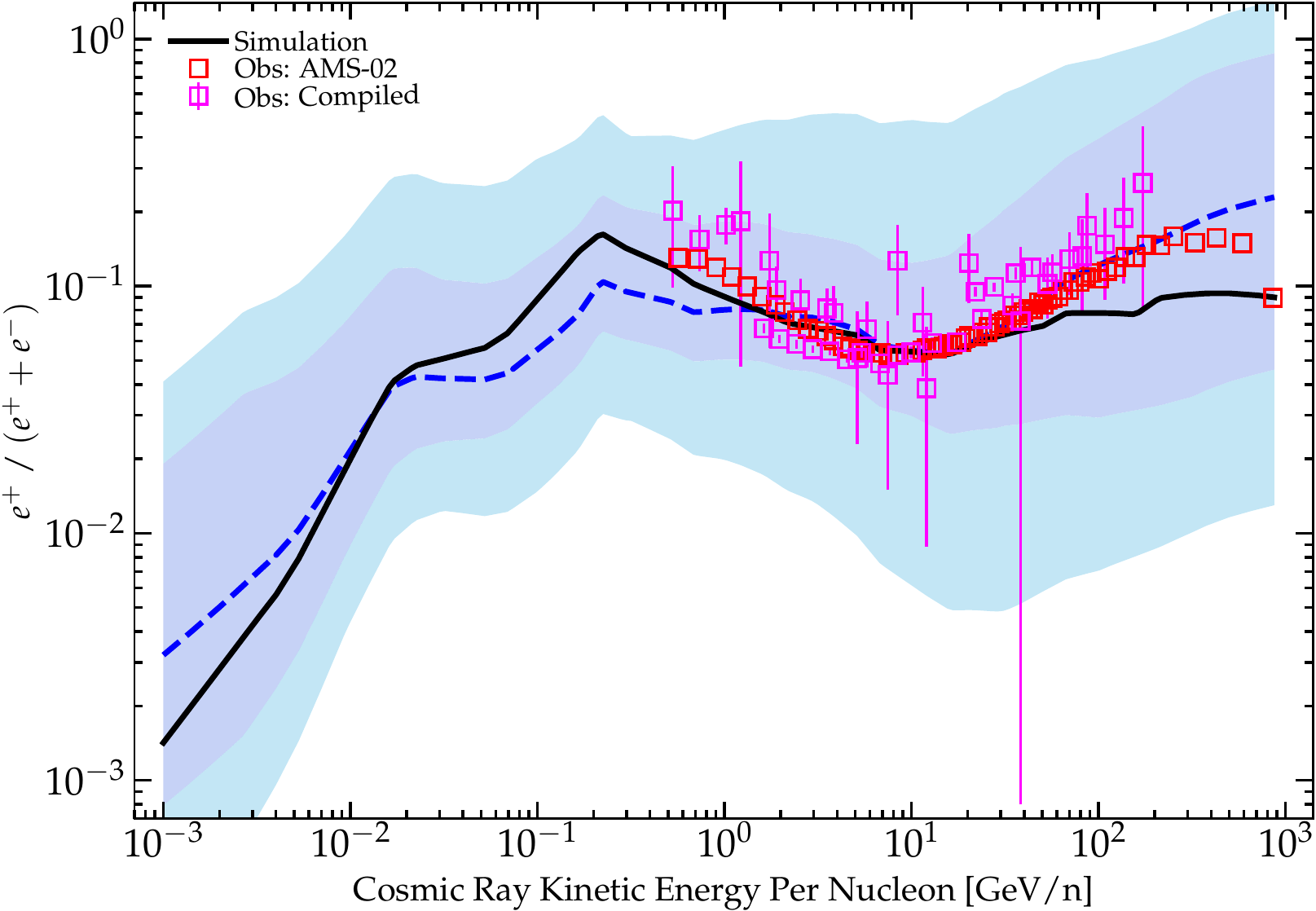} \\
	\vspace{-0.1cm}
	\caption{Example CR spectra and secondary-to-primary ratios from our full live galaxy-formation simulation at redshift $z=0$ in a ``working model'' with a single-power-law injection spectrum (with all abundance ratios at injection determined in-code) and fiducial scaling of the scattering rates $\bar{\nu} \sim 10^{-9}\,{\rm s}^{-1}\,\beta\,R_{\rm GV}^{-0.6}$; see \S~\ref{sec:params} for details. 
	{\em Top:} CR intensity ({\em left}) and kinetic energy density ({\rm right}) spectra, for different species: protons $p$, anti-protons $\bar{p}$, electrons $e^{-}$, positrons $e^{+}$, B, C, radioactive ($^{10}$Be) and stable ($^{7}$Be+$^{9}$Be) Be. 
	Lines show the median ({\em dashed}) and mean ({\em solid}) values in the simulation (small-scale ``features'' are artifacts of our spectral sampling; see \S~\ref{sec:params:simple.fit}), and shaded range shows the $\pm1\,\sigma$ range, for ISM gas in the Solar circle ($r=7-9\,$kpc) with local density $n=0.3-3\,{\rm cm^{-3}}$. Points show observations (colors denote species), from the local ISM (LISM) from Voyager ({\em circles}; \citealt{cummings:2016.voyager.1.cr.spectra}), AMS-02 ({\em squares}; \citealt{2018PhRvL.120b1101A,2019PhRvL.122d1102A,2019PhRvL.122j1101A}, and references therein), and compiled from other experiments including PAMELA, HEAO, BESS, TRACER, CREAM, NUCLEON, CAPRICE, Fermi-LAT, CALET, HESS, DAMPE, ISOMAX ({\em pentagons}; \citealt{engelmann:1990.heao.cosmic.rays,2007APh....28..154S,2000ApJ...532..653B,2011ApJ...742...14O,adriani:2014.pamela.experiment,2017PhRvD..95h2007A,2017ICRC...35.1091B,2017arXiv170906442H,2017ApJ...839....5Y,2017Natur.552...63D,2018PhRvL.120z1102A,2019ARep...63...66A}). For the non-Voyager data we omit observations at energies where the Solar modulation correction is estimated to be important \citep[see][and references therein]{2017AdSpR..60..865B,bisschoff:2019.lism.cr.spectra}.
	{\em Middle Left:} $^{10}$Be/$^{9}$Be ratio, in the same ISM gas; dark purple (light cyan) shaded range shows the $\pm1\,\sigma$ ($\pm2\,\sigma$) range, lines show median ({\em dashed}) and mean ({\em solid}). Points show compiled AMS-02 and other-experiment data (references above). 
	{\em Middle Right:} B/C ratio. For the Voyager data, we show both the directly observed values and the ``modulation-corrected'' value from \citet{2007ARNPS..57..285S} who consider models where modulation could still be important for V1 data (note this would also reduce the value of B/C observed at $\sim1\,$GeV). 
	{\em Bottom Left:} $\bar{p}/p$ ratio. Note the value at the highest-energies is significantly affected by our spectral upper boundary (we do not evolve $p$ or heavier ions with rigidity $\gtrsim1000\,$GV). 
	{\em Bottom Right:} $e^{+}/(e^{+}+e^{-})$ ratio.
	We stress that we have not marginalized over parameters or ``fit'' to any of these observations, but simply survey a few model choices and show one which gives overall agreement.
	\label{fig:demo.cr.spectra.fiducial}}
\end{figure*}

\subsection{Default Input Parameters (Model Assumptions)}
\label{sec:methods:params}

We vary the physics and input assumptions in tests below, but for reference, the default model inputs assumed are as follows.

\subsubsection{Injection}

By default we assume all SNe (Types Ia \&\ II) and fast (OB/WR) winds contribute to Fermi-I acceleration with a fixed fraction $\epsilon_{\rm cr}^{\rm inj} \sim 0.1$ of the initial (pre-shock) ejecta kinetic energy going into CRs (and a fraction $\epsilon_{e} \sim 0.02$ of that into leptons). We adopt a single-power power-law injection spectrum in momentum/rigidity with $j(R) \propto R^{-\psi_{\rm inj}}$ and $\psi_{\rm inj} \sim 4.2$ (i.e.\ a ``canonical'' predicted injection spectrum; discussed in detail in \S~\ref{sec:params}). We will assume most acceleration happens at early stages after a strong shock forms, when the shocks have their highest velocity/Mach number and the dissipation rates are also highest -- this occurs after the reverse shock forms, roughly when the swept-up ISM mass is about equal to the ejecta mass ($M_{\rm swept} \approx M_{\rm ej}$). Equivalently, given that most of the shock energy injected into the ISM, and therefore CR energy, comes from core-collapse SNe, we obtain nearly-identical results if we instead assume that the injection is dominated by shocks with velocity $\gtrsim 2000\,{\rm km\,s^{-1}}$. We show below that slower (e.g.\ ISM or AGB, or late-stage Sedov/snowplow SNe) shocks cannot contribute significant Fermi-I acceleration of the species followed.

\subsubsection{Scattering Rates}

In future studies we will explore physically-motivated models for scattering rates as a function of local plasma properties, pitch angle, gyro-radius, etc. But in this first study we restrict to simple phenomenological models, where we parameterize by default the (pitch-angle-weighted) scattering rates as a single power-law $\bar{\nu} = \bar{\nu}_{0}\,(v/c)\,(R/R_{0})^{-\delta}$ with $R_{0}\equiv1\,$GV (e.g.\ $\bar{\nu} \propto v / \ell_{\rm scattering}$ where $\ell_{\rm scattering} \propto R^{\delta}$ is some characteristic length). In the strong-scattering flux-steady-state limit, this gives a parallel diffusivity $\kappa_{\|}=v^{2}/3\,\bar{\nu}$ or, in the isotropic Fokker-Planck equation $\partial_{t} f = \nabla (D_{x x}\,\nabla f)$, $D_{x x} = v^{2}/9\,\bar{\nu}$, so reduces to the common assumption in phenomenological Galactic CR models that $D_{x x} = \beta\,D_{0}\,(R/R_{0})^{\delta}$. Here our default models (motivated by both historical studies and the comparison to observations discussed below) take $\bar{\nu}_{0} \approx 10^{-9}\,{\rm s^{-1}}$, $\delta=0.5$, equivalent to $D_{0}(R=1\,{\rm GV}) \approx 10^{29}\,{\rm cm^{2}\,s^{-1}}$.

\subsection{Initial Conditions} 
\label{sec:methods:ics}

In a follow-up paper, we will present full cosmological simulations from $z\approx 100$, as in our previous single-bin CR studies (see \citealt{hopkins:2020.cr.outflows.to.mpc.scales,hopkins:cr.transport.constraints.from.galaxies,hopkins:cr.mhd.fire2,hopkins:2020.cr.transport.model.fx.galform,ji:fire.cr.cgm,ji:20.virial.shocks.suppressed.cr.dominated.halos} and \papertwo). These, however, are (a) computationally expensive, and (b) inherently chaotic owing to the interplay of N-body+hydrodynamics+stellar feedback \citep{su:2016.weak.mhd.cond.visc.turbdiff.fx,su:discrete.imf.fx.fire,keller:stochastic.gal.form.fx,genel:stochastic.gal.form.fx}, which makes it difficult if not impossible to isolate the effects of small changes in input assumptions (e.g.\ the form of $\bar{\nu}(R)$) and to ensure that we are comparing to a ``MW-like'' galaxy. Because we focus on Solar-neighborhood observations, we instead in this paper adopt a suite of ``controlled restarts'' as in \citet{orr:ks.law,hopkins:fire2.methods,daa:20.hyperrefinement.bh.growth}. We begin from a snapshot of one of our ``single-bin'' CR-MHD cosmological simulations from \papertwo, which include all the physics here but treat CRs in the ``single-bin'' approximation from \S~\ref{sec:intro}. 
Illustrations of the stars and gas in these systems are shown in Fig.~\ref{fig:images}. Per \S~\ref{sec:intro}, these initial conditions have been extensively compared to MW observations to show that they broadly reproduce quantities important for our calculation like the Galaxy 
stellar and gas mass in different phases \citep{elbadry:HI.obs.gal.kinematics,hopkins:cr.mhd.fire2,gurvich:2020.fire.vertical.support.balance}, 
molecular and neutral gas cloud properties and magnetic field strengths \citep{guszejnov:fire.gmc.props.vs.z,benincasa:2020.gmc.lifetimes.fire}, 
gas disk sizes and morphological/kinematic structure \citep{elbadry:fire.morph.momentum,garrisonkimmel:fire.morphologies.vs.dm}, 
SNe and star formation rates \citep{orr:ks.law,garrison.kimmel:2019.sfh.local.group.fire.dwarfs}, 
$\gamma$-ray emission properties (provided reasonable CR model choices; \citealt{chan:2018.cosmicray.fire.gammaray,hopkins:cr.transport.constraints.from.galaxies}), 
and circum-galactic medium properties in different gas phases \citep{faucher-giguere:2014.fire.neutral.hydrogen.absorption,ji:fire.cr.cgm}, 
suggesting they provide a reasonable starting point here. We take galaxy {\bf m12i} (with the initial snapshot from the ``CR+($\kappa=3e29$)'' run in \papertwo) as our fiducial initial condition, though we show results are similar for galaxies {\bf m12f} and {\bf m12m}. 

We re-start that simulation from a snapshot at redshift $z\approx 0.05$, using the saved CR energy in every gas cell to populate the CR DF for all species, assuming an initially isotropic DF with the initial spectral shape and relative normalization of different species all set uniformly to fits to the local ISM (LISM) spectra \citep{bisschoff:2019.lism.cr.spectra}. The spectra are re-normalized to match the snapshot CR energy density\footnote{Throughout this paper, when we refer to and plot the CR ``energy density'' $e_{\rm cr}$, we will follow the convention in the observational literature and take this to be the kinetic energy density (not including the CR rest mass energy), unless otherwise specified.}  before beginning, to minimize any initial perturbation to the dynamics. We then run for $\approx 500\,$Myr to $z=0$. As discussed in detail and demonstrated in numerical tests in \S~\ref{sec:numerics}, this is more than sufficient for all quantities in the local ISM (LISM; which we use interchangeably with warm-phase ISM gas at Solar-like galacto-centric radii and densities $\sim 0.1-1\,{\rm cm^{-3}}$) to reach their quasi-steady-state values (which should physically occur on the loss/escape timescale in the LISM, maximized at $\sim 1-10\,$Myr around $\sim 1\,$GeV) -- only in the further CGM at $>10\,$kpc from the galaxy do CR transport timescales exceed $\sim\,$Gyr. 
We have confirmed this directly by comparing the simulation results at various times spread over $\sim 100-200\,$Myr before $z=0$; we discuss this below but the variations in the median are generally much smaller than the $\pm1\sigma$ range for different Solar-like locations. To test the independence of our results on the CR ICs, we have also re-started simulations with {\em zero} initial CR energy in all cells. This produces a more pronounced initial transient in the couple disk dynamical times ($\sim 100$\,Myr) owing to the loss of CR pressure but converges to the same equilibrium after somewhat longer physical time, and all our conclusions are identical at $z=0$. This also provides an independent test that the simulations have converged to steady-state behaviors.

\begin{figure*}
\begin{tabular}{r@{\hspace{0pt}}r@{\hspace{0pt}}r}
	\includegraphics[width=0.33\textwidth]{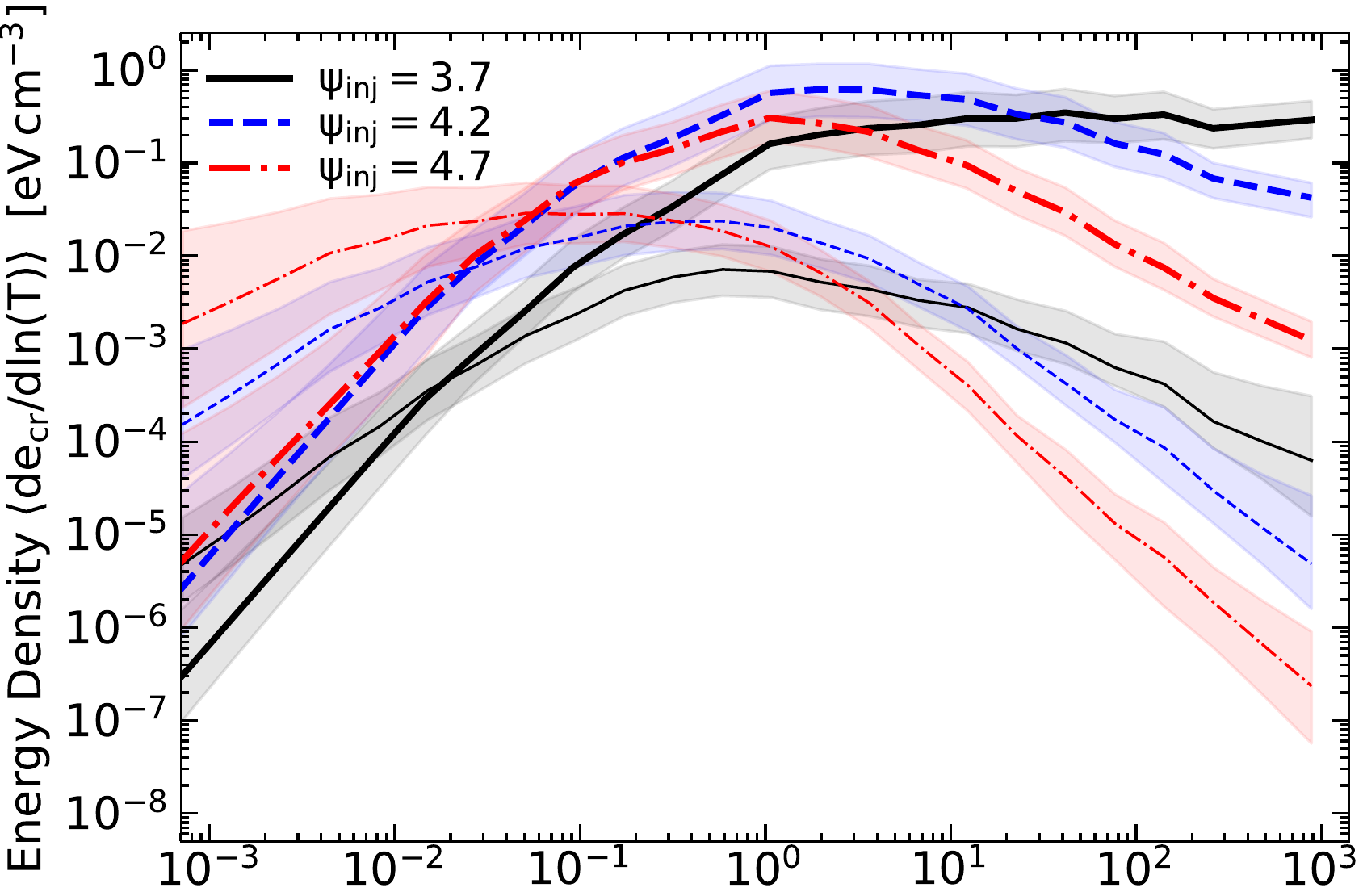}&
	\includegraphics[width=0.33\textwidth]{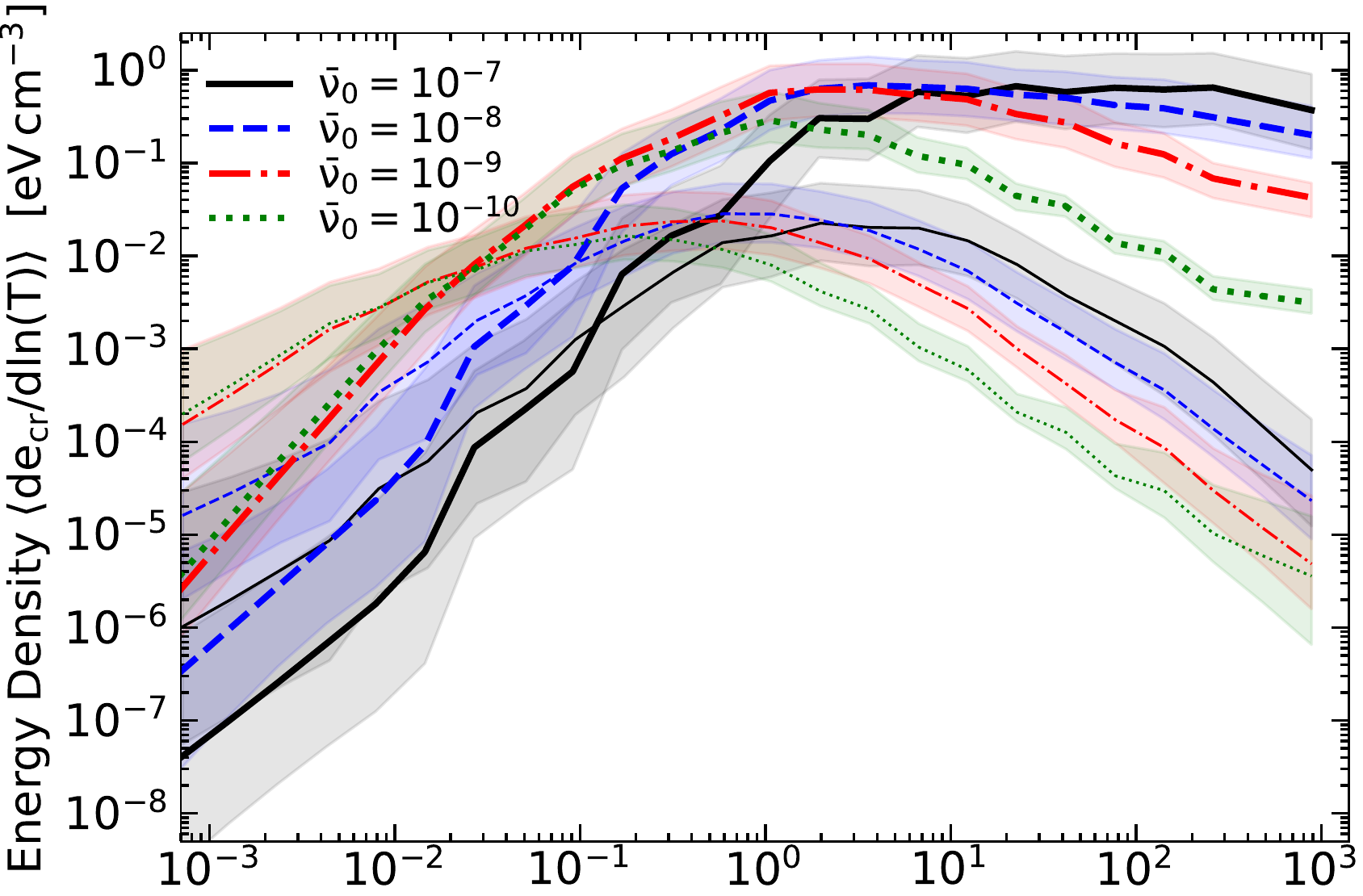} &
	\includegraphics[width=0.33\textwidth]{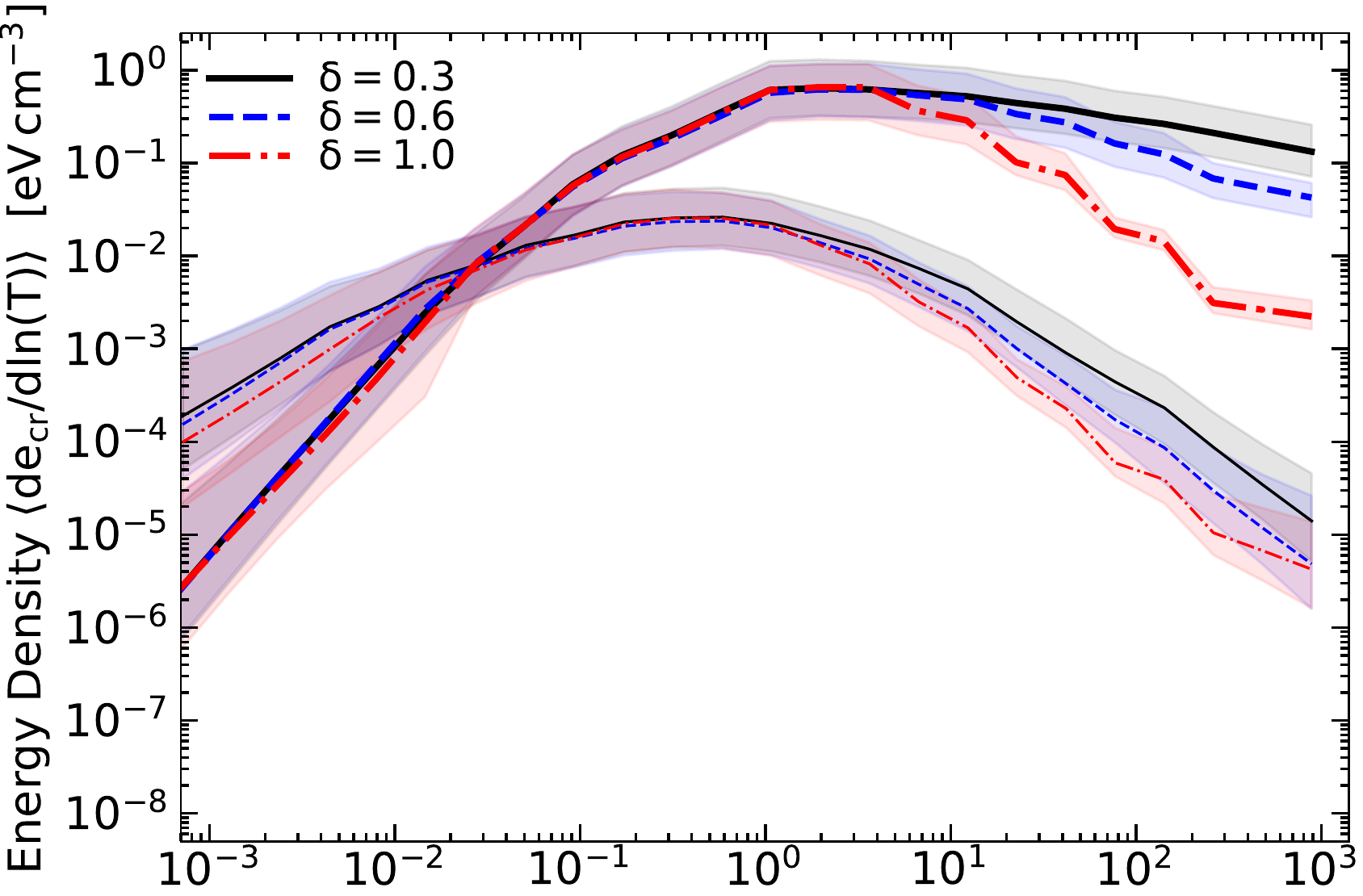} \\
	\includegraphics[width=0.32\textwidth]{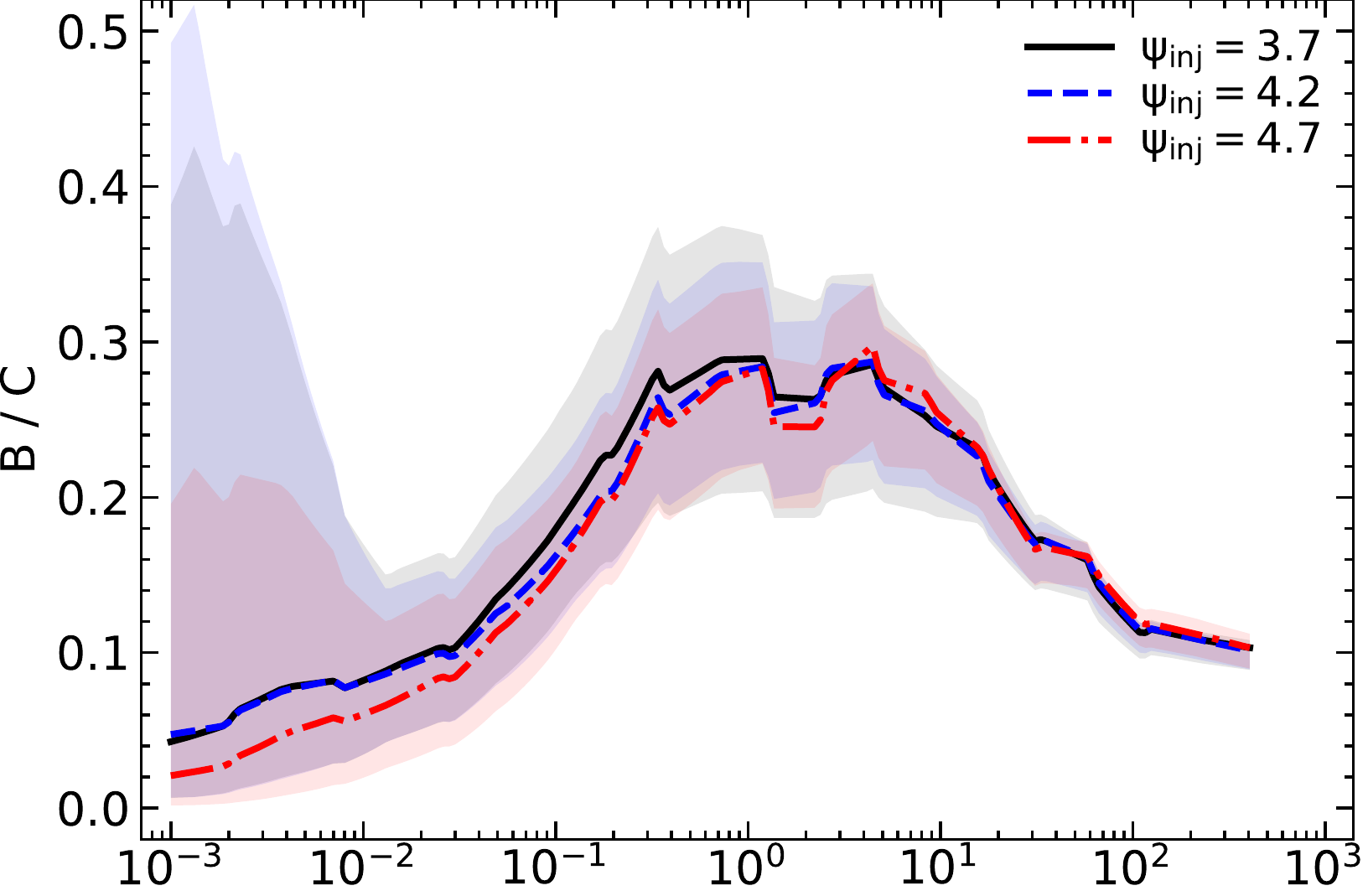}&
	\includegraphics[width=0.32\textwidth]{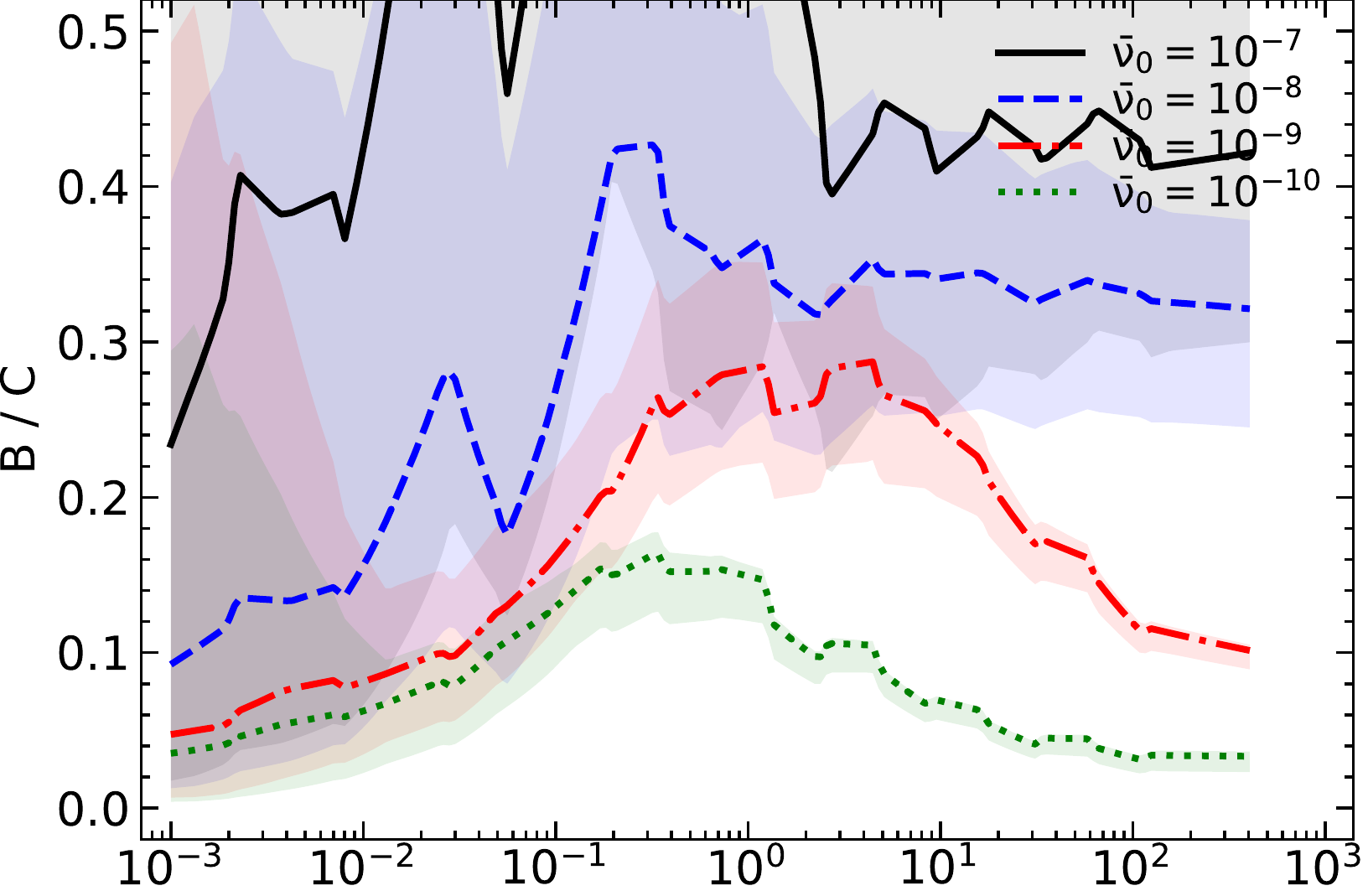}&
	\includegraphics[width=0.32\textwidth]{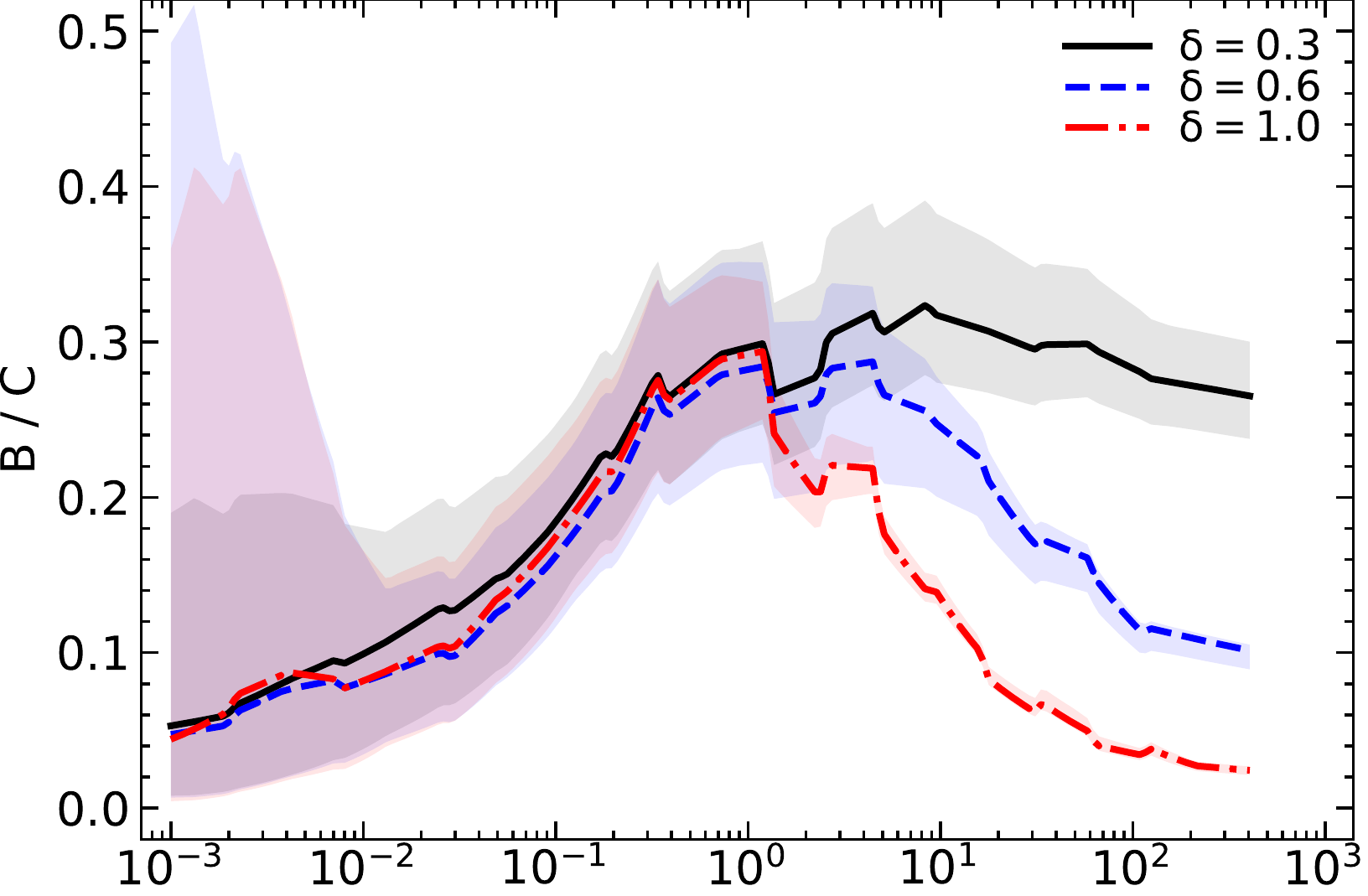} \\
	\includegraphics[width=0.32\textwidth]{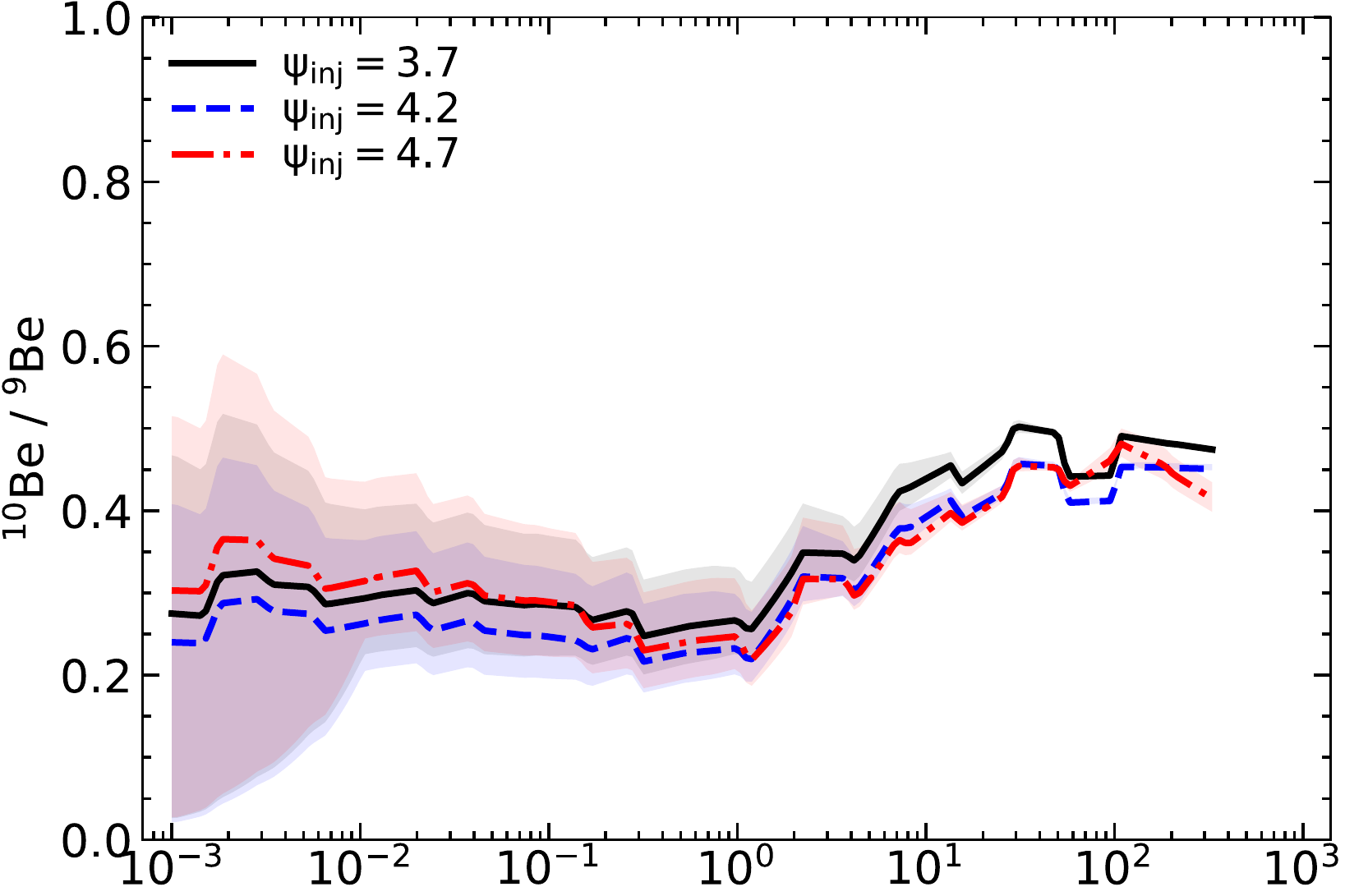}&
	\includegraphics[width=0.32\textwidth]{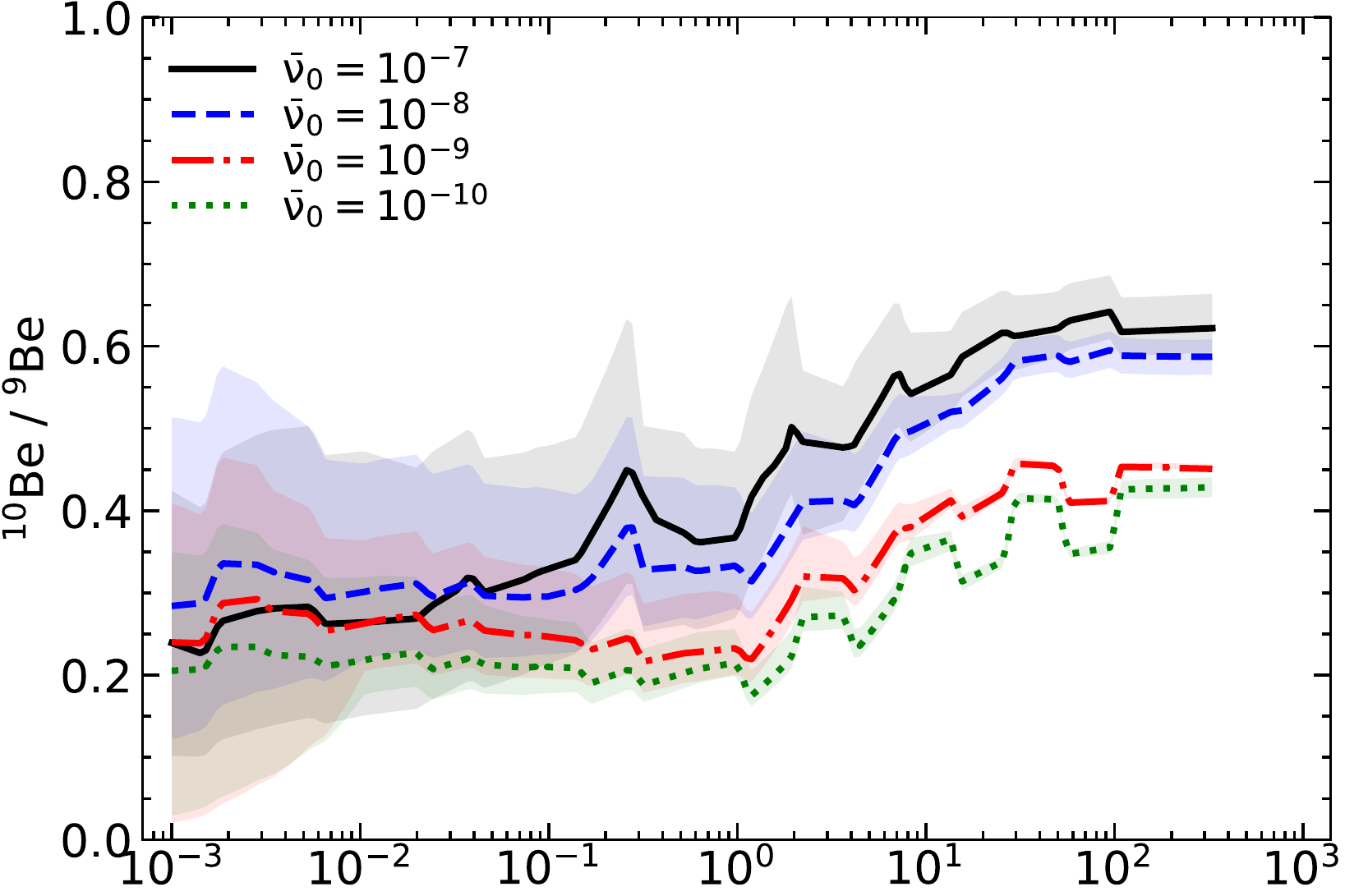}&
	\includegraphics[width=0.32\textwidth]{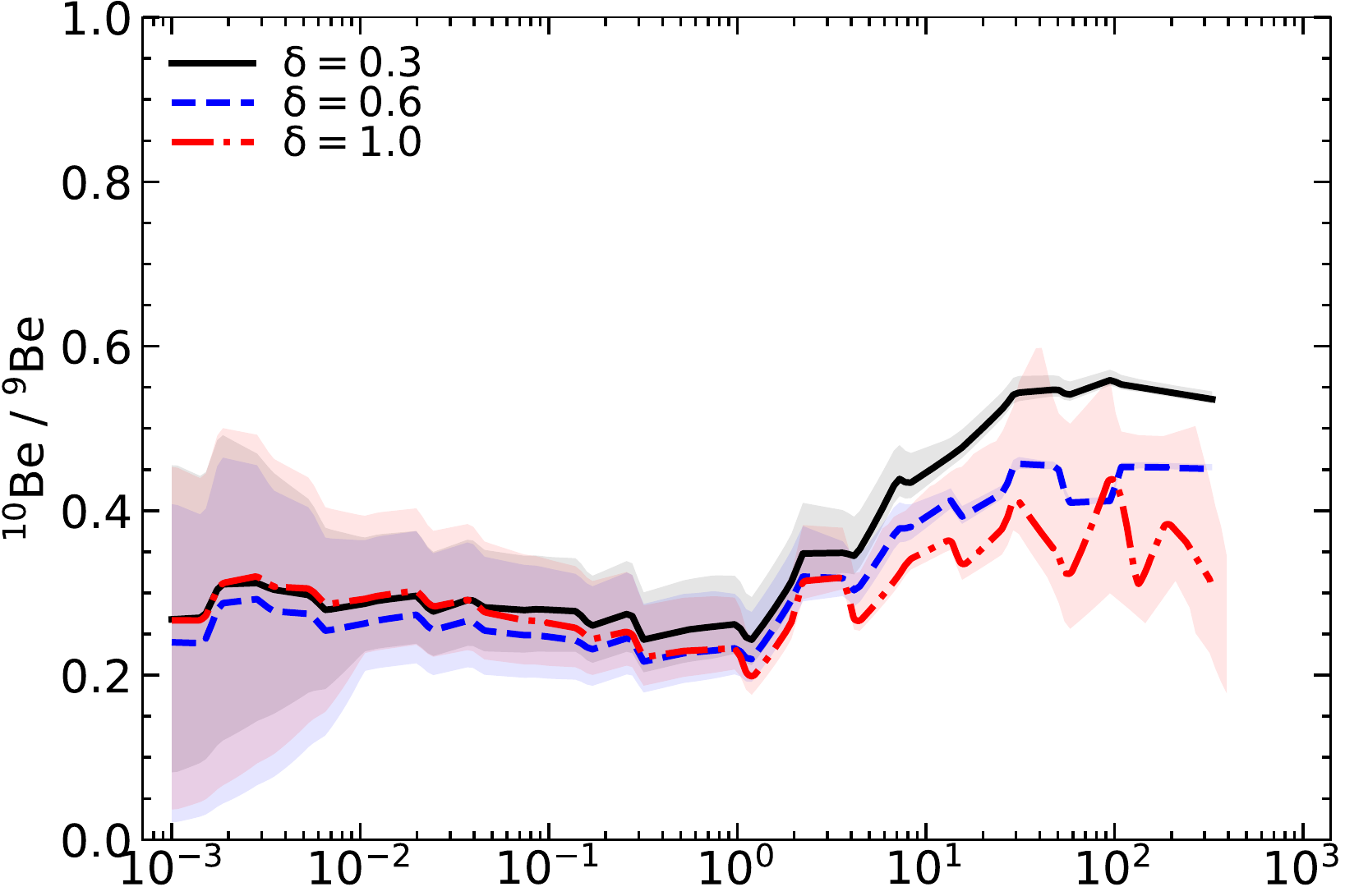} \\
	\includegraphics[width=0.33\textwidth]{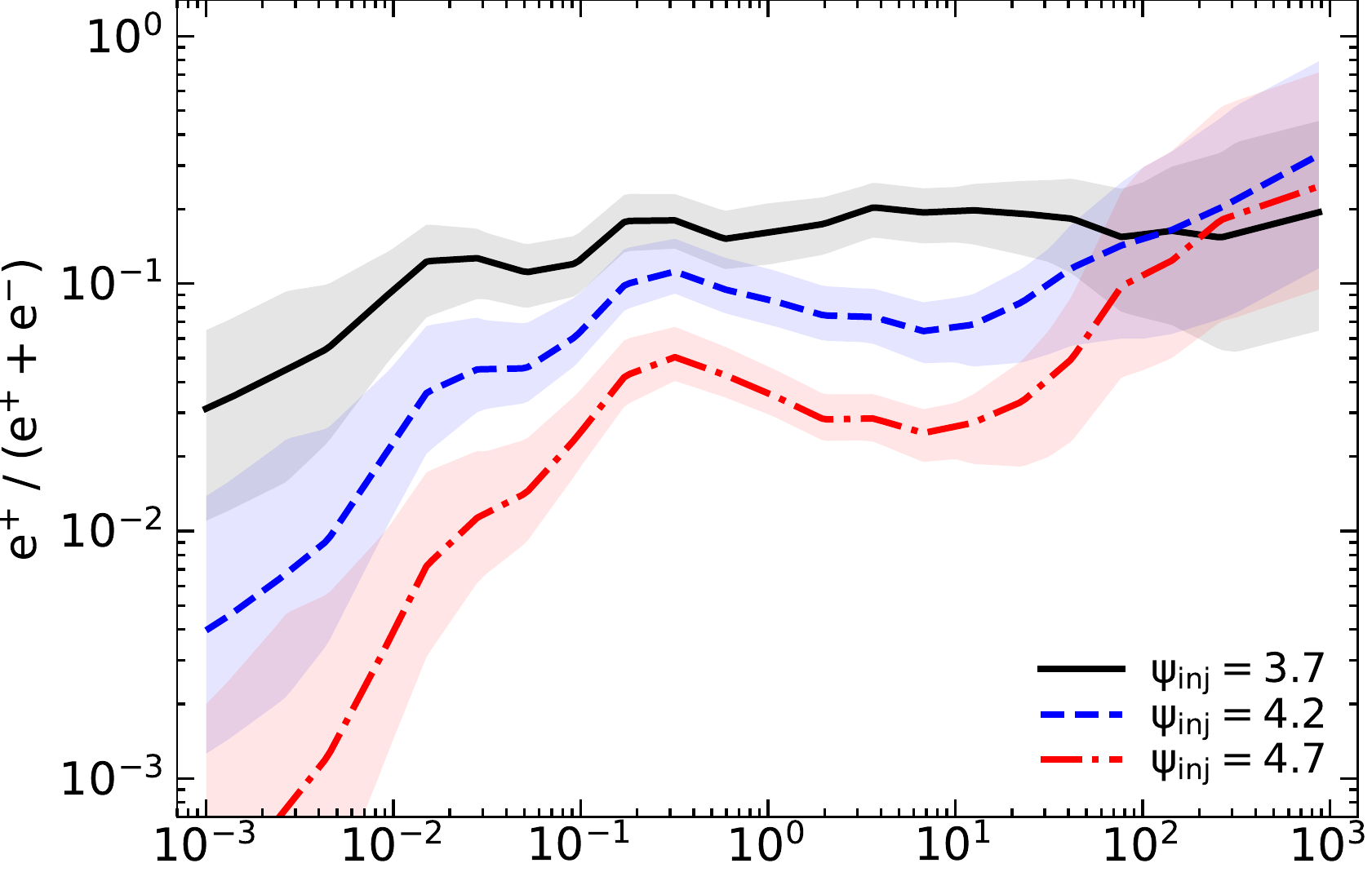}&
	\includegraphics[width=0.33\textwidth]{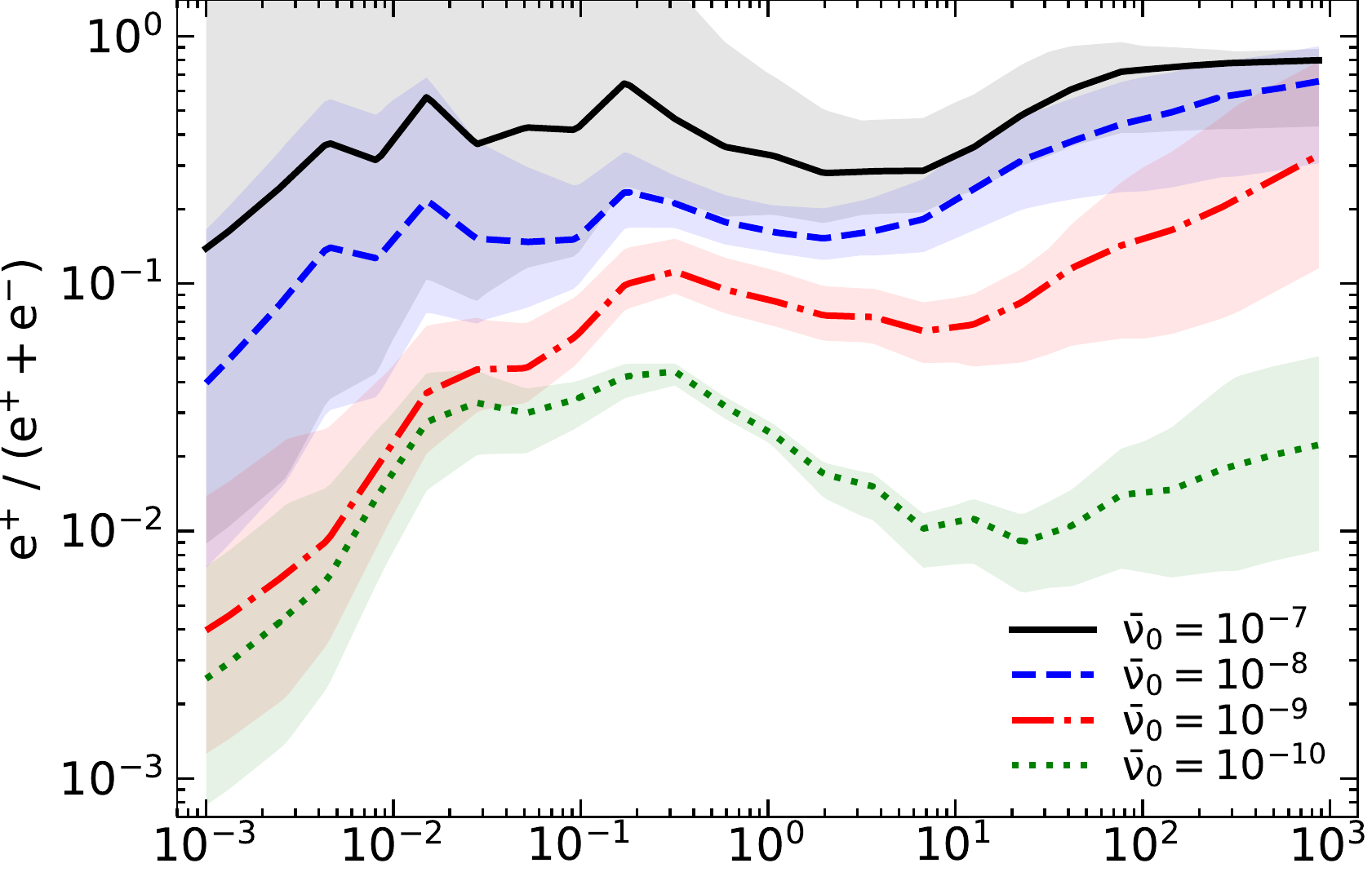}&
	\includegraphics[width=0.33\textwidth]{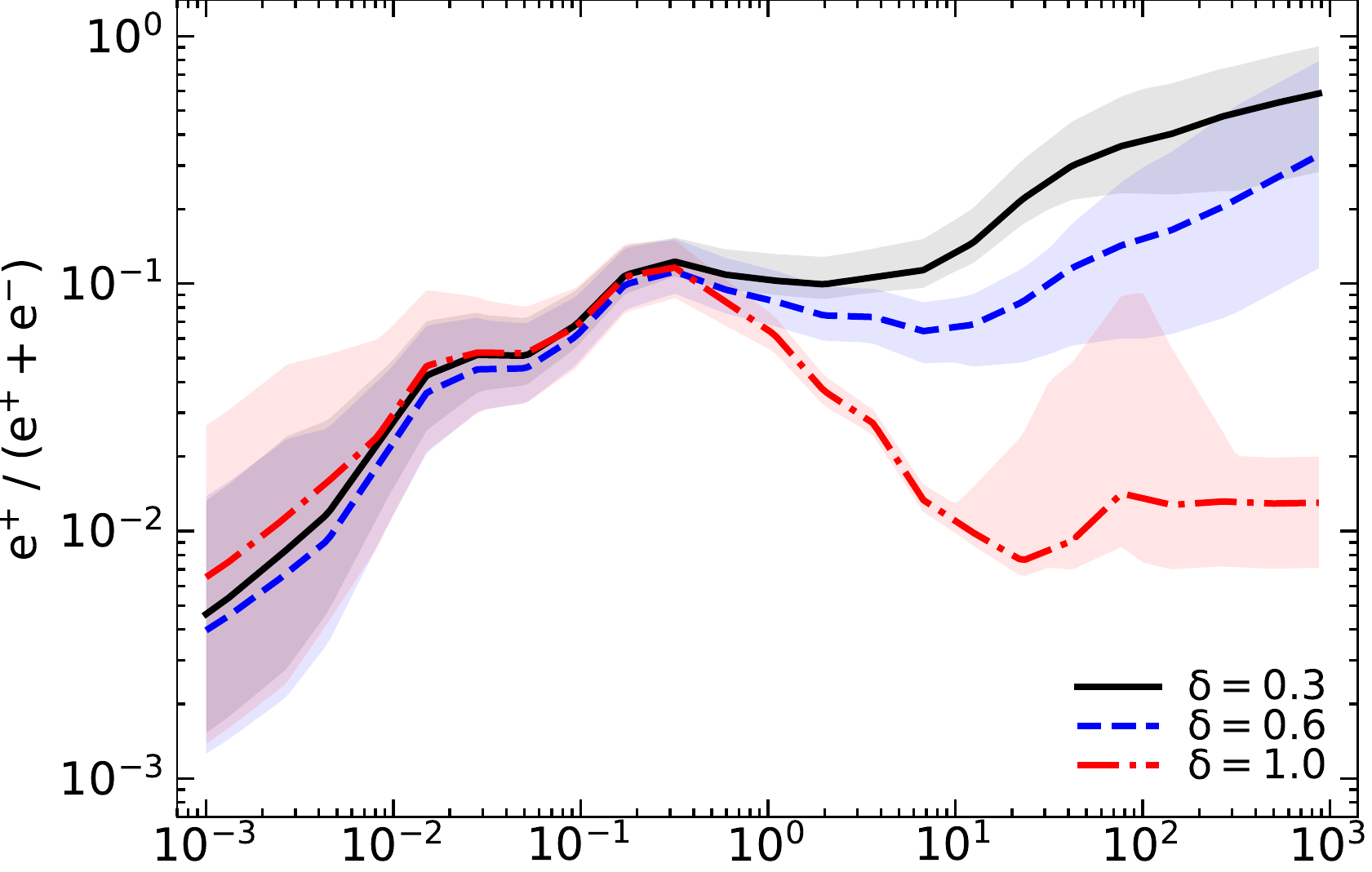} \\
	\includegraphics[width=0.33\textwidth]{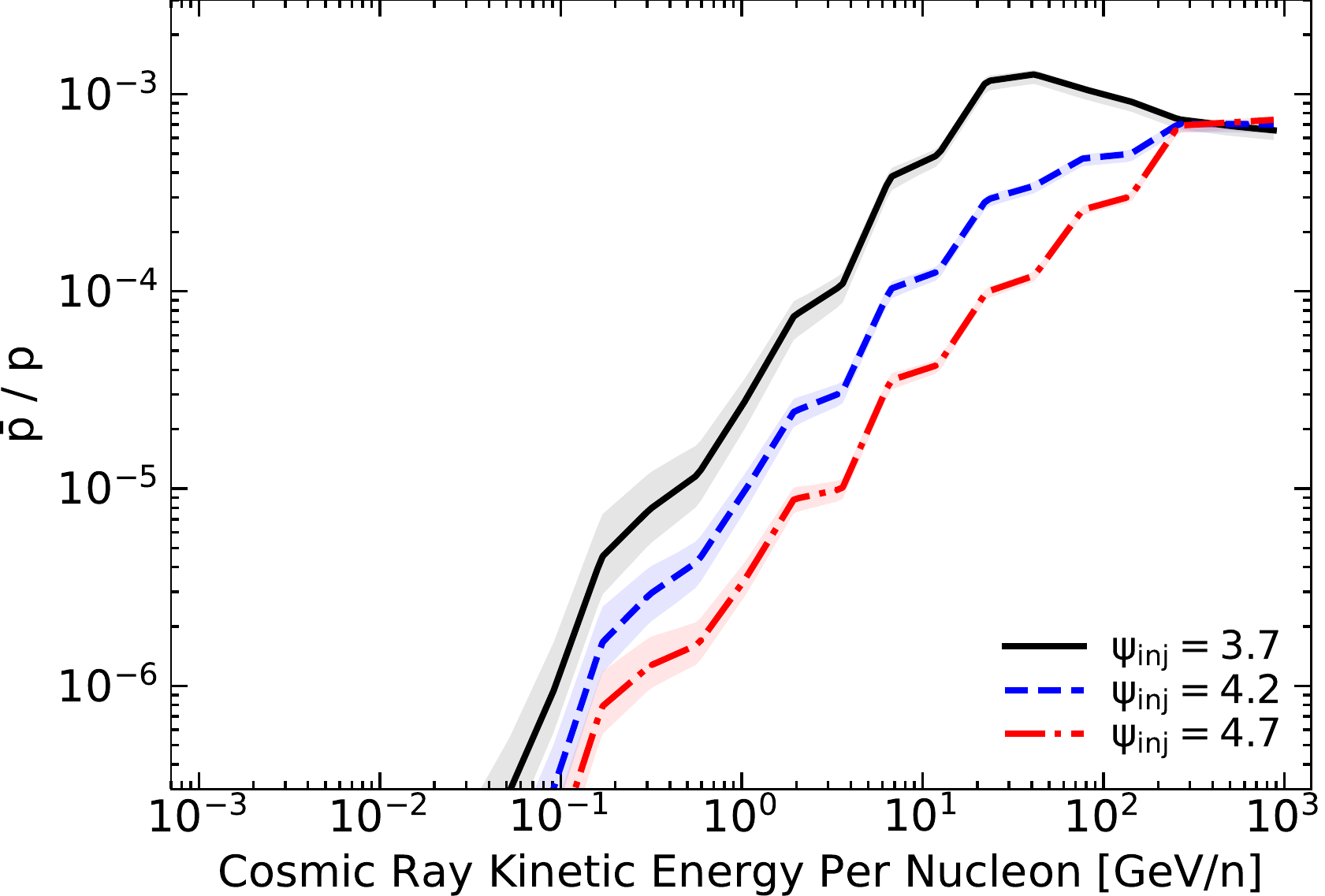}&
	\includegraphics[width=0.33\textwidth]{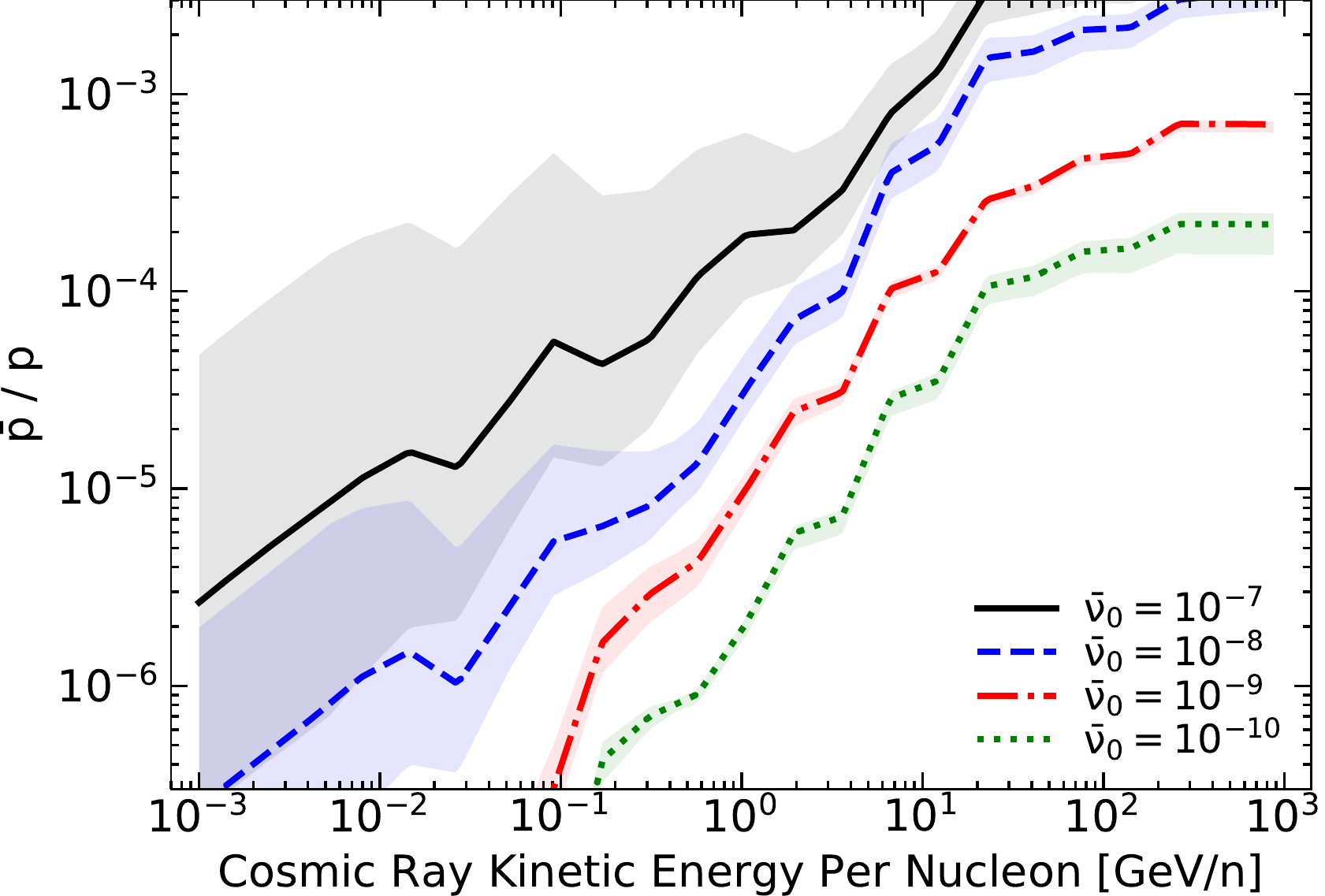}&
	\includegraphics[width=0.33\textwidth]{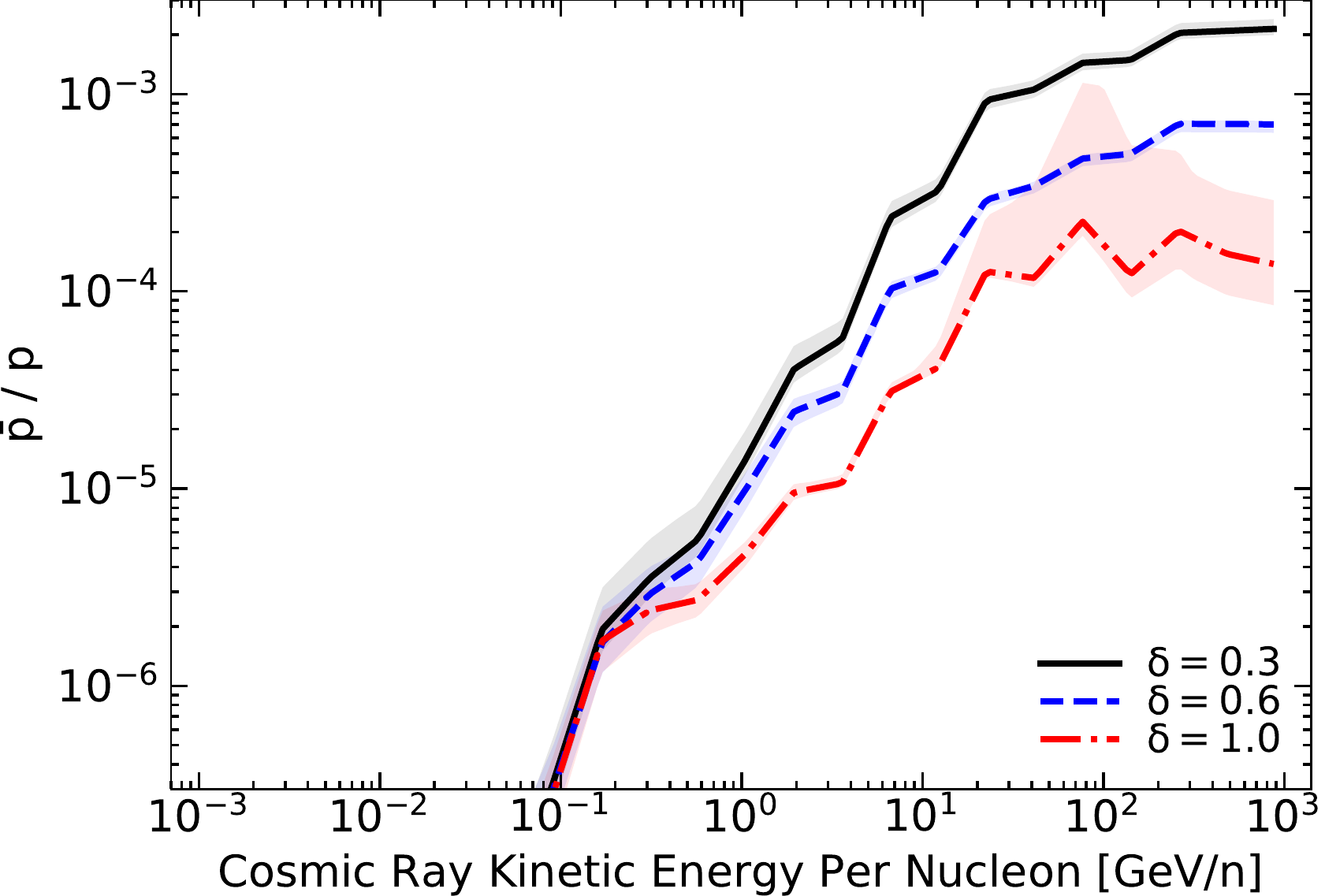} \\
\end{tabular}
	\vspace{-0.1cm}
	\caption{Comparison of CR spectra as Fig.~\ref{fig:demo.cr.spectra.fiducial} from the same galaxy initial condition, in gas with $n=0.3-3\,{\rm cm^{-3}}$, with different parameters of the injection spectrum $j_{\rm inj} \propto p^{-\psi_{\rm inj}}$ and assumed (universal) CR scattering rate $\bar{\nu} = \bar{\nu}_{0}\,\beta\,R_{\rm GV}^{-\delta}$. To reduce clutter, lines+shaded range show just the mean+1$\sigma$ range for each model, we do not overplot the observations, and for the CR kinetic energy density $d e_{\rm cr}/d\ln{T}$ we show just $p$ (thick) and $e^{-}$ (thin). Note the ``reference'' parameters (about which we vary) are those from our best-fit in Fig.~\ref{fig:demo.cr.spectra.fiducial} (Appendix~\ref{sec:appendix:additional} considers variation about an alternative ``reference'' model).
	{\em Left:} Injection slope $\psi_{\rm inj}$. As expected this shifts CR spectral slopes, but it also shifts secondary-to-primary ratios in a manner not trivially predicted by leaky-box type models.
	{\em Middle:} Scattering rate normalization $\bar{\nu}_{0}$. These shift secondary-to-primary ratios qualitatively as expected, but often non-linearly; more surprising, higher $\bar{\nu}_{0}$ (lower effective diffusivity) clearly produces systematically harder/shallower CR spectra. 
	{\em Right:} Dependence of scattering rate on rigidity $\delta$. This shifts the spectral shape and secondary-to-primary dependence on $T$ roughly as expected, though again slightly non-linearly.
	\label{fig:spec.compare.kappa}\vspace{-0.4cm}}
\end{figure*}

\section{Results \&\ Discussion}
\label{sec:results}

\subsection{Working Models}
\label{sec:params}

\subsubsection{Parameters: Single Power-Law Injection \&\ Diffusion Can Fit the Data} 
\label{sec:params:simple.fit}

The first point worth noting is that it is actually {\em possible} to obtain reasonable order-of-magnitude agreement with the Solar neighborhood CR data, as shown in Fig.~\ref{fig:demo.cr.spectra.fiducial}. This may seem obvious, but recall that the models here have far fewer degrees of freedom compared to most historical Galactic CR population models: the Galactic background is entirely ``fixed'' (so e.g.\ \Alf\ speeds, magnetic geometry, radiative/Coulomb/ionization loss rates, convective motions, re-acceleration, etc.\ are determined, not fit or ``inferred'' from the CR observations); we assume a universal single-power-law injection spectrum (with just two parameters entirely describing the injection model for all species) and {\em do not} separately fit the injection spectra for different nuclei but assume they trace the injection of protons given their {\em a priori} abundances in the medium; and we similarly assume a single power-law scattering rate $\nu(R)$ as a function of rigidity to describe all species. 

In our favored model(s), the injection spectrum for {\em all} species is a single power-law with $dj \propto p^{-4.2}\,d^{3}{\bf p}$ (with all heavy species relative abundance following their actual shock abundances), with $\sim10\%$ of the shock energy into CRs and $\sim2\%$ of that into leptons, and the scattering rate scales as $\bar{\nu} \sim 10^{-9}\,{\rm s}^{-1}\,\beta\,R_{\rm GV}^{-(0.5-0.6)}$. Under the assumptions usually made to turn the CR transport equations into an isotropic Fokker-Planck diffusion equation, this corresponds to $D_{xx} \approx \beta\,D_{0}\,R_{\rm GV}^{\delta}$ with $\delta$ in the range $\delta=0.5-0.6$ and $D_{0}\approx 10^{29}\,{\rm cm^{2}\,s^{-1}}$.\footnote{In Appendix~\ref{sec:vdrift.tloss}, we also show that this translates to typical effective CR ``drift velocities'' of roughly $\sim 300\,{\rm km\,s^{-1}}\,(T/{\rm GeV})^{0.3}$ in Solar circle, midplane LISM gas with densities $n\sim 1\,{\rm cm^{-3}}$, but this can vary more significantly with environment.}

Briefly, we note in Fig.~\ref{fig:demo.cr.spectra.fiducial} that the largest statistical discrepancy between the simulations and observations appears to be between the flat values of $\bar{p}/p$ at high energies $\sim 10-300\,$GV, where our model $\pm 1\sigma$ predictions continue to rise by another factor $\sim 2-3$. This is generically the most difficult feature to match, of those we compare, while simultaneously fitting all other observations, and we will investigate in more detail in future work. It is interesting in particular because it runs opposite to the recent suggestion that reproducing $\bar{p}/p$ alongside B/C requires some ``additional,'' potentially exotic (e.g.\ decaying dark matter) source of $\bar{p}$ \citep{2020arXiv201203956H}. But we caution against over-interpretation of our result for several reasons: (1) the systematic detection/completeness corrections in the data and (2) the physical $\bar{p}$ production cross-sections at these energies remain significant sources of uncertainty \citep{2019PhRvD..99j3014C,2020PhRvR...2d3017H}; (3) the observations still remain within the $\pm2\,\sigma$ range, so the LISM may simply be a $\sim 2\sigma$ fluctuation; (4) recalling that the energy of a secondary $\bar{p}$ is $\sim 10\%$ the primary $p$, most of this discrepancy occurs at such high energies that it depends sensitively on the behavior of our highest-energy $p$ and C bins -- i.e.\ our ``boundary'' bins; and (5) we are only exploring empirical models with a constant (in space and time) scattering rate, while almost any physical model predicts large variations in $\bar{\nu}$ with local ISM environment, which can easily produce systematic changes in secondary-to-primary ratios at this level \citep{hopkins:cr.transport.constraints.from.galaxies}.

We also caution (as noted in \S~\ref{sec:methods.spatial.evol} and demonstrated in \citealt{girichidis:cr.spectral.scheme,ogrodnik:2021.spectral.cr.electron.code} and \S~\ref{sec:numerical.tests}) that the small ``step'' features between CR spectral bins (in Fig.~\ref{fig:demo.cr.spectra.fiducial} and our subsequent plots) are a numerical artifact of finite sampling and the ``bin-centered'' approximation for the spatial fluxes. This directly leads to the ``jagged'' small-scale features evident in B/C and $^10$Be/$^9$Be. There, we follow standard convention and take the intensity ratio of e.g.\ B-to-C at fixed CR kinetic energy per nucleon ($T/A$). But recall (\S~\ref{sec:methods:momentum.maintext}, Table~\ref{tbl:intervals}, our spectral bins for different species are aligned in rigidity, not necessarily in kinetic-energy-per-nucleon, so when taking the ratios the bin edges are offset from one another for different nucleons, producing the ``jagged'' or ``odd-even'' type features spaced at semi-regular fractions of the bin widths. Obviously these features should not be over-interpreted; fortunately these effects are small compared to the full dispersion seen in Fig.~\ref{fig:demo.cr.spectra.fiducial} and to the systematic differences between different Galactic environments or scattering rate parameterizations shown below.

\subsubsection{Comparison to Idealized, Static-Galaxy Analytic CR Transport Models}

The ``favored'' parameters (those which agree best with the observations) above in \S~\ref{sec:params:simple.fit} are completely plausible. The injection spectrum ($\psi_{\rm inj} \approx 4.2$) is essentially identical to the ``canonical'' theoretically-predicted injection spectrum and efficiency for first-order Fermi acceleration in SNe shocks \citep{bell:1978.shock.cr.accel,malkov:2001.shock.cr.dsa,spitkovsky:2008.particle.accel.shock.pic.sims,caprioli:2012.sne.cr.shock.accel}. Considering how different the models are in detail, the favored scattering rate in \S~\ref{sec:params:simple.fit} and its dependence on rigidity are remarkably similar to the values inferred from most studies in the past decade which have fit the CR properties assuming a simple toy model analytic Galaxy model and isotropic Fokker-Planck equation model for CR transport, provided they allow for a CR ``scattering halo'' with size $\sim 5-10\,$kpc \citep{blasi:cr.propagation.constraints,vladimirov:cr.highegy.diff,gaggero:2015.cr.diffusion.coefficient,2016ApJ...819...54G,2016ApJ...824...16J,cummings:2016.voyager.1.cr.spectra,2016PhRvD..94l3019K,evoli:dragon2.cr.prop,2018AdSpR..62.2731A}. Consider e.g.\ \citealt{delaTorre:2021.dragon2.methods.new.model.comparison} , who compare the most recent best-fit models from both GALPROP and DRAGON, which both favor a CR scattering halo of scale-height $\sim 7\,$kpc with a very-similar $D_{xx} \sim 0.6\times10^{29}\,{\rm cm^{2}\,s^{-1}}$ for $\sim1$\,GV protons and $\delta \sim 0.4-0.5$. \citet{korsmeier:2021.light.element.requires.halo.but.upper.limit.unconfined} reached similar conclusions.\footnote{In detail \citet{korsmeier:2021.light.element.requires.halo.but.upper.limit.unconfined} more broadly considered an extensive survey of GALPROP model variations with various statistical modeling methods to show that the combination of Li, Be, B, C, N, O requires halo heights $z_{h} \gtrsim 5\,$kpc across models, in turn requiring $\delta\approx0.5$ and $D_{xx} \sim 0.6\times10^{29}\,{\rm cm^{2}\,s^{-1}}$ at $\sim 1\,$GV. But they note that larger halo heights (with correspondingly larger diffusivities scaling as $D_{xx} \propto z_{h}^{0.8-1.0}$) are also allowed, as e.g.\ $^{10}$Be/$^9$Be becomes weakly dependent on height once that height is sufficiently large. It is primarily smaller halo heights, and correspondingly lower diffusivities, that are strongly ruled out by the analytic Galaxy models.} This is also consistent with a number of recent studies using ``single-bin'' $\sim $\,GeV-CR transport models in cosmological galaxy formation simulations of a wide range of galaxy types \citep{chan:2018.cosmicray.fire.gammaray,su:turb.crs.quench,hopkins:cr.transport.constraints.from.galaxies,hopkins:cr.mhd.fire2,hopkins:2020.cr.transport.model.fx.galform}, compared to observational constraints from $\gamma$-ray detections and upper limits showing all known dwarf and $L^{\ast}$ galaxies lie well below the calorimetric limit \citep{lacki:2011.cosmic.ray.sub.calorimetric,fu:2017.m33.revised.cr.upper.limit,lopez:2018.smc.below.calorimetric.crs}, which inferred that a value of $\bar{\nu} \approx 10^{-9}\,{\rm s}^{-1}$ at $E\approx 1\,$GeV was required to reproduce the $\gamma$-ray observations. 

This is by no means trivial, however. Some recent studies using classic idealized analytic CR transport models have argued that features such as the ``turnover'' in B/C at low energies or minimum in $e^{+}/e^{-}$ require strong breaks in either the injection spectrum or dependence of $\bar{\nu}(p)$ (e.g.\ favoring a $D(p)$ which is non-monotonic in momentum $p$ and rises very steeply with lower-$p$ below $\sim$\,GeV; \citealt{strong:2011.strong.break.in.electron.injection.spectra}), or artificially strong re-acceleration terms (much larger than their physically-predicted values here) which would imply (if true) that most of the CR energy observed actually comes from diffusive reacceleration, not SNe or other shocks \citep{drury.strong:power.req.for.diffusive.reaccel}, or some strong spatial dependence of $\bar{\nu}$ in different regions of the galaxy \citep{2018ApJ...869..176L}. Other idealized analytic transport models \citep{2010A&A...516A..67M,2011ApJ...729..106T,2017MNRAS.471.1662B,2020JCAP...11..027Y} have argued for $\delta$ in the range $\sim 0.3-1$ and some for $\bar{\nu}$ as large as $\sim 10^{-7}\,{\rm s}^{-1}$ at $\sim1\,$GeV (equivalent to $D_{0} \sim 10^{27}\,{\rm cm^{2}\,s^{-1}}$). These go far outside the range of models which we find could possible reproduce the LISM observations.

The fundamental theoretical uncertainty driving these large degeneracies in previous studies is exactly what we seek to address in this study here: the lack of a well-defined galaxy model. In the studies cited above, quantities like the halo size, source spatial distribution, Galactic magnetic field structure and \Alf\ speeds, key terms driving different loss processes (ionization, Coulomb, synchrotron, inverse Compton), adiabatic/convective/large-scale turbulent re-acceleration, are all either treated as ``free'' parameters, or some ad-hoc empirical model is adopted. For example, it is well-known that if one neglects the presence of any ``halo''/CGM/thick disk (and so effectively recovers a classic ``leaky box'' model with sources and transport in a thin $\lesssim 200\,$pc-height disk), then one typically infers a best-fit with much lower $D_{0} \sim 10^{27}\,{\rm cm^{2}\,s^{-1}}$ and ``Kolmogorov-like'' $\delta \sim 0.3$ \citep{2010A&A...516A..67M}. At the opposite extreme, assuming the ``convective'' term has the form of a uniform vertical disk-perpendicular outflow everywhere in the disk (neglecting all local turbulent/fountain/collapse/inflow/bar/spiral and other motions, and assuming a vertically-accelerating instead of decelerating outflow) -- the inferred $\delta$ can be as large as $\sim 1$ \citep{2010A&A...516A..67M}. Similarly,  in these analytic models one can make different loss and/or re-acceleration terms as arbitrarily large or small as desired by assuming different \Alf\ speeds, densities, neutral fractions, etc., in different phases; so e.g.\ models which effectively ignore or artificially suppress ionization \&\ Coulomb losses will require a break in the injection or diffusion versus momentum $p$, to reproduce the correct observed spectra.

Briefly, it is worth noting that in some analytic models, especially the classic ``leaky box'' models, it is common to refer to the residence or loss or escape times of CRs. We discuss these further below, but readers interested in more details can see Appendix~\ref{sec:vdrift.tloss}, where we explicitly present the CR drift velocities and loss timescales in our fiducial simulation as a function of species and energy for LISM conditions.

\subsubsection{On the Inevitability of the ``Halo'' Size}
\label{sec:halo.size}

One particular aspect requires comment here: in cosmological galaxy formation models, a very large ``halo'' is inevitable. Indeed, as noted in \S~\ref{sec:intro}, in modern galaxy theory and observations, the region within $< 10\,$kpc above the disk would not even be called the ``halo'' but more often the thick disk or corona or disk-halo interface. It is well-established that most of the cosmic baryons associated with galaxies are located in the CGM reaching several hundred kpc from galaxy centers, distributed in a slowly-falling power-law-like (not exponential or Gaussian) profile with scale lengths $\sim 20-50\,$kpc \citep{maller:2004.hvc.accretion,steidel:2010.outflow.kinematics,martin:2010.metal.enriched.regions,werk:2014.cos.halos.cgm,sravan.2015:metal.emission.line.cgm.fire,tumlinson:2017.cgm.review}. This is visually obvious in Fig.~\ref{fig:images}.

Thus, from a galaxy-formation point of view, it is not at all surprising that models with a large ``CR scattering halo'' are observationally favored and agree better with realistic galaxy models like those here. What is actually surprising, from the galaxy perspective, is {\em how small} the best-fit halo sizes in some analytic Galactic CR transport models (e.g.\ $\sim 7-8\,$kpc, in the references above) actually are. These $\sim 5-10\,$kpc sizes are actually {\em much smaller} than the scale length for the free-electron density or magnetic field strength inferred in theoretical and observational studies of the CGM \citep[see references above and e.g.][]{lan:2020.cgm.b.fields.rm}. However, there is a simple explanation for this: as parameterized in most present analytic models for CR transport, the ``halo size'' does not really represent the scale-length of e.g.\ the magnetic energy or free-electron density profile; rather, the ``halo size'' in these models is more accurately defined as the volume interior to which CRs have an order-unity probability of scattering ``back to'' the Solar position. In the CGM (with sources concentrated at smaller radii), for any spatially-constant diffusivity, the steady-state solution for the CR kinetic energy density is a spherically-symmmetric power-law profile with $e_{\rm cr} \propto 1/\kappa\,r$ \citep{hopkins:cr.mhd.fire2}, so the characteristic length-scale for scattering ``back into'' some $r=r_{0}$ is just $\approx r_{0}$. In other words, in any slowly-falling power-law-like medium with spatially-constant diffusivity, the {\em inferred} CR scattering ``halo scale length'' at some distance $R_{0} \approx 8\,$kpc from the source center (e.g.\ the Solar position) will always be $L_{\rm halo} \approx R_{0}$ to within a factor of $\sim 2$ depending on how the halo and its boundaries are defined (and indeed this is what models infer), more or less independent of the actual CGM $n_{e}$ or $B$-field scale-length (generally $\gg R_{0}$). 

Empirically, \citet{korsmeier:2021.light.element.requires.halo.but.upper.limit.unconfined} argue for a similar conclusion from a comparison of parameterized analytic CR scattering models to LISM data. They show that so long as the assumed scattering halo volume is sufficiently extended ($z_{h} \gtrsim 5\,$kpc, in their models), the CR observables at the Solar position become essentially independent of its true size ($z_{h}$) -- i.e.\ the ``effective'' scattering halo size becomes constant.

\begin{figure*}
\begin{tabular}{r@{\hspace{0pt}}r@{\hspace{0pt}}r}
	\includegraphics[width=0.33\textwidth]{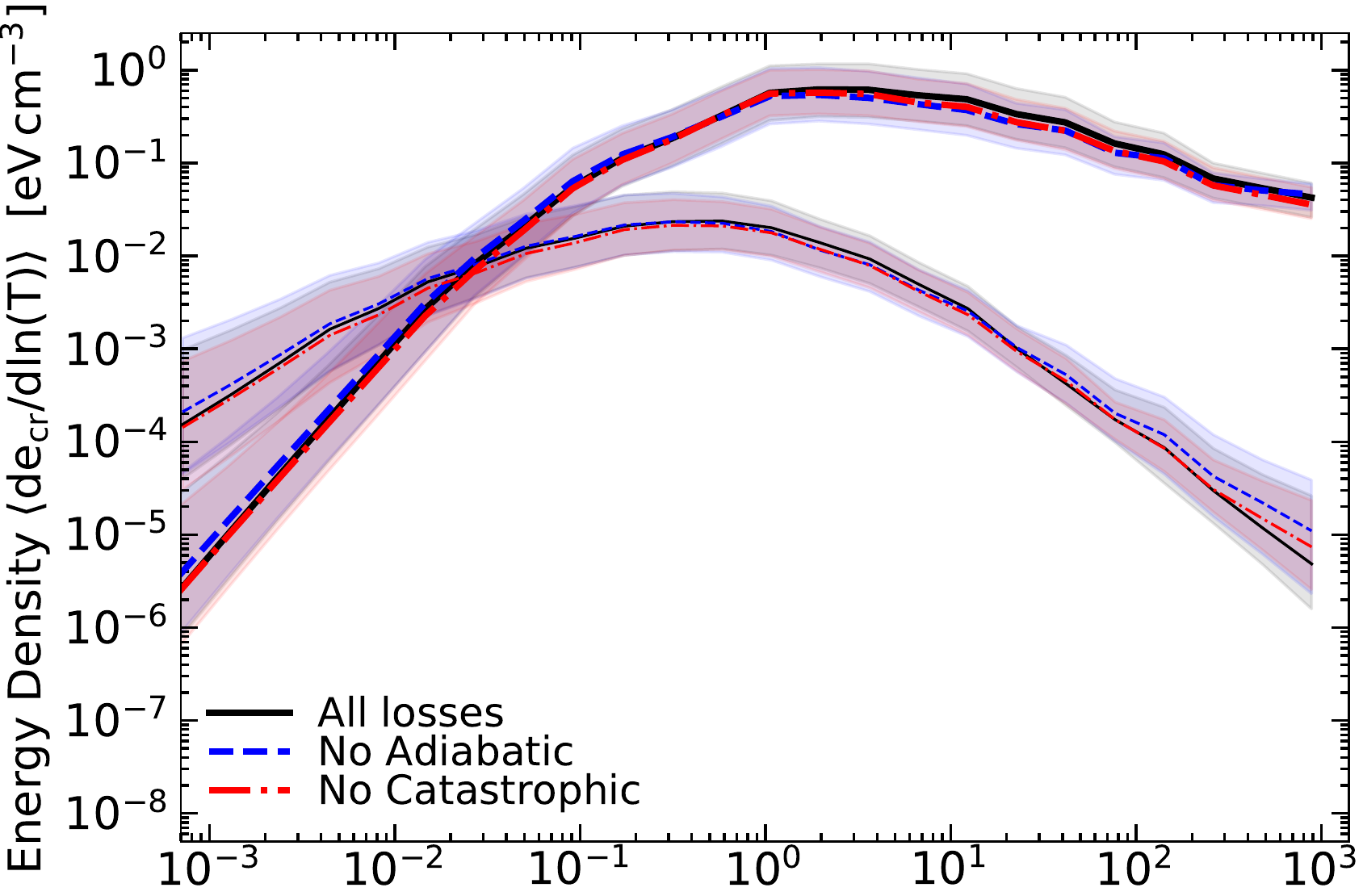}&
	\includegraphics[width=0.32\textwidth]{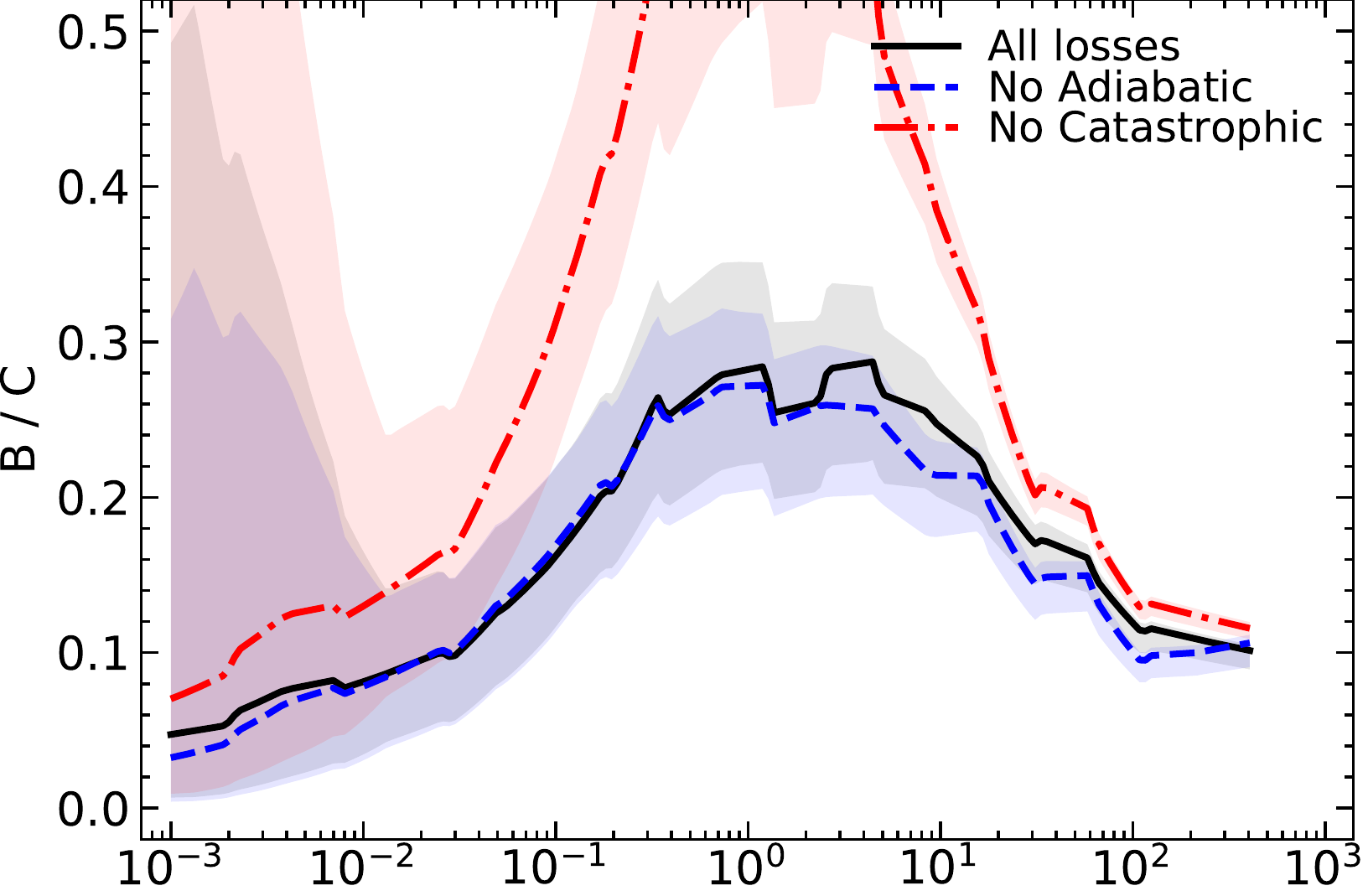}&
	\includegraphics[width=0.33\textwidth]{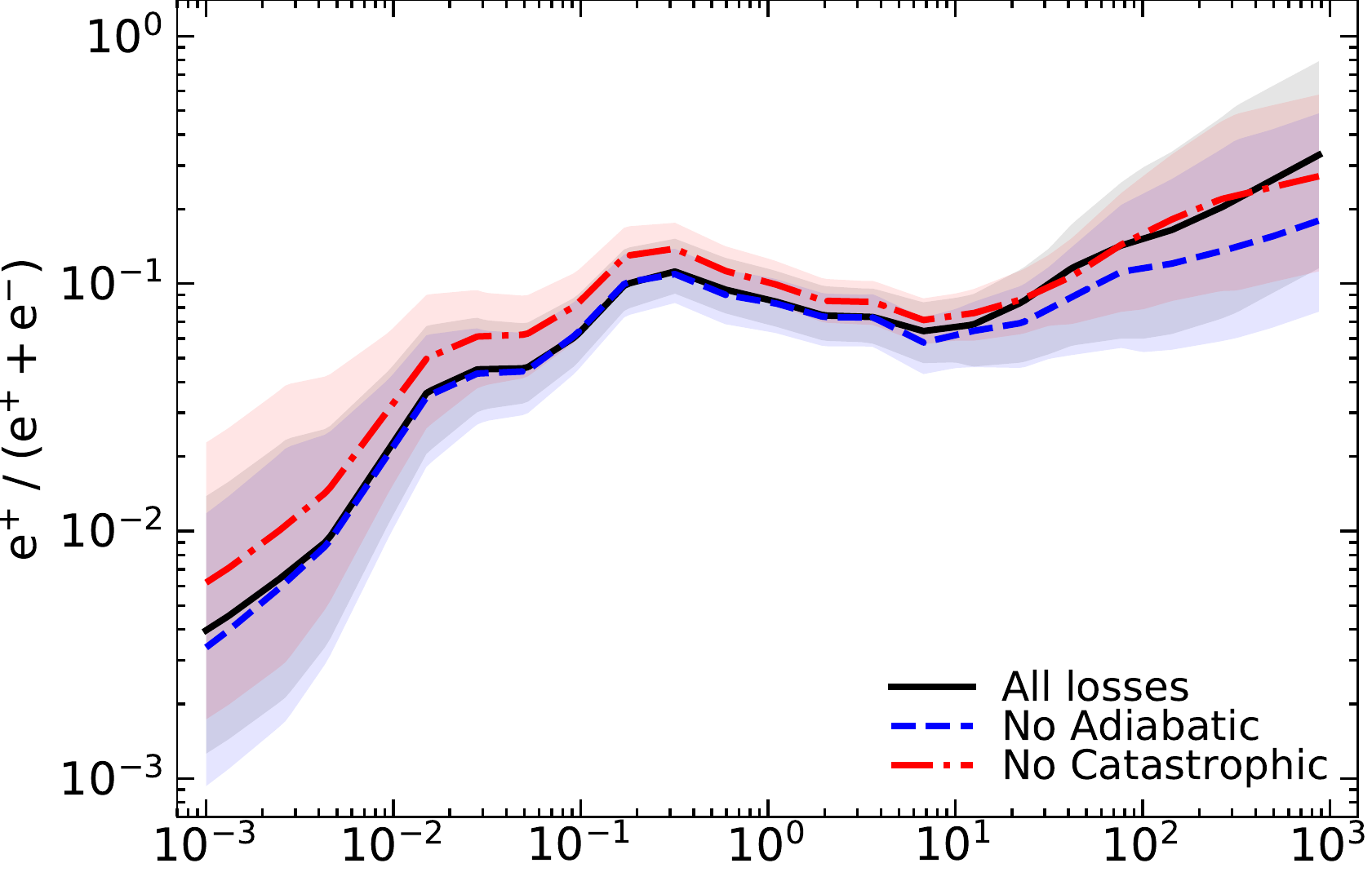}
	\\
	\includegraphics[width=0.33\textwidth]{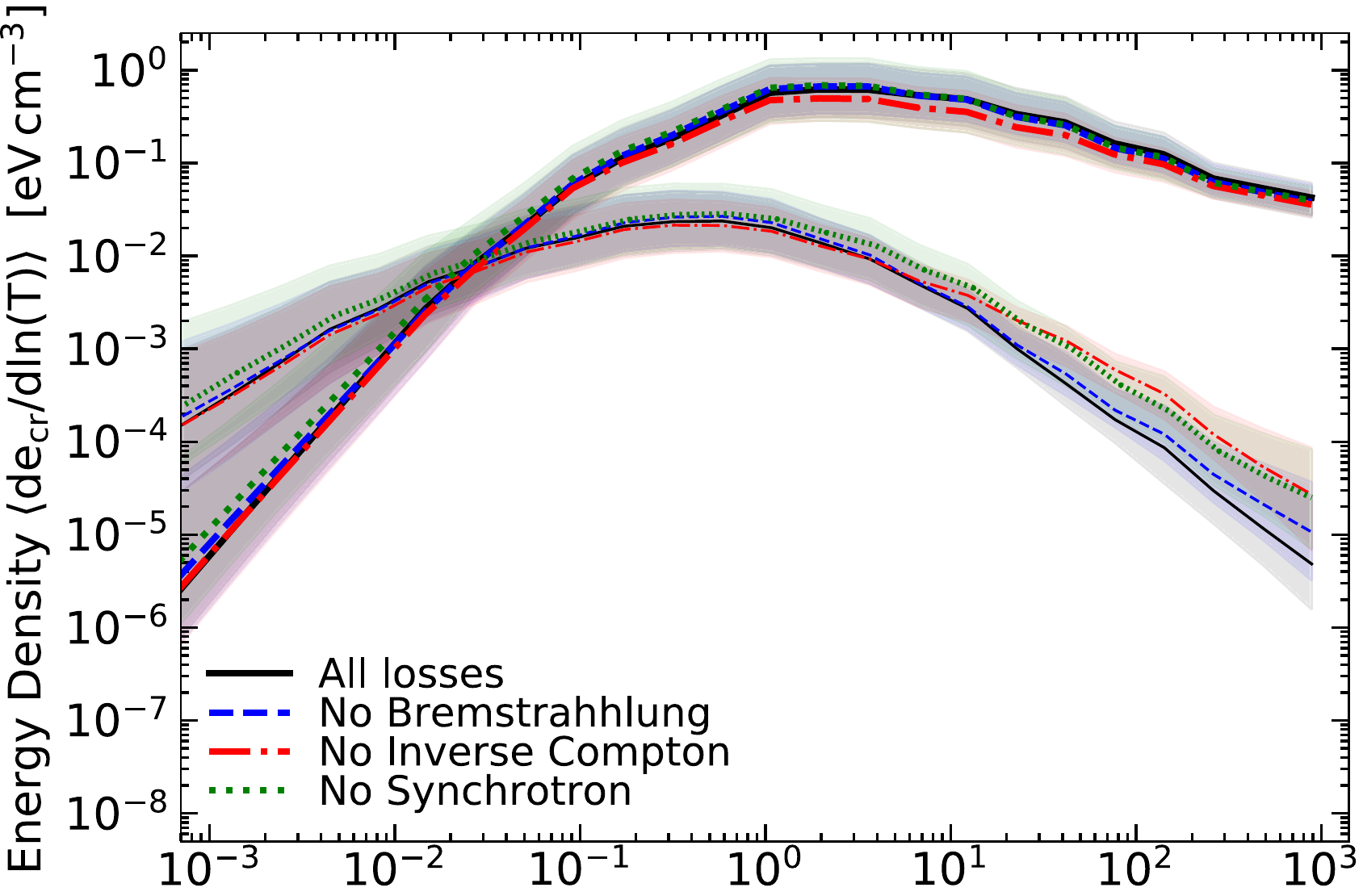}&
	\includegraphics[width=0.32\textwidth]{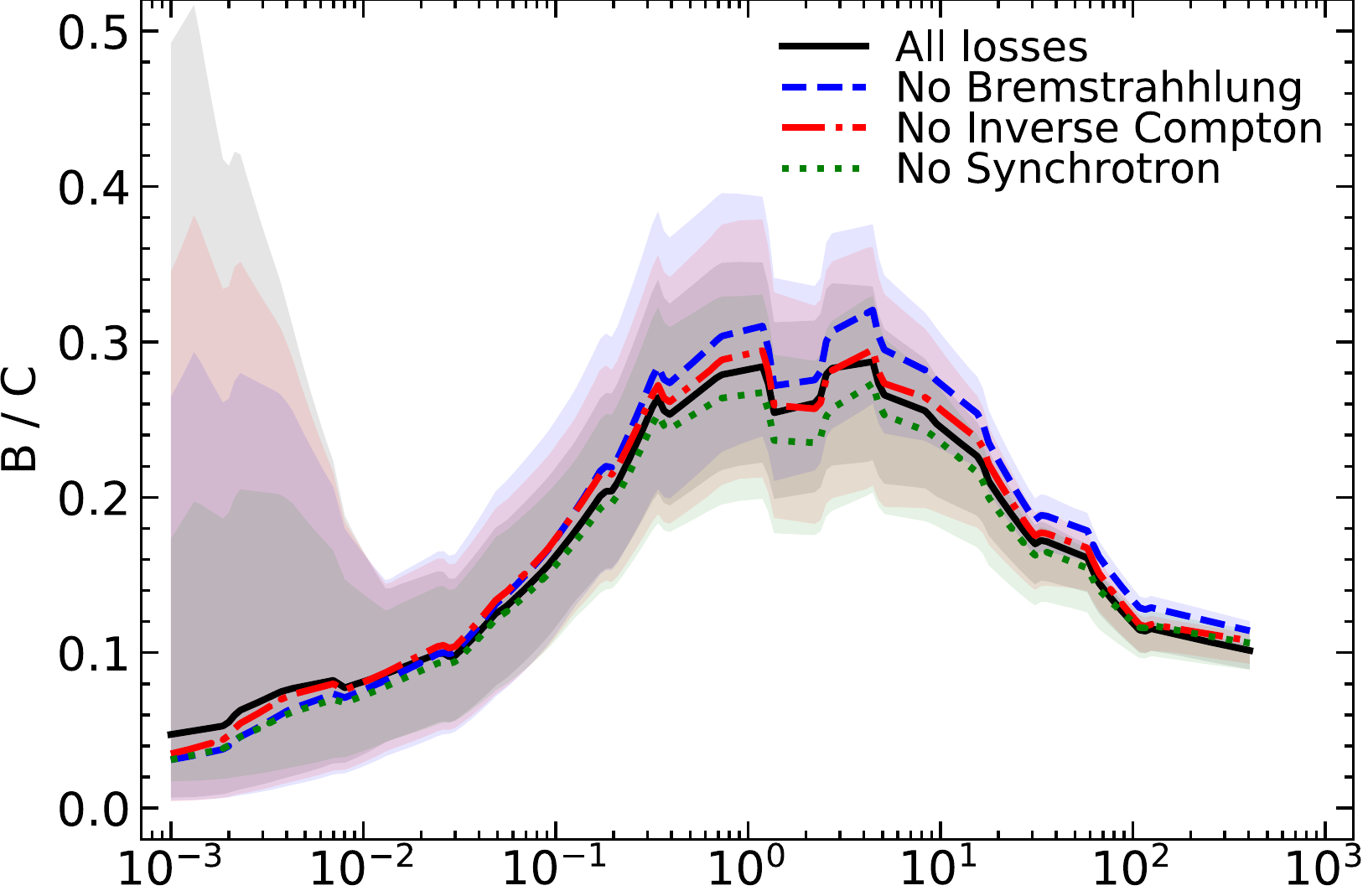}&
	\includegraphics[width=0.33\textwidth]{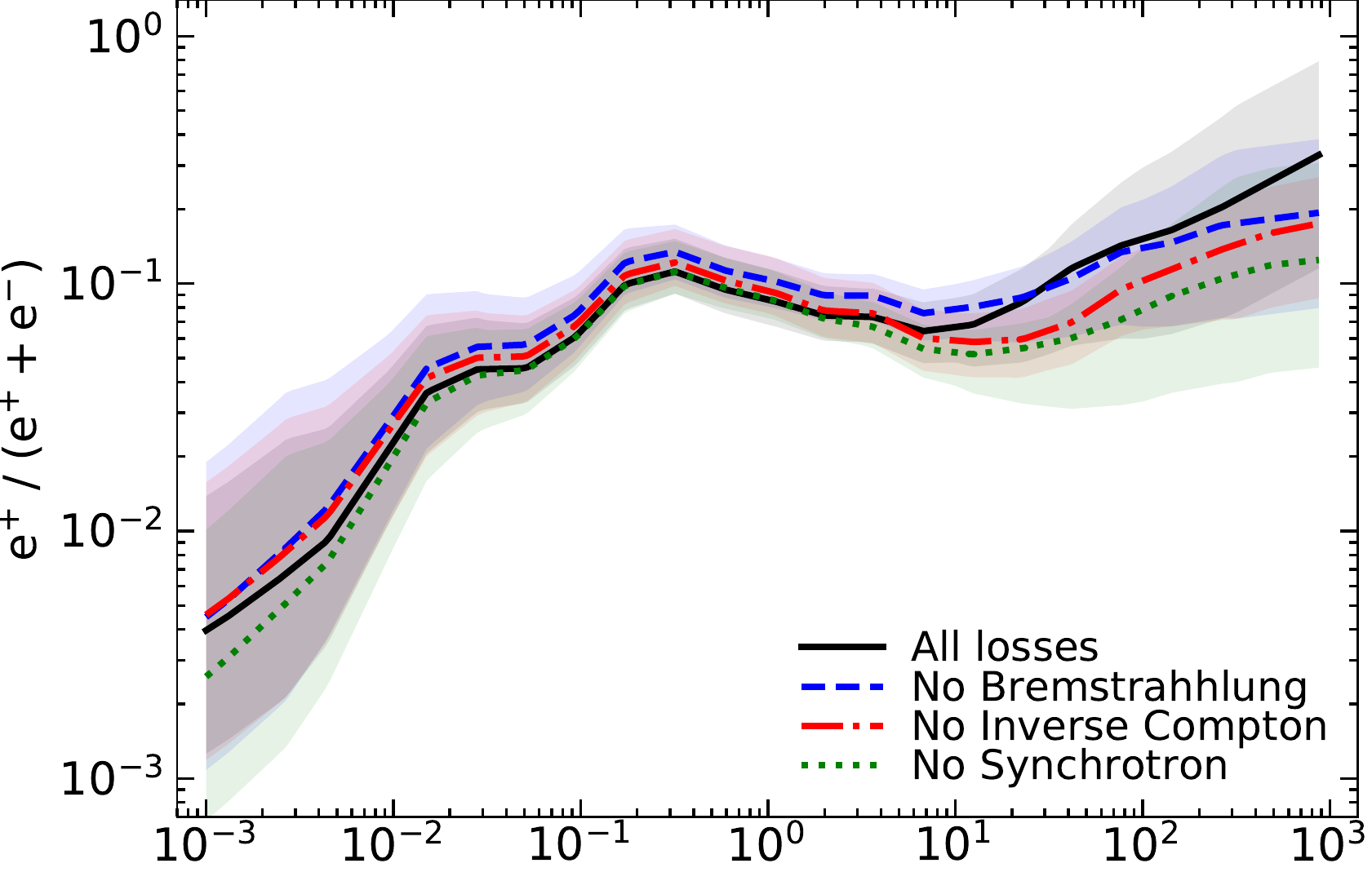}
	\\
	\includegraphics[width=0.33\textwidth]{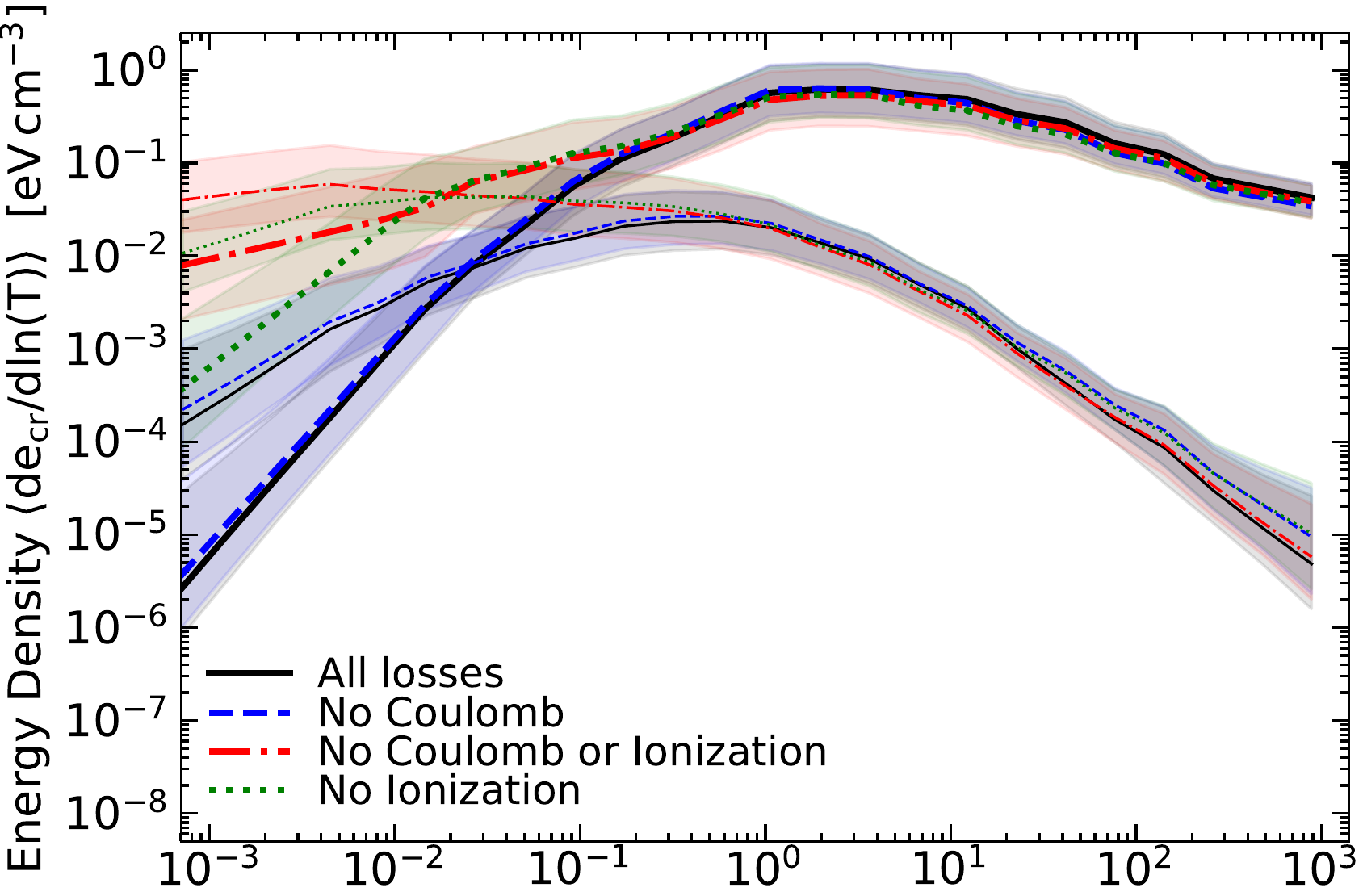}&
	\includegraphics[width=0.32\textwidth]{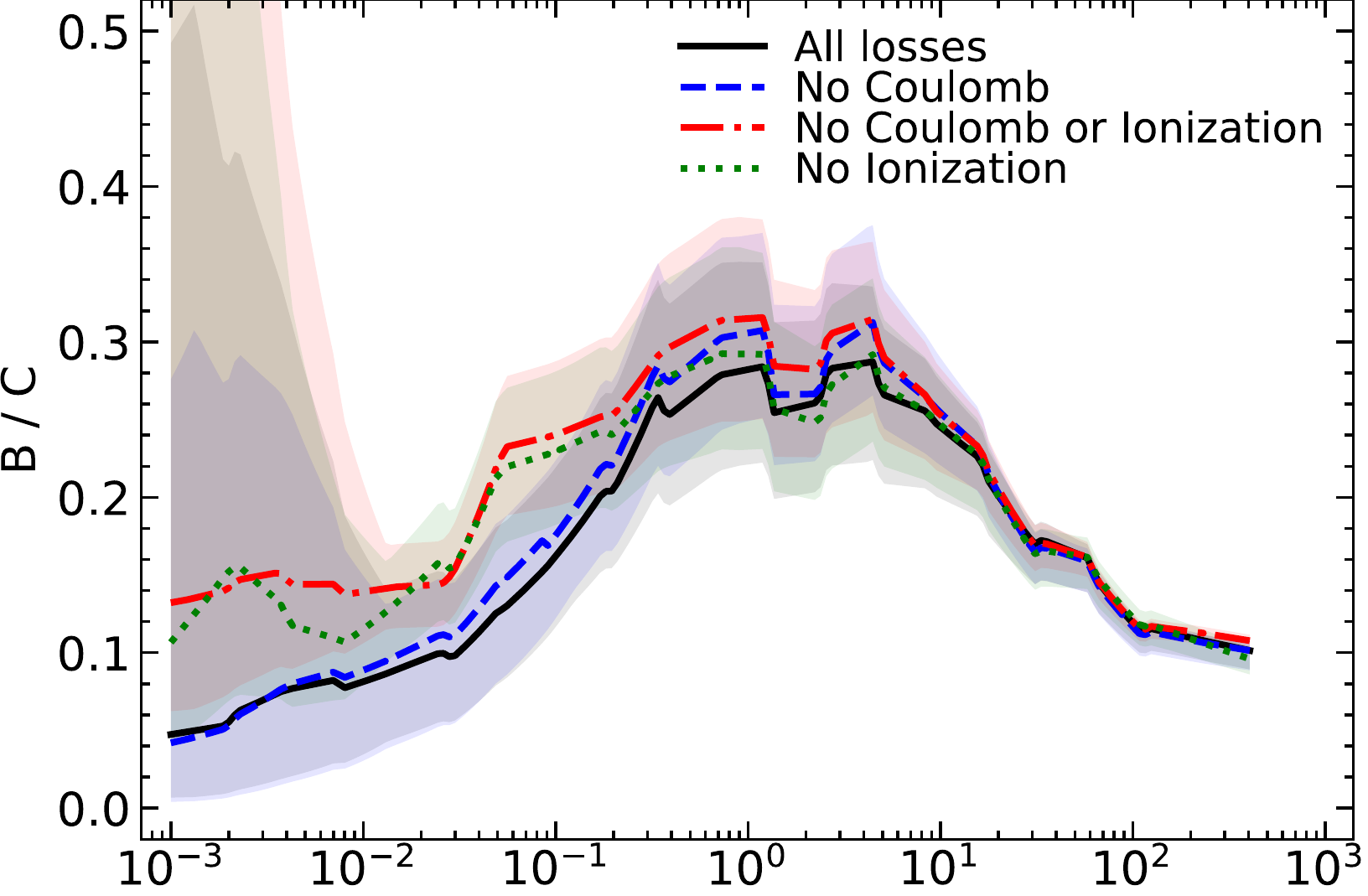}& 
	\includegraphics[width=0.33\textwidth]{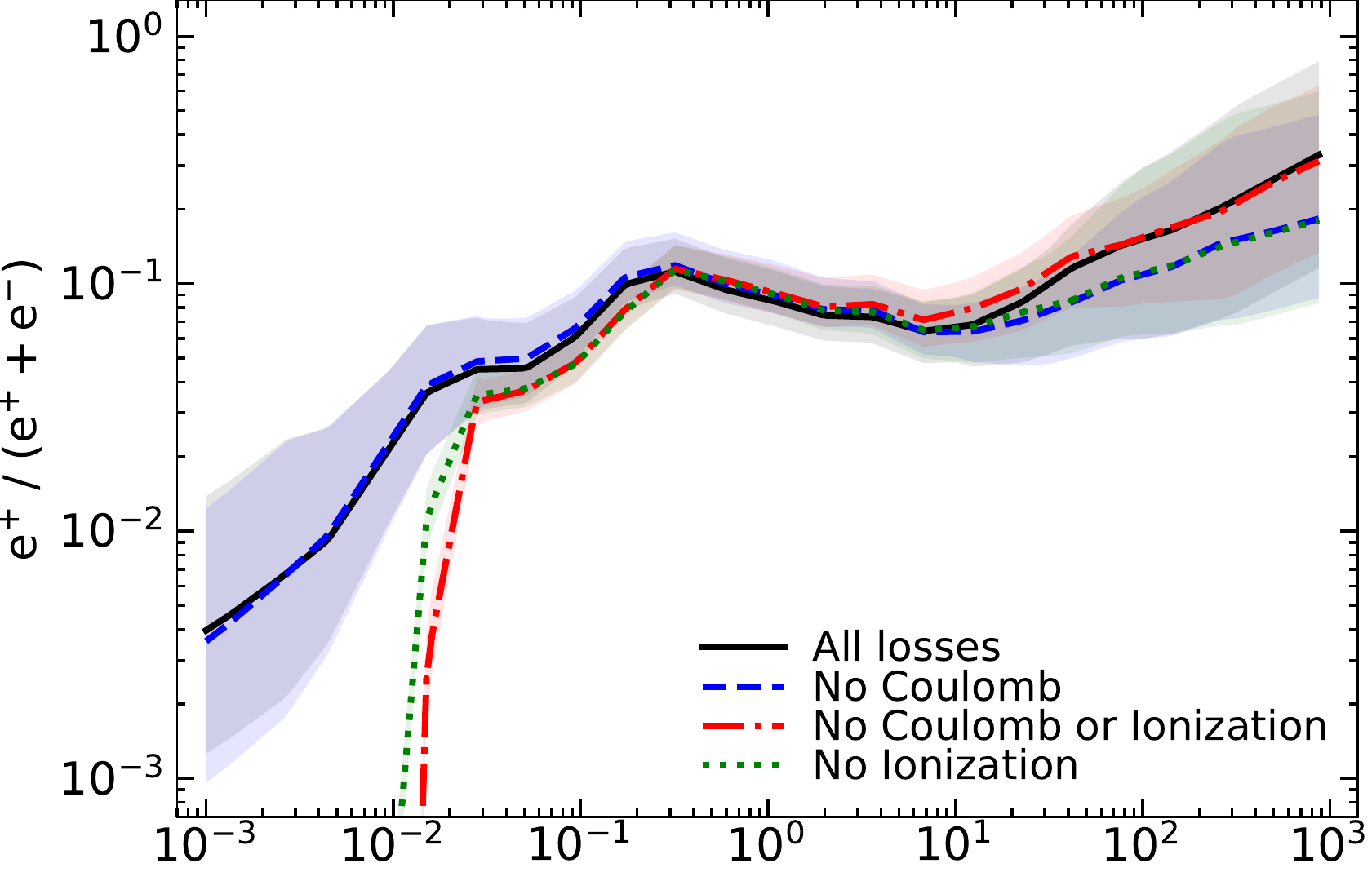} 
	\\
	\includegraphics[width=0.33\textwidth]{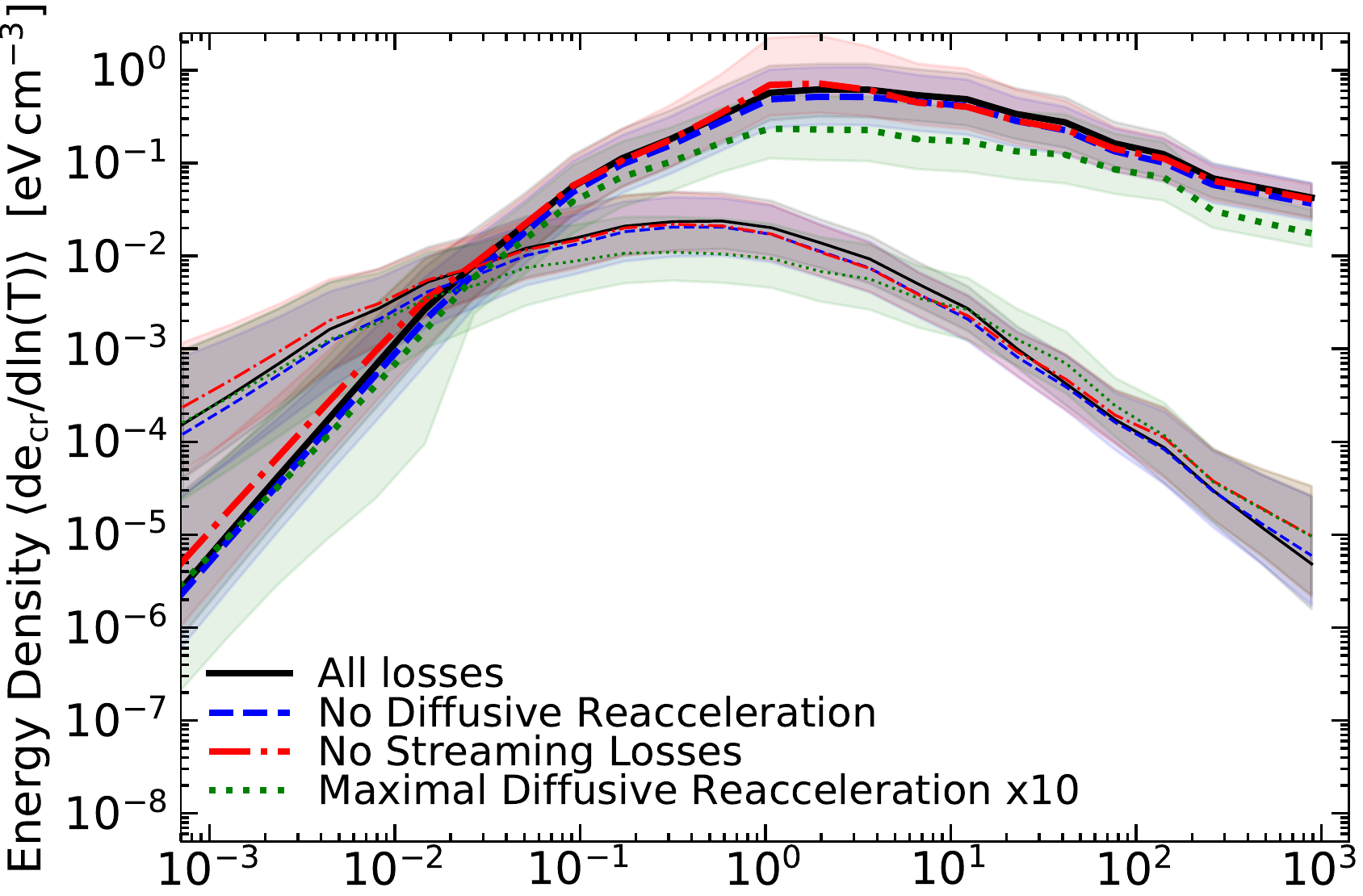}&
	\includegraphics[width=0.32\textwidth]{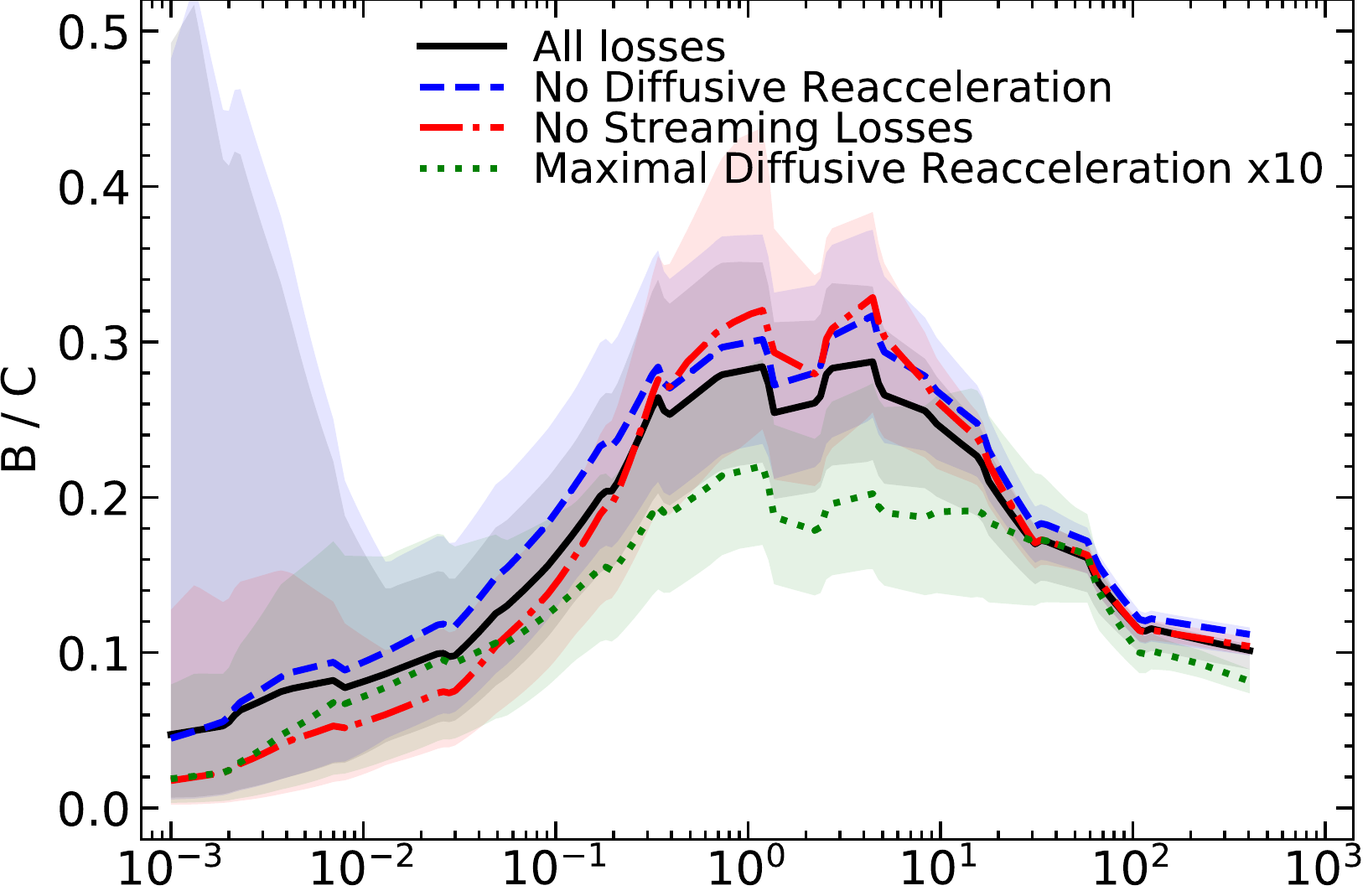}&	
	\includegraphics[width=0.33\textwidth]{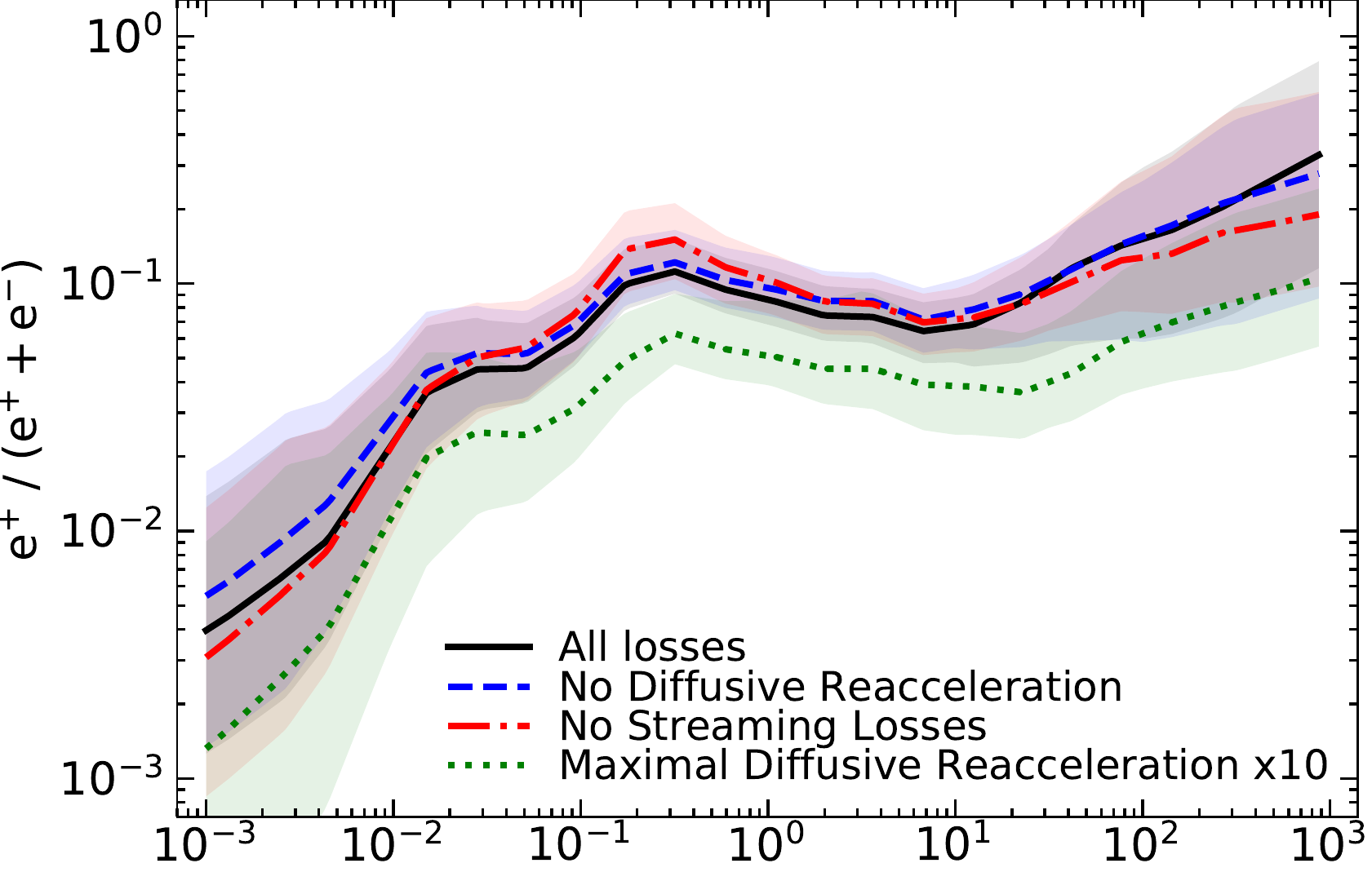}
	\\
	\includegraphics[width=0.33\textwidth]{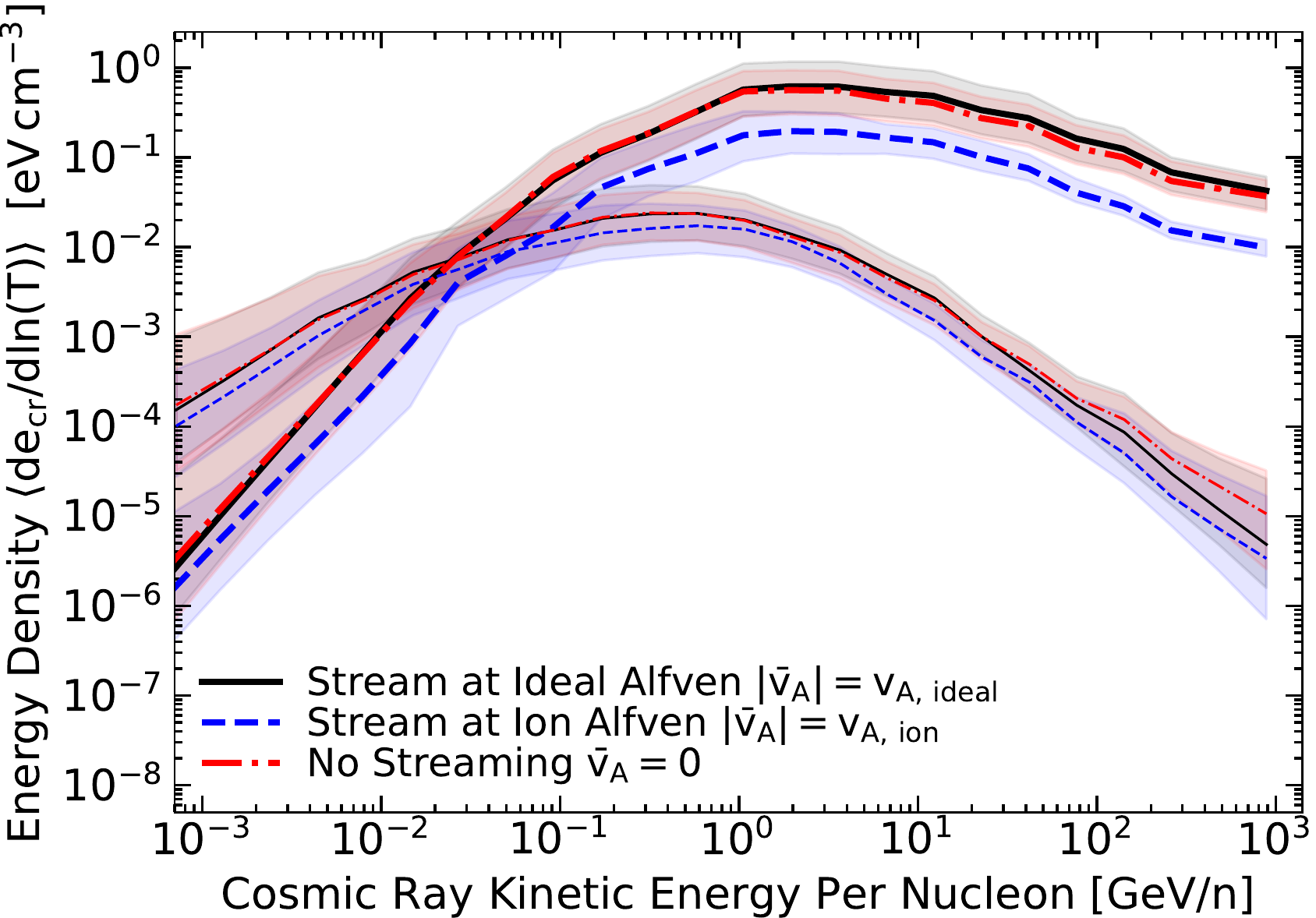}&
	\includegraphics[width=0.32\textwidth]{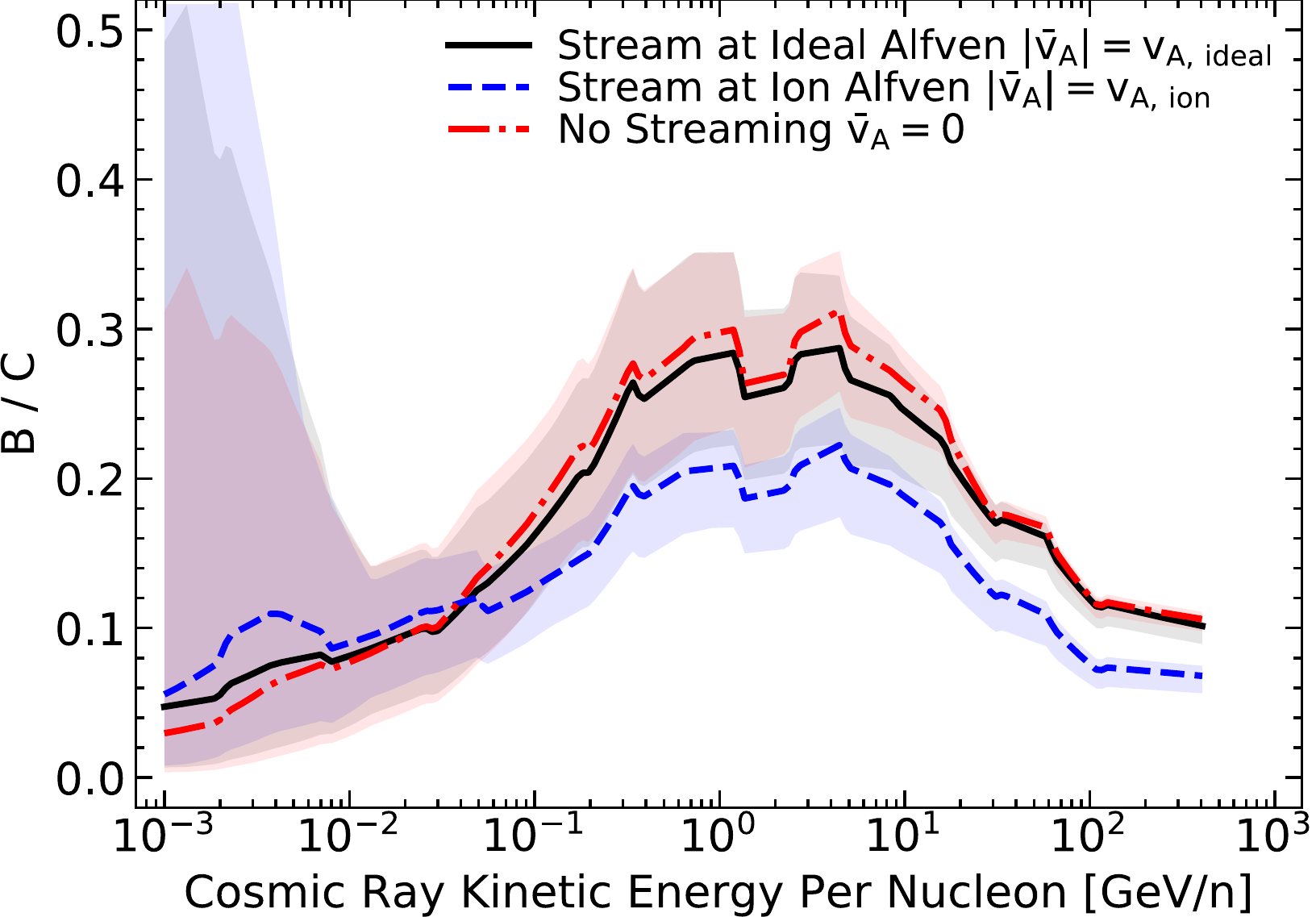}&
	\includegraphics[width=0.33\textwidth]{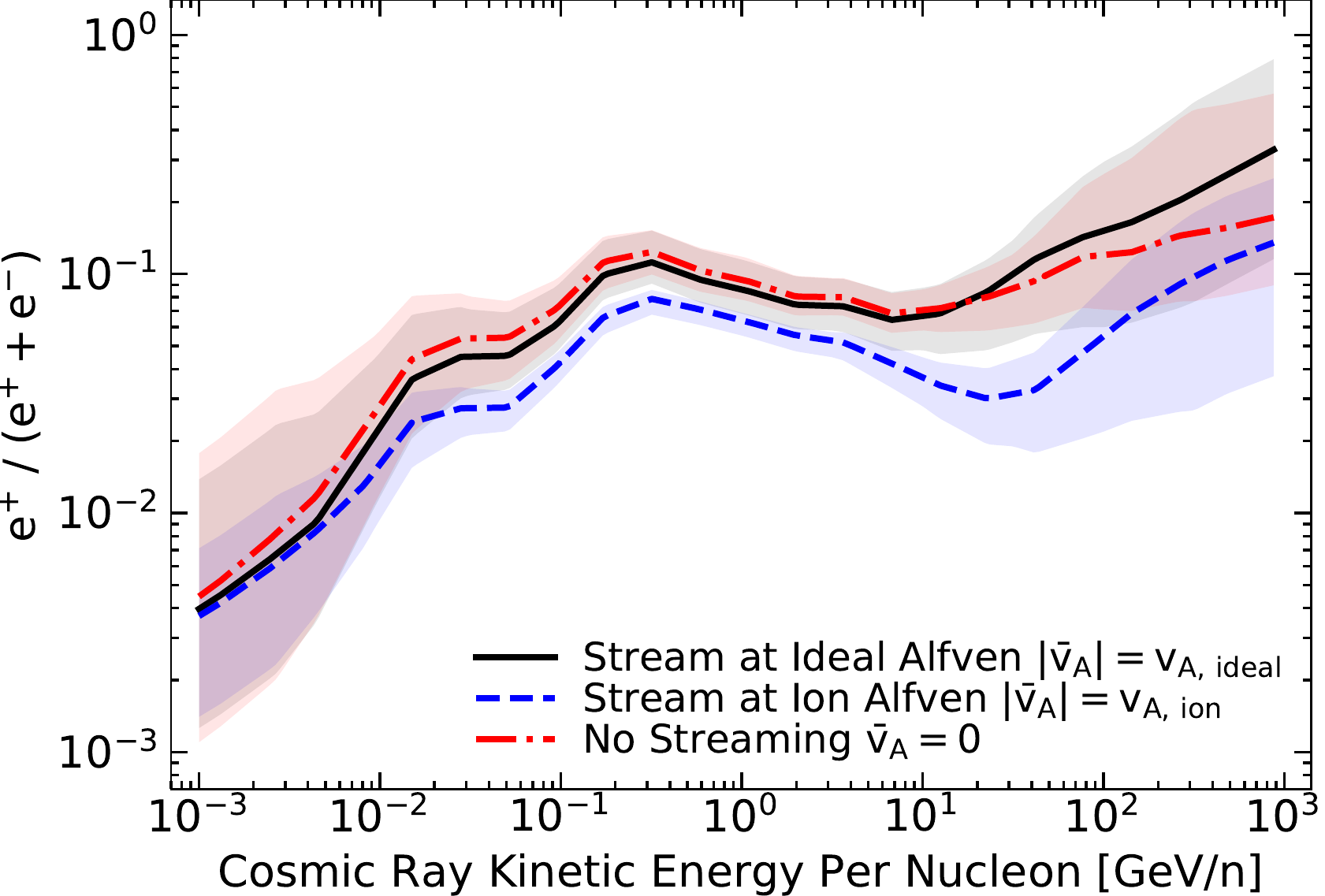}
\end{tabular}
	\vspace{-0.3cm}
	\caption{CR spectra (as Fig.~\ref{fig:spec.compare.kappa}), removing different loss processes to see their effects. 
	We simplify by focusing on just the spectra, B/C, and $e^{+}/e^{-}$, which summarize the key effects. 
	We consider removing, each in turn: 
	{\em First Row:} 
	(1) the ``adiabatic'' $\mathbb{D}:\nabla{\bf u}$ gain/loss term in Eq.~\ref{eqn:f0}; or 
	(2) catastrophic losses (still allowing secondary production, but no primaries are ``destroyed'' in the process; 
	{\em Second:} 
	(3) Bremstrahhlung; (4) inverse Compton; (5) synchrotron; 
	{\em Third:}
	(6) Coulomb; (7) ionization; (8) Coulomb and ionization; 
	{\em Fourth:} 
	(9) the ``diffusive reacceleration'' $\tilde{D}_{p p}$ term in Eq.~\ref{eqn:f0}; 
	(10) the ``streaming loss'' $\tilde{D}_{p\mu}$ term in Eq.~\ref{eqn:f0}. 
	We also consider (11) arbitrarily increasing the diffusive re-acceleration term to a value much larger than physical. 
	{\em Fifth:} Altering the ``streaming speed'' $\bar{v}_{A}$ to (a) $\bar{v}_{A} = v_{A,\,{\rm ideal}} = (|{\bf B}|^{2}/4\pi\rho)^{1/2}$ (our default, the ideal-MHD \Alf\ speed), (b) $\bar{v}_{A} = v_{A,\,{\rm ion}} = (|{\bf B}|^{2}/4\pi\rho_{\rm ion})^{1/2}$ (the ``ion \Alf\ speed,'' much faster in mostly-neutral gas, and favored in self-confinement models), and (c) $\bar{v}_{A}=0$ (assumed in older extrinsic turbulence models). All these changes are discussed in \S~\ref{sec:physics}. Generically inverse Compton+synchrotron alter high-energy lepton spectra, Coulomb+ionization alter low-energy spectra, and the effect of the ``re-acceleration'' terms are modest.
	\label{fig:spec.compare.losses}\vspace{-0.4cm}}
\end{figure*}

\subsection{Effects of Different Physics \&\ Parameters}
\label{sec:physics}

\begin{figure*}
\begin{tabular}{r@{\hspace{0pt}}r@{\hspace{0pt}}r}
	\includegraphics[width=0.33\textwidth]{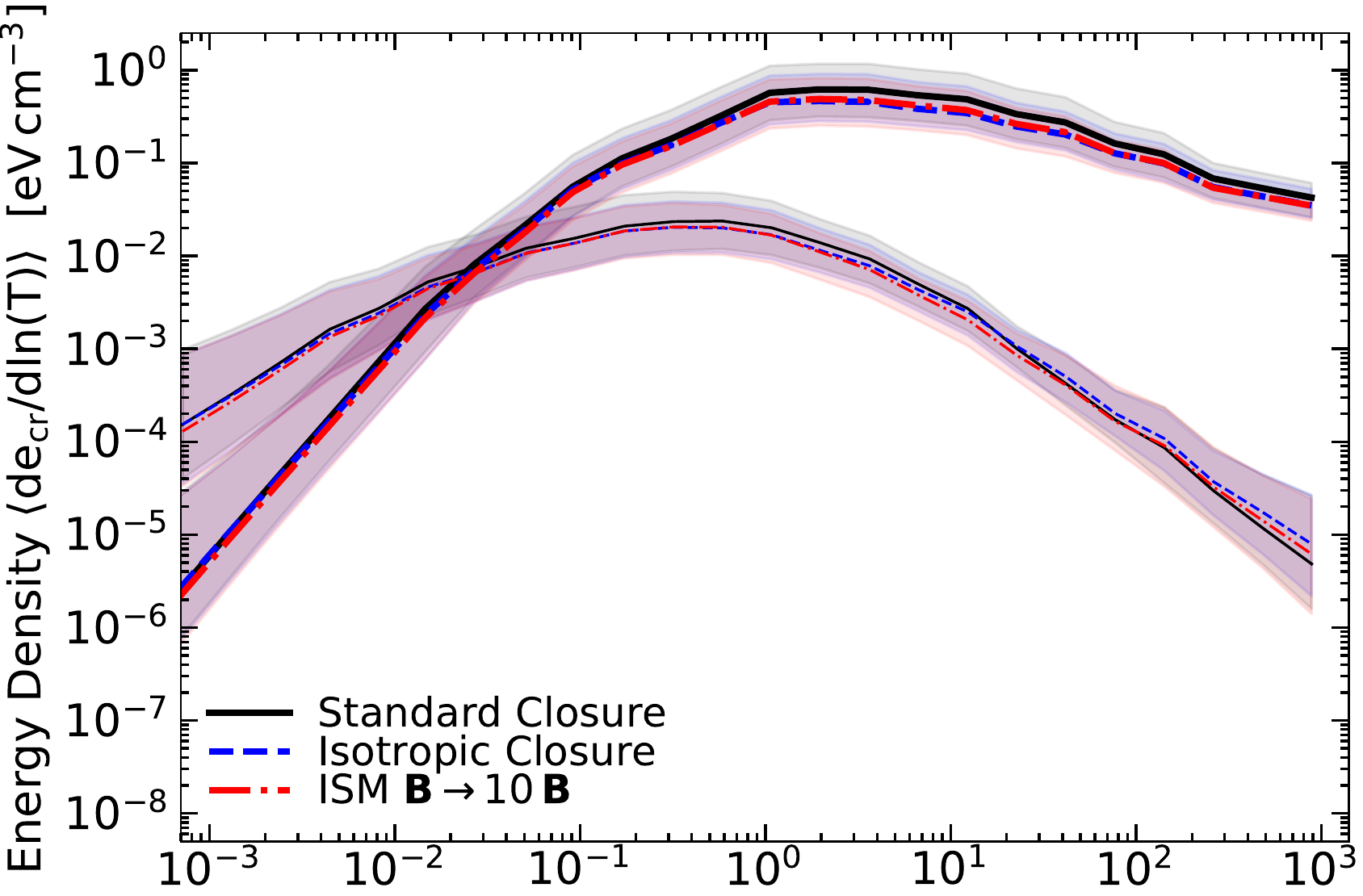}&
	\includegraphics[width=0.32\textwidth]{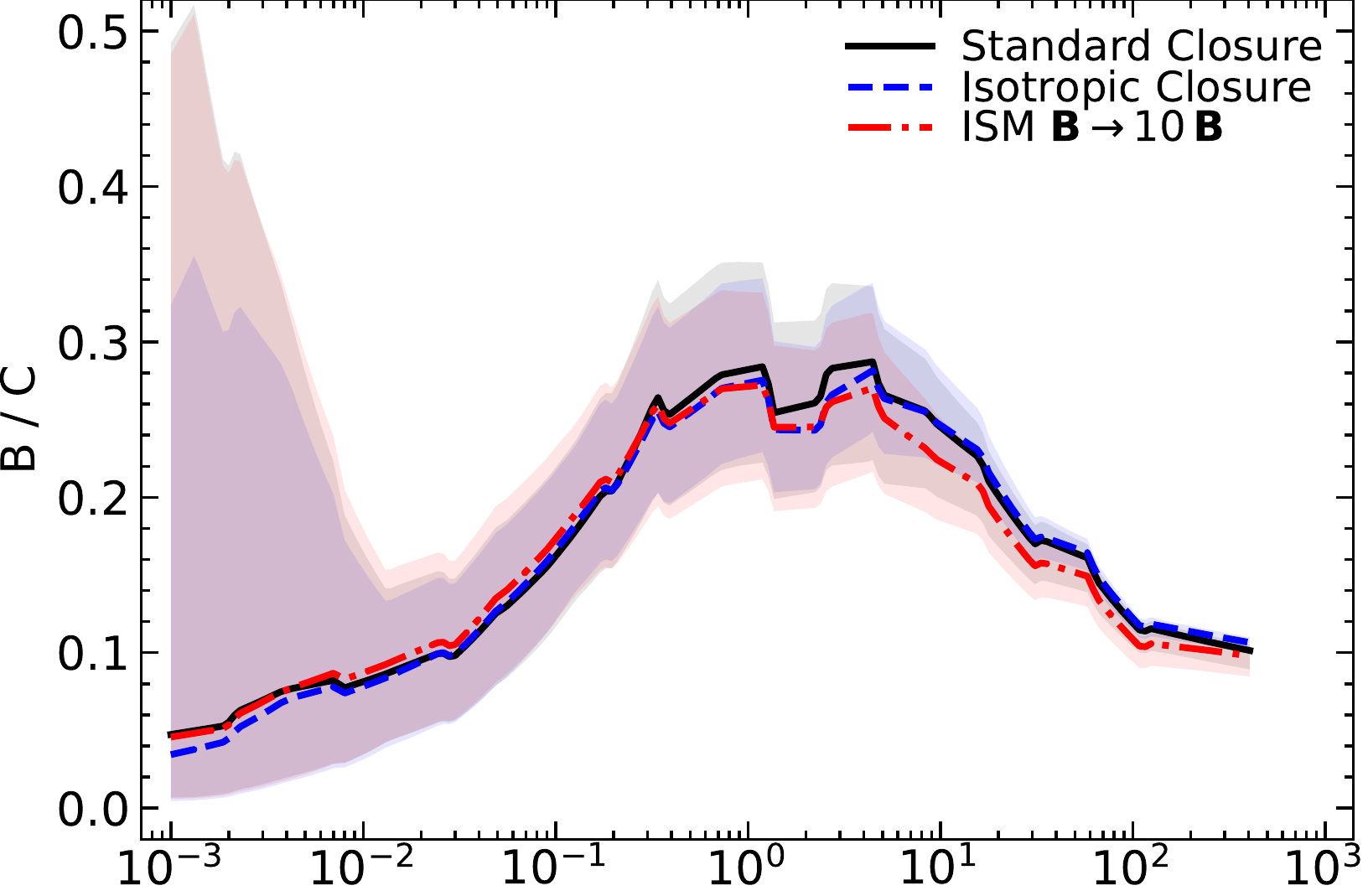}&
	\includegraphics[width=0.33\textwidth]{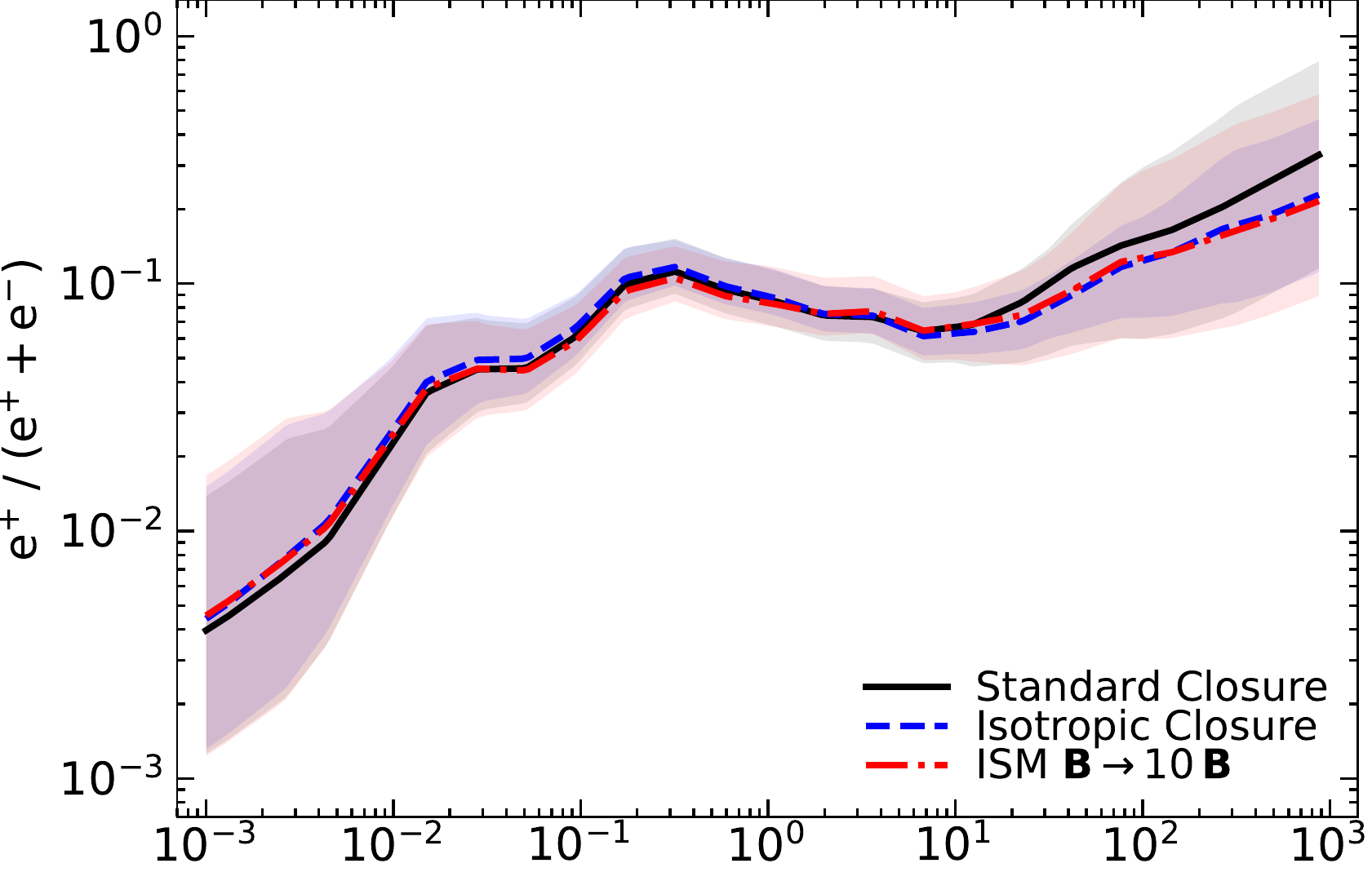}
	\\
	\includegraphics[width=0.33\textwidth]{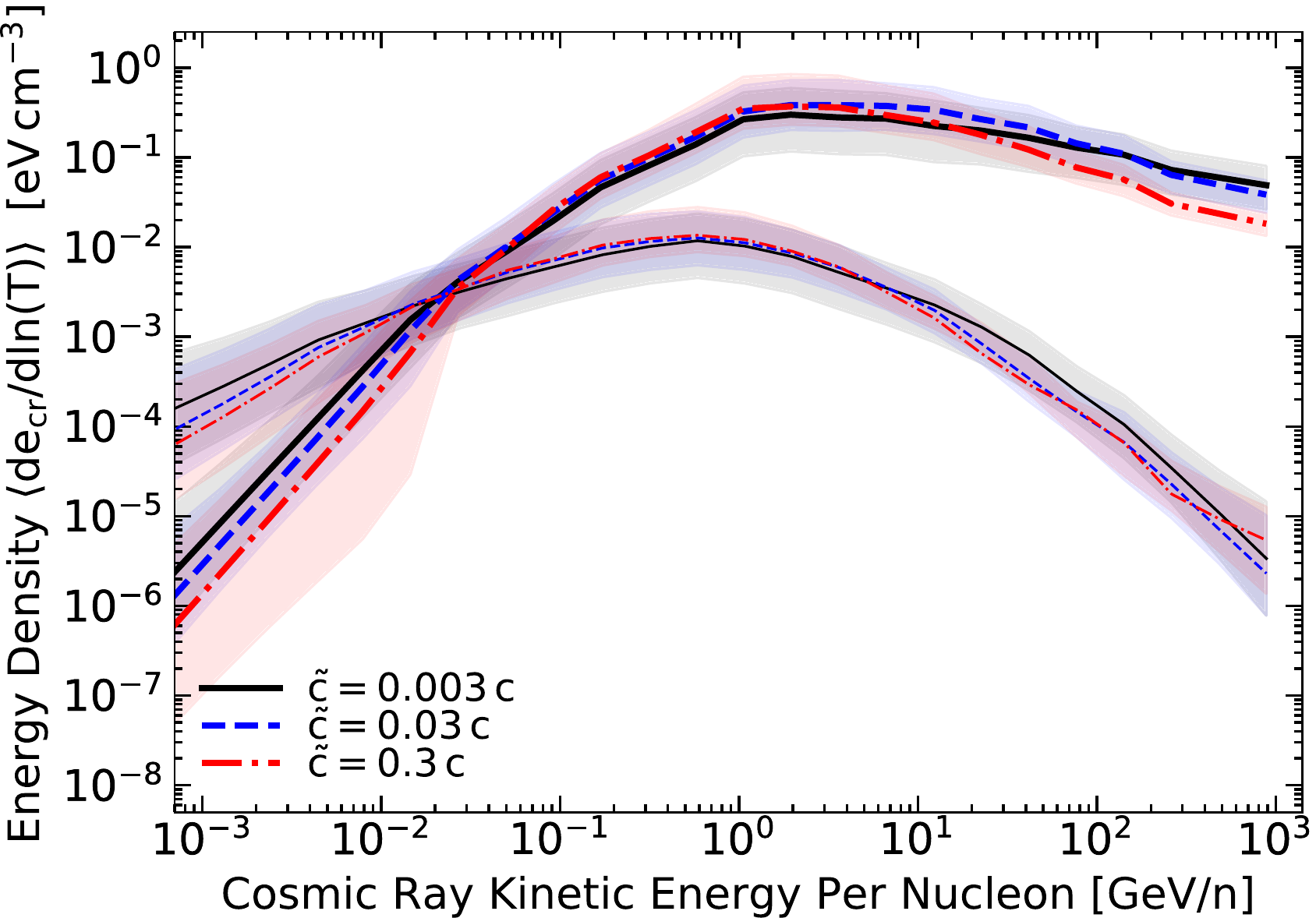}&
	\includegraphics[width=0.32\textwidth]{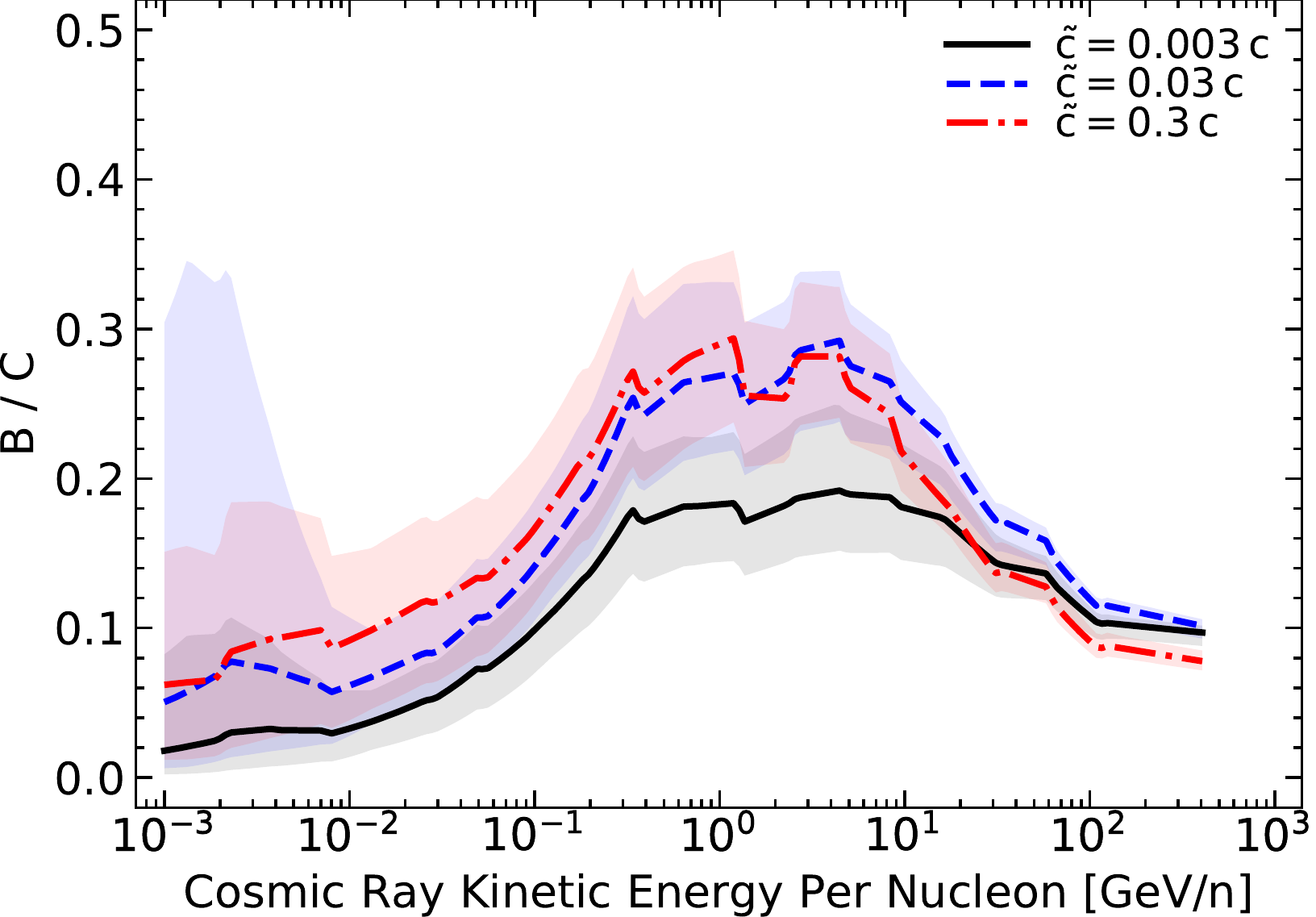}& 
	\includegraphics[width=0.33\textwidth]{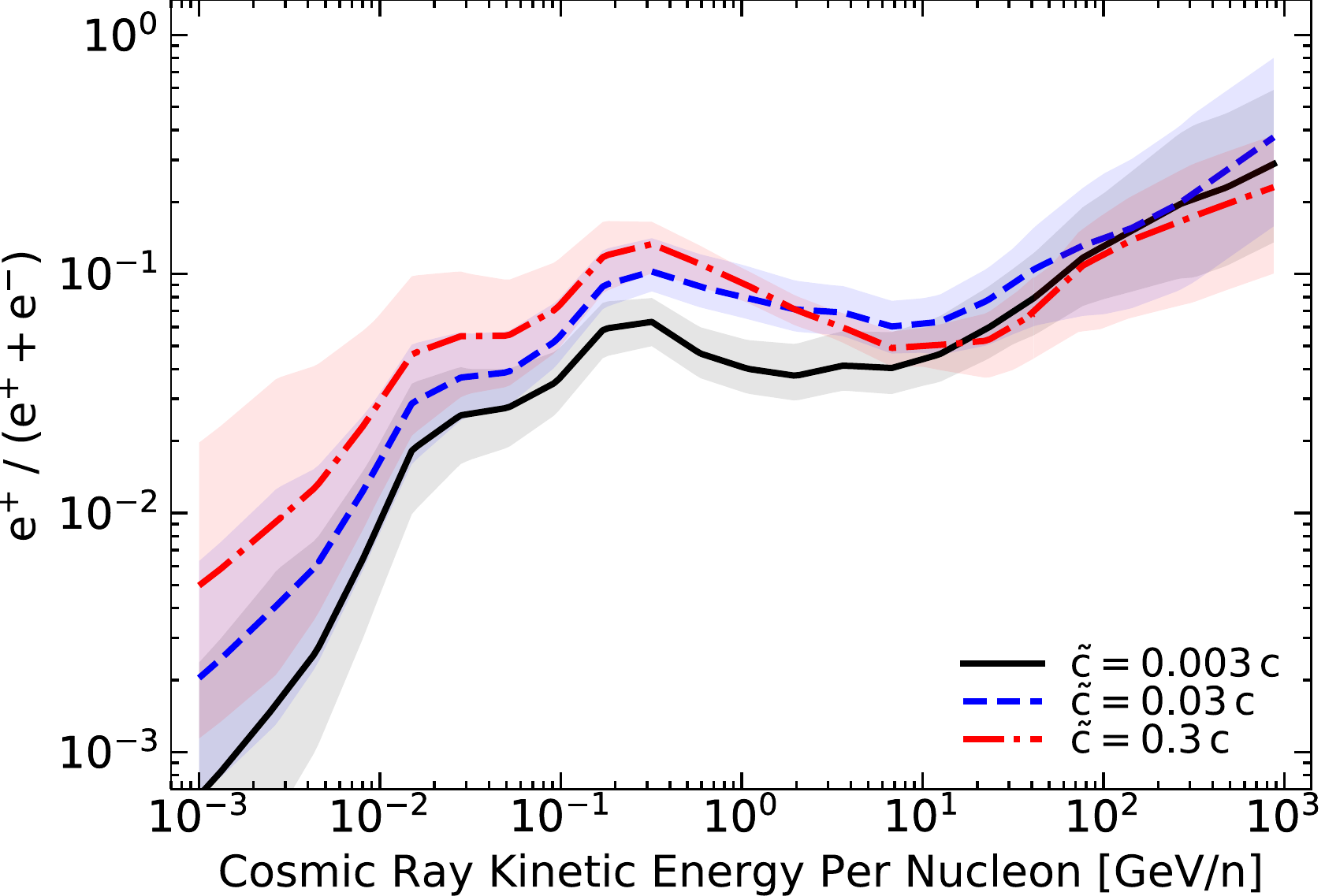} 
\end{tabular}
	\vspace{-0.3cm}
	\caption{As Fig.~\ref{fig:spec.compare.losses}, considering purely-numerical parameters. 
	{\em Top:} We compare our default closure model for the CR pressure tensor $\mathbb{D}(\chi[\mu])$ (Eq.~\ref{eqn:f1}) to a simpler model assuming an always-near-isotropic CR distribution ($\chi=1/3$, $\mathbb{D}\rightarrow \mathbb{I}/3$), or to a model where we arbitrarily multiply the magnetic fields everywhere in our initial conditions (a cosmological simulation snapshot at $z\sim 0.05$) by $10$. This multiplies magnetic energy by $100$, much larger than allowed by observations (see \S~\ref{sec:bfields} and \citealt{su:fire.feedback.alters.magnetic.amplification.morphology,guszejnov:fire.gmc.props.vs.z}), but illustrative of how weak the effects of uncertainty in the true Galactic magnetic field are. 
	{\em Bottom:} Varying the reduced speed of light $\tilde{c}$ in Eqs.~\ref{eqn:f0}-\ref{eqn:f1} to allow larger timesteps from $0.003-0.3\,c$. At the lowest $\tilde{c}$ we see some artifacts: as shown in \citealt{hopkins:m1.cr.closure} reducing $\tilde{c}$ means it takes longer by a factor $\sim c/\tilde{c}$ to converge to the correct steady-state secondary abundances, so at for very-low $\tilde{c}$ we simply have not run our simulation for sufficient time to reach convergence. But for $\tilde{c}>0.01\,c$, the convergence is good.
	\label{fig:spec.compare.numerics}\vspace{-0.4cm}}
\end{figure*}

We now briefly discuss the qualitative effects of different variations on CR spectra, using tests where we fix all parameters and physics but then ``turn off'' different physics or adjust different parameters each in turn, with resulting spectra shown in Figs.~\ref{fig:spec.compare.kappa}, \ref{fig:spec.compare.losses}, \&\ \ref{fig:spec.compare.numerics}. Here, our ``reference'' model is that in Fig.~\ref{fig:demo.cr.spectra.fiducial}. We have considered a set of simulations varying other parameters simultaneously, and in Appendix~\ref{sec:appendix:additional}, we repeat the exercise in Figs.~\ref{fig:spec.compare.kappa}, \ref{fig:spec.compare.losses}, \&\ \ref{fig:spec.compare.numerics}, but for variations with respect to a different reference model with larger scattering rate and different dependence of scattering on rigidity. This allows us to confirm that all of our qualitative conclusions here are robust. 

It is useful to define some reference scalings, by reference to a toy leaky-box type model: if the CR injection rate in some $p$ interval were $dj = j_{0}\,(p/p_{0})^{-\psi_{\rm inj}}\,d^{3}{\bf p}$, and the CR ``residence time'' (or escape time) were $\Delta t_{\rm res} = \Delta t_{0}\,(p/p_{0})^{-\psi_{\rm res}}$, then the observed number density would scale as $ d N_{\rm obs} = \Delta t\,dj = \Delta t_{0}\,j_{0}\,(p/p_{0})^{-\psi^{\rm N}_{\rm obs}}\,d^{3}{\bf p}$. For the more usual units of intensity we have $dI \propto d N/dt\,dA\,dT \propto p^{-\psi_{\rm obs}}$ with $\psi_{\rm obs} =  \psi_{\rm inj} + \psi_{\rm res} - 2$. Again, the explicit loss or escape timescales calculated in LISM conditions in our reference simulation from Fig.~\ref{fig:demo.cr.spectra.fiducial}, for each of the processes discussed below, are presented in Appendix~\ref{sec:vdrift.tloss}, to which we refer for additional details.

\begin{itemize}

\item{Injection Spectra:} As expected, the CR spectral shapes scales with the injection spectrum, shown in Fig.~\ref{fig:spec.compare.kappa}. However, the scaling is not perfectly linear as the above toy model would imply: changing the injection $\psi_{\rm inj}$ by some $\Delta \psi_{\rm inj}$, we obtain $\Delta \psi_{\rm obs} \sim (0.7-0.9)\,\Delta \psi_{\rm inj}$, depending on the CR energy range, species, etc. The issue is that part of the change in assumed slope is offset by (a) losses, (b) non-linear effects of CRs on the medium, and (c) non-uniform source distributions (where e.g.\ the effective ``volume'' of sources in a realistic disk sampled by a given $\Delta t_{\rm res}$ is not $p$-independent, so one needs to convolve over the source distribution at each $p$). Shallower slopes (smaller $|\psi_{\rm inj}|$) produce a B/C ratio which is shallower (drops off more slowly) at low energies. More dramatically, in e.g. $e^{+}/e^{-}$, because $e^{-}$ and $p$ are injected with the same slope and the $e^{+}$ secondaries have energy $\sim 0.1$ times their $p$ progenitors, a steeper $\psi_{\rm inj}$ gives a lower value of $e^{+}/e^{-}$ at a given $R$ or $E$, and a sharper ``kink'' in the distribution, while shallower $\psi_{\rm inj}$ gives a higher $e^{+}/e^{-}$ (rising more continuously to low-$E$). The CR kinetic energy density (normalization of the spectra) is slightly sub-linear in the injected CR fraction $\epsilon_{\rm cr}$, as lower CR pressure allows slightly more rapid gas collapse and star formation, raising the SFR and CR injection rate \citep[see][]{hopkins:cr.mhd.fire2}. The lepton-to-hadron ratio injected translates fairly closely to the $e^{-}/p$ ratio at $\sim 1-10\,$GeV, for realistic diffusivities where losses are not dominant at $\sim 1\,$GeV.

\item{Scattering Coefficients:} Parameterizing the scattering coefficient as: $\bar{\nu} = \bar{\nu}_{0}\,\beta\,R_{\rm GV}^{-\delta}$, recall this corresponds to $D_{x x} \approx \beta\,D_{0}\,R_{\rm GV}^{\delta}$ (with $D_{0} \approx c^{2}/9\,\bar{\nu}_{0}$) in the often-assumed isotropic strong-scattering flux-steady-state negligible-streaming limit. Our preferred model has (in cgs units) $\bar{\nu}_{0} \sim 10^{-9}$ ($D_{0}\approx10^{29}$), $\delta \sim 0.5-0.6$. As shown in Fig.~\ref{fig:spec.compare.kappa}, lowering $\delta$ produces a ``flatter'' (nearly energy-independent) B/C ratio and systematically higher $e^{+}/e^{-}$ and $\bar{p}/p$ ratio at energies $\gtrsim 1\,$GeV, as well as flatter CR spectral slopes $\psi_{\rm obs}$ for high-$E$ hadrons (where the residence time is primarily determined by diffusive escape), as expected. Larger $\delta$ has the opposite effects (as expected), but also large-enough $\delta \gtrsim 1$ (see also Appendix~\ref{sec:appendix:additional}) at low energies increases B/C and makes hadronic and leptonic slopes more shallow, by increasing the effect of losses via slower transport. Even a modestly-lower $\delta \sim 0.3$ is strongly disfavored, given the fact that we cannot ``remove'' the halo here to compensate for the flatter B/C predicted. A much-higher $\delta > 1$ is also clearly ruled out, and these limits are robust even after marginalizing over the assumed injection spectra.

Changing the normalization $\bar{\nu}_{0}$ has the obvious effects of e.g.\ increasing/decreasing the secondary-to-primary ratio and normalization\footnote{Briefly, at lower $\bar{\nu}_{0}$ in steady-state with all else equal, the CR energy density should increase $\propto \bar{\nu}_{0}^{-1}$. But as we decrease $\bar{\nu}_{0}$ (1) the size of the CR scattering halo also decreases (making this dependence weaker) and (2) losses become important even for $\sim$\,GeV protons, so the CR energy density cannot continue to increase.} of the spectra, but more interestingly also has a strong effect on the {\em shape} of the CR spectra (and scaling of secondary-to-primary ratios with $p$), where larger $\bar{\nu}_{0}$ (lower diffusivity) produces shallower slopes for hadrons. This arises from the non-linear competition between the various loss terms (which become stronger at lower-$\bar{\nu}_{0}$) and escape, and is generally a larger effect for hadrons (where the loss timescales are systematically shorter at lower-$E$) as compared to leptons (except for the very lowest $\bar{\nu}_{0}$ considered, e.g.\ $D_{0} \lesssim 10^{27}\,{\rm cm^{2}\,s^{-1}}$).\footnote{It is worth commenting on the behavior of $^{10}$Be/$^{9}$Be with varying $\nu$ in Fig.~\ref{fig:spec.compare.kappa} (and Appendix~\ref{sec:appendix:additional}). Naively we would expect that, all else equal, $^{10}$Be/$^{9}$Be should decrease with increasing CR ``residence time'' between secondary production and arrival at the Solar system, hence be lower for higher $\bar{\nu}$. And at low CR energies, we often see behavior consistent with this (but the effects are weak and somewhat non-linear, owing to the non-zero effects of streaming and losses controlling the residence time, instead of diffusive escape). At high-energies, however, we clearly see $^{10}$Be/$^{9}$Be increase with  $\bar{\nu}$ (either from increasing $\bar{\nu}_{0}$, or increasing $\delta$ at $T\gg$\,GeV). While some of this owes to lower-$\bar{\nu}$ runs sampling an effectively smaller CR scattering halo and source region, most of the effect owes to the fact that the runs with larger $\bar{\nu}$ {\em also} produce much higher B/C at these energies. At $\sim100\,$GeV, for B/C\,$\gtrsim0.3$ (much higher than observed, but predicted in these models if we artificially increase $\bar{\nu}$), B actually dominates over C in producing $^{10}$Be \citep{moskalenko:2003.galprop.cx}, with a significantly higher ratio of $^{10}$Be to $^{9}$Be production factors. So what we see is effectively that tertiary $^{10}$Be production from B becomes important (though we caution that many of the relevant cross sections are not well-calibrated at these energies).}

\item{Ionization \&\ Coulomb Losses:} Fig.~\ref{fig:spec.compare.losses} shows that if we artificially disable ionization losses, the low-$E$ spectra of $p$, $e^{-}$, CNO, and many other species are significantly more shallow, and the B/C ratio also becomes flat below $\sim 100-200\,$MeV (in conflict with the Voyager data). At these densities and diffusivities, the effect of disabling Coulomb losses alone is relatively weak compared to ionization,  however if we either consider the spectrum in much more tenuous gas (a poorer match to observations overall) or higher diffusivities, then the relative role of Coulomb losses increases until both are comparable. The Coulomb or ionization loss time at low energies is $\sim 1\,{\rm Myr}\,(T/{\rm 10\,MeV})\,(n/{\rm cm^{-3}})^{-1}\,Z^{-2}$, so this is easily shorter than CR diffusive lifetimes at low-$E$ (see also \S~\ref{sec:vdrift.tloss}, Figs.~\ref{fig:tloss} \&\ \ref{fig:tloss.weights}), and they (Coulomb \&\ ionization losses) scale almost identically, the only difference is whether they act in neutral or ionized gas. So if CRs are spread uniformly in volume (e.g.\ owing to efficient diffusion) then the ratio of losses integrated over CR trajectories or volume is just the ratio of total ISM+inner CGM gas mass in ionized vs neutral phases \citep[see e.g.][for a derivation of this]{hopkins:cr.transport.constraints.from.galaxies}, which is $\mathcal{O}(1)$ in the ISM (with modestly more gas in neutral phases, but not by a large factor). However as shown below, the lowest-energy CRs are not infinitely-diffusive, so the CR energy density and loss rates at low rigidities are higher in denser gas, which tends to be neutral (explaining why ionization losses have a larger integrated effect at low rigidities than Coulomb losses). In either case, for low-energy hadrons (with $\gamma \sim 1$, i.e.\ not ultra-relativistic), this gives $\Delta t_{\rm res} \propto p^{2}$ (mildly non-relativistic) or $\Delta t_{\rm res} \propto p^{3}$ (highly non-relativistic; at $T \ll 100\,{\rm MeV} $), giving $\psi_{\rm obs} \sim 0$ (assuming the usual injection spectrum), i.e.\ a ``flat'' intensity at intermediate energies turning over to $\psi_{\rm obs} \sim -0.8$, i.e.\ an intensity $\propto T^{0.4}$ in the low-energy/sub-relativistic regime, as observed. For electrons $e^{-}$ (with $\beta \approx 1$ and $\gamma \gg 1$) this gives $\Delta t_{\rm res} \propto p^{1}$, so $\psi_{\rm obs} \sim 1$ (intensity $\propto T^{-1}$), also as observed. 

\item{Hadronic/Catastrophic/Spallation/Pionic/Annihilation Losses:} Obviously, we cannot get the correct secondary-to-primary ratios if we do not include these processes; our question here is whether these processes strongly modify the primary spectrum. Annihilation serves to ``cut off'' the spectrum of $\bar{p}$ and $e^{+}$ around their rest-mass energies. Radioactive losses here only shape the $^{10}$Be ratios. As for the spectra of CRs, at LISM conditions, the $e^{-}$ and $p$ population (as required by the $e^{+}/e^{-}$ and $\bar{p}/p$ ratios and $\gamma$-ray luminosity) is mostly primary, with relatively modest catastrophic losses, so we see in Fig.~\ref{fig:spec.compare.losses} that such losses do not dramatically reshape the spectra of these primaries (of course, they can do so in extreme environments like starbursts, which reach the proton calorimetric limit). Nonetheless removing the actual losses from e.g.\ pionic+hadronic processes does produce a non-negligible increase in the $p$ spectrum, and artificially boosts B/C owing to the ``retained'' primaries producing more B, and the lack of losses of B from spallation, which are actually significant under the conditions where B/C would normally be maximized.

\item{Inverse Compton \&\ Synchrotron Losses:} Fig.~\ref{fig:spec.compare.losses} also shows that if we disable inverse Compton (IC) \&\ synchrotron losses, the high-$E$ $e^{-}$ and $e^{+}$ spectra become significantly more shallow, basically tracing the shape of the $p$ spectrum (set by injection+diffusion). The magnitude of the change to the spectrum therefore depends on the assumed $\bar{\nu}(p)$ scaling (compare e.g.\ Appendix~\ref{sec:appendix:additional}, where we consider a reference model with $\delta \sim1$, where the effect is somewhat smaller). For high-energy leptons, IC+synchrotron loss times are $\sim 1\,{\rm Myr}\,(T/100\,{\rm GeV})^{-1}\,[(u_{\rm B}+u_{\rm rad})/{\rm 3\,eV\,cm^{-3}}]^{-1}$, so shorter than diffusive escape times (again see \S~\ref{sec:vdrift.tloss}, Figs.~\ref{fig:tloss} \&\ \ref{fig:tloss.weights}), and this $\Delta t_{\rm res} \propto p^{-1}$ produces $\psi_{\rm obs} \sim 3$, as observed. Since IC \&\ synchrotron scale identically with the radiation \&\ magnetic energy density, respectively, whichever is larger on average dominates (volume-weighted, since CR transport is rapid at these $p$). 

Even in the MW, it is actually not always trivial that the synchrotron losses should be comparable to IC losses, since in many Galactic environments, $u_{\rm B} \ll u_{\rm rad}$. Consider some basic observational constraints in different regions, noting $u_{\rm B} = 0.02\,{\rm eV\,cm^{-3}}\,B_{\rm \mu G}^{2}$. First, e.g.\ the CGM, where $B \ll 1\,\mu\,G$ \citep{farnes:2017.sub.microgauss.dla.bfields.from.rms,prochaska:2019.weak.magnetization.low.Bfield.rm.massive.gal.frb,vernstrom:2019.40.nanogauss.igm.field.limits.from.rms,lan:2020.cgm.b.fields.rm,malik:2020.mgII.absorber.rm.Bfield.microgauss.in.galaxies,osullivan:2020.nanoGauss.upper.limits.galactic.B.at.mpc}, but $u_{\rm rad}$ cannot be lower than the CMB value $\approx 0.3\,{\rm eV\,cm^{-3}}$; or at the opposite extreme consider typical star-forming complexes or OB associations or superbubbles (where most SNe occur) with observed upper limits from Zeeman observations in e.g. \citet{crutcher:cloud.b.fields,crutcher:2012.zeeman} of $\langle |{\bf B}|\rangle \lesssim 10\,{\rm \mu G}\,(n/{\rm 300\,cm^{-3}})^{2/3} \sim 5\,{\rm \mu G}\,(M_{\rm GMC}/10^{6}\,M_{\odot})^{-1/3}$ (inserting the GMC size-density relation; \citealt{bolatto:2008.gmc.properties}) compared to observed $u_{\rm rad}\sim 300\,{\rm eV\,cm^{-3}}$ averaged over the entire regions out to $\sim 200\,$pc and $\sim 10^{4}\,{\rm eV\,cm^{-3}}$ in the central $\sim 40\,$pc \citep{lopez:2010.stellar.fb.30.dor,pellegrini:2011.30.dor.structure.feedback.effects,barnes:2020.fb.comparison.cmz.rad.sne.winds,olivier:2020.hii.region.feedback.mechanism.breakdown}.\footnote{This can also be derived taking the observed nearly-constant MW cloud surface density and star formation efficiency and convolving with the IMF for a young SSP, see \citet{lee:2020.hopkins.stars.planets.born.intense.rad.fields}.} But the ratio $u_{\rm B}/u_{\rm rad}$ is maximized in the WIM phases with $n\sim 0.1-1\,{\rm cm^{-3}}$, $u_{\rm rad} \sim 1.3\,{\rm eV\,cm^{-3}}$ (the ISRF+CMB; \citealt{draine:ism.book}) and $B \sim 1-10\,{\rm \mu\,G}$ \citep[$u_{\rm B} \sim 0.02-2\,{\rm eV\,cm^{-3}}$][]{sun:2010.galactic.bfield.halo.strong.upper.limits.microGauss,jansson:2012.galactic.bfield.new.models,2015ASSL..407..483H,beck:2016.galactic.random.bfields.strong,mao:2018.ska.Galactic.bfield.constraints,mao:galactic.bfields.revisions}.
In Fig.~\ref{fig:egy.corr.density}, we show a quantitative plot of this for the same simulation as Fig.~\ref{fig:spec.compare.losses}, comparing the energy density in different (radiation, magnetic, CR, thermal) forms, as a function of local gas density, just for gas in the Solar circle. This agrees well with the broad observational constraints above, and indeed shows that $u_{\rm B}/u_{\rm rad}$ is maximized in the WIM phases. The fact that this is a volume-filling phase, and that CRs diffuse effectively (so that the total synchrotron emission is effectively a volume-weighted integral) ensures the synchrotron losses are not much smaller than the inverse Compton in the integral, allowing for the standard arguments \citep[e.g.][]{voelk:1989.fir.radio.calorimetric} to explain the observed far infrared (FIR)-radio correlation (at least within the $>1\,$dex observed $90\%$ inclusion interval; \citealt{magnelli:2015.fir.radio,delhaize:2017.fir.radio,wang:2019.fir.radio.corr}).\footnote{It is worth noting that other authors have shown that even if IC losses are significantly larger than synchrotron, the FIR-radio correlation is not strongly modified, when one accounts for secondary processes, radiation escape, and other effects \citep{lacki:2010.fir.radio.conspiracy,lacki:2010.fir.radio.highz.synchrotron,werhahn:2021.cr.calorimetry.simulated.galaxies}.} 

As a consequence of this, in Fig.~\ref{fig:spec.compare.losses}, we see that the effect of removing synchrotron losses on the $e^{-}$ spectrum is generally comparable to the effect of removing IC losses, but the synchrotron losses are somewhat larger at energies $\lesssim 20\,$GeV which contain most of the $e^{-}$ energy (thus in a ``bolometric'' sense synchrotron dominates over IC losses), while IC losses slightly dominate at even larger energies. This owes to the fact that higher-energy CRs (being more diffusive) sample an effectively larger CR scattering halo, therefore with loss rates reflecting lower-density CGM gas where $u_{\rm rad} \gtrsim u_{\rm B}$.

\item{Re-Acceleration (Convective, Streaming/Gyro-Instability, and Diffusive):} We discuss the different ``re-acceleration'' terms in detail below in \S~\ref{sec:reaccel}. In Fig.~\ref{fig:spec.compare.losses}, we see that removing each of the three re-acceleration terms in turn has relatively small effects. The convective term can have either sign, while the ``streaming'' term is almost always a loss term, and the ``diffusive reacceleration'' term is a gain term; on average for CRs we see the sum of the three (usually dominated by the convective term) results in a weak net loss term on average. 

For the sake of comparison with historical Galactic CR transport models which usually only include the ``diffusive re-acceleration'' term with an ad-hoc or fitted coefficient, we run one more test (``Maximal Diffusive Reacceleration x10'') where we artificially (1) turn off both the convective and streaming re-acceleration/loss terms, which are generally larger and have the opposite sign; (2) adopt $\bar{\nu} = 10^{-8}\,{\rm s^{-1}}\,R_{\rm GV}^{-1}$, so $D_{p p}$ is a factor $\sim 10\times$ larger at $\sim 1\,$GeV and $\sim100\times$ larger at $\sim$\,MeV compared to our ``preferred'' values (closer to what would be inferred in a ``leaky box'' model with no halo); (3) further replace our expression for the $D_{p p}$ terms derived directly from the focused CR transport equation with the more ad-hoc expression $\dot{p}/p = (1/9)\,(v_{A}^{2}/\kappa_{\|})\,|\partial\ln f/\partial \ln p| \sim (4/9)\,v_{A}^{2}/\kappa_{\|}$, about $\sim 5$ times larger than the value we would otherwise obtain. With this (intentionally un-realistic) case we find noticeable effects with a steeper low-$E$ slope and a more-peaked B/C, reproducing the very large implied role of diffusive re-acceleration for CR energy in some previous models.

\item{Streaming Terms (Non-Symmetric Scattering):} Per \S~\ref{sec:methods:crs}, we assume by default (motivated by SC models) that scattering is anisotropic in the fluid frame such that $\nu_{+} \ne \nu_{-}$, giving $\bar{v}_{A}\,\bar{f}_{1} \approx v_{A}\,|\bar{f}_{1}|$. Although this is almost always expected, if somehow the scattering were perfectly isotropic in that frame and the \Alf\ frame, we would have $\bar{v}_{A}\rightarrow0$, so the $D_{\mu p}$ term (which gives rise to CR ``streaming'' motion at $v_{\rm eff} \rightarrow v_{A}$ in the strong-scattering $\bar{\nu} \rightarrow \infty$ limit) and $D_{p \mu}$ term (the ``streaming loss''; \S~\ref{sec:reaccel}) vanish. Since we do not predict $\bar{\nu}_{\pm}$ here, in Fig.~\ref{fig:spec.compare.losses} we compare a run where we simply set $\bar{v}_{A}=0$. This makes only very small differences. Even for the scaling adopted in Fig.~\ref{fig:spec.compare.losses}, $\bar{\nu} \approx 10^{-8}\,{\rm s}^{-1}\,R_{\rm GV}^{-1}$, and reasonable $v_{A} \sim 10\,{\rm km\,s^{-1}}$, the streaming velocity only dominates over the diffusive velocity ($\sim \kappa\,|\nabla e|/e \sim c^{2}/(\bar{\nu}\,\ell_{\rm grad})$) at $E \lesssim 100\,$MeV. For our preferred model with smaller and more-weakly-$R$-dependent $\bar{\nu} \approx 10^{-9}\,{\rm s}^{-1}\,R_{\rm GV}^{-0.5}$, streaming only dominates at $\lesssim$\,MeV. 

Note, however, that in this study $v_{A}^{2}\equiv|{\bf B}|^{2}/4\pi\,\rho$ is the ideal MHD \Alf\ speed. As discussed in \citet{hopkins:cr.transport.constraints.from.galaxies}, for low-energy CRs where the frequency of gyro-resonant \Alf\ waves is much higher than the ion-neutral collision frequency, the CR streaming speed in a partially-ionized gas is the ion \Alf\ speed $v_{A,\,{\rm ion}}^{2}\equiv|{\bf B}|^{2}/4\pi\,\rho_{\rm ion}$, which can be very large in molecular clouds with typical $\rho_{\rm ion} \lesssim  10^{-7}\,\rho$. If we simply use this everywhere, we find in Fig.~\ref{fig:spec.compare.losses} that it has a significant effect, making the low-energy slopes shallower in $p$ and $e$ and lowering the peak B/C, as the CRs escape neutral gas nearly immediately without losses. However properly treating this regime requires a self-consistent model for self-confinement including the damping terms acting on gyro-resonant waves, which we defer to future work.

\item{Numerics:} For extensive tests of the numerical implementation of the spatial CR transport, we refer to \citet{chan:2018.cosmicray.fire.gammaray,hopkins:cr.transport.constraints.from.galaxies,hopkins:cr.mhd.fire2}. Briefly, however, we have also considered some pure-numerical variations here in Fig.~\ref{fig:spec.compare.numerics}, including: (1) replacing the generalized closure relation in Eqs.~\ref{eqn:f0}-\ref{eqn:f1} with the simpler ``isotropic'' closure from \citet{hopkins:m1.cr.closure}, which assumes the CR DF is always close-to-isotropic, closer to e.g.\ the formulation in \citet{thomas.pfrommer.18:alfven.reg.cr.transport}, or going further and using the older (less accurate) formulation of the CR flux equation from \citet{chan:2018.cosmicray.fire.gammaray}. This makes very little difference, as CRs are indeed close-to-isotropic and the timescale for the flux equation to reach steady-state (where the formal differences in these formulations vanishes) is short ($\sim \nu^{-1}$) compared to other simulation timescales \citep[as argued in e.g.][]{zweibel:cr.feedback.review,hopkins:m1.cr.closure,hopkins:2021.sc.et.models.incompatible.obs,thomas:2021.compare.cr.closures.from.prev.papers}. (2) We have also considered the effect of simply assuming the ultra-relativistic limit always in the spatial transport terms including the ``re-acceleration'' terms, instead of correctly accounting for $\beta$. This is purely to test how the more accurate formulation alters the results; removing the $\beta$ dependence in these terms artificially makes the low-energy spectra more shallow and lowers the low-energy ($\ll 1\,$GeV) B/C while raising $^{10}$Be/$^{9}$Be. So it is important to properly include these terms. (3) We have re-run our fiducial and several parameter-variation models with both the FIRE-2 and FIRE-3 \citep{hopkins:fire3.methods} versions of the FIRE code, which utilize the same fundamental physics and numerical methods, but differ in that FIRE-2 adopts somewhat older fits to quantities like stellar evolution tracks and cooling physics. This has no significant effects on any CR quantities we examine in this paper. Finally (4) we have tested various ``reduced speed of light'' (RSOL) values (which limit the maximum free-streaming speed of CRs to prevent extremely small timesteps). As extensively detailed in \citet{hopkins:m1.cr.closure} our numerical formulation is designed so that when the system is in steady-state, the RSOL has no effect at all on solutions, so long as is faster than other relevant speeds in the problem. Our default tests here adopt an RSOL of $\tilde{c} = 10^{4}\,{\rm km\,s^{-1}}$, which is more than sufficient for convergence, but in several model variants including raising/lowering $\bar{\nu}(1\,{\rm GV})$ by $\pm 1$\,dex, and changing the slope $\bar{\nu} \propto R^{-\delta}$ from $\delta=0.3-1$, we have tested values $\tilde{c} = 300 - 3\times10^{5}\,{\rm km\,s^{-1}} = (0.001 - 1)\,c$. We find that at $\sim 1\,$GV, we can reduce $\tilde{c}$ as low as $\sim 500-1000\,{\rm km\,s^{-1}}$ and obtain converged results; but for the highest-energy CRs (which can reach diffusive speeds $\sim \kappa\,|\nabla e_{\rm cr}|/e_{\rm cr} \sim 0.1\,c$) we require $\tilde{c} \gtrsim 3000\,{\rm km\,s^{-1}}$ for converged results with this particular RSOL formulation (Eq.~50 from \citealt{hopkins:m1.cr.closure}, as compared to the formulation from Eq.~51 therein which converges more rapidly, but potentially less-robustly in some conditions).\footnote{As shown in \citet{hopkins:m1.cr.closure}, with the RSOL formulation used here, if the background conditions are in steady-state then the steady-state CR predictions are strictly RSOL-independent. However, the time required to come into this steady state is increased by a factor $\sim c/\tilde{c}$. For this reason, we must (with this formulation) run our $\tilde{c}=3000\,{\rm km\,s^{-1}}$ simulation somewhat longer (beginning at $z=0.075$) to ensure it converges to steady state. If the characteristic CR escape or loss times (which normally set this timescale) with the true $\tilde{c}=c$ are $\sim 1-10$\,Myr, then for $\tilde{c} \lesssim 300\,{\rm km\,s^{-1}}$, this becomes $\sim 1-10$\,Gyr, timescales over which we cannot treat the galaxy as being in ``steady state,'' thus we do not expect our results with this method to converge for such low $\tilde{c}$.}

\end{itemize}

\begin{figure}
	\includegraphics[width=0.98\columnwidth]{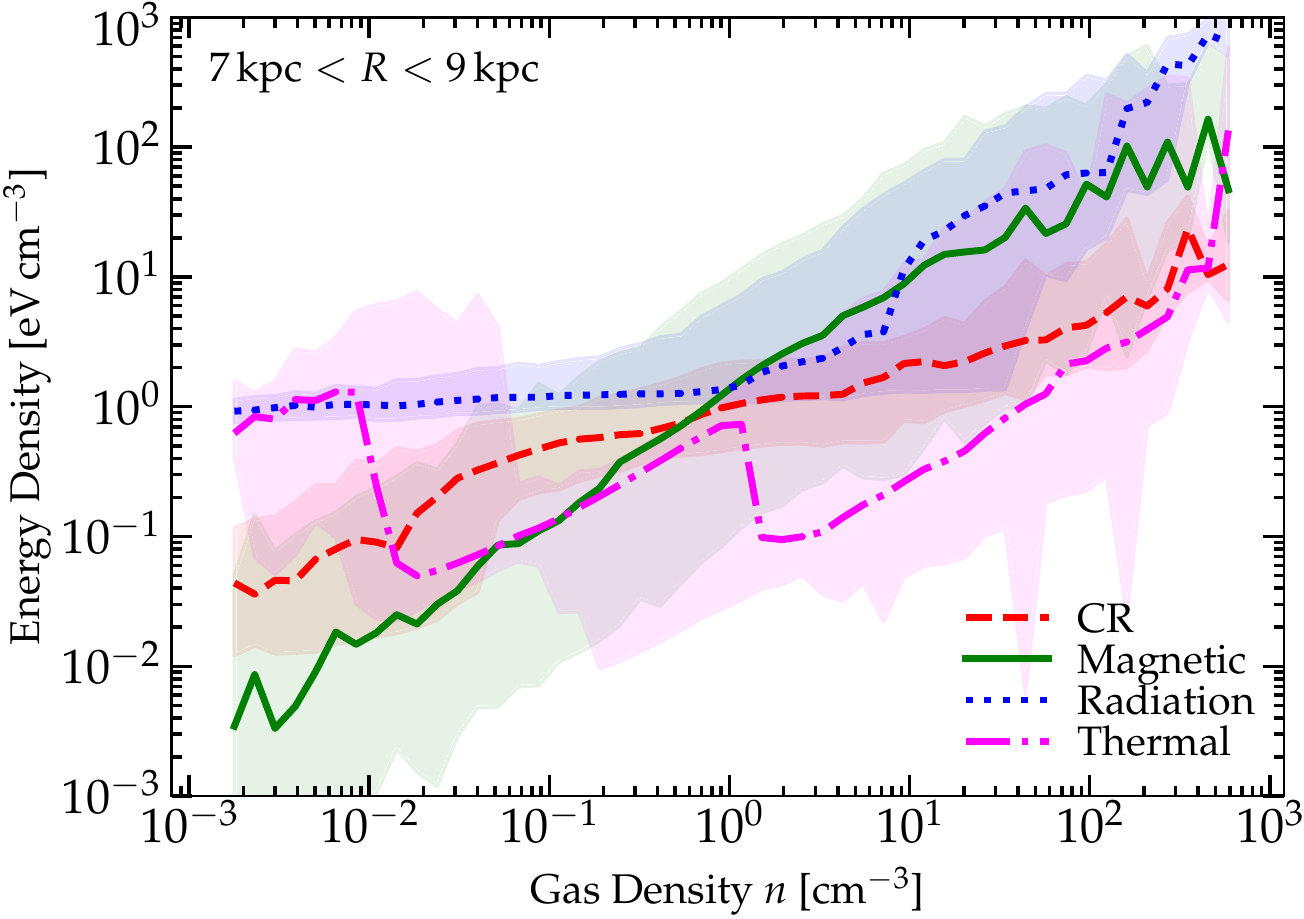}  
	\vspace{-0.15cm}
	\caption{Correlation of different energy densities with gas density in gas around the Solar circle (galacto-centric radius $R=7-9\,$kpc). Radiation $e_{\rm rad}$ is roughly constant at low densities owing to the semi-uniform ISRF+CMB, and rises in dense star-forming regions, but is sub-dominant or comparable to magnetic $e_{\rm B}$ at intermediate volume-filling densities, so synchrotron losses for leptons are dominant or comparable to inverse Compton. CRs are diffusive but not infinitely so, so follow a power-law $e_{\rm cr} \propto n^{0.4-0.5}$ between uniform ($n^{0}$) and pure-adiabatic ($n^{4/3}$). The ``jumps'' in thermal $e_{\rm th}$ arise from phase transitions (primarily hot to warm at $n \gg 0.01\,{\rm cm^{-3}}$, warm to cold at $n \gg 1\,{\rm cm^{-3}}$). These dependencies translate to different loss processes dominating with different rates in different ISM phases. It also produces systematic differences in CR spectra, losses, and ionization rates with density and galacto-centric radius.
	\label{fig:egy.corr.density}}
\end{figure}

\begin{comment}
\begin{figure}
	\includegraphics[width=0.98\columnwidth]{figures/output_12_16_20_newrsol_m10k_coeff700_anisoclosure_Dcoeff0pt5/corr_egy_vs_n_cen}  
	\includegraphics[width=0.98\columnwidth]{figures/output_12_16_20_newrsol_m10k_coeff700_anisoclosure_Dcoeff0pt5/corr_egy_vs_n_cgm}  
	\vspace{-0.15cm}
	\caption{Correlation of different energy densities with gas density, as Fig.~\ref{fig:egy.corr.density}, at the galactic center ($R<2\,$kpc) and disk-halo interface/inner CGM ($10<R<20\,$kpc).
	\label{fig:egy.corr.density.location}}
\end{figure}
\end{comment}

\subsection{On Re-Acceleration, ``Adiabatic,'' and ``Streaming Loss'' Terms}
\label{sec:reaccel}

We generically find that re-acceleration plays a modest to minimal role (see Fig.~\ref{fig:spec.compare.kappa}). But there are three different ``re-acceleration'' terms, per Eq.~\ref{eqn:reaccel}, and contradictory conclusions in the literature. We therefore discuss the physics of each in turn, to assess their relative importance. We will discuss physical behaviors in both self-confinement (SC) and extrinsic-turbulence (ET) limits.

\subsubsection{``Adiabatic'' Term}

First, consider the ``adiabatic'' term, $\dot{p}/p = \dot{p}_{\rm ad}/p= -\mathbb{D}:\nabla{\bf u} \sim \mathcal{O}(\nabla\cdot {\bf u})$. Despite its simplicity, in a complicated flow there are contributions to $\nabla{\bf u}$ from modes on all scales $\lambda$, which we can decompose as $\sim \delta u(\lambda)/\lambda$. In a standard turbulent cascade, $\delta u(\lambda)/\lambda \sim 1/t_{\rm eddy}(\lambda) \sim \lambda^{-(1/2-2/3)}$ (depending on the cascade model) is larger on small scales. Galactic fountains, pure gravitational collapse/fragmentation cascades, etc., all produce similar results in  this respect \citep[see][]{elmegreen:sf.review,vazquez-semadeni:2003.turb.reg.sfr,krumholz.schmidt,klessen:2010.accretion.turb,kim:disk.self.reg,krumholz:2010.instab.turb.in.disks,ballesteros-paredes:2011.core.collapse.sizemass,2015ApJ...815...67K}. However, the quantity of interest (what actually determines the net effect on the CR spectrum and energy) is not actually $\dot{p}_{\rm ad}/p$, but a mean (volume-averaged) time-integrated (over the CR travel time) $\langle \dot{p}_{\rm ad} \rangle$, which it is well-known from many galactic/ISM theoretical and observational studies \citep{stanimirovic:1999.smc.hi.pwrspectrum,elmegreen:2002.fractal.cloud.mf,decamp:2002.turb.concentration.ism.conditions.chemistry.fx.make.stronger,mac-low:2004.turb.sf.review,block:2010.lmc.vel.powerspectrum,bournaud:2010.grav.turbulence.lmc,hopkins:frag.theory,squire.hopkins:turb.density.pdf} is dominated by the largest-scale modes which are coherent over $\lambda \sim H$, the disk scale height.\footnote{For our purposes, the largest modes with $\lambda\sim H$ where $H$ is the disk scale-height or Toomre length have, by definition in a trans or super-sonically turbulent ISM, $|\nabla{\bf u}| \sim V_{c}/r_{\rm disk} \sim 1/t_{\rm dyn}$, where $t_{\rm dyn}$ is the galactic dynamical time \citep{elmegreen:1997.open.closed.cluster.same.mf.form,gammie:2001.cooling.in.keplerian.disks,hopkins:2012.intermittent.turb.density.pdfs,hopkins:2013.turb.planet.direct.collapse}.} Briefly, this can be understood with a simple toy model. Since $\nabla{\bf u}$ has either sign, and modes on small scales $\lambda$ compared to the total CR travel length $\ell$ along $\bhat$ are un-correlated, then averaging over CR  paths (assuming a diffusive 3D random walk in space with $\lambda \ll \ell$) or averaging over volume $d^{3}{\bf x}$ (equivalent if the CRs are in steady-state or we assume ergodicity), the {\em coherent} effect of the modes is reduced by a factor of $\sim N_{\rm modes}^{-1/2} \sim (\ell/\lambda)^{-3/2}$. So for any realistic spectrum the largest coherent modes dominate the integral, and for any realistic disk structure these must have $\lambda \sim {\rm MIN}(\ell,\,H) \sim H$ (for the energies of interest), giving $\mathcal{O}(\langle \mathbb{D}:\nabla{\bf u} \rangle) \sim \mathcal{O}(t_{\rm dyn}^{-1})$ with the disk dynamical time $t_{\rm dyn} \sim 35\,$Myr at the Solar position.

The magnitude of the coherent effect of this term can then be estimated as $\Delta p/p \sim \mathcal{O}(\Delta t_{\rm res}\,\langle \mathbb{D}:\nabla {\bf u}\rangle) \sim \mathcal{O}(\Delta t_{\rm res}/t_{\rm dyn})$. Since at $R\gtrsim 1\,$GV, the residence time $\Delta t_{\rm res}$ decreases with $R$, this term is most important at lower energies, as expected. The sign is not {\em a-priori} obvious, however. But again note the averaging above: if CRs diffuse efficiently, so the CR density is not strongly dependent on the local gas density, then the CR travel time integral above is dominated by the most {\em volume-filling} phases of the ISM and inner halo/corona traversed. These diffuse phases are the ones most strongly in outflow, so more often than not, the appropriately-weighted $\langle \mathbb{D}:\nabla{\bf u} \rangle > 0$ \citep[for detailed discussion of how the adiabatic term depends on ISM phases, see][]{2017ApJ...847L..13P,chan:2018.cosmicray.fire.gammaray}, and the net effect of this term is usually to {\em decrease} CR energies. Because the effect is weaker at higher CR energies, in a volumetric sense this has the net effect of making the CR spectra more {\em shallow} (i.e.\ if $J_{\rm obs} \propto p^{-\alpha}$, this decreases $\alpha$). But we stress, again, that the sign of the effect will be different in different Galactic environments.

\subsubsection{``Streaming'' Term}

Next consider the $D_{p\mu}$ or ``streaming'' term $\dot{p} = \dot{p}_{\rm st} = -\langle \mu\rangle\,\bar{D}_{p\mu}$ or $\dot{p}_{\rm st}/p =  -\bar{\nu}\,(\bar{v}_{A}\,\bar{f}_{1})/(v\,\bar{f}_{0}) = -\bar{\nu}\,(\bar{v}_{A}/c)\,F^{\prime}_{e}/(3\,c\,P_{0}^{\prime})$, where $F^{\prime}_{e}$ and $P_{0}^{\prime}$ are the CR energy flux and pressure in a narrow interval in $p$. In SC-motivated models, as discussed in detail in \citet{hopkins:m1.cr.closure} and noted above, the asymmetry in $\nu_{+}$ and $\nu_{-}$ gives $\bar{v}_{A}\,\bar{f}_{1} \approx v_{A}\,\bar{f}_{1}$, so the ratio of the ``streaming'' $D_{\mu p}$ to ``diffusive'' $D_{p p}$ terms is always $\ge F_{e}/v_{A}\,e \sim v_{\rm eff}/v_{A}$ (where $v_{\rm eff}$  is the effective bulk transport speed of CRs) -- i.e.\ it dominates whenever the CRs move at trans or super-\Alf{ic} speeds, which is usually true. Moreover, as shown in \citet{hopkins:m1.cr.closure} or seen by plugging into Eq.~\ref{eqn:f0}-\ref{eqn:f1}, in flux-steady-state the sum of these two terms becomes $\dot{p}_{\rm st}/p = -\bar{\nu}\,[\bar{f}_{1}/\bar{f}_{0}\,\bar{v}_{A}/v - \chi\,\psi\,v_{A}^{2}/v^{2}] \rightarrow -[v_{A}\,|\bhat\cdot \nabla P_{0}|/3\,P_{0} + \bar{\nu}\,|\psi/(2+2\,\beta^{2}))|\,(v_{A}/\gamma\,\beta\,c)^{2}]$. This term is negative-definite, representing the ``streaming losses'' (energy lost to gyro-resonant instability as the CRs move), and the leading term is $\sim v_{A}/\ell_{\rm grad,\,cr}$ where $\ell_{\rm grad,\,cr} = 3\,P_{0}/|\bhat \cdot \nabla P_{0}|$. Comparing this to the magnitude of the adiabatic term, we have $|\dot{p}_{\rm ad}/\dot{p}_{\rm st}| \sim (|{\bf u}|/v_{A})\,(\ell_{\rm grad,\,cr} / \ell_{\rm grad,\,u}) \gtrsim 1$. Thus the ``adiabatic'' term is almost always larger than the ``streaming'' term, because (a) on galaxy scales the bulk turbulent and convective/fountain/inflow/outflow motions are trans or super-\Alf{ic} ($|{\bf u}|\gtrsim v_{A}$), and (b) CR diffusion/streaming means that the CR pressure profile is almost always smoother than the local gas velocity structure ($\ell_{\rm grad,\,cr} \gg \ell_{\rm grad,\,u}$). 

In ET-motivated historical Galactic CR transport models, this $D_{p\mu}$ term is often neglected, implicitly assuming that the scattering modes are exactly symmetric in the co-moving and \Alf\ frames ($\nu_{+}(\mu)=\nu_{-}(\mu)$, so $\bar{v}_{A}\rightarrow0$). In reality, multiple effects break this degeneracy: for example gyro/streaming instabilities (both resonant and non-resonant) excite modes in one direction and damp in the other, giving $\nu_{+}\gg \nu_{-}$ or $\nu_{+}\ll \nu_{-}$ (depending on the sign of $\langle \mu\rangle$), giving the SC behavior above. Even if these instabilities are negligible, symmetry is broken by advective/transport terms, potentially at the order-unity level \citep[see][]{zweibel:cr.feedback.review} -- the symmetry-breaking would have to be smaller than $\mathcal{O}(v_{A}/c)$ for the $D_{p\mu}$ term to be much smaller than the $D_{pp}$ term.

\subsubsection{``Diffusive'' Term}

Now consider the $D_{p p}$ or ``diffusive'' term $\dot{p} = \dot{p}_{\rm di} \sim 4\,\bar{D}_{p p}/p^{2} \sim \bar{\nu}\,4\,\chi\,(v_{A}/v)^{2}$. First note, as shown in \citet{hopkins:m1.cr.closure}, that this term vanishes entirely when CRs approach a ``free-streaming'' or highly-anisotropic distribution function limit ($\chi\rightarrow 0$), and in any weak-scattering (small $\nu$) limit $\mathbb{D}:\nabla{\bf u}$ (which does not depend on $\nu$) trivially becomes the dominant re-acceleration term. 

The $D_{p p}$ term is also, as noted above, guaranteed to be sub-dominant in steady-state to the ``streaming'' term if streaming instabilities are significant (e.g.\ in SC-motivated models). So in order to estimate the maximum possible importance of the term, let us assume ET-type models, with a nearly-isotropic DF,  and relatively low-energy CRs (where $\nu$ is larger) giving $\dot{p}_{\rm ad}/p \approx (2/3)\,\bar{\nu}\,v_{A}^{2}/v^{2} \approx (2/27)\,v_{A}^{2}/D_{x x}$. Using the fact that in this limit, the diffusive bulk  transport speed is $v_{\rm eff} \sim D_{x x} / 2\,\ell_{\rm grad,\,cr}$, the ratio of the adiabatic $\dot{p}_{\rm ad}$ term to this diffusive $\dot{p}_{\rm di}$ term is $|\dot{p}_{\rm ad} / \dot{p}_{\rm di}| \sim (10)\, (|{\bf u}|/v_{A})\,(v_{\rm eff}/v_{A})\,(\ell_{\rm grad,\,cr}/\ell_{\rm grad,\,u}) \gg 1$, as each of the four $(...)$ terms is $\gtrsim 1$. Even comparing this value of the diffusive term in the ET limit to the ``streaming'' term, if we include the minimal advective symmetry-breaking terms above, we see $|\dot{p}_{\rm di}| \lesssim |\dot{p}_{\rm st}|$ for reasonable coefficients. 

Finally, even if we ignore these other re-acceleration terms, it is difficult for $\dot{p}_{\rm di}$ to have a large effect: since the magnitude scales as $\propto v_{A}^{2}/D_{x x} \propto \bar{\nu}\,v_{A}^{2}/c^{2}$, the re-acceleration time is $(\dot{p}_{\rm di}/p)^{-1} \sim 30\,{\rm Gyr}\,(\beta^{2}\,\bar{\nu}/10^{-9}\,{\rm s^{-1}})\,(10\,{\rm km\,s^{-1}}/v_{A})^{2}$, so it requires unrealistically small diffusivities (large $\bar{\nu}$) and/or large $v_{A}$ to reach the regime where it could have an order-unity effect on the CR spectrum, and even in that regime would require residence times longer than hadronic or ionization loss timescales, leading to net losses.

\subsubsection{Summary}

In summary, for almost any internally-consistent (let alone physically-plausible) ET or SC models for $\bar{\nu}$ and large-scale motions, there is a robust ordering of the three re-acceleration terms: $|\dot{p}_{\rm ad}| \gtrsim |\dot{p}_{\rm st}| \gtrsim |\dot{p}_{\rm di}|$, i.e.\ the ``adiabatic'' term $\sim \nabla{\bf u}$, which is itself dominated by the driving-scale motions with $\lambda\sim H$ and $\mathcal{O}(\nabla{\bf u}) \sim 1/t_{\rm dyn}$ is largest, with the ``streaming'' or gyro-instability loss ($D_{p\mu}$) term intermediate, and the ``diffusive'' or micro-turbulent ($D_{p p}$) re-acceleration term smallest. In the LISM, $t_{\rm dyn} \sim 35\,$Myr, generally longer then CR residence/loss timescales, though not completely negligible near the peak of the CR spectrum, so this can have a $\mathcal{O}(1)$ effect on the peak amplitude (which is energetically plausible for convective/adiabatic terms, as the energy in ISM turbulence and bulk outflows is generally larger by a factor of at least several than CR energies) but weak effects elsewhere in the CR spectrum. We also see this directly plotting the relevant loss timescales in \S~\ref{sec:vdrift.tloss}.

Previous studies which inferred a dominant role of diffusive CR re-acceleration gains generally (1) require un-realistic energetics in this component \citep[as cautioned by][]{drury.strong:power.req.for.diffusive.reaccel}; (2) ignored both convective and streaming loss terms which, for any realistic model, offset the diffusive term as described above; (3) adopted un-realistically low diffusivities (high $\bar{\nu}$) appropriate for e.g.\ leaky-box models but not models with a more realistic halo, so $D_{p p}$ is much larger than it should be; and (4) treat the normalization of the diffusive-reacceleration as arbitrary, e.g.\ through ``fitting'' the value of $v_{A}$ that appears in $D_{pp}$, and adopt order-of-magnitude larger values than physically allowed here. This can easily be seen from Eq.~\ref{eqn:reaccel}: the diffusive-only re-acceleration timescale is $p/\dot{p} \sim v^{2}/4\,\chi\,\bar{\nu}\,v_{A}^{2} \sim (c/v_{A})^{2}\,(2/\bar{\nu})\sim 10^{6}\,{\rm Myr}\,(n/{\rm cm^{-3}})\,(B/{\rm \mu G})^{-2}\,(\bar{\nu}/10^{-9}\,{\rm s}^{-1})^{-1}$. Without invoking orders-of-magnitude lower $\bar{\nu}$ or larger $B$, this cannot compete with the streaming loss and adiabatic terms (timescale $\sim 10-100\,$Myr, as noted above) let alone other CR loss/escape terms.

\begin{figure}
	\hspace{-0.2cm}\includegraphics[width=1.0\columnwidth]{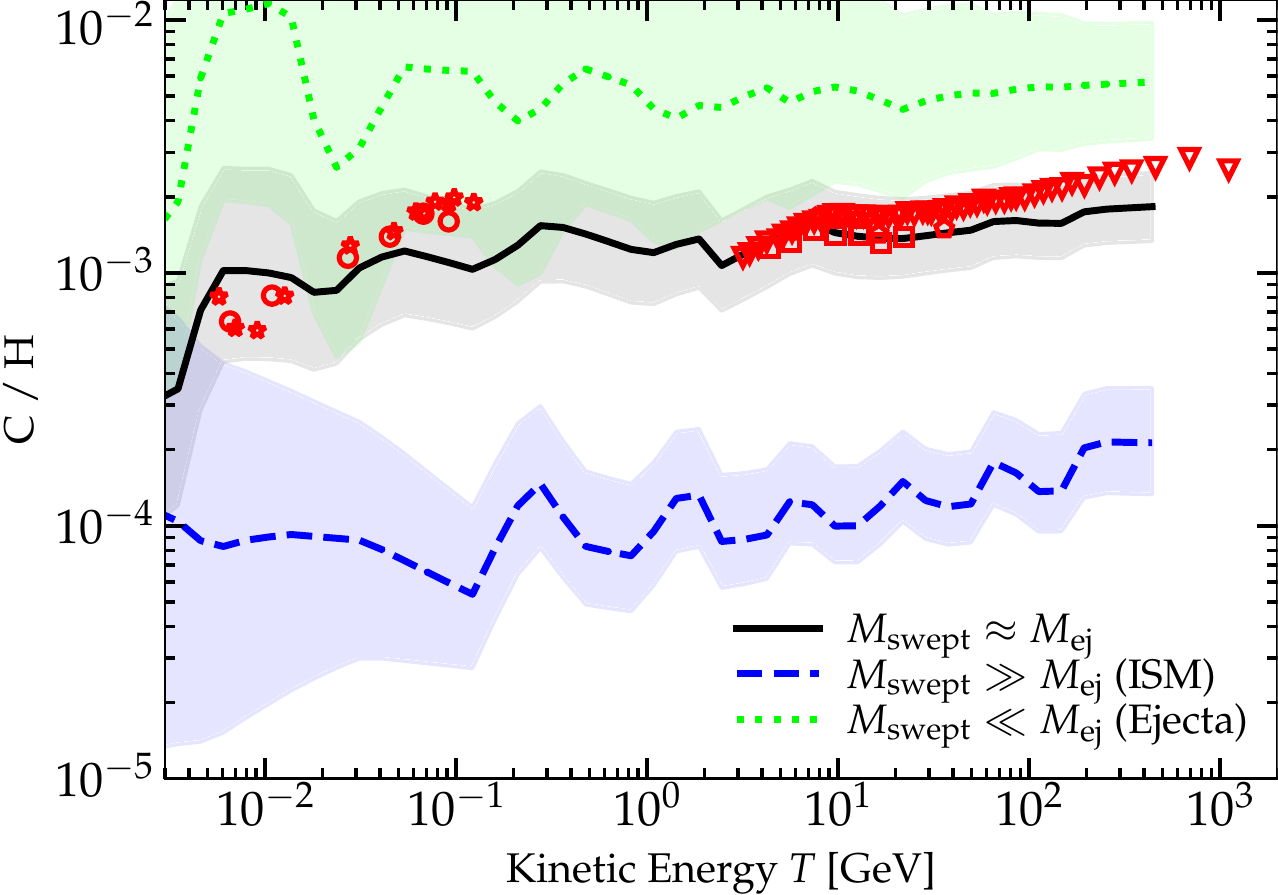} \\ 
	\includegraphics[width=0.98\columnwidth]{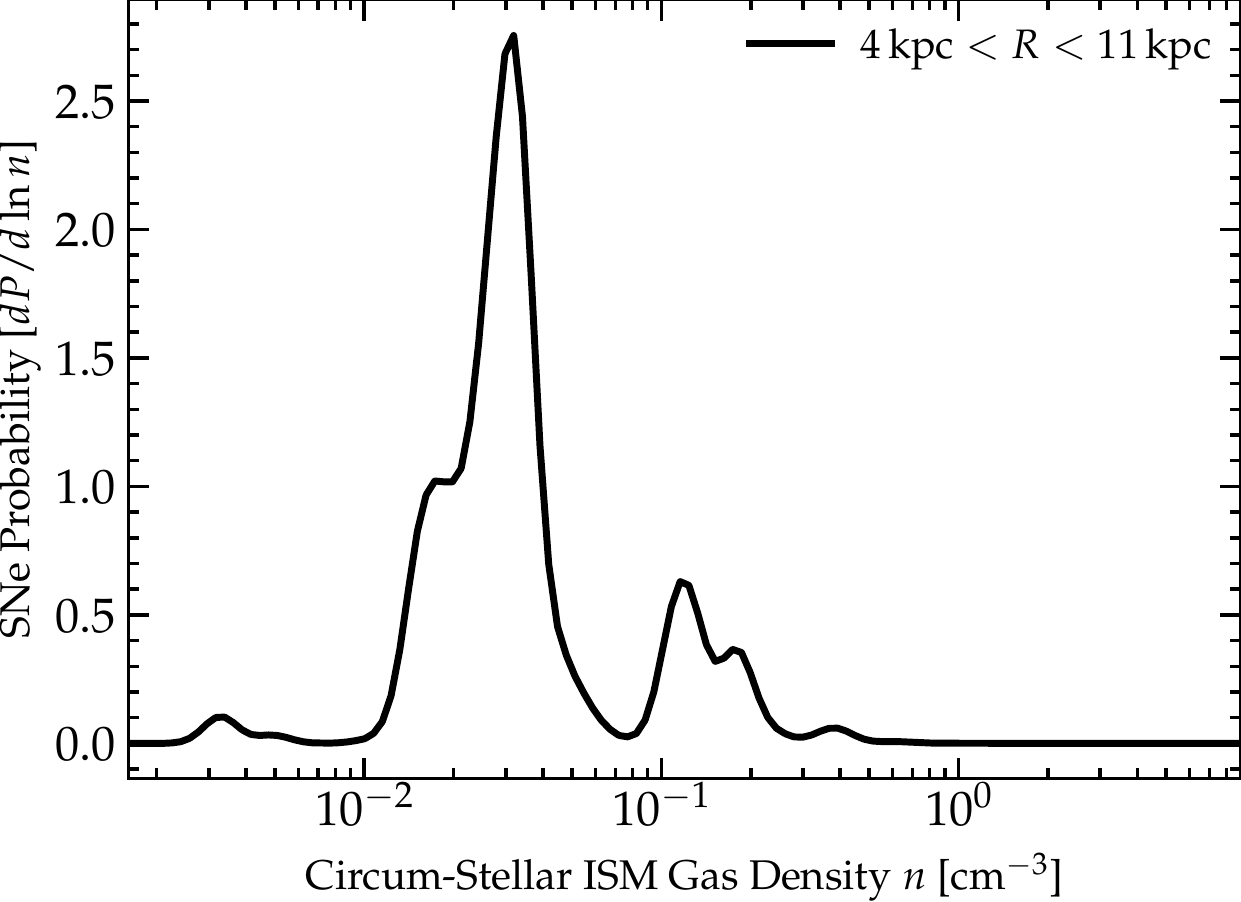} \\ 
	\vspace{-0.4cm}
	\caption{{\em Top:} Ratio of observed C/H in LISM CRs (at Solar-circle with gas $n=0.3-3\,{\rm cm^{-3}}$) vs.\ CR energy in simulations (lines are median+shaded ranges $\pm2\,\sigma$) and observations (points), for three different assumptions about where CRs are accelerated, which determines the injection abundance ratios (see \S~\ref{sec:methods:injection}). Our default ($M_{\rm swept} \approx M_{\rm ej}$) assumes CRs accelerate when the swept-up mass from SNe or stellar wind injection equals the ejecta mass (i.e.\ around reverse-shock formation or the onset of the Sedov-Taylor phase). For SNe (which dominate the sources) this is similar to assuming most acceleration occurs in shocks with velocity $\gtrsim 2000\,{\rm km\,s^{-1}}$. Blue/dashed line assumes efficient acceleration occurs through the end of the Sedov-Taylor phase such that most of the total CR energy is produced in the final stages when $M_{\rm swept} \sim 3000\,M_{\odot} \gg M_{\rm ej}$ and the initial CR abundances essentially trace the ambient ISM abundances (roughly Solar). This gives C/H an order of magnitude lower than observed. Green/dotted shows the result assuming acceleration with ``pure'' ejecta abundance ratios. The results for N/H and O/H are almost identical, and do not depend strongly on e.g.\ CR transport parameters.
	{\em Bottom:} Distribution of injection sites: PDF of ISM circum-SNe gas densities just before explosion in the outer disk (radii labeled), where most LISM CRs are accelerated. Despite stars forming at resolved densities $n \gtrsim 1000\,{\rm cm^{-3}}$ in these simulations, most SNe (by number) explode in evacuated (super)bubble-like regions.
	\label{fig:sne.abundances.sites}}
\end{figure}

\subsection{Where are Most CRs Accelerated?} 

There is an extensive literature using the abundance patterns of CRs to constrain their acceleration sites; most of these studies have argued that most of the $\sim\,$MeV-TeV CRs followed here must come from sites relatively near SNe, probably in ``super-bubbles,'' with $\sim 10\%$ of the initial ejecta kinetic energy ending up in CRs \citep[see e.g.][and references therein]{higdon:crs.accel.in.superbubbles.in.sne.ejecta.mass.dominated.regions,parizot:2004.superbubble.cr.accel.bulk,becker.tjus:2020.cr.multi.messenger.accel.regions}. As illustrated in Fig.~\ref{fig:sne.abundances.sites}, we find the same, and below we argue this must be the case from abundance, energetic, and gravito-turbulent considerations.

\subsubsection{Abundances: Most CRs Must Come From Initial SNe Shocks}

It has been argued by a number of authors that significant CR Fermi-I acceleration could occur in shocks with modest Mach number of just $\sim 5$ or even lower \citep{ryu:2003.low.mach.shocks.most.of.dissipation.for.cluster.cr.accel,schure:2012.diffusive.cr.shock.accel.needs.bfield.amplification.in.shock,vink:2014.critical.shock.mach.number.cr.accel,guo:2014.low.mach.shock.accel.electrons}. If this alone were sufficient to produce an order-unity fraction of the ``new'' CRs/Fermi-I acceleration to $\sim $\,GeV (we stress that Fermi-II acceleration from shocks {\em is} included in our methods), then CRs could be continuously created almost everywhere throughout the ISM, as it is well-established observationally that the majority of all the gas in the ISM is super-sonically turbulent, with e.g.\ typical Mach numbers in the massive atomic and molecular clouds that contain $\sim 1/2$ of the galaxy gas mass typically ranging from $\mathcal{M} \sim 10-100$ \citep{evans:1999.sf.gmc.review,mac-low:2004.turb.sf.review,elmegreen:2004.obs.ism.turb.review,mckee:2007.sf.theory.review}. If these were the {\em dominant} sites of initial CR acceleration, then the primary CR O/H ratio would just trace the ISM abundance, with e.g.\ $N_{\rm O}/N_{\rm H} \sim 0.0005$ \citep{lodders:updated.solar.abundances.review}. But in the LISM, the ratio of $N_{\rm O}/N_{\rm H}$ in CRs (where almost all O is primary) is $\sim 0.01$ \citep{cummings:2016.voyager.1.cr.spectra}. The qualitative discrepancy is the same if we consider any of CNO or any heavier species from Ne through Fe. Most other processes (re-acceleration, spallation, violations of the test-particle limit in acceleration) make the discrepancy more, not less, dramatic. This also rules out stellar winds as the source of most CR acceleration: while AGB winds (which are very low-energy) are mildly enhanced in CNO, even this is nowhere near sufficient, and faster OB/WR winds are so weakly enhanced that they give essentially identical results to ``pure ISM'' acceleration. 

If we assume acceleration near SNe, we can constrain the total ratio of ``entrained'' mass per SNe at the time of acceleration, to the initial ejecta mass. Again, if the only requirement for efficient CR acceleration were a Mach number $\mathcal{M} \gtrsim 5$, then acceleration would be efficient throughout the end of the Sedov-Taylor and well into the snowplow phase of remnant evolution -- we test this directly in Fig.~\ref{fig:sne.abundances.sites} by running a model where the swept-up mass $M_{\rm swept}$ is given by the mass at the end of the Sedov-Taylor phase. In general, at the end of the Sedov-Taylor phase for a clustered group of $N_{\rm SNe}$, the shock velocity is $v_{\rm shock} \sim 200\,{\rm km\,s^{-1}}$, and the swept-up ISM mass is $\sim 3000\,N_{\rm SNe}\,M_{\odot}$ \citep{cioffi:1988.sne.remnant.evolution,walch.naab:sne.momentum,kim.ostriker:sne.momentum.injection.sims,hopkins:sne.methods}. The initial ejecta metallicity has been completely diluted at this point, predicting $N_{\rm O}/N_{\rm H} \sim 0.0005$, again. Per \S~\ref{sec:methods:injection}, if the acceleration occurs from a mix of ambient gas and ejecta when the shock has entrained a mass $M_{\rm swept}$, then the CR $N_{\rm O}/N_{\rm H} \approx (N^{\prime}_{\rm O,\,ej}\,M_{\rm ej} + N^{\prime}_{\rm O,\,ISM}\,M_{\rm swept}) / (N^{\prime}_{\rm H,\,ej}\,M_{\rm ej} + N^{\prime}_{\rm H,\,ISM}\,M_{\rm swept})$. If we instead assume $M_{\rm swept} \approx M_{\rm ej}$ (our default model), then we obtain $N_{\rm O}/N_{\rm H} \sim 0.008$ for {\em both} core-collapse and Ia SNe, in excellent agreement with the CR observations, as shown for C/H in Fig.~\ref{fig:sne.abundances.sites} (we do not show O/H but the conclusions are essentially identical). Theoretically, this is a particularly interesting value, since it corresponds to the time when the reverse shock fully-forms and propagates through the ejecta, essentially to the ``onset'' of the shock and end of the ejecta free-streaming phase. In any model where the CR acceleration efficiency to $\sim\,$GeV is an increasing function of Mach number or an increasing function of the shock kinetic energy dissipation rate, this will be the phase which dominates acceleration.\footnote{For the usual definition of CR acceleration efficiency $\eta(\mathcal{M})$, the flux of accelerated CRs in a strong shock is $F_{\rm cr} \sim \eta(\mathcal{M})\,(1/2)\,\rho\,V_{\rm sh}^{3}$ (for shock velocity $V_{\rm sh}$ and upstream density $\rho$), so the contribution to the total CR energy in some time interval $dt$ is $d E_{\rm cr} \propto \oint F_{\rm cr}\,dt \propto \eta(\mathcal{M})\,r_{\rm sh}^{2}\,V_{\rm sh}^{3}\,dt \sim \eta(\mathcal{M})\,r_{\rm sh}^{2}\,V_{\rm sh}^{2}\,dr_{\rm sh}$ (with $r_{\rm sh}$ the shock radius). But in the Sedov-Taylor phase ($r_{\rm sh} \propto V_{\rm sh}\,t \propto t^{2/5}$) this is just $\propto \eta(\mathcal{M})\,d\ln{r_{\rm sh}}$ with $\mathcal{M} \propto r_{\rm sh}^{-3/2}$. Thus, any model where $\eta(\mathcal{M})$ is an even weakly increasing function of $\mathcal{M}$ (as expected qualitatively in most theories of diffusive shock acceleration, see e.g.\ \citealt{blandford:1987.cr.acceleration.review,amato:2005.cr.shock.accel.analytic.toy.model,bell:2013.cr.acceleration.review}) will produce most CR acceleration at the smallest $r_{\rm sh}$ possible (i.e.\ the onset of the Sedov-Taylor phase, when the reverse shock forms).} For completeness, in Fig.~\ref{fig:sne.abundances.sites} we also consider a model where CRs are accelerated with ``pure ejecta'' abundances, i.e.\ $M_{\rm swept} \rightarrow 0$, which in contrast {\em over} predicts the abundance of intermediate elements.

\subsubsection{Energetics: Most CR Energy Comes From SNe Energy}

If we assume that CR acceleration imparts a constant fraction $\epsilon \sim 0.1$ of the thermalized/dissipated shock kinetic energy to CRs, then we have directly verified that in our simulations most of the CR energy comes from SNe, even if we allow CR injection at ISM shocks of arbitrarily low Mach number. This is expected: integrated over time and the stellar IMF, the kinetic energy input from stellar mass-loss (dominated by fast O/B winds) is $\sim 10\%$ that of core-collapse SNe \citep{starburst99,2014ApJS..212...14L,2014ARA&A..52..487S,rosen:2014.xray.energy.wind.clusters,BPASS}. The input from proto-stellar radiation and jets is only $\sim 1\%$ of that from SNe, while the energy from winds accelerated around remnants (e.g.\ PWNe, XRBs, etc.) is even smaller still \citep{federrath:subgrid.jet.model,bally:2016.protostellar.outflows,guszejnov:2020.starforge.jets}. From ISM shocks, our simulations reproduce the usual result that the ISM turbulent dissipation rate is $\sim 1-5\%$ of the SNe energy input rate \citep{hopkins:fb.ism.prop,cafg:sf.fb.reg.kslaw,kim.ostriker:sne.momentum.injection.sims,martizzi:sne.momentum.sims,orr:ks.law}, which also follows from the trivial order-of-magnitude expectation for super-sonic turbulence, $\dot{E} \sim (1/2)\,M_{\rm swept}\,\sigma_{\rm turb}^{2}/t_{\rm dyn}$, where $t_{\rm cross} = t_{\rm eddy}(H)=H/\sigma=t_{\rm dyn}$ for any disk with Toomre $Q\sim 1$, given canonical MW-like values for $M_{\rm swept} \sim 10^{10}\,M_{\odot}$ and $\sigma_{\rm turb} \sim 10\,{\rm km\,s^{-1}}$ with $t_{\rm dyn}=r/V_{c}$ at the effective radius $\sim 5\,$kpc. What this cannot tell us is ``how close'' to SNe CRs are accelerated (e.g.\ at the onset or later in the shock), since by definition during e.g.\ the Sedov-Taylor phase the shock energy is conserved -- for this we refer to the abundance argument.

\subsubsection{Environment: Most SNe Explode in Super-Bubbles}

If most CRs are accelerated ``near'' SNe (before they sweep up a mass $\gg M_{\rm ejecta}$), it follows trivially in simulations like ours that most CRs are accelerated in super-bubble environments, simply because the majority of SNe explode in such environments. Note this is weighted by number or energy, so most SNe come from $\sim 10\,M_{\odot}$ stars that explode $\sim 30\,$Myr after they reach the main sequence, well after more massive stars in the complex have exploded and destroyed their natal GMCs; see e.g.\ \citealt{grudic:sfe.cluster.form.surface.density}. We have shown that the fact that most SNe energy goes into super-bubble type structures (overlapping SNe shocks during the energy-conserving phase) is true for FIRE simulations in a number of studies \citep{hopkins:stellar.fb.winds,hopkins:2013.merger.sb.fb.winds,muratov:2015.fire.winds,escala:turbulent.metal.diffusion.fire}, and many other simulation and observational studies have shown the same \citep{walch.naab:sne.momentum,martizzi:sne.momentum.sims,haid:snr.in.clumpy.ism,2018MNRAS.481.3325F,gentry:sne.momentum.boost,2019MNRAS.487..364L,2019MNRAS.483.4707G}. And we also confirm this directly in Fig.~\ref{fig:sne.abundances.sites}. Indeed, the fact that star formation (and therefore Type-II SNe) are strongly clustered in both space and time are not just observational facts, but are generic consequences of any reasonable model where gravitational collapse and hierarchical fragmentation (e.g.\ from ISM to clouds to clumps to cores to stars) plays an important role \citep{hopkins:frag.theory,guszejnov:protostellar.feedback.stellar.clustering.multiplicity,guszejnov:universal.scalings,grudic:cluster.properties}. 

If superbubble type environments have typical densities $n\equiv 0.01\,n_{0.01}\,{\rm cm^{-3}}$ (per Fig.~\ref{fig:sne.abundances.sites}), and the majority of the CR acceleration occurs early in the Sedov-Taylor phase when the shock forms and the dissipation rate is maximized, i.e.\ when the entrained mass is $\sim M_{\rm ejecta} \sim 10\,M_{\odot}$, then the characteristic shocks occur at radii $\sim 20\,n_{0.01}^{-1/3}\,{\rm pc}$, after $\sim 10^{4}\,n_{0.01}^{-1/3}\,{\rm yr}$ of free-expansion, with the ``ambient'' mass being primarily SNe-enriched material and a characteristic delay time between events (i.e.\ delay time seen by the {\em ambient} medium before the accelerating shock) of $\Delta t_{\rm SNe} \sim  2\times10^{5}\,{\rm yr}\,(M_{\rm cl}/10^{4}\,M_{\odot})^{-1}$ in terms of association stellar mass $M_{\rm cl}$. All of these properties agree well with the constraints from CR observations (e.g.\ detailed isotopic ratios) argued for in \citet{higdon:crs.accel.in.superbubbles.in.sne.ejecta.mass.dominated.regions,parizot:2004.superbubble.cr.accel.bulk,becker.tjus:2020.cr.multi.messenger.accel.regions}.

\begin{figure}
	\includegraphics[width=0.98\columnwidth]{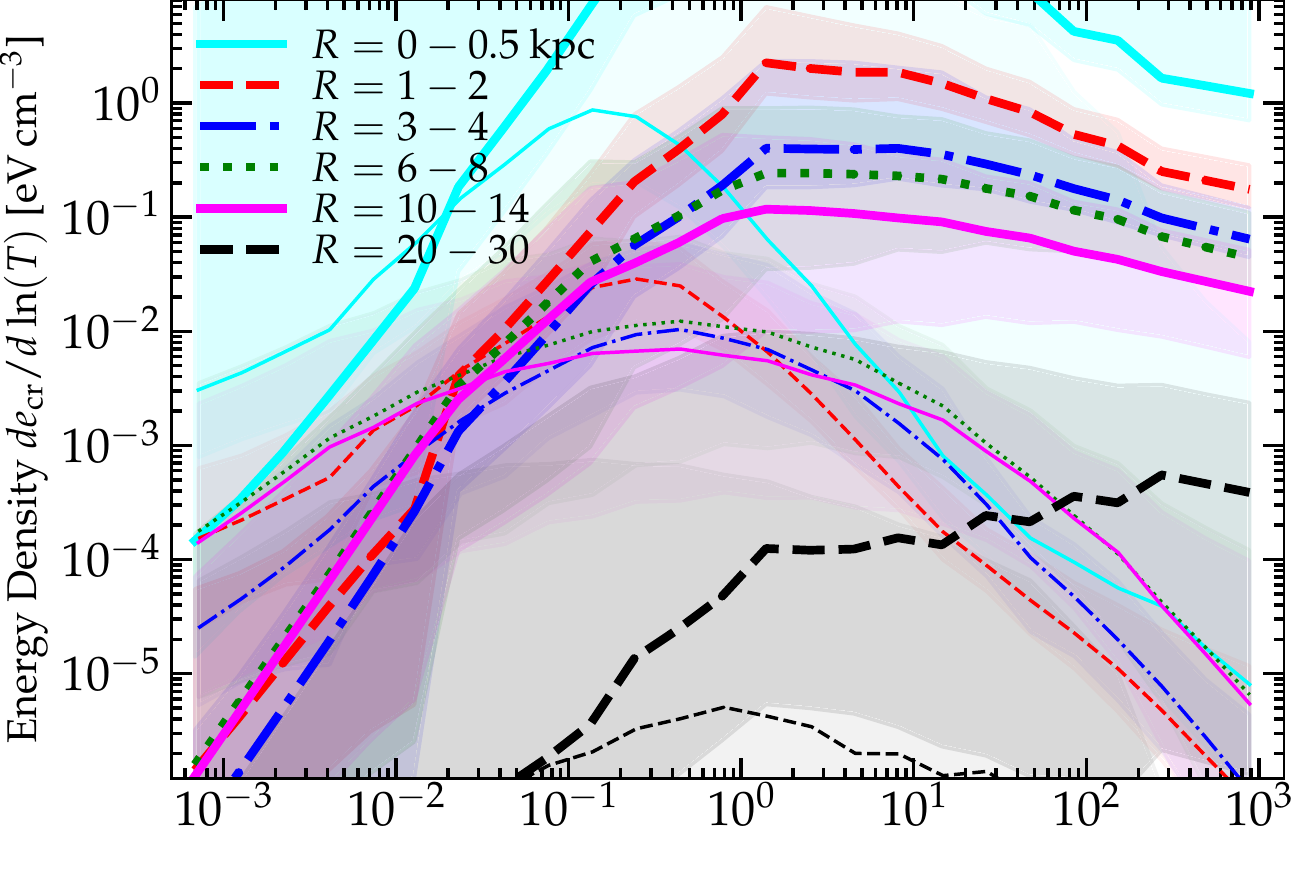} \\ 
	\includegraphics[width=0.98\columnwidth]{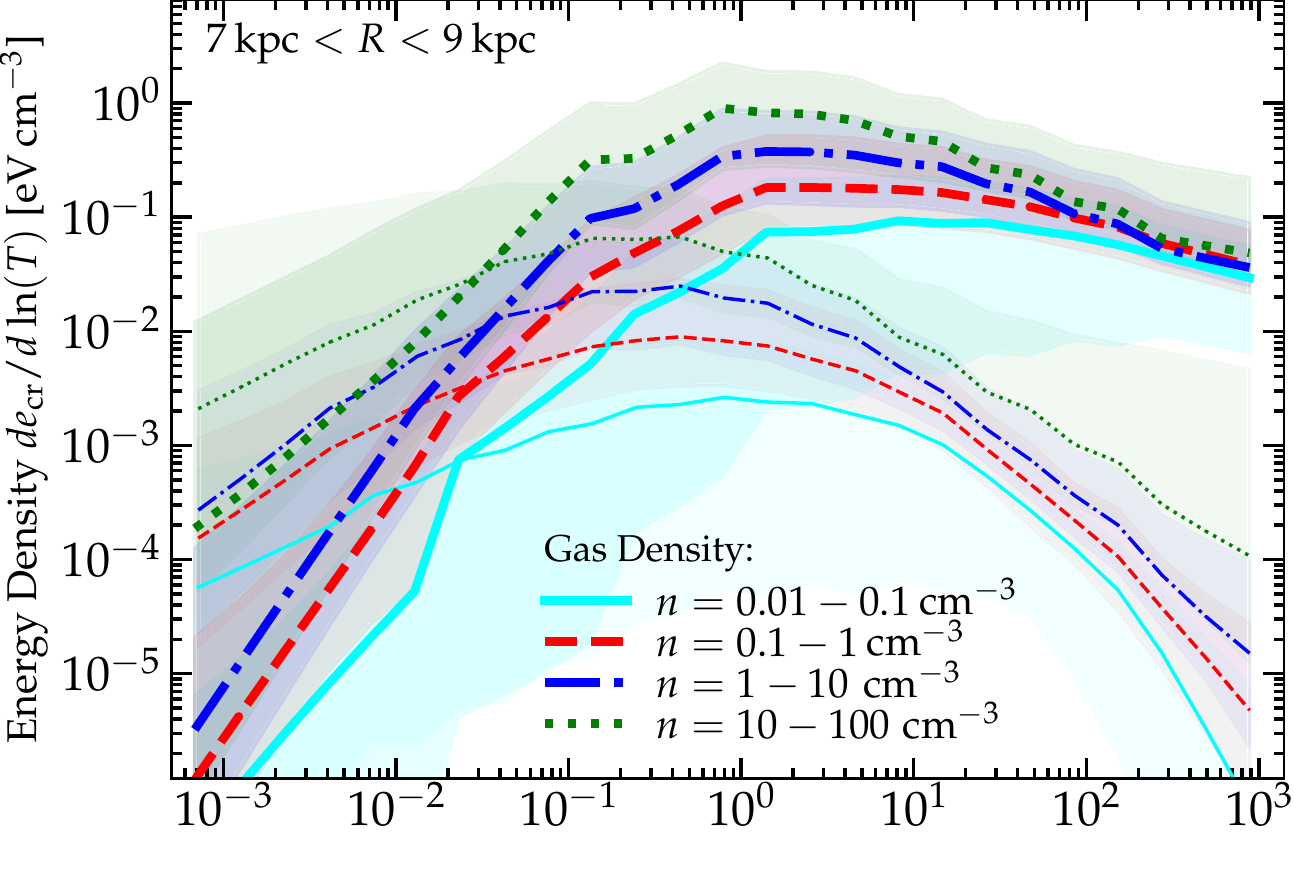} \\ 
	\includegraphics[width=0.98\columnwidth]{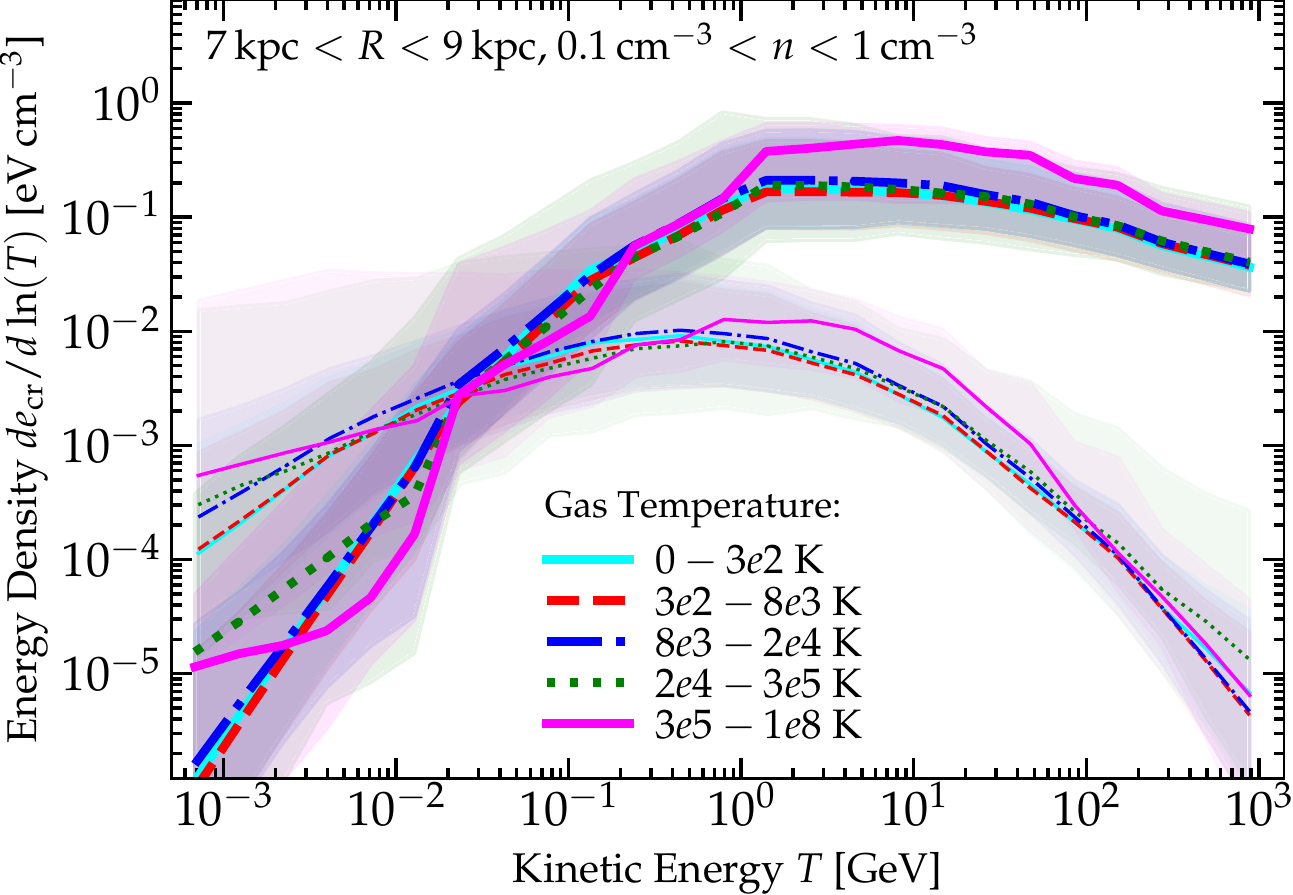} \\ 
	\vspace{-0.4cm}
	\caption{Comparison of CR energy spectra ($p$, {\em thick}, and $e$, {\em thin}) vs.\ environmental properties, as Fig.~\ref{fig:demo.cr.spectra.fiducial}.
	{\em Top:} Spectra vs.\ galacto-centric radius (allowing for the entire range of gas densities at each $R$). The CR spectra depend strongly on $R$, with higher normalization (tracing higher source density+proximity) as $R\rightarrow 0$, but also harder intermediate-energy hadronic spectra \&\ shifted leptonic, owing to more rapid losses.
	{\em Middle:} Spectra vs.\ gas density $n$ at fixed galacto-centric radii $R=7-9\,$kpc. 
	At fixed $R$, there is still significant dependence on $n$. High-energy CRs, which diffuse rapidly, exhibit weaker variations. Low-energy CRs exhibit dramatic systematic dependence, as their lower diffusivity leads to ``trapping'' in dense gas.
	{\em Bottom:} Spectra vs.\ gas temperature at fixed $n=0.1-1\,{\rm cm^{-3}}$ and $R=7-9\,$kpc. After controlling for $n$ and $R$, there is little {\em systematic} dependence on temperature or other phase properties (ionization state, magnetic field strength), but there is still substantial variation in low-energy CR spectra, especially in hot gas, reflecting the stochastic nature of SNe super-bubbles.
	\label{fig:env.compare.spectra}}
\end{figure}

\begin{figure}
	\includegraphics[width=0.98\columnwidth]{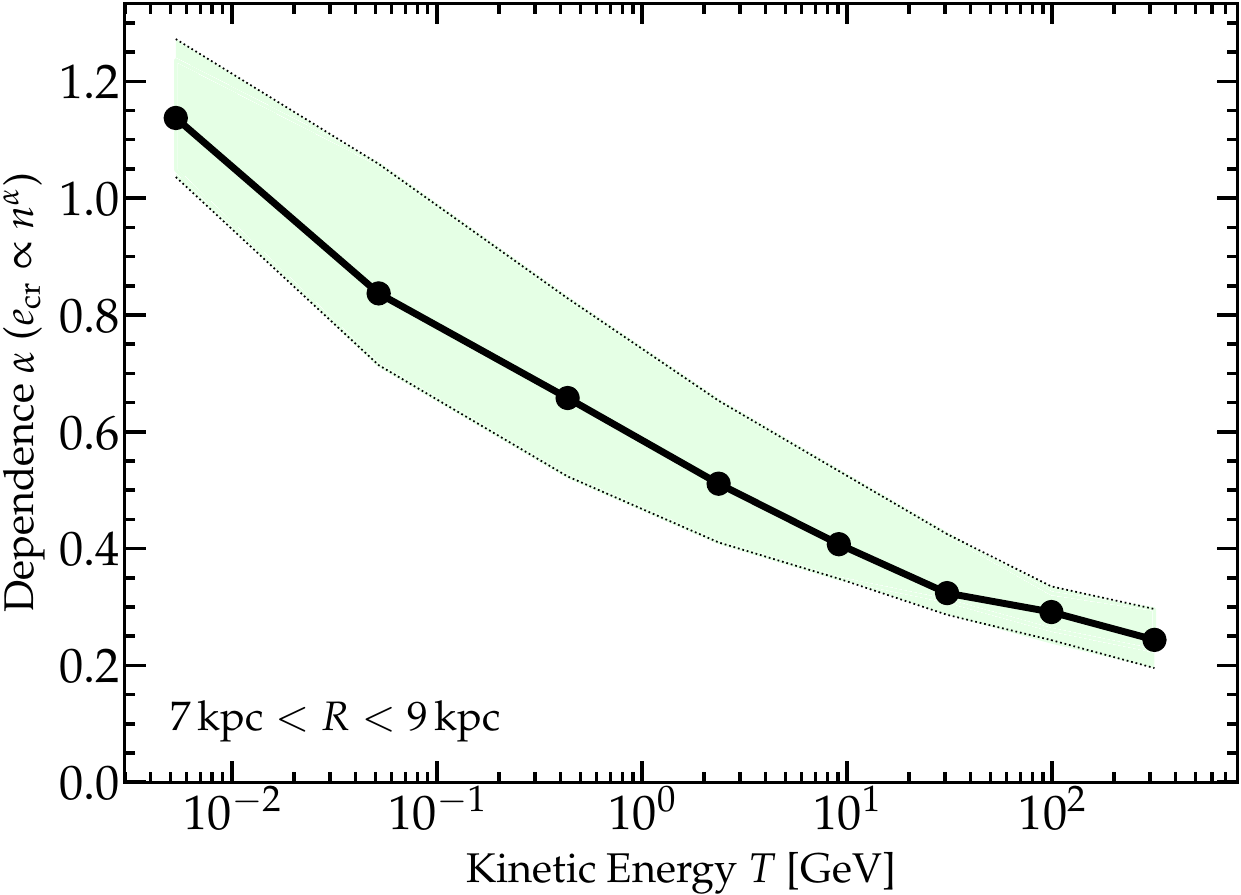}  
	\vspace{-0.15cm}
	\caption{Correlation between CR kinetic energy density in bins of CR energy/momentum/rigidity ($de_{\rm cr}/d\ln{T}$) and gas density ($n$), in gas at the Solar circle ($R=7-9$\,kpc). For CR protons in a narrow interval of energy at these radii, we fit the correlation $de_{\rm cr}/d\ln{T} \propto n^{\alpha}$ (as in Fig.~\ref{fig:egy.corr.density}) to a power-law slope $\alpha$, and show the best fit $\alpha$ (line) with the $\pm1\sigma$ range (shaded), as a function of CR energy $T$. This can also be considered an ``effective adiabatic index'' $d P_{\rm cr}/d\ln{T} \propto \rho^{\alpha}$ at each $T$. Low-energy CRs approach the adiabatic relativistic tight-coupling limit $\alpha \rightarrow 4/3$ (slightly lower, as losses also increase at high-$n$ for low-energy protons, and offset the adiabatic increase in $e_{\rm cr}$). High-energy CRs being more diffusive become more spatially-uniform with $\alpha\rightarrow0$.
	\label{fig:egy.corr.density.slope.vs.rigidity}}
\end{figure}

\begin{figure}
	\includegraphics[width=0.98\columnwidth]{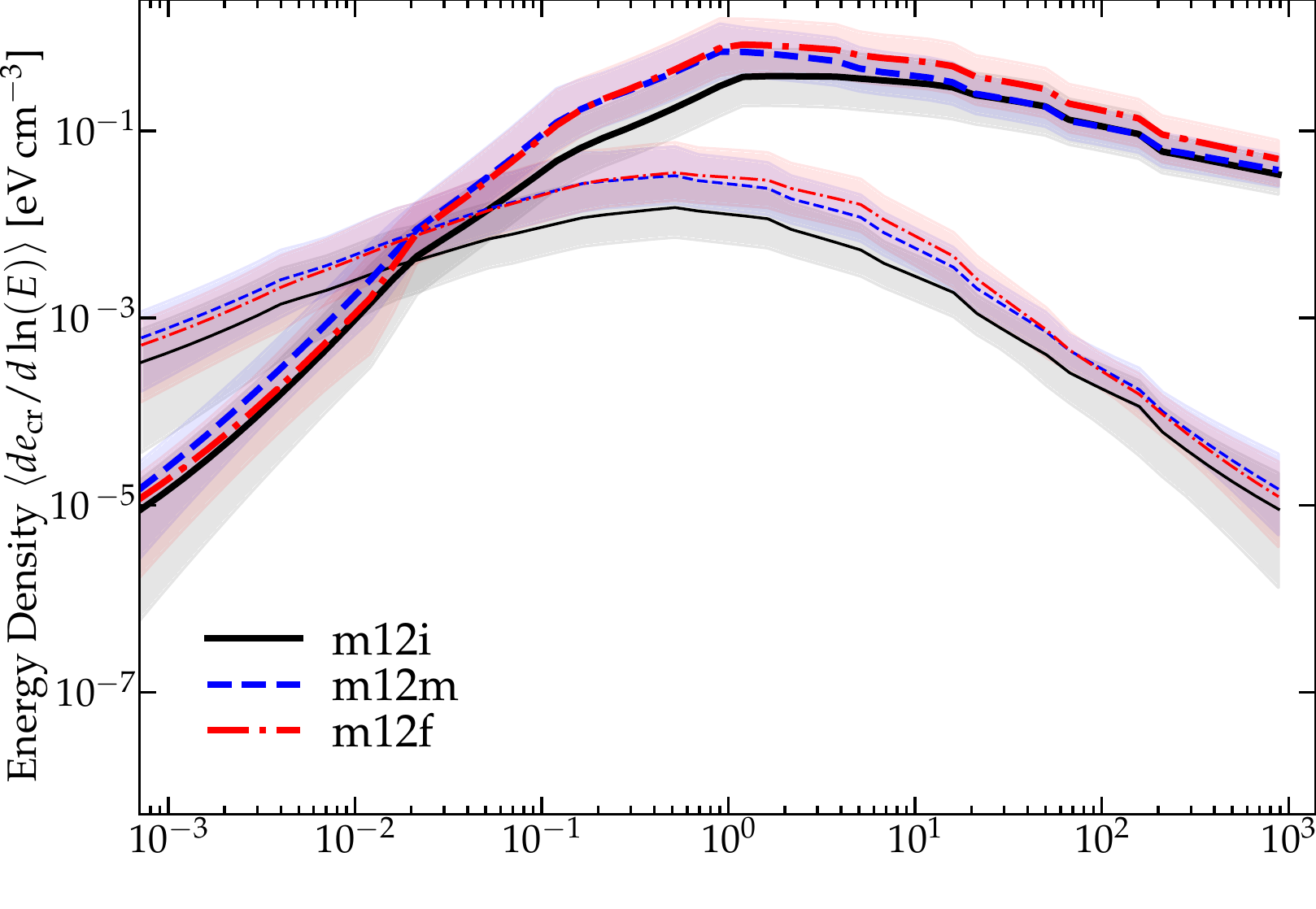} \\
	\hbox{\hspace{0.18cm}\includegraphics[width=0.96\columnwidth]{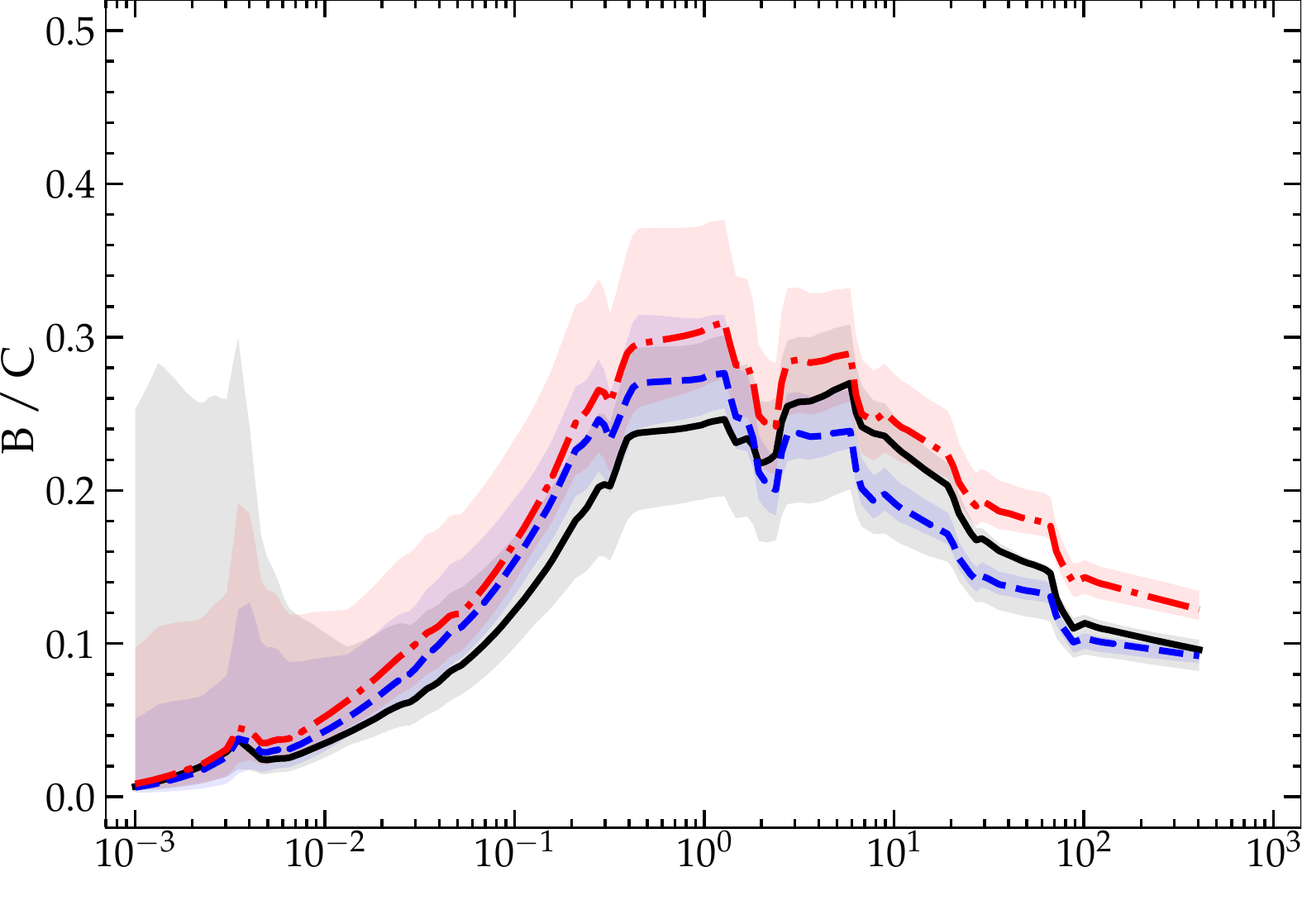}} \\ 
	\includegraphics[width=0.98\columnwidth]{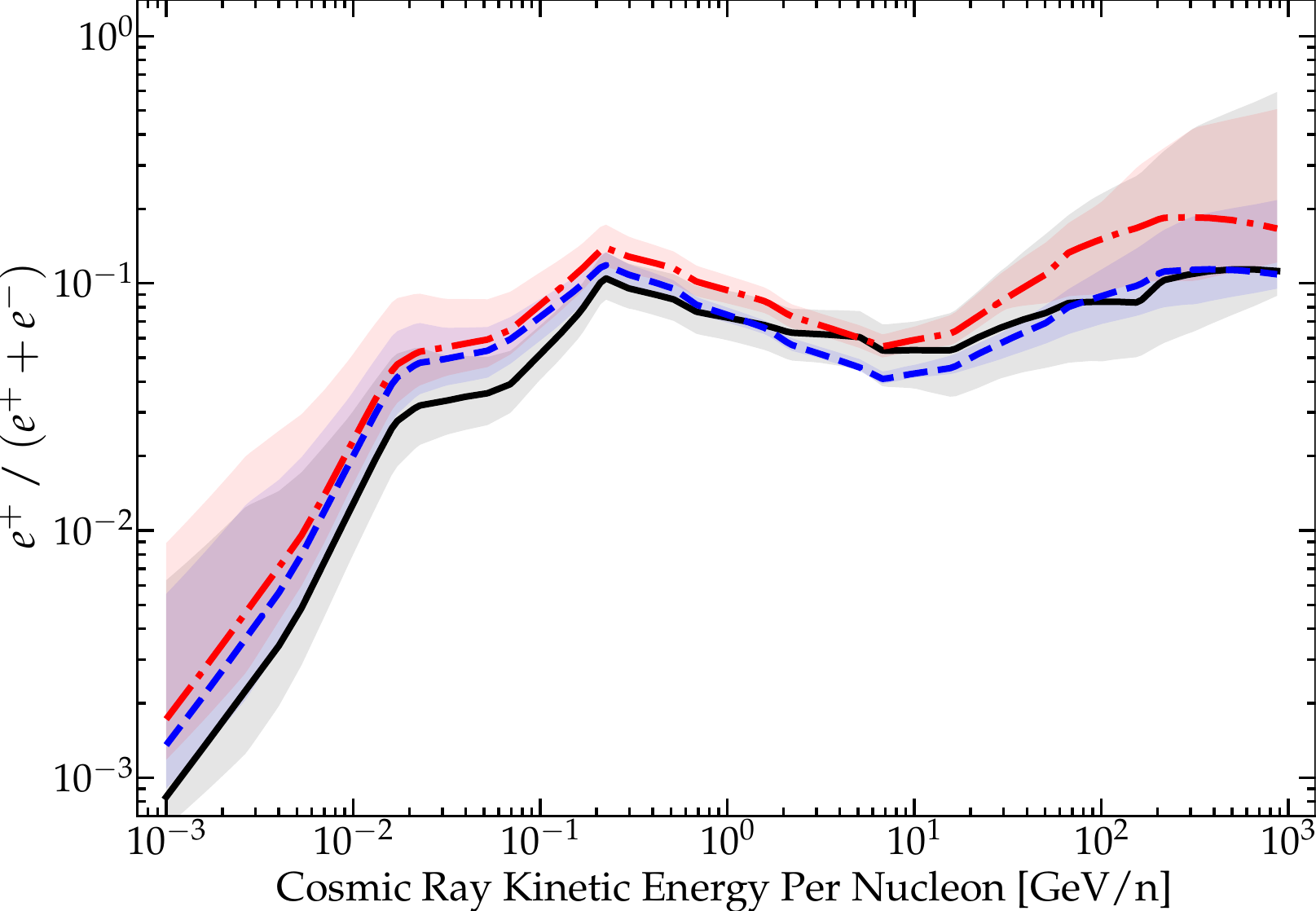} \\
	\vspace{-0.1cm}
	\caption{CR spectra in different Milky Way-like galaxies (as Fig.~\ref{fig:spec.compare.kappa}). We select three of the most ``Milky Way-like'' galaxies ({\bf m12i}, {\bf m12f}, {\bf m12m}) from cosmological simulations in our large FIRE suite \citep{garrisonkimmel:local.group.fire.tbtf.missing.satellites}, and compare CR spectra for each, restricted to Solar circle ($R=7-9\,{\rm kpc}$) and LISM-like densities ($n=0.1-1$). Differences in detailed galaxy properties (e.g.\ exact spatial distribution of star formation, disk thickness, etc) do not qualitatively change any of our conclusions, but can quantitatively shift our predicted spectra by amounts similar to their $\sim1\,\sigma$ scatter at a given $(R,\,n)$ -- an amount much larger than the LISM observations' statistical error bars (compare Fig.~\ref{fig:demo.cr.spectra.fiducial}). We therefore caution against over-interpreting detailed features at this level.
	\label{fig:spec.compare.galaxies}}
\end{figure}

\subsection{Variations Across and Within MW-Like Galaxies}
\label{sec:variation}

\subsubsection{Variation With Galactic Environment}
\label{sec:variation:env}

In Fig.~\ref{fig:env.compare.spectra} we note that the {\em variation} in the CR spectra can be significant, and quantify some of these variations and their dependence on the local Galactic environment. We have examined how the spectra vary as a function of local ISM properties including: galacto-centric radius, height above the midplane, gas density, temperature, ionization fraction, local inflow/outflow velocity, magnetic energy density, radiation energy density, turbulent dissipation rate, star formation or SNe rate per unit volume, plasma $\beta$, and other properties. For most of these, there is some correlation with CR spectra, but it is important to remember that all of these parameters are themselves mutually correlated within a galaxy; as a result most of the systematic variation with the properties above can be captured by the dependence on galacto-centric radius $r$ and local gas density $n$, shown explicitly in Fig.~\ref{fig:env.compare.spectra}.

The dependence on galacto-centric $r$ in Fig.~\ref{fig:env.compare.spectra} (even at fixed $n$, $T$, $|{\bf B}|$, etc.) is easily understood: towards the galactic center, the source density is higher, and assuming CRs are efficiently escaping, steady-state requires that the kinetic energy density be some declining function of galacto-centric radius $r$ \citep[see][]{hopkins:cr.mhd.fire2}. Specifically, $e_{\rm cr} \propto \dot{E}_{\rm cr}(<r)/2\pi\,D_{xx}\,r$ for a spatially-constant isotropic-equivalent diffusivity $D_{xx} \sim \kappa_{\|}/3 \sim v^{2}/9\,\bar{\nu}$ (where $\dot{E}_{\rm cr}$ is the injection rate, proportional to the SNe rate) if losses are negligible. This scaling provides a reasonable description of what we see outside of a few hundred pc, in fact, and the CR kinetic energy density therefore drops by an order of magnitude (or more) between the equivalent of the Galactic Central Molecular Zone ($r \lesssim 400\,$pc) and Solar neighborhood/LISM ($r\sim 8\,$kpc). The spectra towards the Galactic center are also shallower/harder in hadrons (at intermediate rigidities from $\sim 0.01-10\,$GV), while being steeper/softer in leptons: this owes to the fractional importance of losses. The Galactic center has much higher neutral gas/nucleon/radiation/magnetic energy densities overall, so the loss rates are all enhanced: this makes the hadronic spectra (where loss rates increase at low energies) shallower/harder and leptonic spectra (where loss rates increase at high energies) steeper/softer.  

At a {\em given} galacto-centric $r$, e.g.\ at the Solar circle ($r\sim 8\,$kpc),\footnote{Note that the galactocentric radii in Fig.~\ref{fig:env.compare.spectra} are spherical radii, so some of the variation with density reflects gas at different scale-heights; we consider in more detail the difference between the variation along the disk midplane cylindrical radius versus vertical height above the disk below in \S~\ref{sec:gamma.rays}.} there is much less variation, but there is still a significant systematic dependence on the local gas density $n$, seen also in Fig.~\ref{fig:env.compare.spectra}. Higher-density environments have higher CR energy density: again this is qualitatively unsurprising, since (a) in the ``tight coupling'' limit $e_{\rm cr} \propto n^{4/3}$, (b) even without tight-coupling, the ``adiabatic'' re-acceleration term is typically positive in denser regions, and typically negative in low-density regions, and (c) denser regions are positively correlated with CR sources (e.g.\ SNRs). Figure~\ref{fig:egy.corr.density.slope.vs.rigidity} quantifies how CR energy density/pressure scales with gas density, as a function of CR energy or rigidity, giving a quantitative indication of ``how tightly coupled'' CRs are to gas (which is also crucial for understanding how CRs do or do not modify thermal behaviors of the gas such as the classical thermal instability; see \citealt{2020arXiv200804915B}). At the highest energies for hadrons, the effect is negligible, owing to fast diffusion (the CR density is basically un-coupled from the gas density), while at the lowest energies the effect is strongest (the small diffusivity produces tight coupling). Thus the net effect is that lower-density regions {\em at a given galacto-centric radius} have slightly harder CR spectra with lower total energy density (opposite the effect with galacto-centric radius). The effect at the highest densities for leptons is a bit more complicated owing to the non-linear effects of IC and synchrotron losses. As discussed below, this has important implications for CR ionization.

Even controlling for galactocentric radius and gas density, there is still large variation in the total CR kinetic energy density, with the $90\%$ inclusion interval  ($-2\,\sigma$ to $+2\,\sigma$ range) spanning nearly $\sim 3\,$dex ($\sim 0.7\,$dex $1\,\sigma$ range) at the lowest energies or for certain species (with smaller $\sim 0.2\,$dex variation in the energy density of e.g.\ high-energy protons where diffusion is rapid and losses minimal). This scatter does not primarily come from a systematic dependence on a third variable that we can identify (e.g.\ any of the variables noted above) -- the residual dependence on e.g.\ gas temperature, etc., in Fig.~\ref{fig:env.compare.spectra} is minimal. Rather, this appears to owe mostly to effectively stochastic variations in space and time: the combination of e.g.\ multiple second-order/weak correlations, how many SNe recently occurred in a given vicinity and how strongly-clustered they were in both space and time, the recent turbulent or gravitational collapse history (how strong and with what sign and on what spatial scale the net recent re-acceleration terms acted), the local turbulent magnetic field geometry relative to the large-scale CR gradients, etc.

\begin{figure*}
\begin{tabular}{r@{\hspace{2pt}}r@{\hspace{0pt}}r}
	\includegraphics[width=0.48\textwidth]{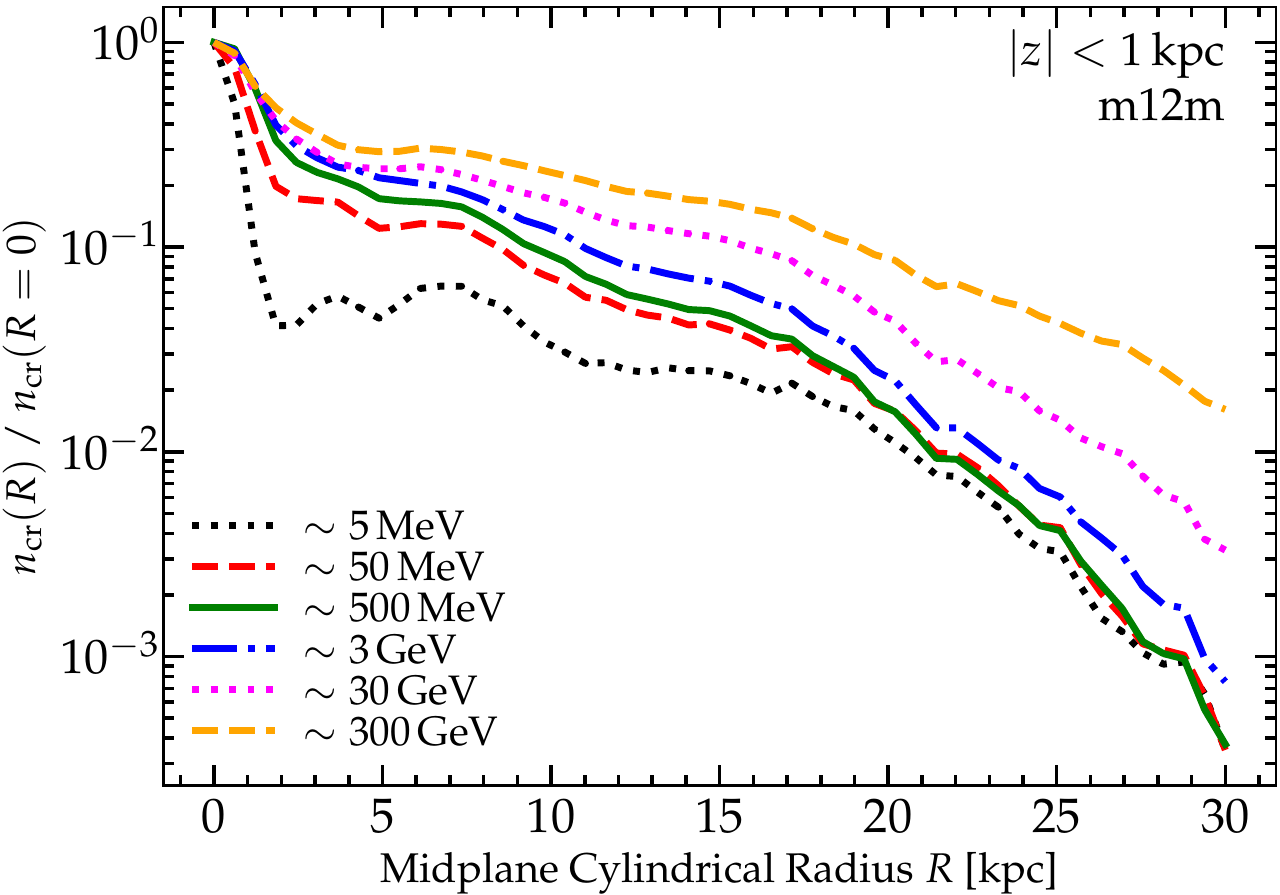}&
	\includegraphics[width=0.48\textwidth]{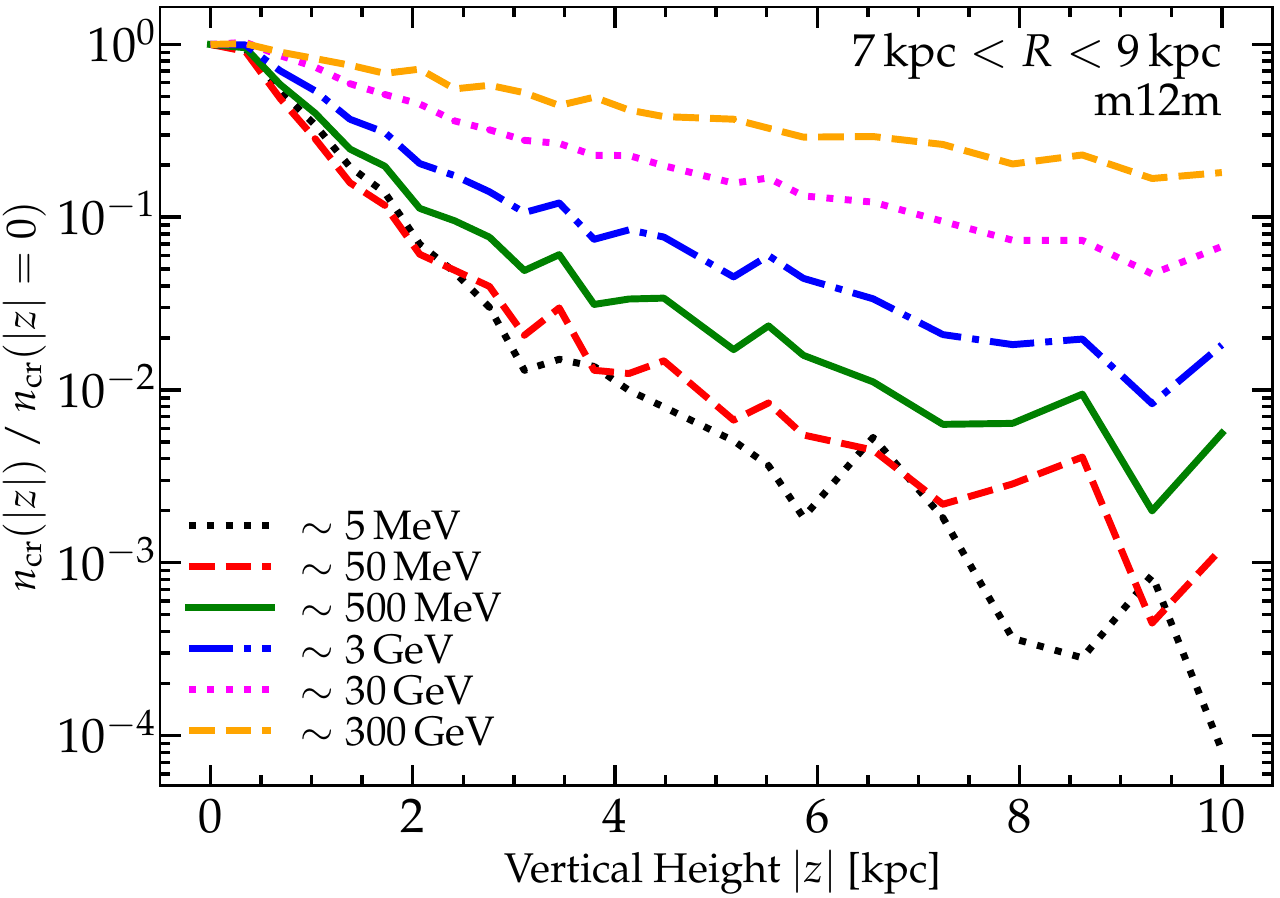}
	\\
	\includegraphics[width=0.48\textwidth]{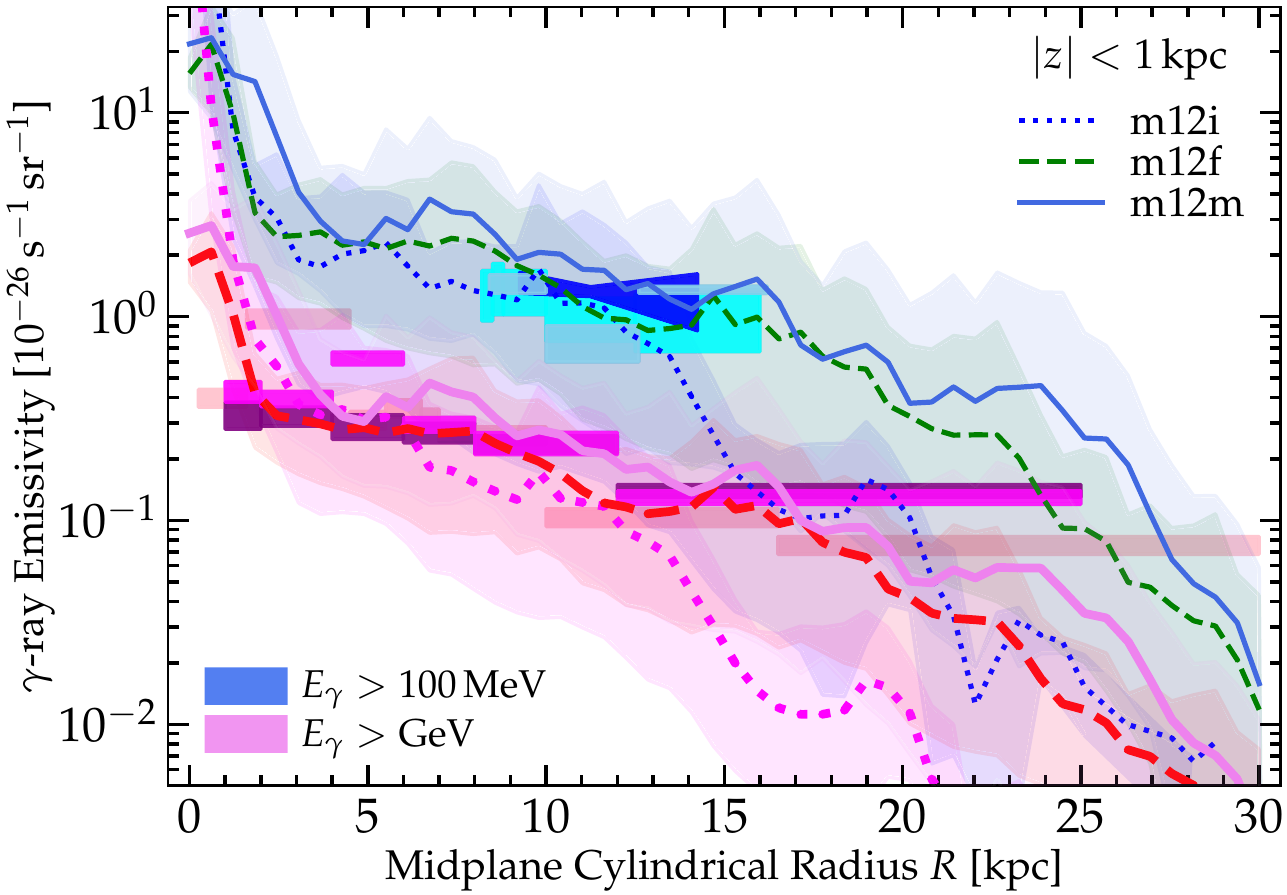} &
	\includegraphics[width=0.48\textwidth]{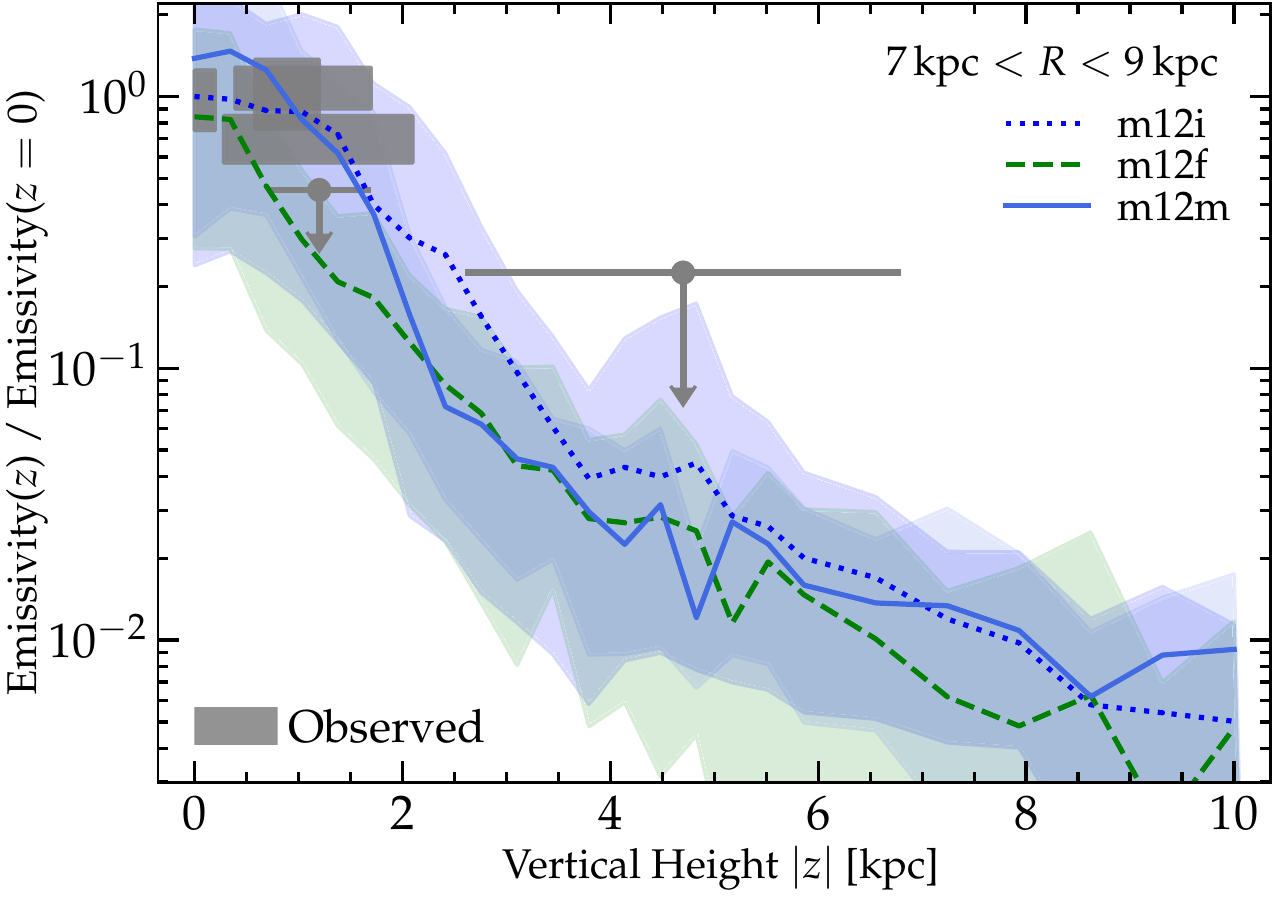}
\end{tabular}
	\vspace{0.05cm}
	\caption{CR properties versus Galactic spatial position in detail. 
	{\em Top:} Mean (volume-averaged) number density of CR protons in cylindrical midplane (height $|z|<1$\,kpc) radial annuli of radius $R$ ({\em left}), and as a function of vertical height within the Solar circle ($7\,{\rm kpc}<R<9\,{\rm kpc}$; {\em right}), in narrow bins of CR energy or momentum (labeled). The profiles are normalized at $(R,\,z)=0$ to compare all energies easily. We consider galaxy {\bf m12m} because the large bar/disk warp in {\bf m12i} at $R\sim 10\,$kpc (see Fig.~\ref{fig:images}) complicates the geometric structure, but trends are similar in {\bf m12i} and {\bf m12f}. Vertical profiles within a few kpc of the Solar circle are close to exponential with scale-height increasing from $\sim 1-10\,$kpc for $\sim5$\,MeV to $\sim1$\,TeV, owing to increasing CR diffusivity. The midplane profile has more complicated structure and less clear separation owing to the extended source distribution but is qualitatively similar. 
	{\em Bottom:} CR $\gamma$-ray emissivity (calculated from the CR spectra; $5-95\%$ range light shaded, with lines showing the $\gamma$-ray emission-weighted mean) versus midplane $R$ ({\em left}) or height at the Solar circle ({\em right}). We compare our three different Milky Way-like galaxies (Fig.~\ref{fig:spec.compare.galaxies}) to observations compiled from EGRET \citep{digel.2001:diffuse.gamma.ray.cr.profile.constraints} and Fermi-LAT \citep{ackermann.2011:diffuse.gamma.ray.cr.profile.constraints,tibaldo.2014:diffuse.gamma.ray.cr.profile.constraints,tibaldo.2015:diffuse.gamma.ray.cr.profile.constraints,tibaldo.2021:diffuse.gamma.ray.cr.profile.constraints,acero:2016.gamma.ray.constraints.cr.emissivity,yang.2016:diffuse.gamma.ray.cr.profile.constraints}, shown as the dark shaded bands (with $1\sigma$ upper limits in the vertical profile denoted with an arrow). We compare predictions and measurements both for lower photon energies $>100\,$MeV (thin blue/green higher values at {\em left}; grey at {\em right}) and $>1\,$GeV (thick pink/red lower values at {\em left}). 
	\label{fig:profiles.r.z}\vspace{-0.4cm}}
\end{figure*}

\subsubsection{Variation with Time}
\label{sec:variation:time}

We can also examine the variations of CR spectra with time in a given galaxy. We can do this either by simply taking the CR spectra in different regions as above (\S~\ref{sec:variation:env}) at different snapshots near $z\sim 0$ (separated by some $\Delta t$ -- we typically output $\sim 100$ snapshots spaced over the time  range simulated), or, because our code is quasi-Lagrangian, we can explicitly ask how the CR spectrum seen by an individual Lagrangian gas parcel varies in time. Comparing simulation snapshots at different times before $z=0$, we find that on sufficiently small timescales $\Delta t \ll t_{\rm dyn}$, smaller than the galaxy dynamical time ($\sim 100\,$Myr), the variations are small as expected: the galaxy and source distribution and bulk ISM properties by definition evolve on slower timescales and even if the CR escape/loss timescales are more rapid, they simply converge to quasi-steady-state (for fiducial simulation, the differences in median spectra between snapshots are comparable to or smaller than the differences between e.g.\ the median and mean curves shown in Fig.~\ref{fig:demo.cr.spectra.fiducial}). On very large timescales $\Delta t \sim t_{\rm H} \gtrsim {\rm Gyr}$ of order the Hubble time, we are asking an effectively different question (how CR spectra vary with cosmic time, or as a function of redshift), and the galaxy is fundamentally different: this is a key question for understanding e.g.\ the FIR-radio correlation but is qualitatively different from what we seek to understand here and we require fully-cosmological simulations run for a Hubble time to address it, so we defer this to future work. On intermediate timescales $t_{\rm dyn} \lesssim \Delta t \lesssim {\rm Gyr}$, the variations in time at a given $r$, $n$ are largely effectively stochastic, with a similar amplitude to the differences between different galaxies in Fig.~\ref{fig:spec.compare.galaxies} or differences in position within a galaxy -- i.e.\ the statistics of the CR spectra are effectively ergodic. In principle there are some events (e.g.\ a starburst, large galaxy merger, etc.) which could cause a substantial deviation from this, but recall our galaxies our chosen for their MW-like properties which means we specifically selected systems without such an event between $z\sim 0-0.1$ (so we would require dedicated simulations of different galaxy types to explore this).

\subsubsection{Variation Across MW-Like Galaxies}
\label{sec:variation:galaxy}

Next compare variation across different MW-like galaxies. Fig.~\ref{fig:spec.compare.galaxies} specifically compares our {\bf m12i}, {\bf m12f}, and {\bf m12m} halos, at the same cosmic time: these three halos produce arguably the three ``most Milky Way-like'' galaxies in the default FIRE suite, and all have been extensively compared to each other and to different MW properties in previous studies \citep[see][]{ma:2015.fire.mass.metallicity,elbadry:fire.morph.momentum,elbadry:HI.obs.gal.kinematics,sanderson:stellar.halo.mass.vs.fire.comparison,bonaca:gaia.structure.vs.fire,gurvich:2020.fire.vertical.support.balance,guszejnov:fire.gmc.props.vs.z,garrison.kimmel:2019.sfh.local.group.fire.dwarfs,benincasa:2020.gmc.lifetimes.fire,samuel:2020.plane.of.satellites.fire}. There are some non-negligible differences in detail between the systems, many apparent in Fig.~\ref{fig:images}: for example, in this particular set of runs, {\bf m12f} has a slightly higher mass and more extended gas+stellar disk, with a hotter CGM halo; {\bf m12i} has a slightly more rising star formation history in its outskirts (making them bluer) and a strong bar which induces a warp and a slight central ``cavity'' in the SFR, akin to the MW central molecular zone \citep[see][]{orr:2021.fire.cmz.analog}; {\bf m12m} has a less centrally-peaked rotation curve, and a pseudobulge driven by bar-buckling \citep{debattista:fire.bar.galaxy.m12m}. But given the large uncertainties in characterizing the MW's star formation history and present-day spatial distribution of star formation over the entirety of the Galactic disk (let along the disk+halo gas \&\ magnetic field structure across the entire galaxy, not just the Solar neighborhood), these are all broadly ``equally-plausible'' MW analogs \citep[for more detailed discussion, see][]{sanderson:gaia.mocks}. 

We see systematic differences between the three galaxies which are comparable to the dispersion in CR spectra within a galaxy. Even though these are all very similar (relative to e.g.\ other $\sim L_{\ast}$ galaxies of much earlier or later type), the detailed differences above produce some systematic differences in the spatial distribution and degree of clustering of SNe, the local PDF of gas and radiation and magnetic field energy densities (and therefore losses and secondary production rates), and other related quantities. This goes further to show that differences in the {\em details} of the local structure of the ISM are crucial for interpreting CR spectra at better than order-of-magnitude level.

\subsubsection{Comparison with $\gamma$-ray Observations}
\label{sec:gamma.rays}

Fig.~\ref{fig:profiles.r.z} further compares the variation of CR properties with Galactic environment to constraints from diffuse Galactic  $\gamma$-ray emission. First, we show the spatial dependence from Fig.~\ref{fig:env.compare.spectra} in more detail, specifically plotting the CR proton number or energy density in narrow bins of CR energy or rigidity as a function of Galactic position. We decompose the spherical radial trend from Fig.~\ref{fig:env.compare.spectra} into cylindrical coordinates, comparing the CR profile in the disk midplane versus cylindrical galactocentric radius, and at the Solar circle ($\sim 8$\,kpc cylindrical) as a function of vertical height above the disk. We see the expected trend: owing to less-efficient diffusion, lower-energy CR protons have shorter radial and vertical scale-lengths. In the vertical direction (again, {\em at the Solar circle}), the CR profiles are approximately exponential ($n_{\rm cr}(E_{\rm cr},\,|z|) \propto \exp{(-|z|/h[e_{\rm cr}])}$) within a few kpc of the disk,\footnote{In Fig.~\ref{fig:profiles.r.z} we see that there is, especially at low energies, a steeper initial falloff at small $|z|$ followed by a somewhat shallower vertical profile. This reflects the fact that the quasi-exponential vertical profile transitions to a more quasi-spherical, power-law profile in the CGM, at sufficiently large radii where the disk/source geometry is no longer important.} and the vertical scale-height increases systematically with CR energy, from $\sim 0.5-1$\,kpc at $\sim1-5\,$MeV to $\sim 2\,$kpc at  $\sim 1-3\,$Gev to $\sim 6-10\,$kpc at $\sim 0.3-1\,$TeV. In the midplane radial direction the qualitative trends are similar but the profiles deviate more strongly from a single exponential and differ from one another more weakly as a function of energy, owing to the continuous distribution of sources populating the disk and much larger range of galactocentric scales considered. This confirms, however, that there is an extended CR halo, whose size increases as a function of CR energy. The dependence of profile shape on energy is generally weaker at low energies ($\lesssim$\,GeV), owing to the fact that low-energy hadrons (with low diffusivity) have residence times increasingly dominated by losses.

Fig.~\ref{fig:profiles.r.z} compares to observations of the  inferred CR emissivity in $\gamma$-rays at energies $E_{\gamma} > 100\,$MeV and $E_{\gamma} > 1\,$GeV. We calculate the emissivity directly from the CR spectra (including pion production from protons and heavier nuclei, as well as $e^{+}$ and $\bar{p}$ annihilation), with the same cross sections used in-code (see \S~\ref{sec:methods:crs} and e.g.\ \citealt{dermer:cr.gamma.rays.vs.protons}). We caution that there can be spatial variations in emissivity from e.g.\ nearby clouds  or Galactic structures even at a given galactocentric radius (see \citealt{ackermann:2012.fermi.obs.cr.emissivity.variation} and note that detections and upper limits in Fig.~\ref{fig:profiles.r.z} at the same distance range often differ in emissivity by much more than their statistical error bars), and that there are often very large distance uncertainties (which are themselves model-dependent) regarding where observed emission actually originates, so it is important to  compare the observed points and uncertainty range to the full range (shaded area) predicted. With that in mind, the simulations agree quite well with both the radial and vertical trends, at a range of different $\gamma$-ray/CR energies. Comparing our three different galaxy models, we note that there are some appreciable differences in the profiles especially in the galactic nucleus (which is sensitive to the instantaneous state of the galaxy, e.g.\ whether there has been a recent nuclear starburst, while the Milky Way appears to be in a period of quiescence; \citealt{orr:2021.fire.cmz.analog}), and at large radii (where the less-extended star forming disk in {\bf m12i} owing to its bar and warped disk structure noted in Fig.~\ref{fig:images}  leads to a more rapid falloff at $\gtrsim15\,$kpc).

Note that the above applies to protons (other hadrons are similar). The electrons (leptons) behave differently, however, owing to the more complicated role of losses. Considering the electron vertical profiles at the Solar circle, for example, we see the $e^{-}$ scale-height decreases with increasing energy weakly (by $\sim 20\%$) from $1\rightarrow 50\,$MeV where diffusivities are very low so transport is dominated by advection and streaming, then increases with energy (by a factor of $\sim2$ from $50\,$MeV to $50\,$GeV) at intermediate energies where losses are not dominant (outside the disk) and diffusivity increases with energy, then at $\gtrsim 50-100\,$GeV the scale height starts to drop significantly with increasing energy owing to rapid inverse Compton and synchrotron losses even in the halo.

We have also compared the variation of the mean predicted spectral index ($\alpha_{\gamma}$ in $I \propto E^{-\alpha_{\gamma}}$) of the emissivity as a function of midplane radius $R$ to FERMI observational estimates from \citet{acero:2016.gamma.ray.constraints.cr.emissivity} and \citet{yang.2016:diffuse.gamma.ray.cr.profile.constraints}. Interestingly, at energies $\sim 3-30\,$GeV which dominate the fitted indices in those observational studies, our {\bf m12i} run at $z=0$ produces a trend very similar to that observed (wherein $\alpha$ peaks with a very steep/soft value at $\lesssim 2\,$kpc in the galactic center, then falls rapidly by $\Delta \alpha_{\gamma} \sim 0.4-0.6$ with increasing $R$ to shallower/harder values from $R\sim 2-6\,$kpc, then gradually increases/steepens again by $\Delta\alpha_{\gamma} \sim 0.3$ out to $R\sim 10-20\,$kpc). However, our {\bf m12m} and {\bf m12f} runs at $z=0$ do not show the same trend (they show a weaker trend of $\alpha_{\gamma}$ with $R$, with occasionally opposite sign); moreover analyzing different snapshots shows this varies in time, as well. This is because the $\gamma$-ray emission (scaling as $\sim n_{\rm gas}\,n_{\rm cr}$) is sensitive to the densest emitting regions in each annulus, which have different spectral slopes at intermediate energies as shown in Fig.~\ref{fig:env.compare.spectra}; moreover as discussed in \citet{acero:2016.gamma.ray.constraints.cr.emissivity} this can also depend on variations in losses (as some dense regions reach calorimetric losses) and the structure of outflows (with advection modifying transport speeds). As such, these higher-order trends, while possible to reproduce, are sensitive to the instantaneous dynamical state of the galaxy.

\begin{figure}
	\includegraphics[width=0.98\columnwidth]{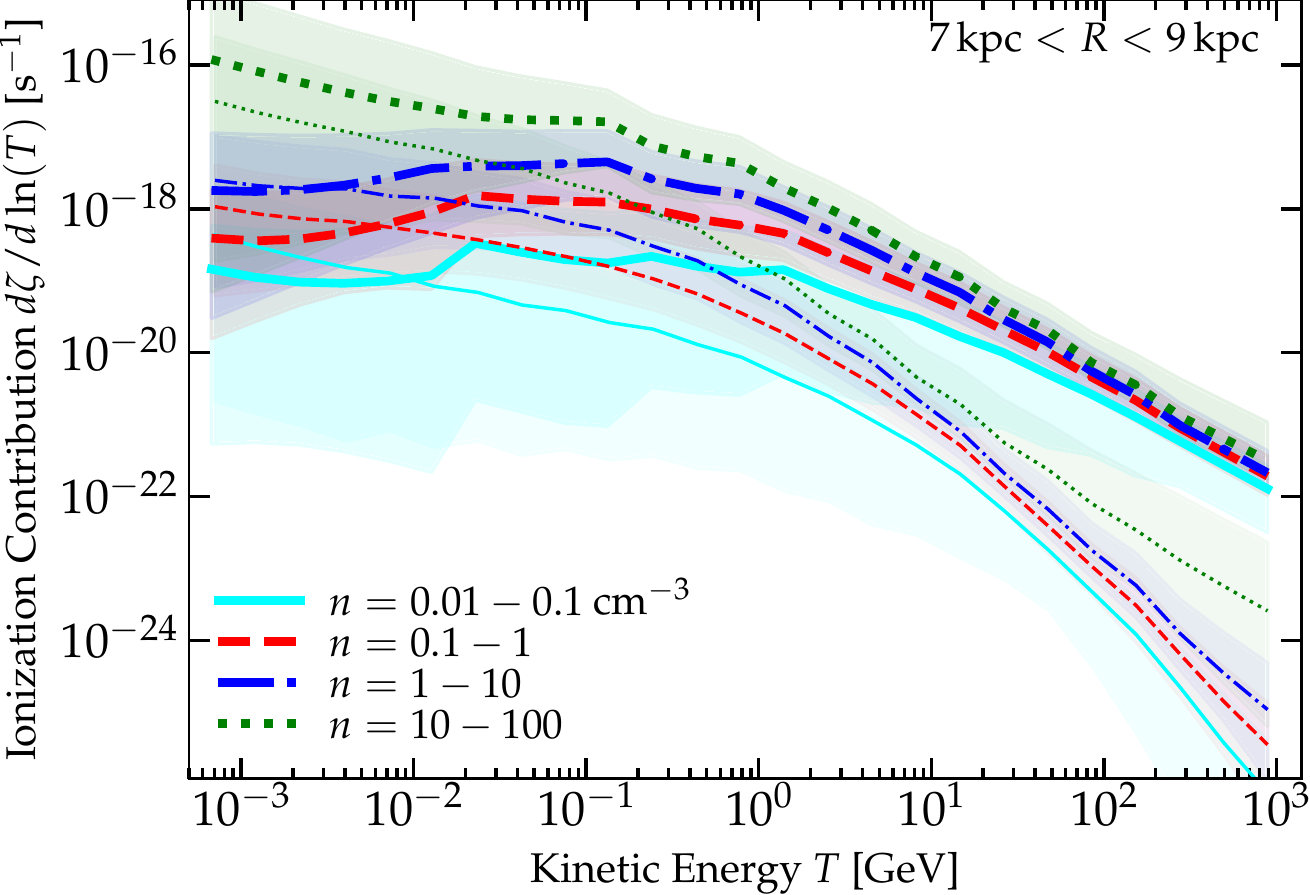} \\ 
	\includegraphics[width=0.98\columnwidth]{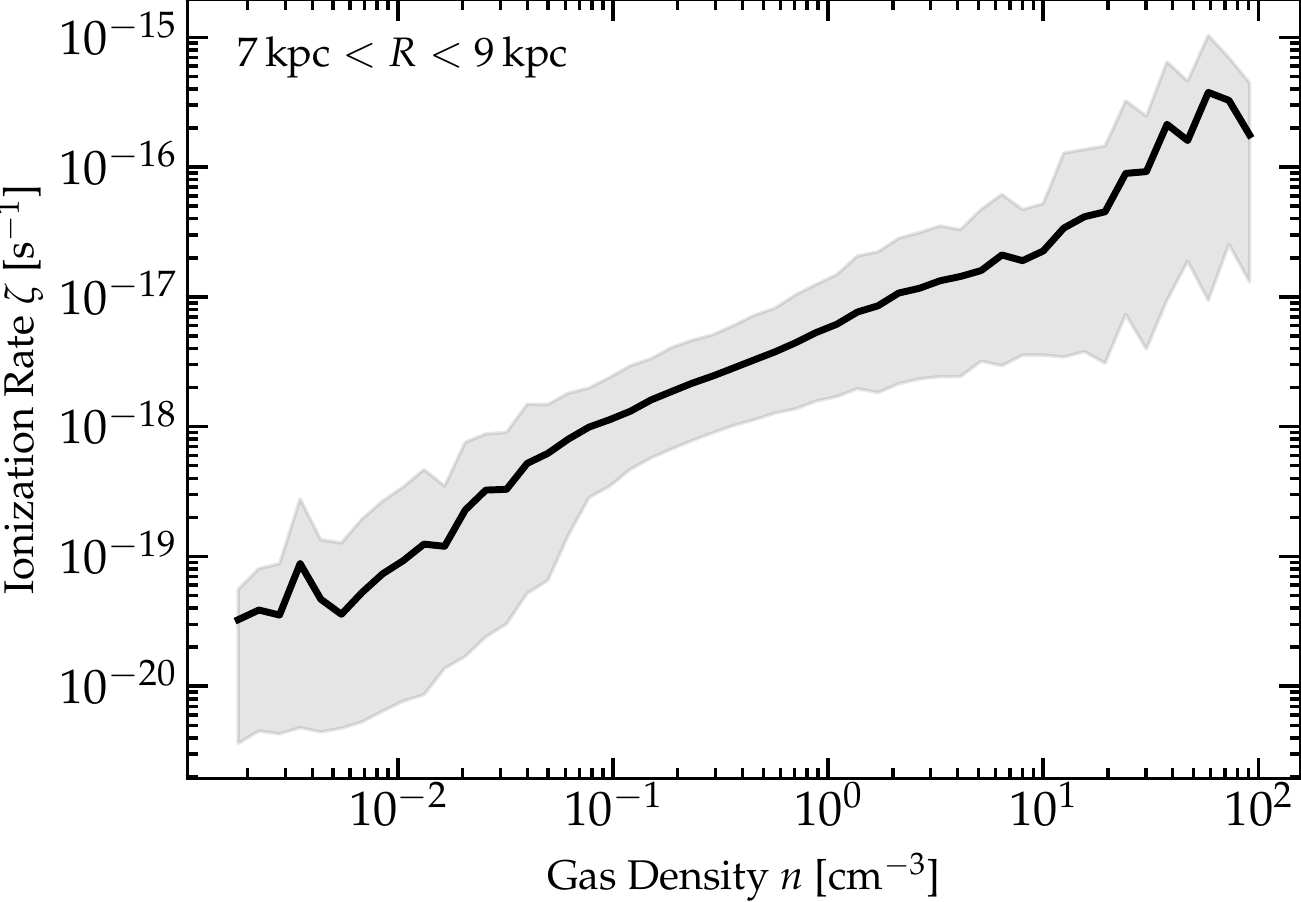} \\ 
	\includegraphics[width=0.98\columnwidth]{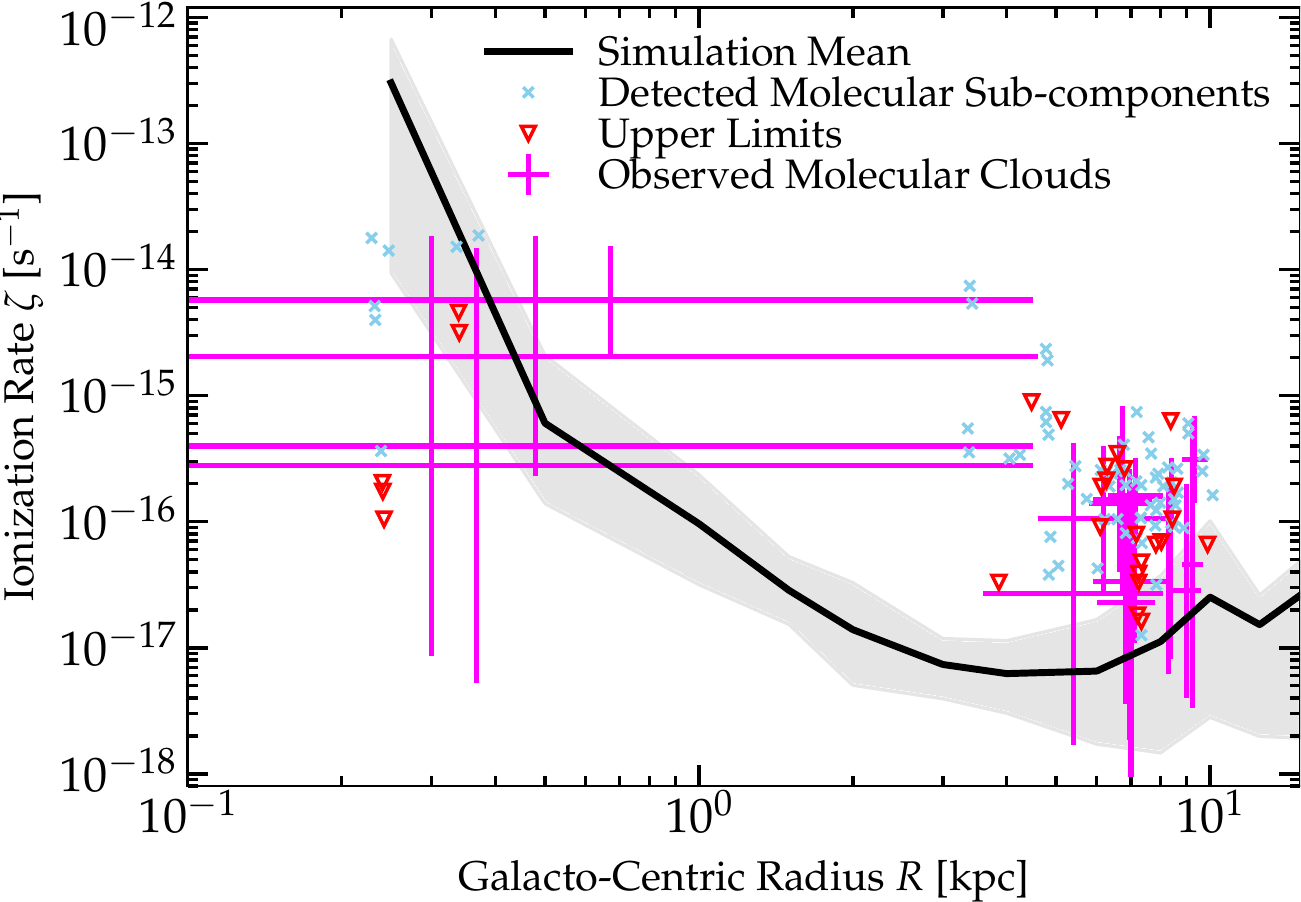} \\ 
	\vspace{-0.4cm}
	\caption{Comparison of CR ionization rates vs.\ environment. 
	{\em Top:} Differential contribution to the total CR ionization rate $\zeta$ from CRs with different kinetic energies, from $p$ ({\em thick}) and $e$ ({\em thin}; all other species are sub-dominant) in different gas densities $n$ at the Solar circle ($R=7-9$\,kpc). A broad range of energies contribute, with the lowest-energy CRs dominant in the most-dense gas. Usually $p$ dominate.
	{\em Middle:} Total ionization rate $\zeta$ (summing all CR energies and species) vs.\ gas density at fixed galacto-centric radius $R=7-9\,$kpc. Enhanced ionization in GMC environments (high densities, compared to LISM $n\sim 0.1-1\,{\rm cm^{-3}}$) arises naturally.
	{\em Bottom:} Ionization rate $\zeta$ vs.\ galacto-centric radius (weighted by total ionization rate of neutral gas). The enhanced nuclear CR densities lead to strongly-enhanced $\zeta$ as $R\rightarrow 0$. We compare observational estimates in \citet{indriolo:2015.cr.ionization.rate.vs.galactic.radius}, from dense GMC molecular line tracers. Error bars show the range of distance and inferred $\zeta$ for each cloud; points show individual upper limits and detections without error bars treating each velocity sub-channel of each cloud as a separate system and using kinematic models to place it at its own distance.
	\label{fig:env.compare.ion}}
\end{figure}

\subsubsection{Implications for the CR Ionization Rate}

A number of studies have attempted to compare the CR ionization rate $\zeta$ inferred from the observed molecular line structure of GMCs, to the rate one would get from simply extrapolating the LISM CR spectrum. Although these inferences of $\zeta$ must be taken with some caution as the values are strongly model-dependent and have potentially large systematic errors, a number of independent studies in e.g.\ \citet{indriolo:2009.high.cr.ionization.rate.clouds.alt.source.models,padovani:2009.cr.ionization.gmc.rates.model.w.alt.sources,indriolo:2012.cr.ionization.rate.vs.cloud.column,indriolo:2015.cr.ionization.rate.vs.galactic.radius,cummings:2016.voyager.1.cr.spectra} have concluded that there must be some variation in $\zeta$ between nearby molecular gas in GMCs in the Solar neighborhood (where the inferred $\zeta \sim 10^{-16}\,{\rm s^{-1}}$) and the (predominantly ionized, much-lower-density) LISM (which has an implied $\zeta \sim 10^{-17}\,{\rm s^{-1}}$). Similarly, \citet{indriolo:2015.cr.ionization.rate.vs.galactic.radius} showed there must be significant variation with galacto-centric radius (with larger $\zeta$ towards the Galactic center). 

Recalling that CR ionization is dominated by low-energy CRs with $E\lesssim 100\,$MeV, the variations we have described above immediately provide a potential explanation for all of these observations. We can make this more rigorous by directly calculating the CR ionization rate from our spectra, as shown in Fig.~\ref{fig:env.compare.ion}. Following \citet{indriolo:2009.high.cr.ionization.rate.clouds.alt.source.models},\footnote{Note our definition is slightly different from \citet{indriolo:2009.high.cr.ionization.rate.clouds.alt.source.models}, who also multiply by the parameter $\xi_{s}=0,\,1.5,\,2.3$ in ionized or atomic or molecular gas respectively, to account for secondary ionizations. We correct the observations by this factor so they can be compared directly: i.e.\ we define $\zeta$ such that an exactly identical CR spectrum will produce an identical $\zeta$, regardless of the ambient non-relativistic gas properties.} and consistent with our assumed in-code CR ionization rates (corresponding to the CR ionization losses in \S~\ref{sec:collisional.losses}), we take: $\zeta \equiv \sum_{s} 4\pi\,1.5\,\int_{T_{\rm low}}^{T_{\rm high}}\,J_{s}(T)\,\sigma_{s}(T)\,dT$ with $T_{\rm low}\approx 2\,$MeV/n, $T_{\rm high} \approx 10^{7}\,$MeV/n, and $\sigma_{s}(T)\approx 7.63\times 10^{-20}\,{\rm cm^{2}}\,Z^{2}\,\beta^{-2}\,(1 - 0.069\,\beta^{2} + 0.16\,\log_{10}{[\gamma\,\beta]})$, and calculate $\zeta$ in every cell in our simulations. We can then weight by the actual total ionization rate of molecular gas ($\propto \int d^{3}{\bf x}\,\zeta({\bf x})\,n_{H_{2}}({\bf x})$) to compare to the GMC observations which measure molecular indicators.

First, we examine which CRs contribute primarily to the ionization rate. As expected, these are low-energy CRs, primarily protons, but there is a broad range of energies which contribute similarly to the total ionization rate, and low-energy electrons are not totally negligible.

By definition, since our ``fiducial'' model roughly reproduces the observed low-energy Voyager CR spectra of $e^{-}$ and $p$, it should also approximately reproduces the LISM-inferred $\zeta \sim 10^{-17}\,{\rm s^{-1}}$ in diffuse ionized gas of the appropriate densities, and we see in Fig.~\ref{fig:env.compare.ion} that this is indeed the case. But we also see immediately in Fig.~\ref{fig:env.compare.ion} that this reproduces the observed $\zeta \sim 10^{-16}\,{\rm s^{-1}}$ in dense molecular gas at the Solar neighborhood. The reason is simply the combination of (1) the dependence of {\em low-energy} CR densities on gas density discussed in \S~\ref{sec:variation:env} and shown explicitly in Fig.~\ref{fig:env.compare.ion} for the ionizing spectra in different gas environments, together with (2) the fact that the observations are sensitive to a {\em total-ionization-rate-weighted-mean} in molecular gas, which will always (by definition) give a systematically higher value than the volume-weighted mean or median in a medium with variations. Likewise, for the same reasons detailed above, these models naturally reproduce the observed trend of $\zeta$ with Galacto-centric radius, again shown in Fig.~\ref{fig:env.compare.ion}. Here, to compare the simulations with observations even heuristically, we simply take a total actual-ionization-rate-weighted average, so weighted somewhat towards more dense gas, at various galacto-centric annuli, and compare with compiled observations of dense GMC cloud cores, attempting to account for the large systematic uncertainties in their distances. 

We stress that this is occurring as described above in the low-energy CRs: some of the CR ionization studies above assumed, based on $\gamma$-ray observations, that the CR background must be more smooth; however the observed $\gamma$ rays are dominated by orders-of-magnitude higher-energy CRs, which as shown in \S~\ref{sec:variation:env} are indeed distributed much more uniformly. Of course, we cannot rule out models  which depend on physics not modeled here, such as a different, steep injection population at even lower CR energies ($\lesssim  1\,$MeV) from some sources which are not massive  stars, as explored in e.g.\ \citet{cummings:2016.voyager.1.cr.spectra,gaches:2018.protostellar.cr.acceleration,offner:2019.cr.feedback.accretion.chem.disks}.

\section{Conclusions}
\label{sec:conclusions}

We have presented and studied the first live-MHD galaxy formation simulations to self-consistently incorporate explicitly-evolved CR spectra (as opposed to a single field simply representing the total CR energy). As such we explicitly follow the ISM+CGM gas dynamics (inflows, outflows, fountains, ISM turbulence, super-bubbles, etc), thermal phase structure of the gas, magnetic field structure, alongside spectrally-resolved CR populations from $\sim$\,MeV-TeV, including protons $p$, electrons $e^{-}$, anti-protons $\bar{p}$, positrons $e^{+}$, intermediate primaries like C, N, O, stable secondaries B, $^{7}$Be, $^{9}$Be, and radioactive secondaries $^{10}$Be, with a network that allows for all the major CR evolutionary processes. We also adopt a recently-developed detailed treatment of the CR transport equations which, unlike the commonly-adopted ``isotropic Fokker Planck'' equation does not impose any assumptions about strong scattering, isotropic scattering rates or magnetic field structures, etc, but instead includes all terms from the Vlasov equation to leading order in $\mathcal{O}(u/c)$ (where $u$ is the Galactic fluid velocity), and gives much more general expressions for terms like the ``re-acceleration'' terms commonly invoked.

From the point of view of understanding CR transport physics, the fundamental advantage of these models as compared to traditional historical models which treat the Galaxy with a simplified time-static analytic model (ignoring or making different assumptions for e.g.\ inflows/outflows, turbulence, source distributions, etc), is that many terms which can, in principle, be freely adjusted or varied in those analytic models (e.g.\ presence or absence or ``size'' of the ``halo,'' presence of inflows/outflows, structure or absence of turbulent/fountain motions, structure or neglect of the magnetic field structure, inhomogeneous density/ionization/magnetic field strength/radiation energy density variations in the ISM and therefore rates of all loss terms, ratios of injected primaries, etc.) are determined here by the self-consistent cosmological evolution. This removes tremendous degeneracies and allows us to explore some crucial sources of systematic uncertainty in those models (which remain order-of-magnitude). 

From the point of view of constraining CR models for application in Galaxy formation simulations, the models here allow us to calibrate CR transport assumptions in far greater detail and rigor than is possible with ``single'' bin models (where one can compare e.g.\ the total $\gamma$-ray emission from $p\rightarrow$\,pion production to observations, but this provides only a single, galaxy-integrated data point for a few galaxies). In the future, we will explore detailed synchrotron spectra from these models in external galaxies. This will allow us to explore the consequences of CRs in other galaxies with much greater fidelity. 

\subsection{Key Conclusions}

We show the following: 

\begin{enumerate}

\item{\textit{It is possible to (roughly) match Solar/LISM CR constraints with simple transport and injection models.} Specifically assuming a single-power-law injection spectrum with a standard slope ($\sim 4.2$), single-power-law scaling of the CR scattering rate with rigidity $\bar{\nu} \propto \beta\,R^{-\delta}$ and $\delta \sim 0.5-0.6$, following all the major loss/gain processes with their expected (locally-varying) rates. As expected, in the LISM, we show that the shape of the high-energy hadronic spectra are regulated by injection+escape (dependence of scattering rates on rigidity), while high-energy leptonic spectral shapes are regulated by synchrotron+inverse Compton losses, and low-energy leptonic+hadronic spectra are regulated (primarily) by ionization+Coulomb losses.} 

\item{\textit{``Large'' halo sizes are inevitable \&\ favored.} The normalization of the halo structure is not a free parameter in our models. Indeed, it is now well-established that a majority of the baryons and significant magnetic field strengths extend to hundreds of kpc around galaxies in the CGM, so it is un-avoidable that the ``thin disk'' or ``leaky box'' model would be a poor approximation. In terms of the idealized cylindrical CR scattering halos sometimes adopted analytically, in the limit of diffusive CRs, the correct ``effective'' halo size (defined as the region interior to which a CR has a non-negligible probability of scattering to Earth) will always be (up to an order-unity factor) the same as the Solar circle radius $r_{\odot} \sim 8\,$kpc). This in turn means that relatively low CR scattering rates (giving relatively high effective diffusivities), compared to decades-older ``leaky box'' models which ignored the halo+CGM, are required. Our inferred scattering rate at $\sim1\,$GV, $\bar{\nu} \sim 10^{-9}\,{\rm s^{-1}}$, is in fact in excellent agreement (within a factor of $\sim 2$, despite enormous differences in model details) with most recent analytic Galactic CR transport models, almost all of which have argued that a scattering halo\footnote{As discussed in \S~\ref{sec:halo.size}, the term ``halo'' in the CGM literature generally refers to the extended gas (and cosmic ray) distribution on tens to hundreds of kpc scales (out to or past the virial radius), while in the CR literature it often refers to a region confined to a few kpc above/below the disk (also often called the ``thick disk'' or ``corona'' or ``disk-halo interface''). The ``CR scattering halo'' specifically refers here to the effective volume interior to which CRs have a non-negligible probability of scattering back interior to the Solar circle.} with effective size $\sim 5-10\,$kpc is required to match the LISM observations \citep{blasi:cr.propagation.constraints,vladimirov:cr.highegy.diff,gaggero:2015.cr.diffusion.coefficient,2016ApJ...819...54G,2016ApJ...824...16J,cummings:2016.voyager.1.cr.spectra,2016PhRvD..94l3019K,evoli:dragon2.cr.prop,2018AdSpR..62.2731A,korsmeier:2021.light.element.requires.halo.but.upper.limit.unconfined,delaTorre:2021.dragon2.methods.new.model.comparison}.}

\item{\textit{Re-acceleration terms are not dominant, and obey a generic ordering.} There are three terms which can act as ``re-acceleration'': the ``adiabatic'' or non-inertial frame term $\dot{p}_{\rm ad}=-p\,\mathbb{D}:\nabla{\bf u}$, the ``streaming loss'' term $\dot{p}_{\rm st}=-\langle \mu\rangle \bar{D}_{p \mu} \sim -p\,\bar{\nu}\,(\bar{v}_{A}/c)\,(F/e\,c)$, and the ``diffusive'' or ``micro-turbulent'' re-acceleration term $\dot{p}_{\rm di} = 4\,p^{-1}\,\bar{D}_{p p} \sim p\,\bar{\nu}\,(v_{A}/c)^{2}$. We show that for almost any physically-realistic structure of the ISM in terms of $v_{A}$, ${\bf u}$, etc.\ and allowed values of $\bar{\nu}$, there is a robust ordering with $|\dot{p}_{\rm ad}| \gtrsim |\dot{p}_{\rm st}| \gtrsim |\dot{p}_{\rm di}|$ at $\gtrsim$\,GV, and that these terms have at most modest (tens of percent) effects on the total CR spectrum.}

\item{\textit{Most $\lesssim\,$TeV Galactic CRs are accelerated in SNe shocks, in super-bubbles, early in the Sedov-Taylor phase (after the reverse shock forms).} Observed abundances of intermediate and heavy primary elements in CRs are all consistent, to leading order, with a universal single-power-law acceleration spectrum with all species tracing their in-situ abundances in the test particle limit if we assume CRs are accelerated with an efficiency $\epsilon \sim 10\%$ of strong shock energy when the shock first forms -- i.e.\ when the entrained mass of ambient ISM material is approximately equal to the initial ejects mass ($M_{\rm swept} \approx M_{\rm ej}$). This is naturally predicted if CRs accelerate when the shock first ``forms,'' and therefore the kinetic energy dissipation rate and Mach number are maximized. If instead acceleration occurred primarily in stellar wind/jet shocks, diffusive ISM shocks with Mach number $\gg 1$, or throughout the entire Sedov-Taylor phase of SNe remnants, then the abundances of CNO at $\sim\,$MeV-TeV would be under-predicted by factors of $\sim 20$. Given the favored conditions, most MW acceleration occurs in SNe shocks within super-bubble-type conditions.}

\item{\textit{CR spectra vary significantly in time \&\ space, both systematically and stochastically.} With more realistic Galactic models, substantial variations are expected between and within Galaxies. CR energy densities decrease with increasing galacto-centric radius $r$ ($\propto 1/r$, for constant scattering rates, over a range of radii) and spectra are harder in hadrons, softer in leptons towards the Galactic center, owing to  differences in loss rates and source spatial distributions. We show this naturally reproduces Galactic $\gamma$-ray emissivity observations, though $\gamma$-ray-inferred variations in spectral shape can be sensitive to the local dynamical state of the dense $\gamma$-ray emitting gas in the Galactic center. At the Solar circle, the CR spectra still vary significantly with local environment and gas density $n$, with e.g.\ CR kinetic energy density $\propto n^{0.5}$ (i.e.\ higher in more-dense environments) -- the effect is stronger at lower CR energies owing to tighter coupling with the gas, while becoming negligible at $\gtrsim 100\,$GeV. Even controlling for e.g.\ $r$, $n$ and other variables (temperature, etc.), the scatter in particularly low-energy CR spectra from point-to-point in space or time or between galaxies can vary by orders of magnitude ($\sim 90\%$ interval of $\pm 1.5\,$dex at $\lesssim 10\,$MeV), owing to the enormous inhomogeneities in local source distributions (clustered star formation \&\ SNe), loss rates (orders-of-magnitude variation in local densities, ionization fractions, radiation energy densities, etc.), and local gas dynamics (e.g.\ local inflow/outflow, turbulence structure). This provides a natural explanation for observations which have inferred a different CR ionization rate in local molecular clouds (compared to the diffuse LISM observed by Voyager) and different ionization rates at different Galactic positions.}

\end{enumerate}

\subsection{Future Work}

This is only a first study and there are many different directions in which it can be extended. In future work, we will explore predictions for a wide range of galaxies outside of the MW, from dwarfs through starbursts and massive ellipticals, at both low and high redshifts. With the models here, we can make detailed forward-modeled predictions for spatially-resolved synchrotron spectra in these galaxies, to compare to the tremendous wealth of resolved extragalactic synchrotron studies. We can also forward-model the $\gamma$-ray spectrum, which provides a key complementary constraint, albeit only in a few nearby galaxies. 

We stress that the extremely simple (and constant in space and time) scaling of the scattering rates adopted here, $\bar{\nu} = \nu_{0}\,\beta\,R_{\rm GV}^{-\delta}$, is purely heuristic/empirical. This is quite radically different, in fact, from what is predicted by either traditional ``extrinsic turbulence'' models for CR scattering or more modern ``self-confinement'' motivated models. In both of those model classes the local scattering rates (e.g.\ across different ISM regions at the Solar circle) can vary by {\em many orders of magnitude} in both space and time, on scales smaller than the CR residence time or disk/halo scale height, as a strong function of the local turbulence properties, plasma-$\beta$, neutral fractions, magnetic field strength, gas density, and other parameters \citep{hopkins:cr.transport.constraints.from.galaxies}. These effects simply cannot be captured in standard CR transport models which adopt simplified static analytic models for Galactic structure. The most interesting application of the new methods here, which attempt to combine more detailed CR propagation constraints with detailed, live galaxy simulations that explicitly evolve those parameters, is therefore likely to be exploring and making detailed predictions from those more physically-motivated CR transport/scattering models, in a way which was previously not possible. 

It will also be particularly important, especially with a variable $\bar{\nu}$, to investigate how local variations in plasma properties modify CR loss and other key timescales commonly assumed in analytic models for CR transport or observables such as the FIR-radio or $\gamma$-ray-SFR relations. For example, the ``effective'' or mean synchrotron loss timescale $\langle t_{\rm synch} \rangle$ at some CR energy represents a complicated weighted average over different ISM regions, so can differ significantly from the synchrotron loss timescale estimated using just the volume-averaged mean magnetic field value. Exploring where and when these differences are important in detail will be an important subject for future study.

\acknowledgments{We thank Jonathan Squire for a number of insightful conversations, and Michael Korsmeier for useful comments, as well as our referee for a number of suggestions. Support for PFH was provided by NSF Research Grants 1911233 \&\ 20009234, NSF CAREER grant 1455342, NASA grants 80NSSC18K0562, HST-AR-15800.001-A. Numerical calculations were run on the Caltech compute cluster ``Wheeler,'' allocations AST21010 and AST20016 supported by the NSF and TACC, and NASA HEC SMD-16-7592. CAFG was supported by NSF through grants AST-1715216 and CAREER award AST-1652522; by NASA through grant 17-ATP17-0067; by STScI through grant HST-AR-16124.001-A; and by the Research Corporation for Science Advancement through a Cottrell Scholar Award and a Scialog Award.  EQ's research was supported in part by NSF grants AST-1715070 and AST-2107872 and a Simons Investigator award from the Simons Foundation. GVP acknowledges support by NASA through the NASA Hubble Fellowship grant  \#HST-HF2-51444.001-A  awarded  by  the  Space Telescope Science  Institute,  which  is  operated  by  the Association of Universities for Research in Astronomy, Incorporated, under NASA contract NAS5-26555.}

\datastatement{The data supporting the plots within this article are available on reasonable request to the corresponding author. A public version of the GIZMO code is available at \gizmourl.} 

%\bibliography{/Users/phopkins/Dropbox/Public/ms}
\bibliography{ms_extracted}

\begin{appendix}

\section{CR Drift Velocities and Loss Timescales}
\label{sec:vdrift.tloss}

Here we present and discuss the characteristic CR drift velocities and different loss timescales in LISM conditions. 

\begin{figure}
	\includegraphics[width=0.98\columnwidth]{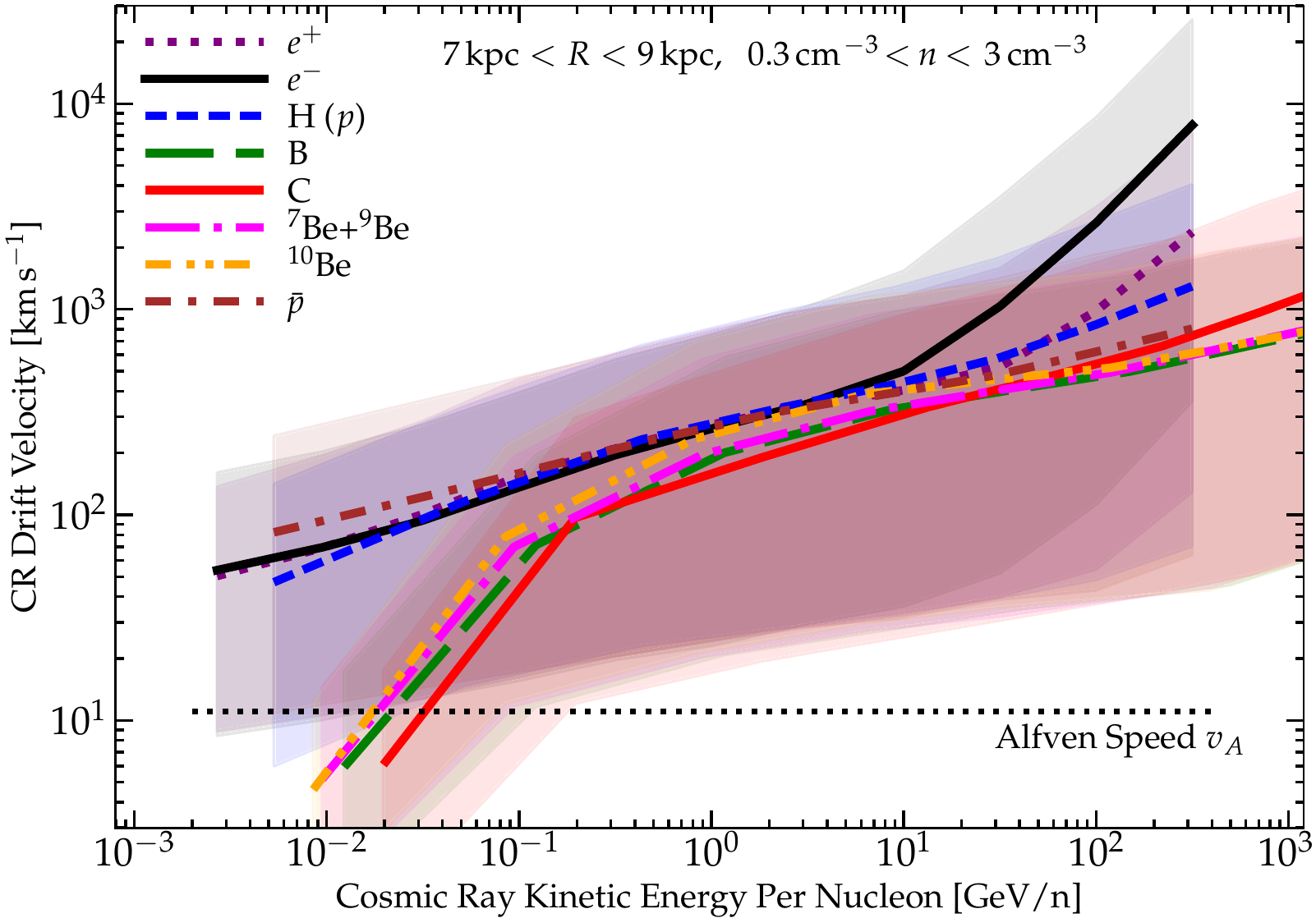}
	\vspace{-0.2cm}
	\caption{Typical CR ``drift velocities'' (\S~\ref{sec:drift.vel}) $v_{\rm drift} \equiv \langle \mu \rangle\,\beta\,c = |v_{\rm cr}\,\bar{f}_{1}|/\bar{f}_{0}$ extracted directly from the simulations for our fiducial model (Fig.~\ref{fig:demo.cr.spectra.fiducial}), in the frame comoving with the gas, measured for Solar-circle ($7\,{\rm kpc}<R<9\,{\rm kpc}$) disk ($|z|<0.5\,$kpc) LISM-like ($0.3\,{\rm cm^{-3}}<n<3\,{\rm cm^{-3}}$) gas, as a function of species ({\em labeled}) and CR energy. We plot the mean ({\em thick line}) and $5-95\%$ interval ({\em shaded range}), weighted by CR energy density at each species/rigidity (evaluated just at the ``bin center'' rigidities). Typical LISM drift velocities scale as $v_{\rm drift} \sim 300\,{\rm km\,s^{-1}}\,(T/{\rm GeV})^{0.3}$, corresponding to expected diffusive drift $\sim \kappa_{\rm eff,\,iso}/\ell_{\rm grad,\,cr}$ for CR gradient scale lengths $\ell_{\rm grad,\,cr} \sim {\rm kpc}\,(T/{\rm GeV})^{0.3}$. The drift speeds scale more weakly with rigidity than the scattering rates, because the CR scale heights also increase (weakly) with CR energy (\S~\ref{sec:gamma.rays}). The steeper scaling for high-energy leptons and low-energy nuclei owes to the dominant role of losses (\S~\ref{sec:drift.vel}). We also compare the mean \Alf\ speed in the same gas ({\em horizontal line}). 
	\label{fig:vdrift}}
\end{figure}

\subsection{Typical LISM Drift Velocities}
\label{sec:drift.vel}

Fig.~\ref{fig:vdrift} presents the typical ``drift velocities'' of the CRs, extracted directly from our fiducial simulation (Fig.~\ref{fig:demo.cr.spectra.fiducial}) at $z=0$. Specifically, we define the drift velocity in standard fashion as $v_{\rm drift} \equiv |F_{\rm cr}^{E}|/e_{\rm cr} = |{\bf F}_{\rm cr}\cdot\bhat|/e_{\rm cr} = |v\,\bar{f}_{1}|/\bar{f}_{0} = |\langle \mu \rangle|\,\beta\,c$ (per Eq.~\ref{eqn:f0}-\ref{eqn:f1} or \S~\ref{sec:numerics:spatial}) measured independently for CRs of each given species $s$ in a narrow range of rigidity (for simplicity, we extract this just at the bin center, for each bin). This is the coherent net drift speed of those CRs along $\bhat$, relative to the gas. We restrict to LISM-like gas (galacto-centric radius $7\,{\rm kpc}<R<9\,{\rm kpc}$, vertical position $|z|<0.5\,$kpc, gas densities $0.3\,{\rm cm^{-3}}<n<3\,{\rm cm^{-3}}$), and weight the mean and distribution of $v_{\rm drift}$ by the contribution to the CR flux, i.e.\ $\langle v_{\rm drift} \rangle = (\int e_{\rm cr}\,v_{\rm drift}\,d^{3}{\bf x}) / (\int e_{\rm cr}\,d^{3}{\bf x})$ (so the mean CR flux is $|F_{\rm cr}^{E}|=\langle v_{\rm drift} \rangle\,\langle e_{\rm cr} \rangle$). 

We see that the drift speeds for different species are broadly similar as a function of CR energy, and (over most of the plotted range) follow a scaling $v_{\rm drift} \sim 300\,{\rm km\,s^{-1}}\,(T/{\rm GeV})^{0.3}$.  This corresponds well to the expected ``diffusive drift speed'' $v_{\rm drift}^{\rm diff} \sim \kappa_{\rm eff}/\ell_{\rm grad,\,cr}$, assuming tangled magnetic fields so the effective isotropically-averaged diffusivity is $\kappa_{\rm eff} \sim v^{2}/(9\,\bar{\nu})$ (with our fiducial $\bar{\nu} = 10^{-9}\,{\rm s^{-1}}\,R_{\rm GV}^{-0.6}$) and a CR gradient scale-length $\ell_{\rm grad,\,cr} \equiv e_{{\rm cr},\,s}/|\nabla e_{{\rm cr},\,s}| \sim {\rm kpc}\,(T/{\rm GeV})^{0.3}$. But that scaling of the gradient scale-length $\ell_{\rm grad,\,cr}$ is very similar to what we found by directly plotting the vertical CR profiles (at energies $\gtrsim 100\,$MeV) in Fig.~\ref{fig:profiles.r.z} (see \S~\ref{sec:gamma.rays}). Importantly, because as we showed there the CR scale height increases (weakly) with CR energy, the resulting CR drift speed ($\propto \kappa_{\rm eff}/\ell_{\rm grad,\,cr} \propto \beta^{2}/\bar{\nu}\,\ell_{\rm grad,\,cr}$) varies less-strongly with energy than the diffusivity or scattering rate. This is a significant, if subtle, distinction between the simulations here and many analytic flat-halo diffusion models for CR transport, which assume a fixed CR gradient scale-length or scale-height by construction in their boundary conditions.

Note that the features in Fig.~\ref{fig:vdrift} which deviate from the trend above are expected. For leptons at the highest energies ($\gtrsim 50\,$GeV) losses (synchrotron and inverse Compton) become so rapid that they regulate the CR scale-height and it begins to {\em decrease} with CR energy (as $\ell_{\rm grad,\,cr} \propto T^{-0.3}$ or so; see \S~\ref{sec:gamma.rays}), so we expect a steeper dependence $v_{\rm drift} \propto \kappa_{\rm eff} / \ell_{\rm grad,\,cr} \propto R^{0.9} \propto T^{0.9}$, as seen. Similarly, for heavier nuclei (B/Be/CNO) at very low energies ($\ll 100\,$MeV), losses become so rapid that they regulate the CR scale height to be essentially energy-independent, and the nuclei are sub-relativistic ($\beta \ll 1$, so $T\propto \beta^{2} \propto R^{2}$), so $v_{\rm drift} \propto \kappa_{\rm eff} / \ell_{\rm grad,\,cr} \propto \beta\,R_{\rm GV}^{-0.6} \propto T^{0.8}$ again becomes steeper. 

For reference we compare the mean (CR-energy-weighted) \Alf\ speed in the same gas, typically $\sim 12\,{\rm km\,s^{-1}}$, corresponding to $|{\bf B}| \sim 6\,{\rm \mu G}$ at these ISM densities ($n\sim1\,{\rm cm^{-3}}$), as expected from Fig.~\ref{fig:egy.corr.density}. For all but the very lowest-energy ($\lesssim 10\,$MeV) heavy nuclei, the streaming is highly super-\Alf{ic}, as we discuss in the main text. Of course, pure \Alf{ic} streaming could not produce the correct scalings of e.g.\ B/C and other features we study even if the typical \Alf\ speed were a factor of $\sim 30-100$ larger, because it (by definition) gives an energy-independent drift velocity.

\begin{figure}
	\includegraphics[width=0.98\columnwidth]{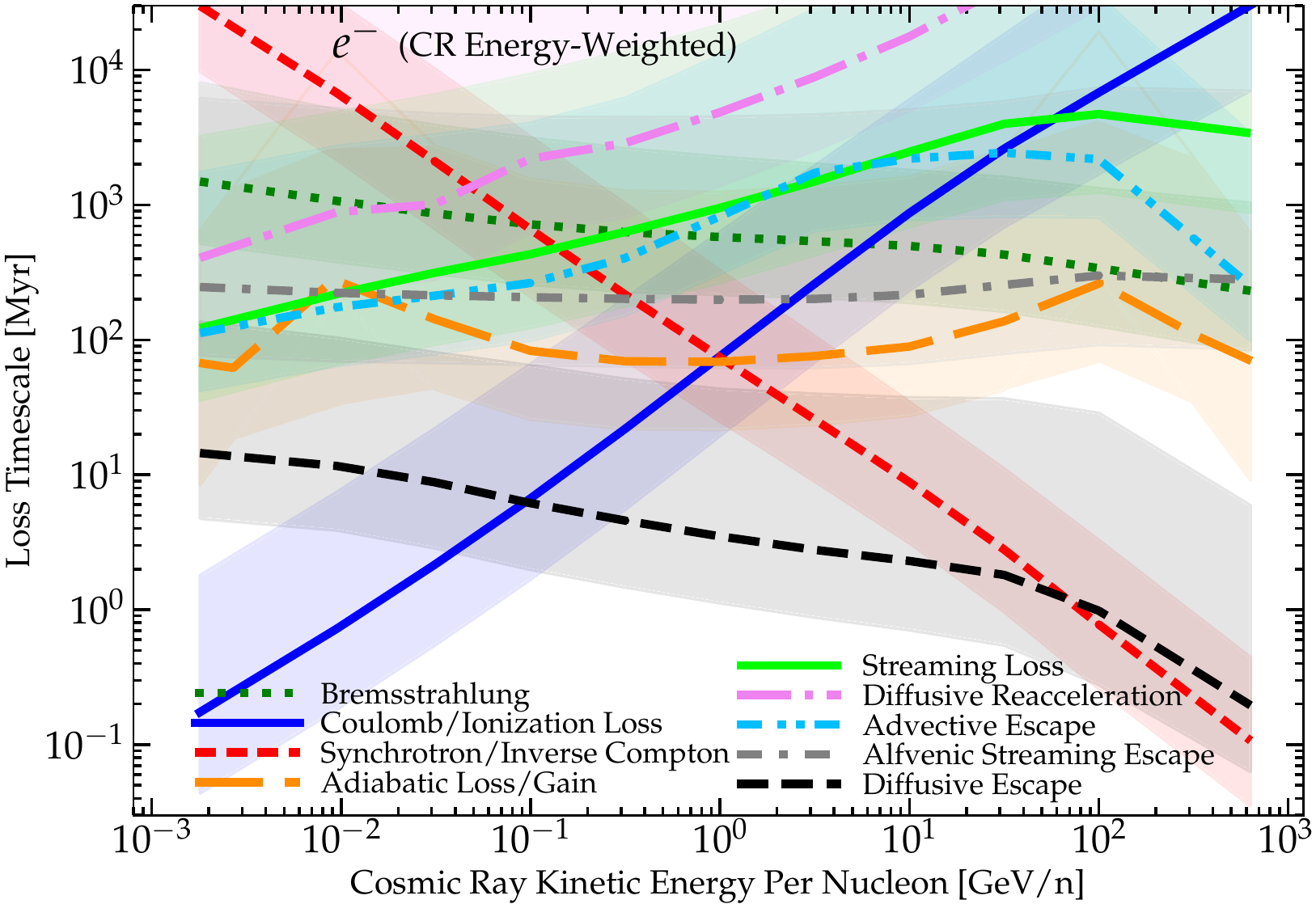}\\
	\includegraphics[width=0.98\columnwidth]{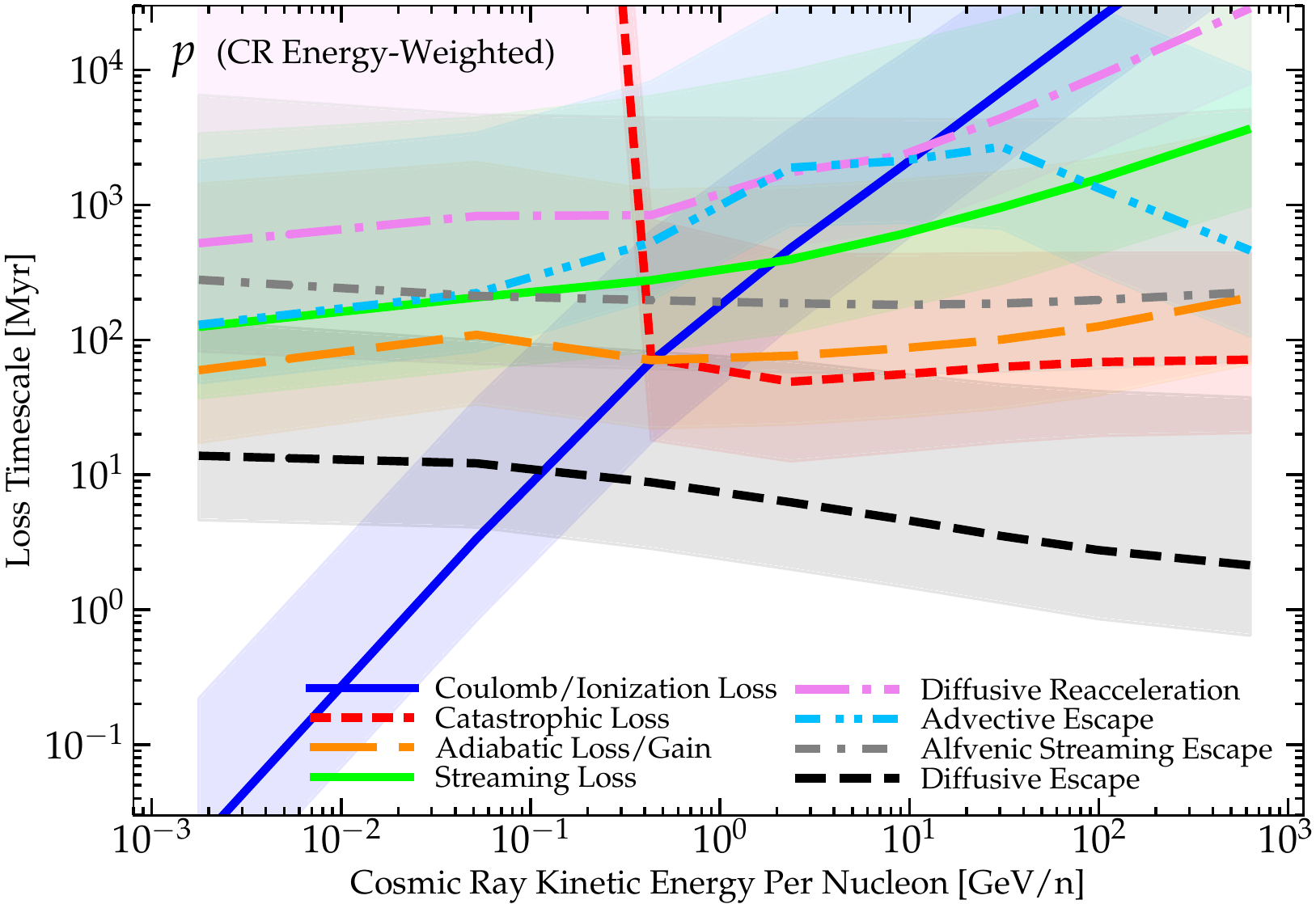}
	\vspace{-0.2cm}
	\caption{CR loss/escape timescales (\S~\ref{sec:loss.timescales}) for Solar-circle ($7\,{\rm kpc}<R<9\,{\rm kpc}$), midplane LISM gas ($|z|<0.5\,{\rm kpc}$, $0.3\,{\rm cm^{-3}}<n<3\,{\rm cm^{-3}}$), for electrons ($e^{-}$; {\em top}) and protons ($p$; {\em bottom}). For each species and energy, we measure the mean loss rate, weighted by total CR kinetic energy density $e_{\rm cr}$ (e.g.\ the contribution to the total CR energy flux/loss; {\em thick lines}), and $5-95\%$ range ({\em shaded}), as a function of CR energy per nucleon. We show the rate corresponding to each process labeled. We confirm the trends discussed in \S~\ref{sec:physics}: at low energies ($\lesssim 100\,$MeV), Coulomb+ionization losses are most important (determining the effective CR residence time); while at high energies for leptons ($\gtrsim 50\,$GeV) synchrotron+inverse Compton losses dominate. At intermediate and (for hadrons) high energies, diffusive escape ($t_{\rm loss} \sim \ell_{{\rm grad,\,cr}}^{2}/\kappa_{\rm eff} \sim (3\,\ell_{{\rm grad},\,cr}/v)^{2}\,\bar{\nu}$) regulates the residence times. Other loss terms (e.g.\ Bremsstrahlung and catastrophic) are less important (though not always negligible); advective and \Alf{ic} streaming/drift are relatively minor effects; and the adiabatic, streaming loss, and diffusive reacceleration terms obey the relative ordering described in \S~\ref{sec:reaccel}.
	\label{fig:tloss}}
\end{figure}

\subsection{CR Loss Timescales}
\label{sec:loss.timescales}

\begin{figure*}
	\includegraphics[width=0.49\textwidth]{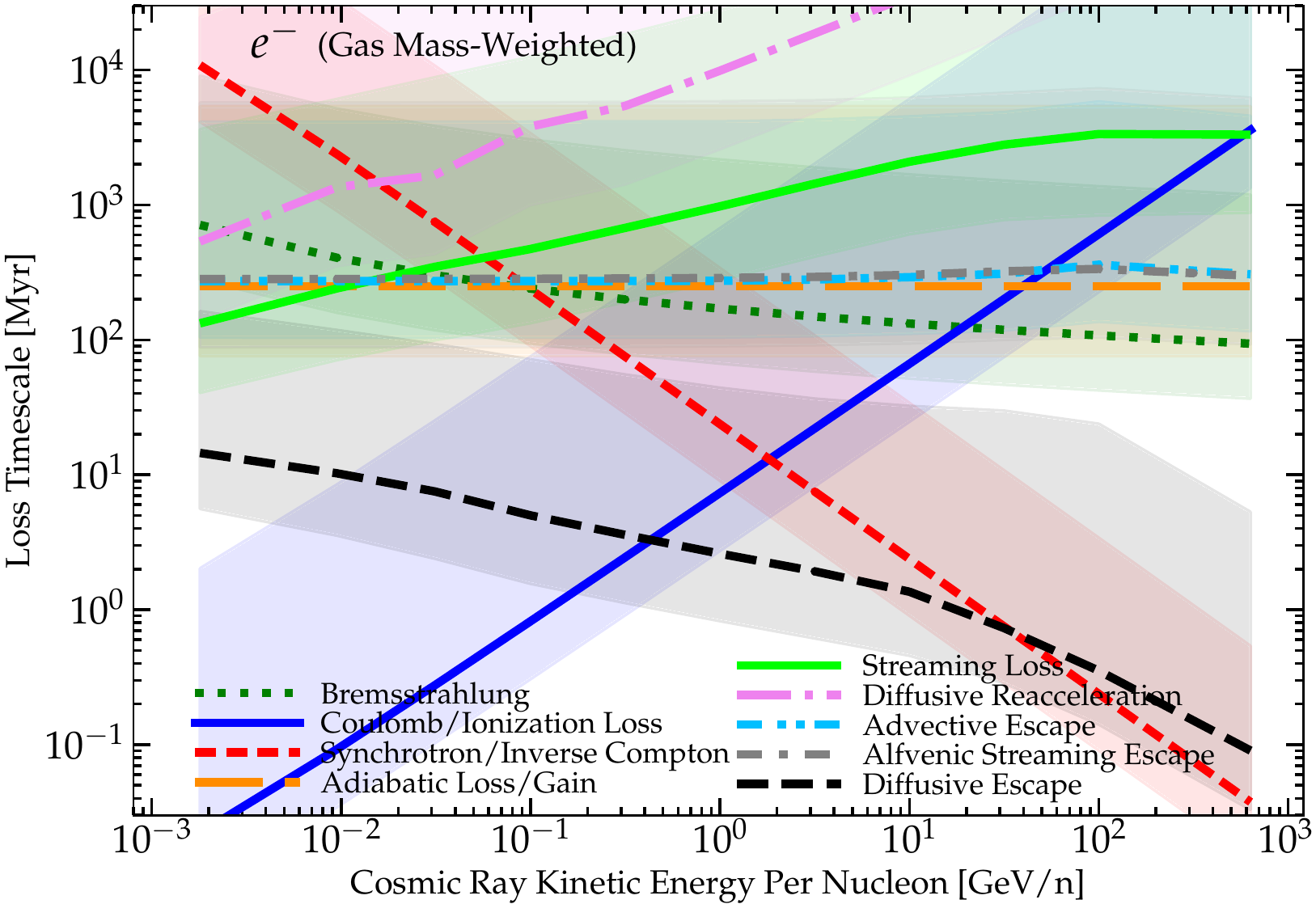}
	\includegraphics[width=0.49\textwidth]{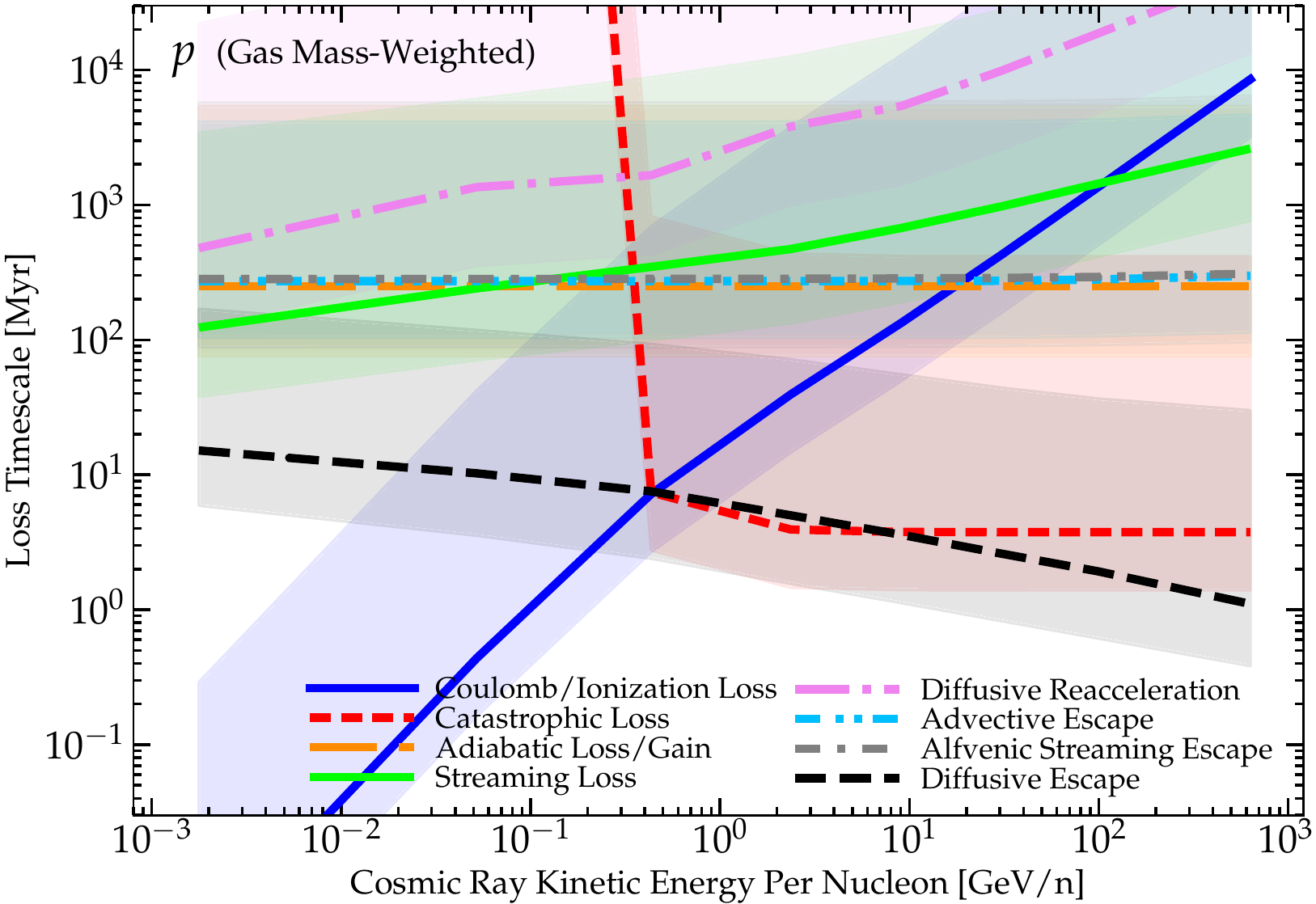} \\
	\includegraphics[width=0.49\textwidth]{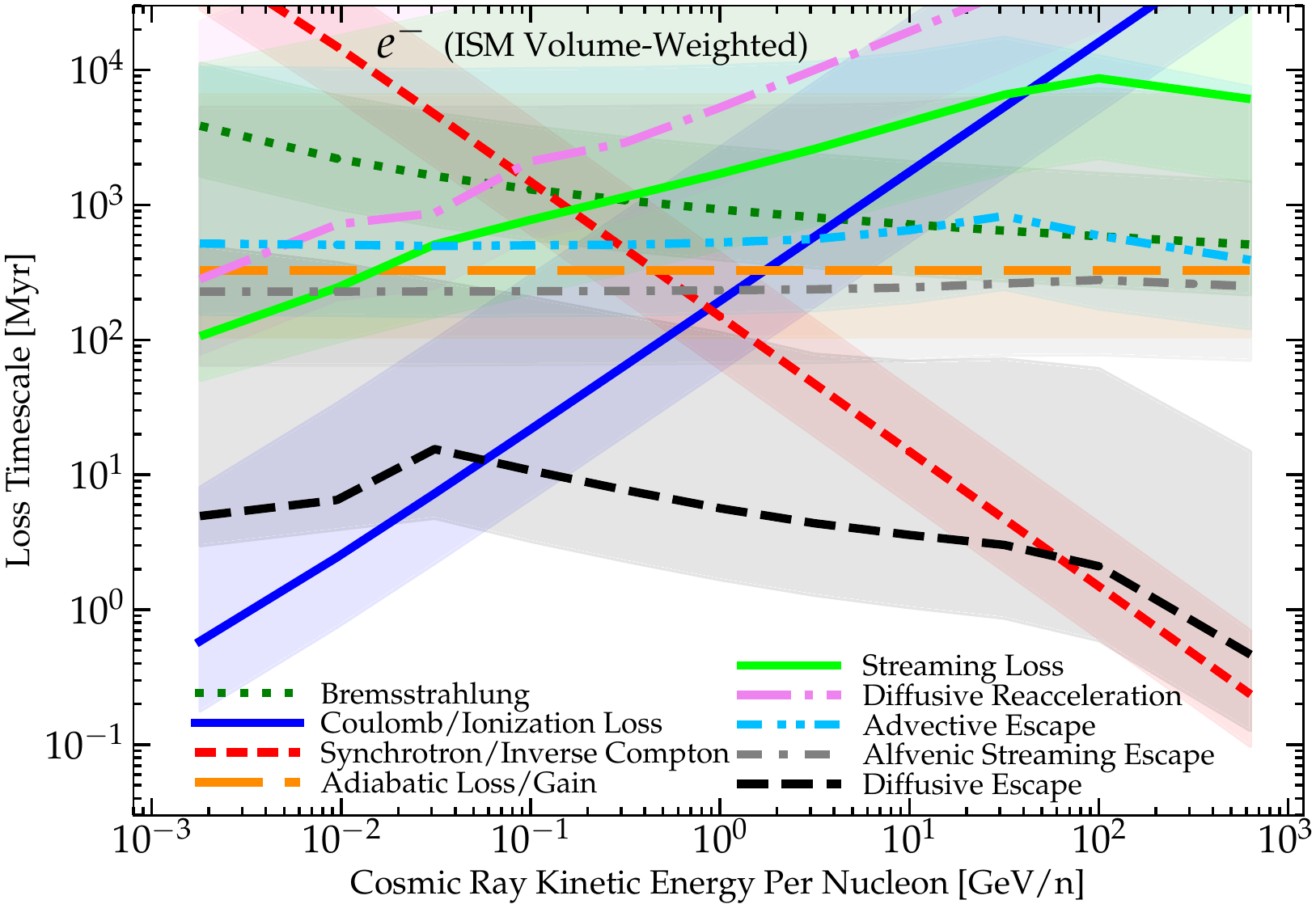}
	\includegraphics[width=0.49\textwidth]{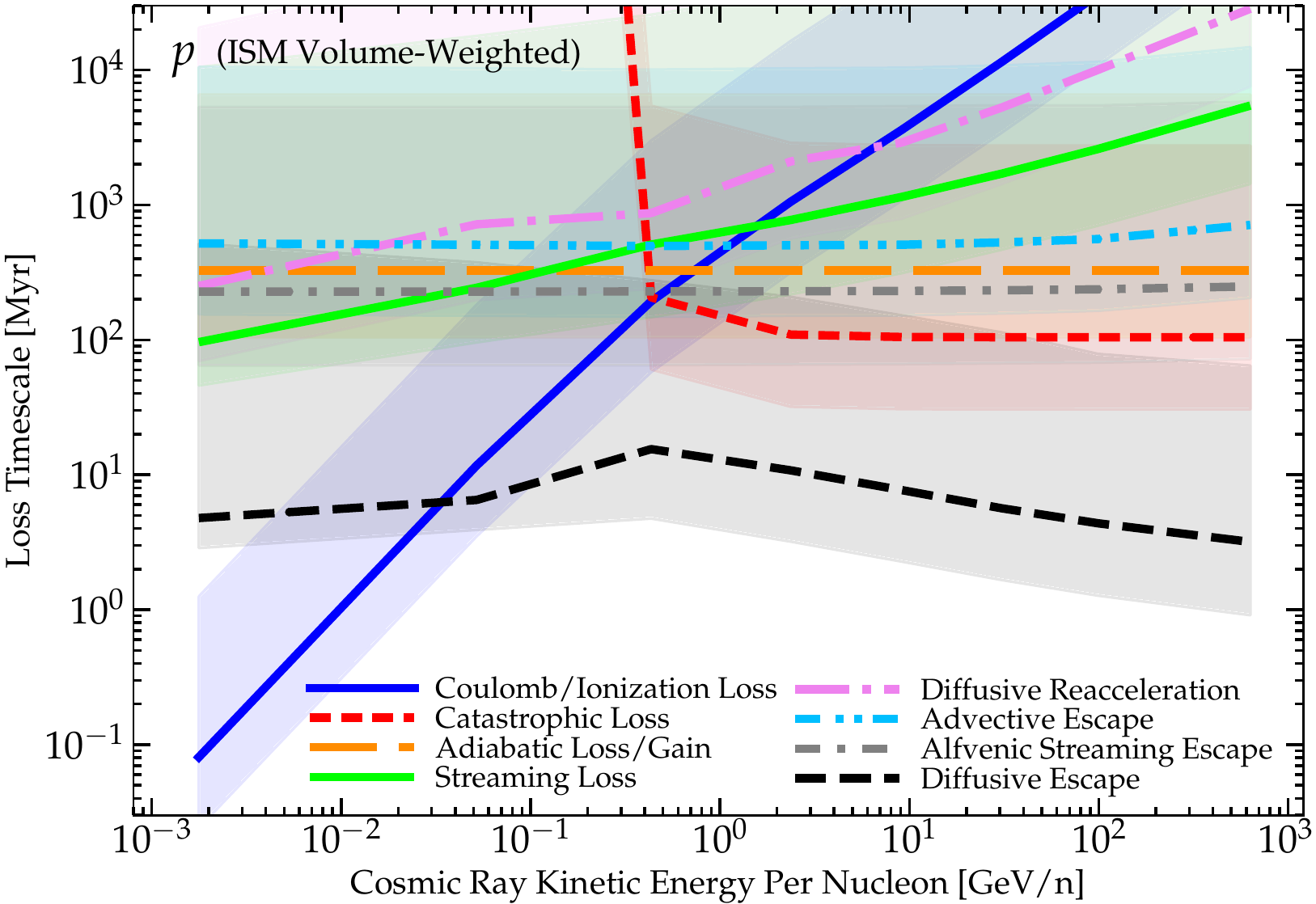}
	\vspace{-0.2cm}
	\caption{CR loss/escape timescales for Solar-circle ISM gas, as Fig.~\ref{fig:tloss}, for electrons ($e^{-}$; {\em left}) and protons ($p$; {\em right}), but weighted by ISM gas mass ({\em top}) or volume ({\em bottom}). Here we include gas in the disk around the Solar circle $7\,{\rm kpc}<R<9\,{\rm kpc}$, $|z|<0.5\,$kpc, but include all gas densities $n$. In a gas-mass-weighted sense the loss timescales are weighted towards higher-density gas so are shorter, particularly from ionization, synchrotron/inverse Compton, and catastrophic processes. Catastrophic loss timescales are comparable to escape in the high-density gas. In a volume-weighted sense the loss timescales are weighted towards low-density regions so are longer, and catastrophic losses are negligible in these regions. But the relative importance of different processes is similar in both cases to that shown in Fig.~\ref{fig:tloss}.  
	\vspace{-0.4cm}
	\label{fig:tloss.weights}}
\end{figure*}

We consider the CR loss timescales in Figs.~\ref{fig:tloss} \&\ \ref{fig:tloss.weights}. We again focus on Solar-circle ($7\,{\rm kpc}<R<9\,{\rm kpc}$) disk ($|z|<0.5\,$kpc) gas, and consider electrons and protons. For each loss or escape process $i$, in cell $j$, for species $s$, at CR momentum $p=p_{j}$, we define the effective loss timescale $t_{{\rm loss},\,j,n,s,i}$ as $=p_{j,n,s} / \dot{p}_{j,n,s,i}$ for continuous CR loss/gain processes, or $=\bar{f}_{0,\,j,n,s} / \dot{\bar{f}}_{0,\,j,n,s,i}$ for catastrophic losses, or $=\ell_{{\rm grad,\,cr},\,j,n,s} / v_{{\rm eff},\,j,n,s,i}$ for different drift/escape terms. Expressions for $\dot{p}$ and $\dot{f}$ are given in \S~\ref{sec:loss.terms.included}. For all species we show the following continuous ($\dot{p}$) terms: Coulomb plus ionization losses (added together because they scale identically with CR properties); adiabatic losses or gains; and the ``streaming loss'' and ``diffusive reacceleration'' terms (\S~\ref{sec:adiabatic.streaming.diffreaccel.terms}). For electrons we also show Bremsstrahlung and synchrotron plus inverse Compton losses (again grouped together given their identical scaling). For protons we show catastrophic losses (\S~\ref{sec:collisional.losses}). 

For the drift/escape terms, we define the loss rate as the time to travel down the CR gradient length scale $\ell_{{\rm grad,\,cr},\,j,n,s} \equiv e_{{\rm cr},\,j,n,s} / |\nabla e_{\rm cr}|_{j,n,s}$ with some effective velocity $v_{{\rm eff},\,j,n,s,i}$. This is defined so that for e.g.\ pure advection/streaming with constant $v_{\rm eff}$, one would simply have $\dot{e}_{\rm cr} = e_{\rm cr} / t_{\rm loss}$ (making it comparable to the other loss timescales defined above), but we stress that this is not generally equal to the true ``escape time'' of CRs from the Galaxy nor to the ``travel time'' between CR sources and the Solar system, both of which are typically longer as for example both $\ell_{{\rm grad,\,cr}}$ and $t_{\rm loss}$ will tend to increase as the CRs travel above the disk through the inner halo. Nonetheless it provides a useful comparison. We define the ``advective escape'' velocity for simplicity as the vertical gas outflow velocity from the disk, $v_{\rm eff}=u_{{\rm gas},\,z}\,z / |z|$ (more complex definitions do not change our qualitative conclusions; see \citealt{muratov:2016.fire.metal.outflow.loading}). This is roughly the time for outflows to advect CRs (with zero drift/streaming speed) out of the disk midplane (though we caution most of this material may not necessarily reach large galactocentric radii). We define the ``\Alf{ic} streaming escape time'' as the timescale for CRs to stream down their gradient at the local \Alf\ speed, giving $v_{\rm eff} = |{\bf v}_{A} \cdot \nabla e_{\rm cr}| / |\nabla e_{\rm cr}|$ where ${\bf v}_{A} = v_{A}\,\bhat$. And we similarly define the ``diffusive escape'' speed as the measured drift speed in excess of the \Alf{ic} streaming speed down the CR gradient, $v_{\rm eff} = (v_{\rm drift} - v_{\rm eff,\,A}) = |(v\,\bar{f}_{1}/\bar{f}_{0} - v_{A})\,\bhat \cdot \nabla e_{\rm cr}| /|\nabla e_{\rm cr}|$ using the values of $\bar{f}_{1}$ (or CR flux) directly from the simulation, which allows for non-steady-state behaviors. As shown in Fig.~\ref{fig:vdrift}, if we just assumed the local-steady-state effective diffusivity so $v_{\rm eff} = \kappa_{\rm eff}/\ell_{\rm grad,\,cr}$ or $t_{\rm loss} \sim \ell_{\rm grad,\,cr}^{2}/\kappa_{\rm eff}$, we would obtain essentially identical results. 

In Fig.~\ref{fig:tloss}, we restrict to LISM gas densities ($n \sim 0.3-3\,{\rm cm^{-3}}$) and weight the distribution by CR energy, defining the mean loss timescale for each process\footnote{We plot the absolute value so some processes which are technically ``gains,'' e.g.\ diffusive re-acceleration, can be compared on the same figure.} $\langle t_{\rm loss} \rangle^{-1} \equiv |\int e_{\rm cr}\,t_{\rm loss}^{-1}\,d^{3}{\bf x}| / \int e_{\rm cr}\,d^{3}{\bf x}$, so that the total CR energy loss rate is just $\dot{E}_{\rm cr} = E_{\rm cr} / \langle t_{\rm loss} \rangle$. Note this means that terms with varying sign (e.g.\ adiabatic or advective terms) are appropriately averaged so that the mean is the {\em net} gain/loss timescale. Also, note that weighting by CR energy density means that even terms which are strictly CR momentum-independent at a given spatial location (e.g.\ the adiabatic term) can have some population-averaged CR momentum/kinetic energy dependence because the averaging weights towards different cells/locations at different CR momenta. 

Fig.~\ref{fig:tloss} confirms all of the general conclusions discussed in \S~\ref{sec:physics}, from our comparison of CR spectra adding/removing these different gain/loss terms and simple analytic scalings. For both electrons and protons, at  low energies ($\lesssim 100\,$MeV), we confirm that Coulomb-plus-ionization losses dominate, with their loss time scaling analytically as expected: $t_{\rm loss} \propto T$ for electrons (which still have $\beta \approx 1$ at these energies) or $\propto T^{1.5}$ for  protons (with $\beta \ll 1$ at $\ll 100\,$MeV). For electrons at the highest energies ($\gtrsim 50\,$GeV), inverse Compton and synchrotron losses dominate, producing a loss timescale that again scales as expected, $t_{\rm loss} \propto T^{-1}$. For high energies in hadrons, and intermediate energies in leptons, diffusive escape dominates, with $t_{\rm loss} \sim \ell_{\rm grad,\,cr}^{2}/\kappa_{\rm eff} \sim 3\,{\rm Myr}\,\beta^{-1}\,R_{\rm GV}^{-0.6}\,(\ell_{{\rm grad,\,cr}}/{\rm kpc})^{2}$ decreasing with CR energy. However, as discussed in \S~\ref{sec:drift.vel} above, because $\ell_{{\rm grad,\,cr}}$ increases with CR energy (at least over the range where diffusive escape is most important), the diffusive escape time is significantly more-weakly dependent on CR energy, compared to the scattering rate or diffusivity alone. The adiabatic, advective escape, and \Alf{ic} streaming escape timescales are order-of-magnitude similar (and nearly energy-independent), as expected in any medium where the characteristic turbulent/fountain motions are broadly trans-\Alf{ic} (like the warm ISM), and do not dominate the loss timescales. We also confirm that the three ``re-acceleration'' terms (adiabatic, streaming loss, and diffusive re-acceleration)  obey the expected hierarchy of relative importance discussed in detail in \S~\ref{sec:reaccel}. Bremsstrahlung (for electrons) and catastrophic losses (for protons) are also generally sub-dominant, though not completely negligible in some environments. Recalling that the total loss/escape rate is given by the sum of all these processes, we see that the loss/escape timescales are maximized for CRs near the peak of the spectrum ($\sim 0.1-1\,$GeV), at a few Myr (see also \S~\ref{sec:numerics:timescales}). 

To get some sense of how these numbers depend on environment, in Fig.~\ref{fig:tloss.weights} we re-calculate the same loss timescales, again in Solar circle disk gas, but include all gas densities and weight by either gas mass (biased to the high-density environments) or volume (biased to low-density environments). As expected, in denser environments (which contain much of the ISM gas mass), the loss timescales are shorter and the relative importance of both radiative and catastrophic losses increases (with catastrophic loss timescales becoming comparable to diffusive escape in e.g.\ GMC environments). Conversely, in lower-density environments (containing much of the ISM volume), loss timescales increase and radiative+catastrophic losses are less important compared to diffusive escape. But the qualitative behaviors and relative importance of different terms, in a broad sense, is similar to what we found in Fig.~\ref{fig:tloss}.

\section{Additional Physics \&\ Parameter Variations}
\label{sec:appendix:additional}

\begin{figure*}
\begin{tabular}{r@{\hspace{0pt}}r@{\hspace{0pt}}r}
	\includegraphics[width=0.33\textwidth]{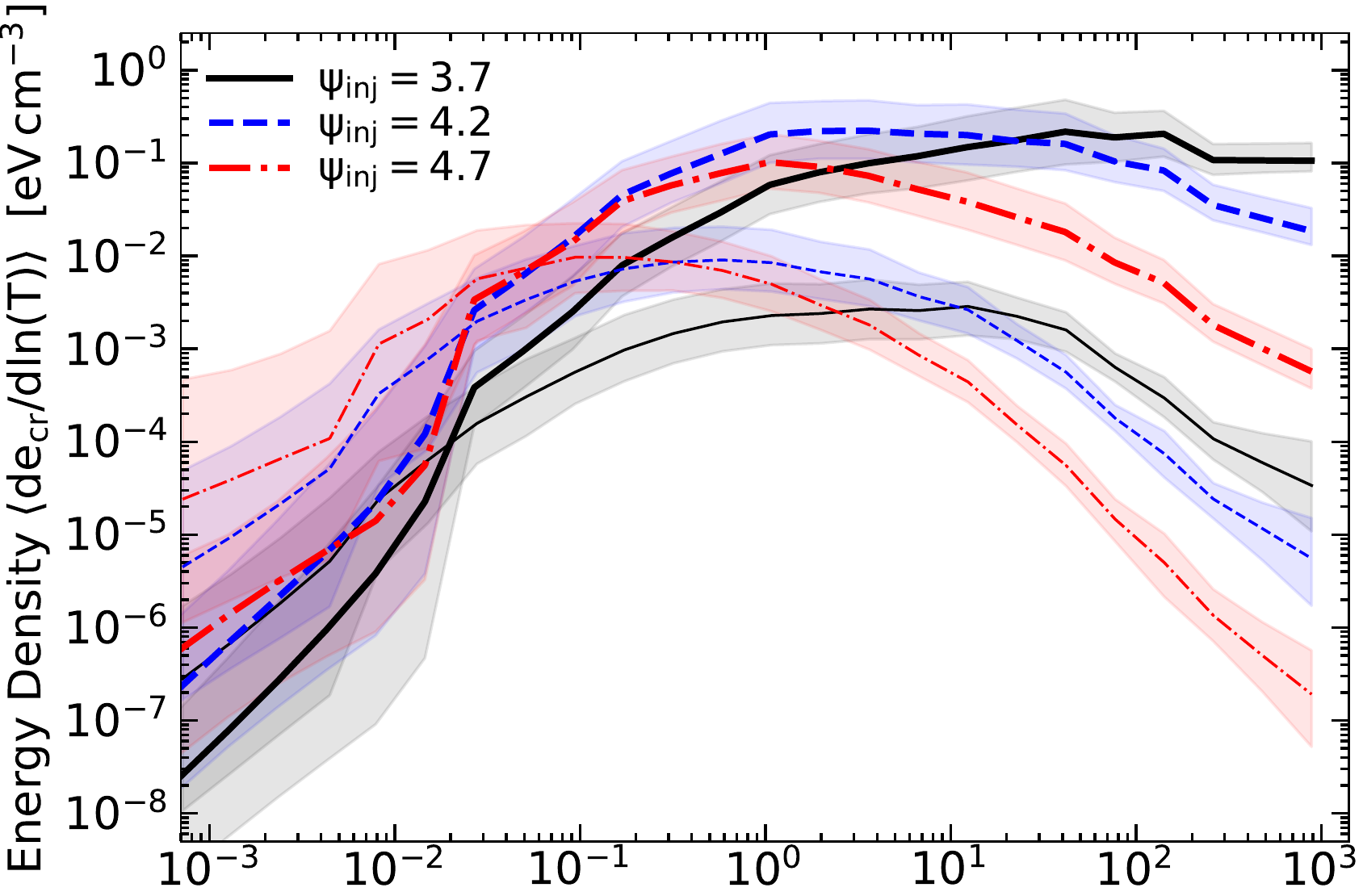}&
	\includegraphics[width=0.33\textwidth]{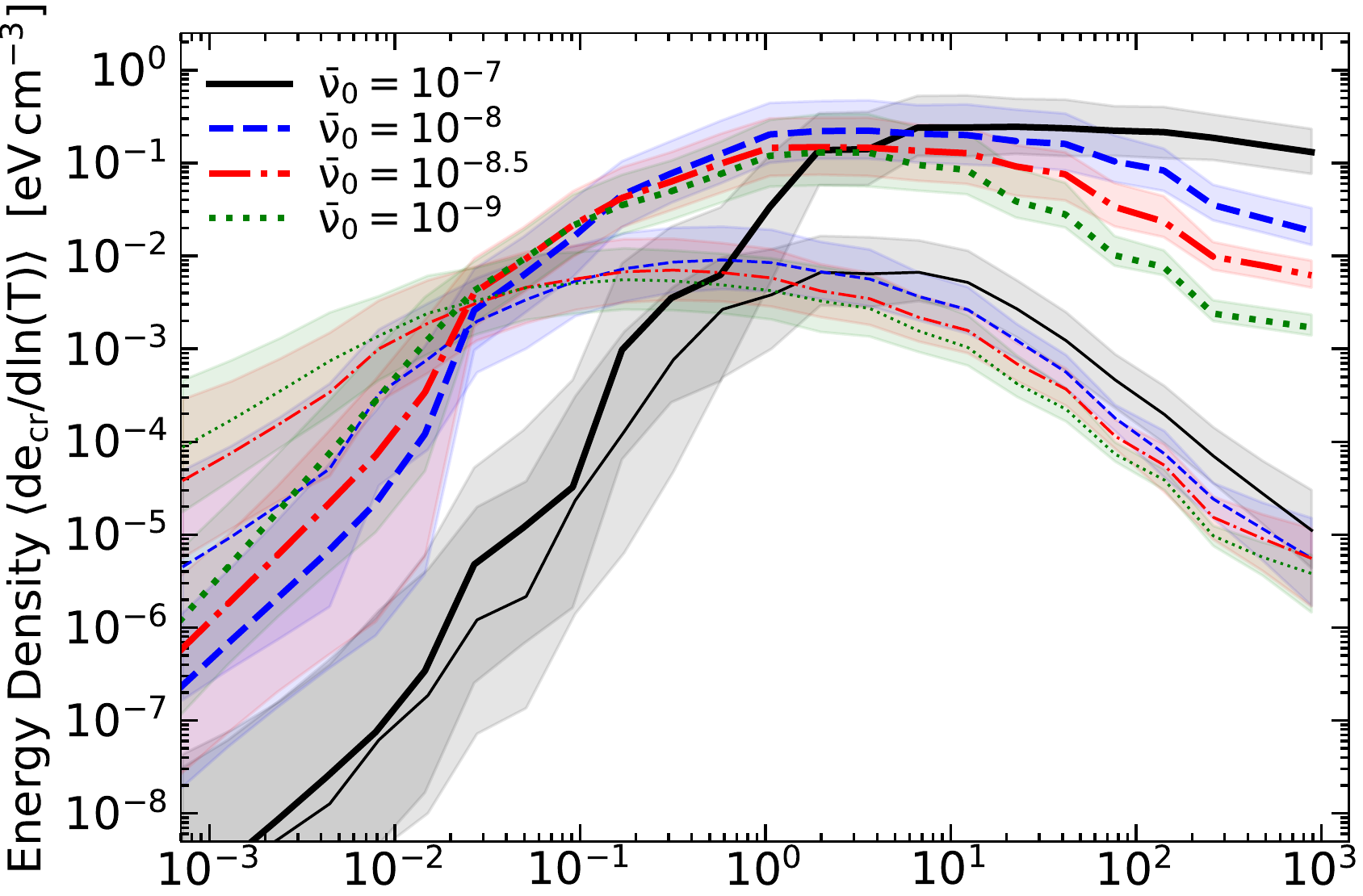}&
	\includegraphics[width=0.33\textwidth]{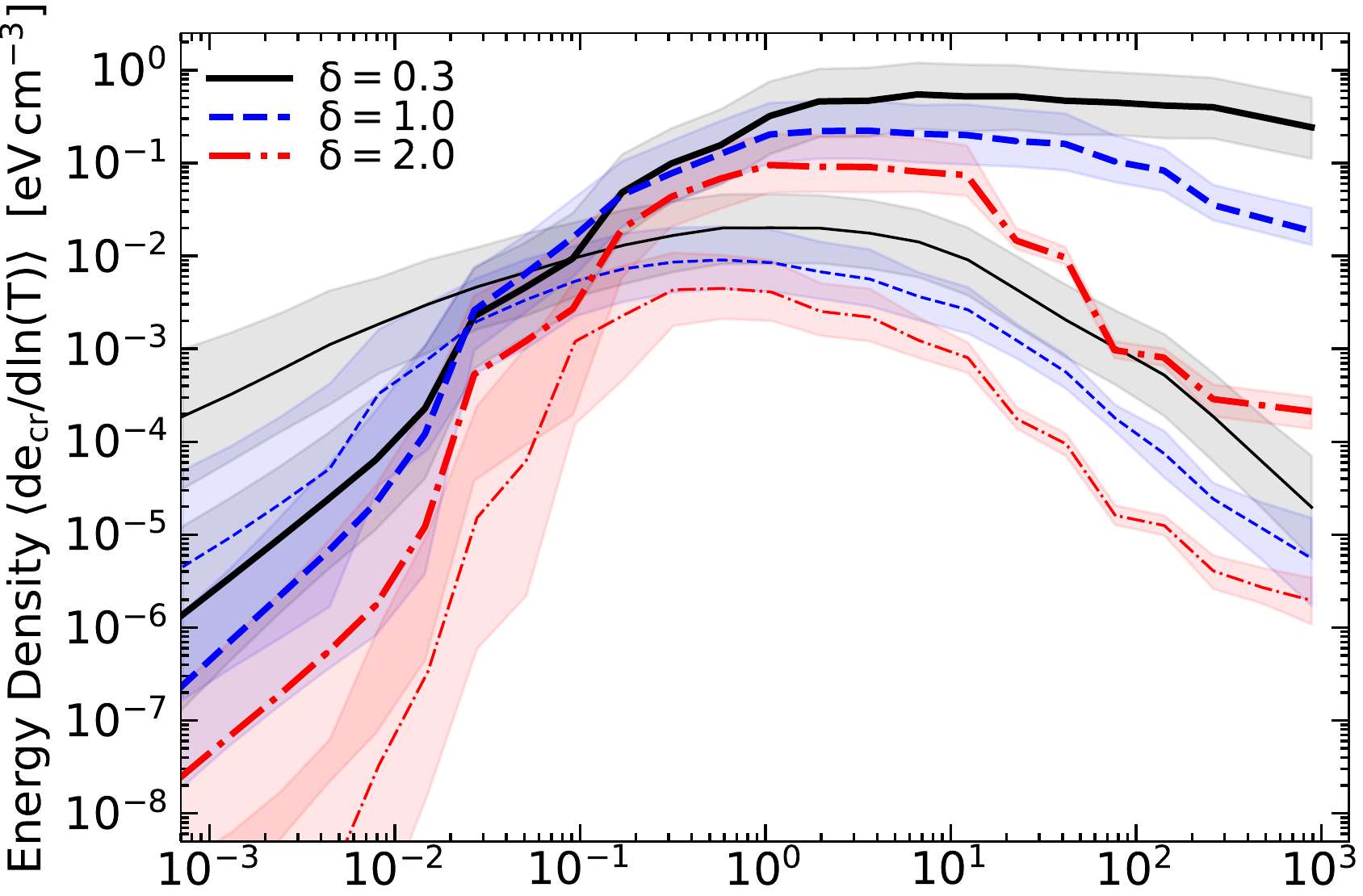} \\
	\includegraphics[width=0.32\textwidth]{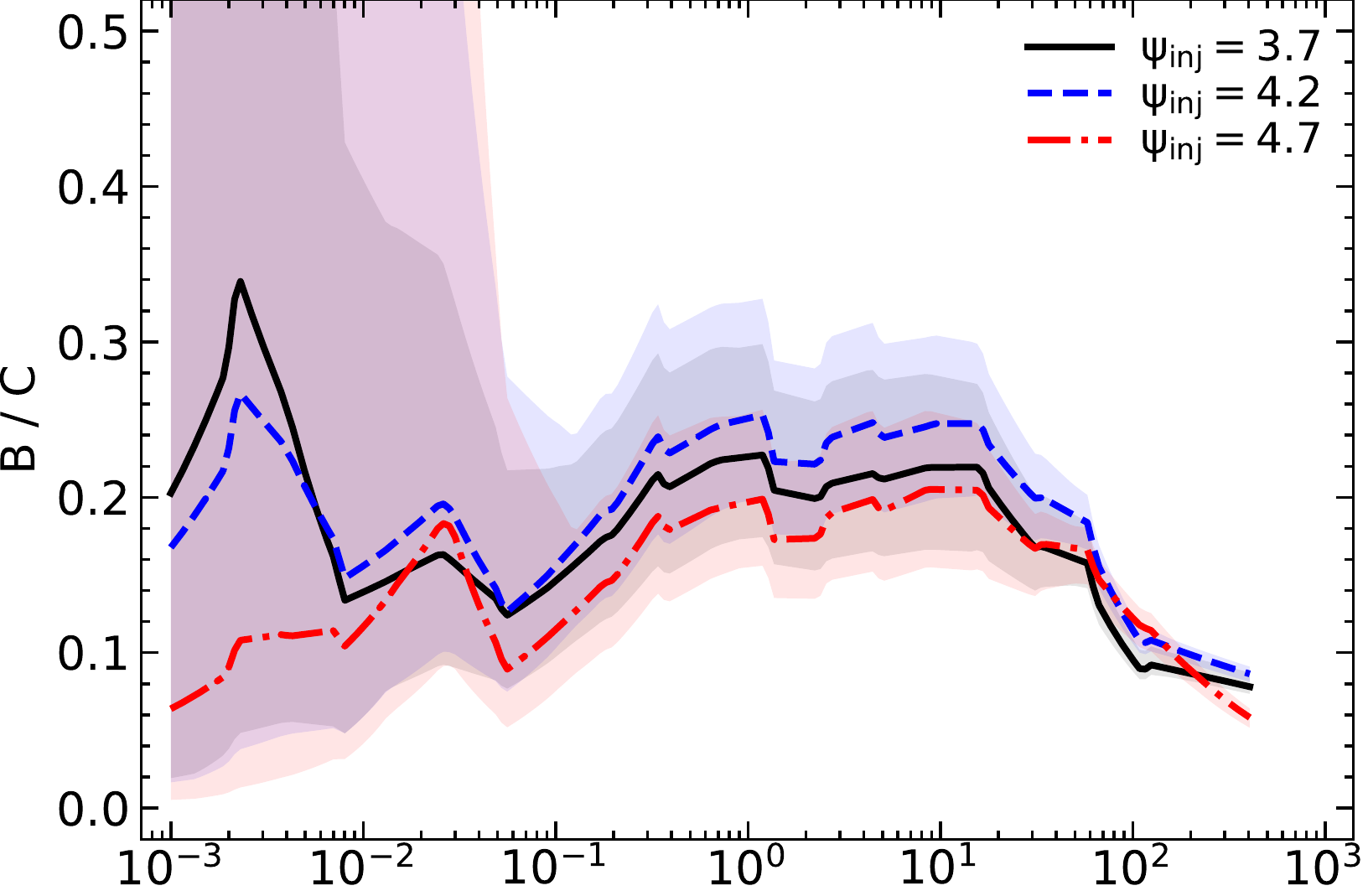}&
	\includegraphics[width=0.32\textwidth]{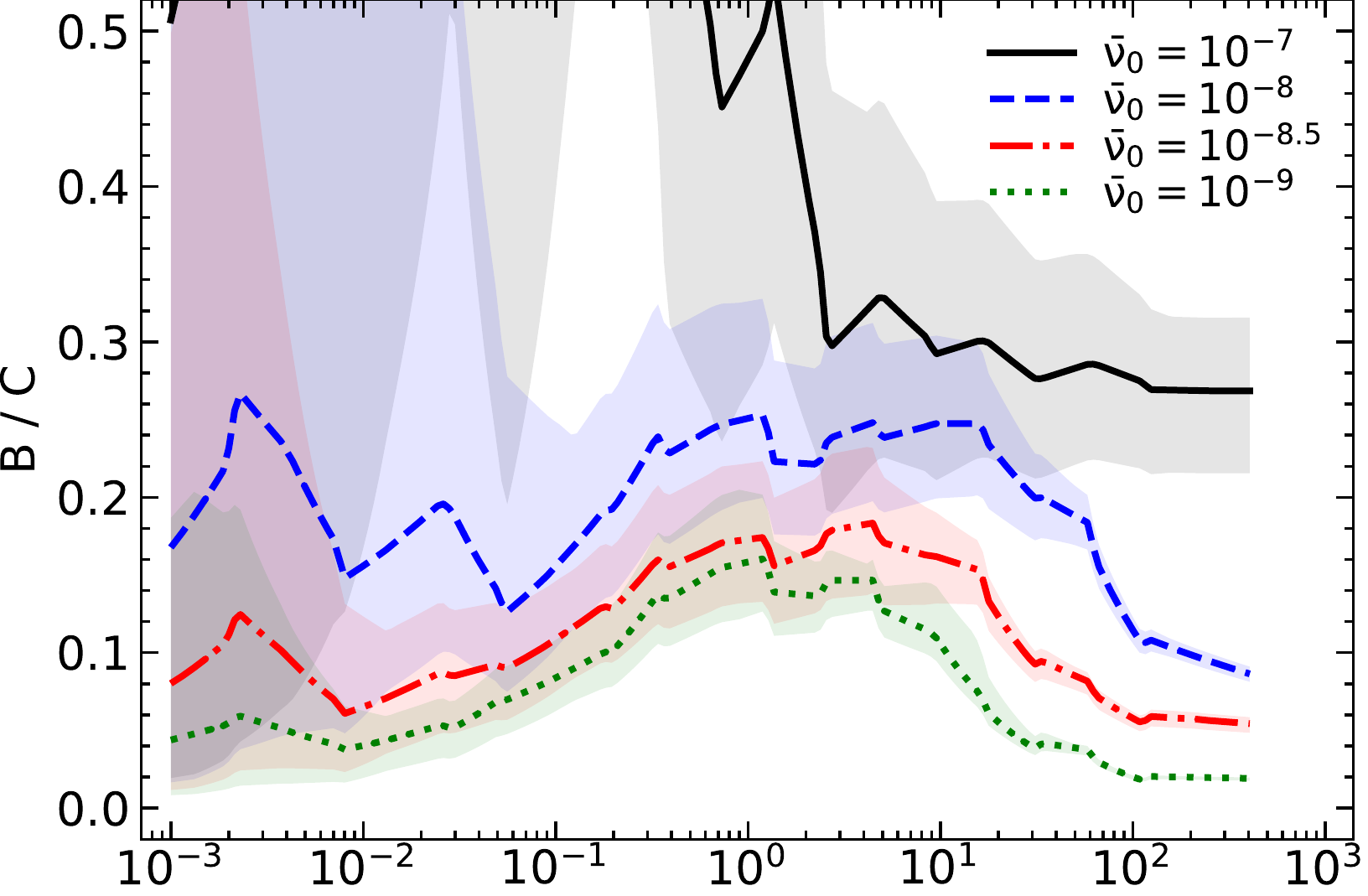}&
	\includegraphics[width=0.32\textwidth]{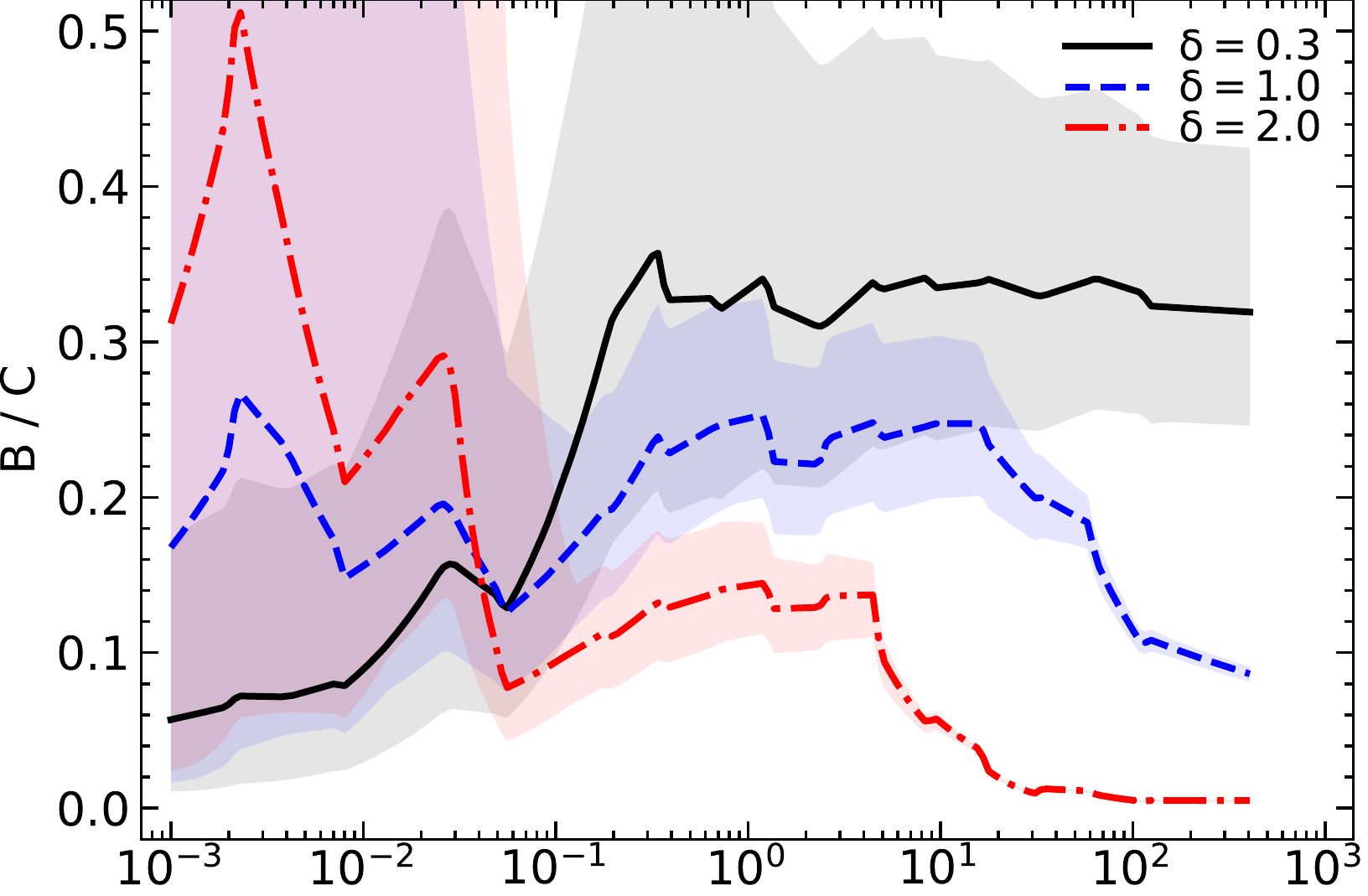} \\
	\includegraphics[width=0.32\textwidth]{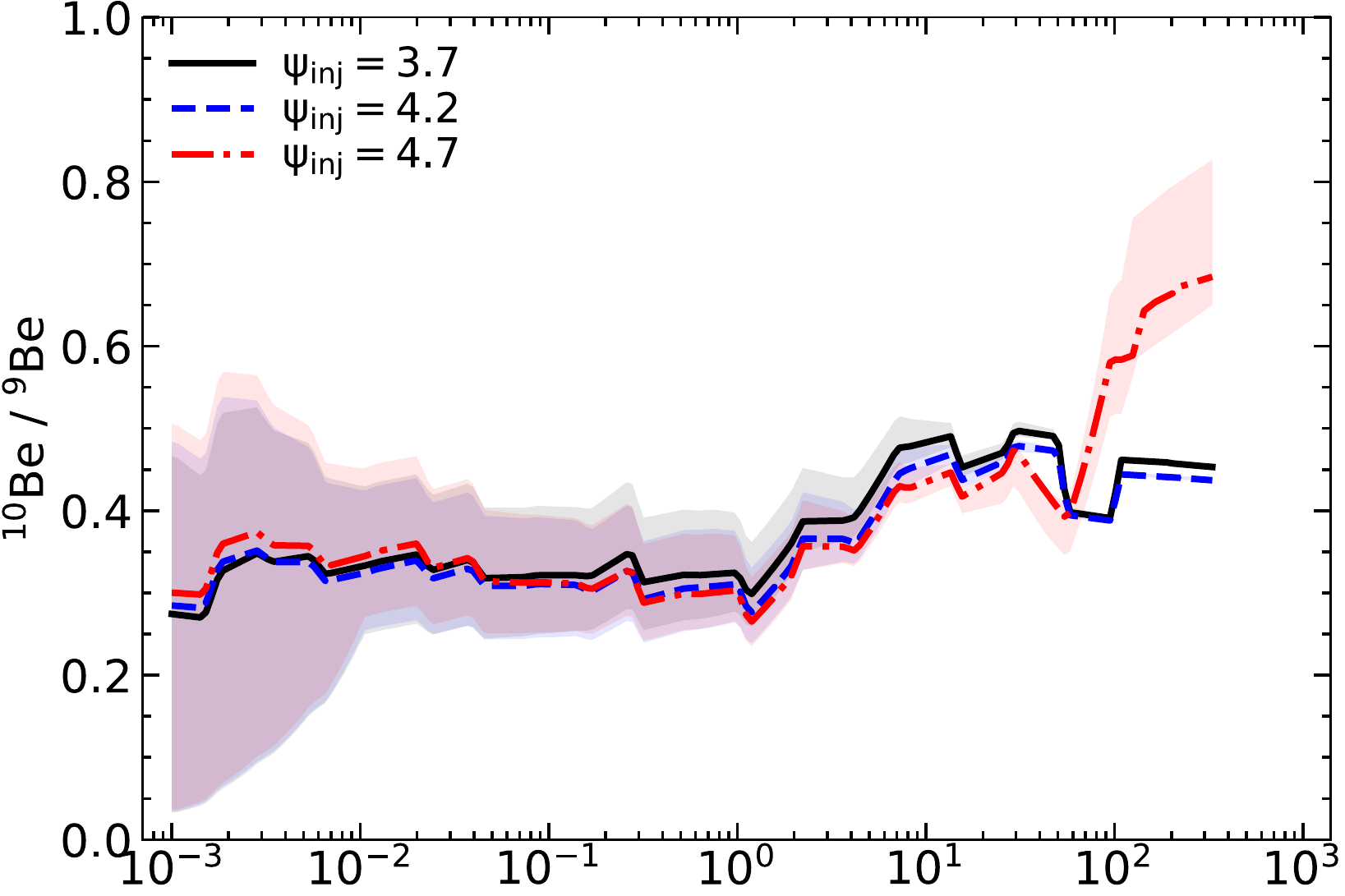}&
	\includegraphics[width=0.32\textwidth]{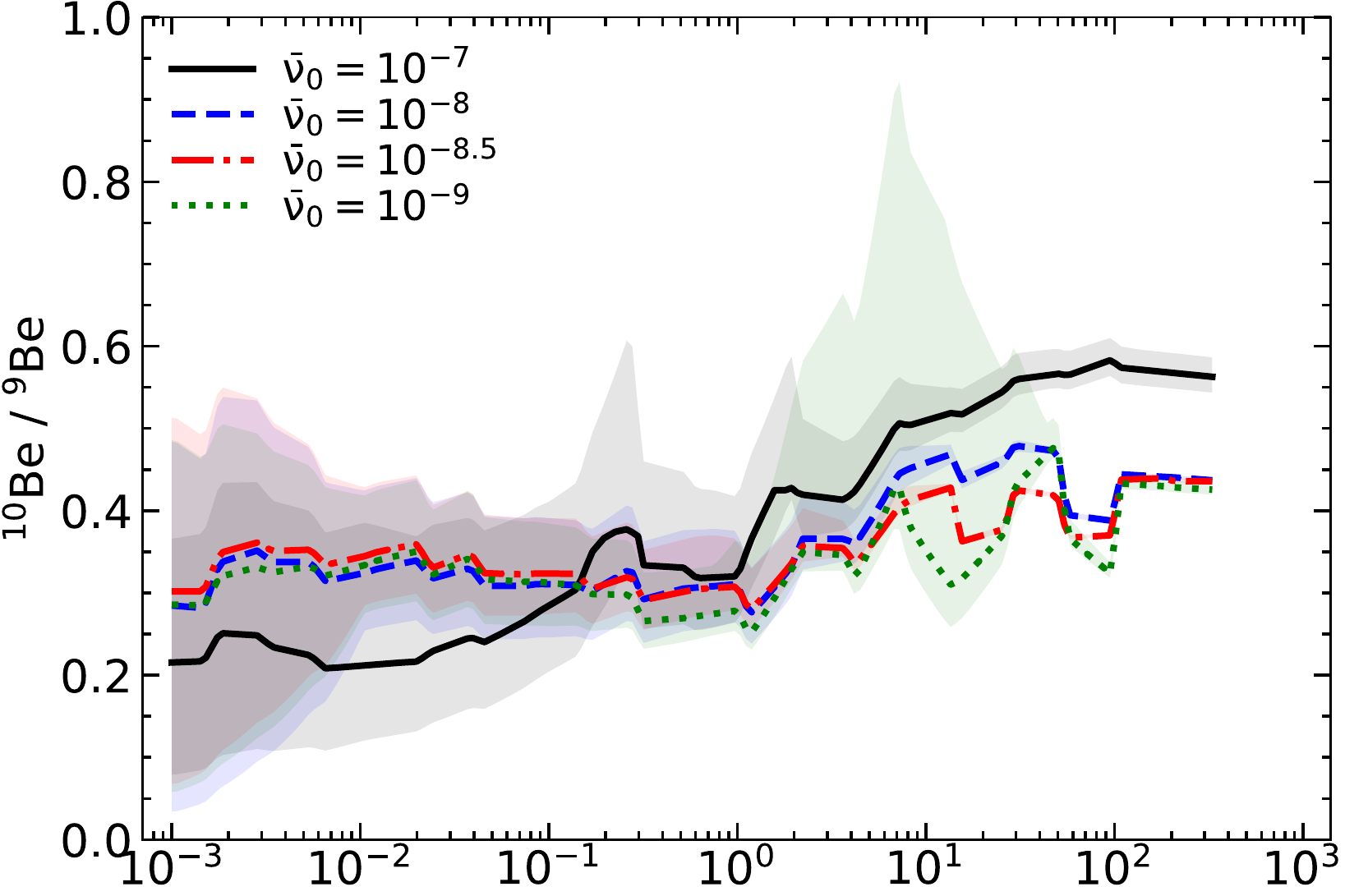}&
	\includegraphics[width=0.32\textwidth]{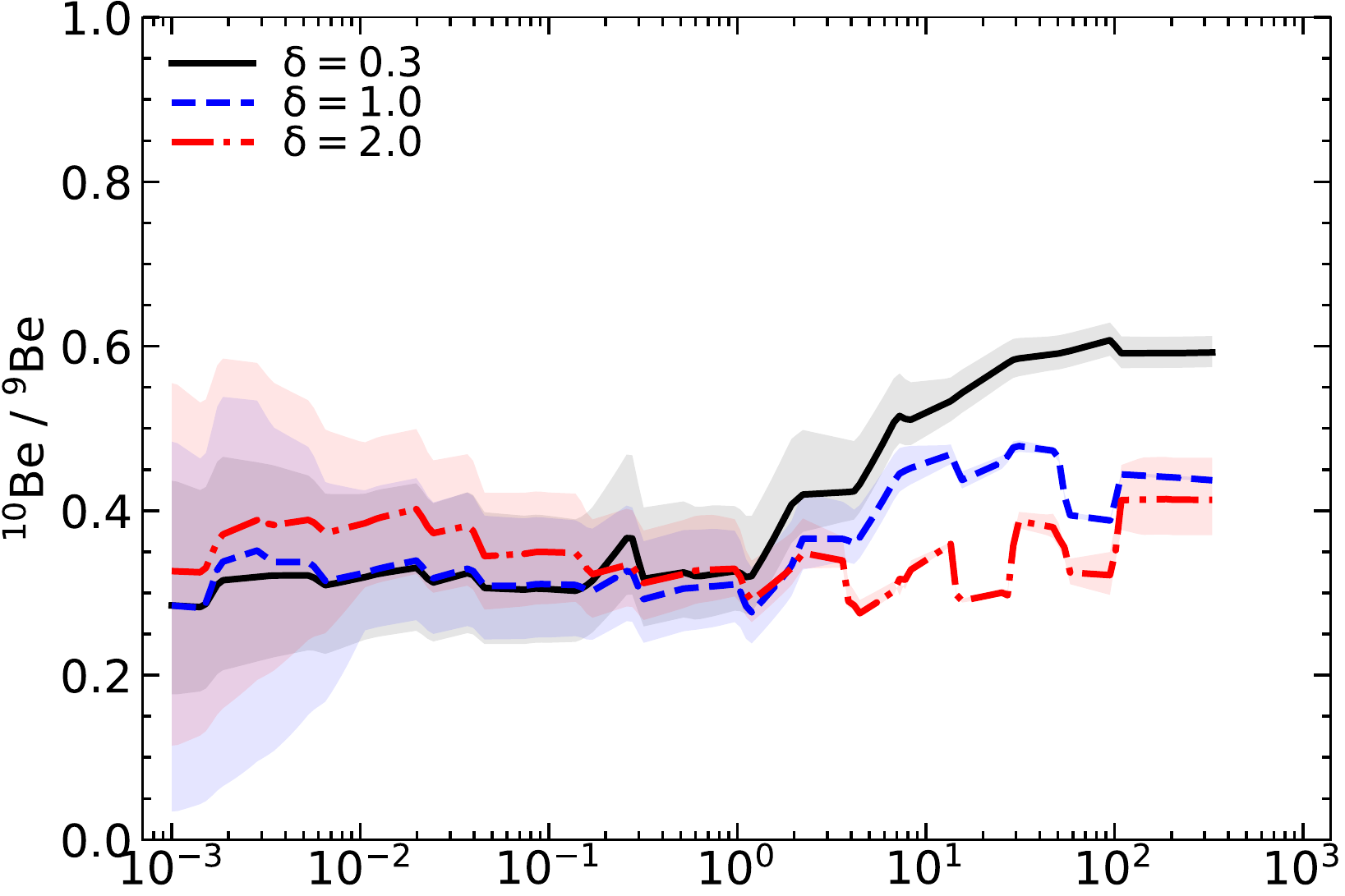} \\
	\includegraphics[width=0.33\textwidth]{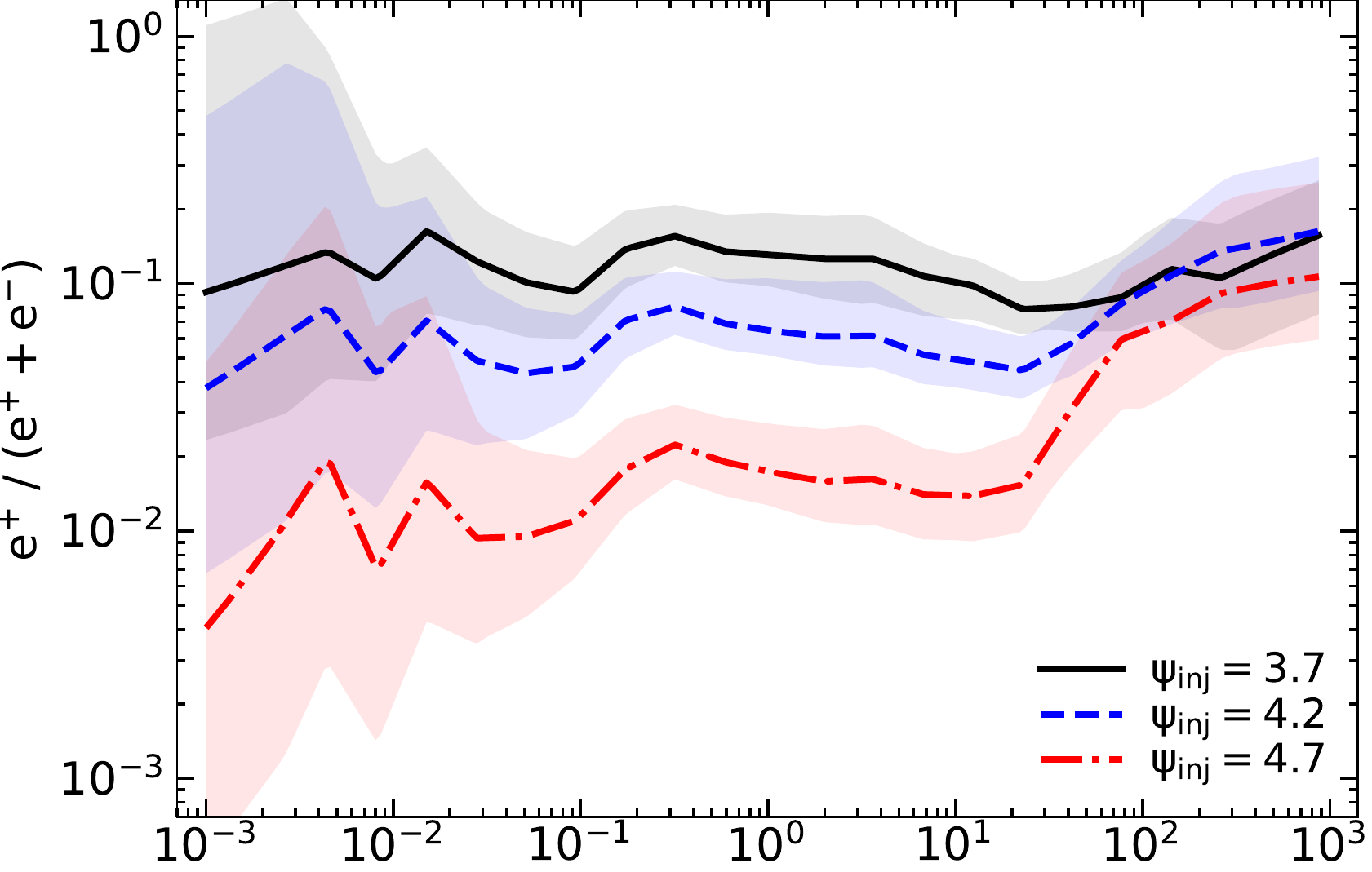}&
	\includegraphics[width=0.33\textwidth]{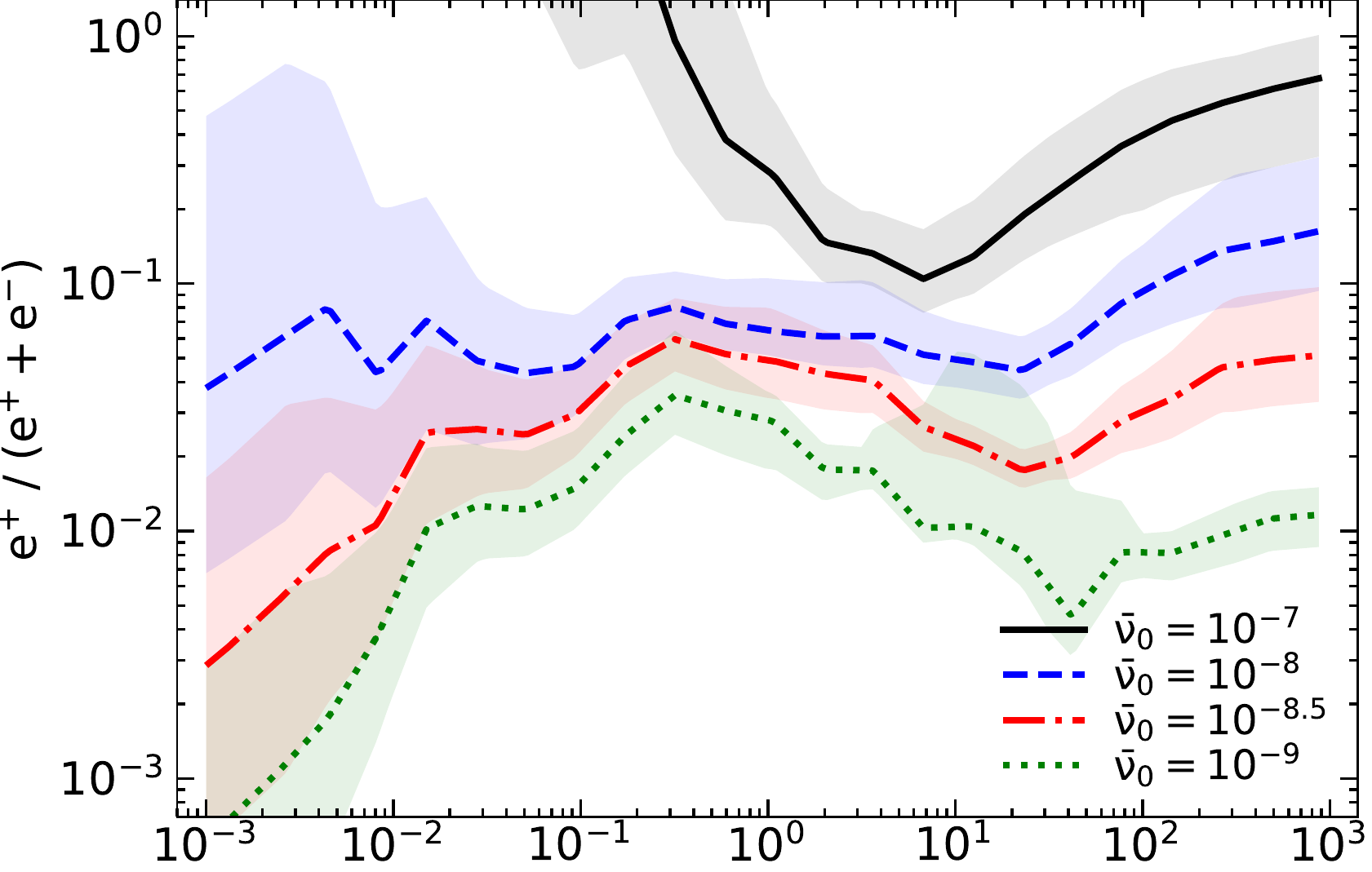}&
	\includegraphics[width=0.33\textwidth]{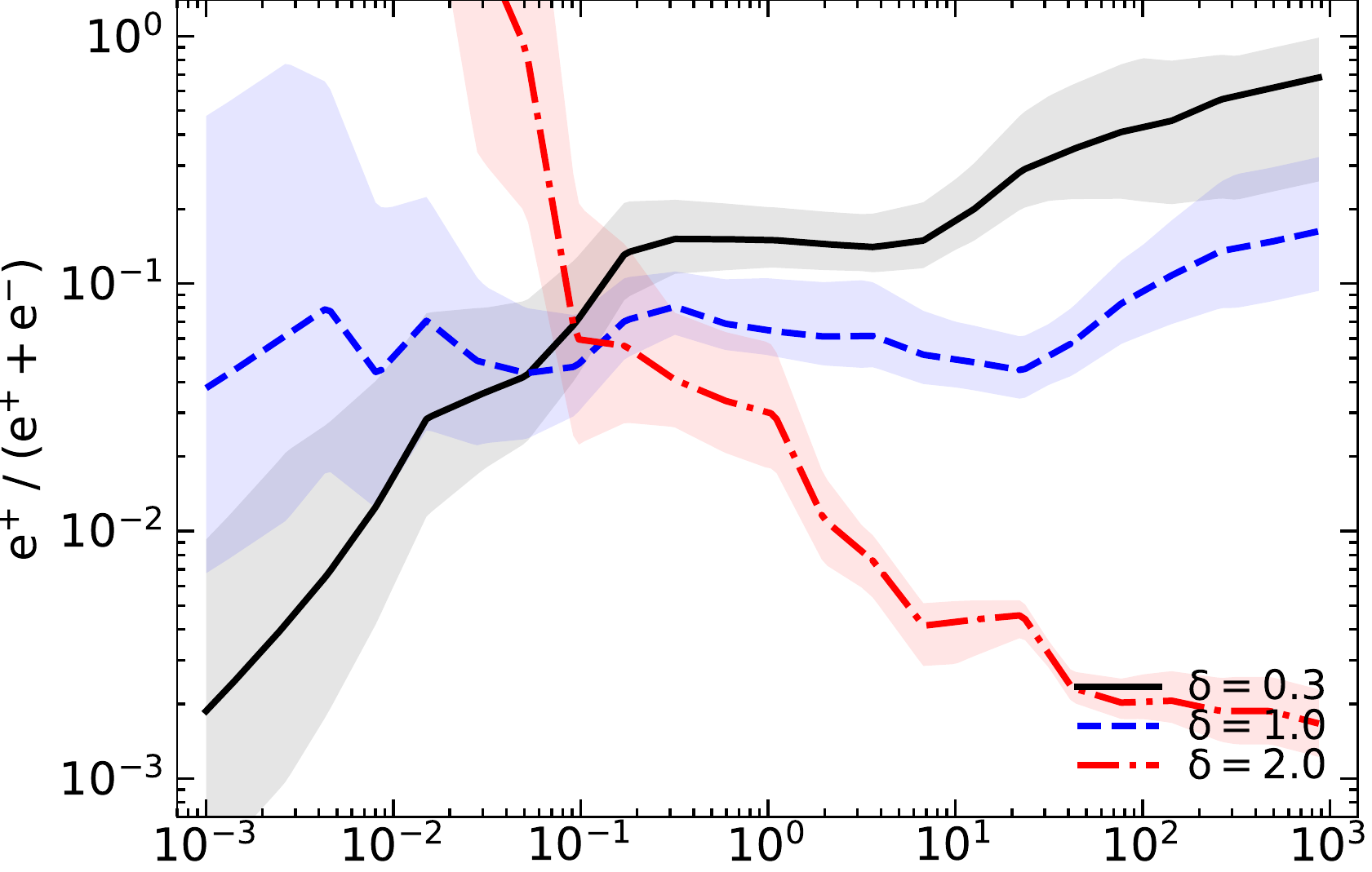} \\
	\includegraphics[width=0.33\textwidth]{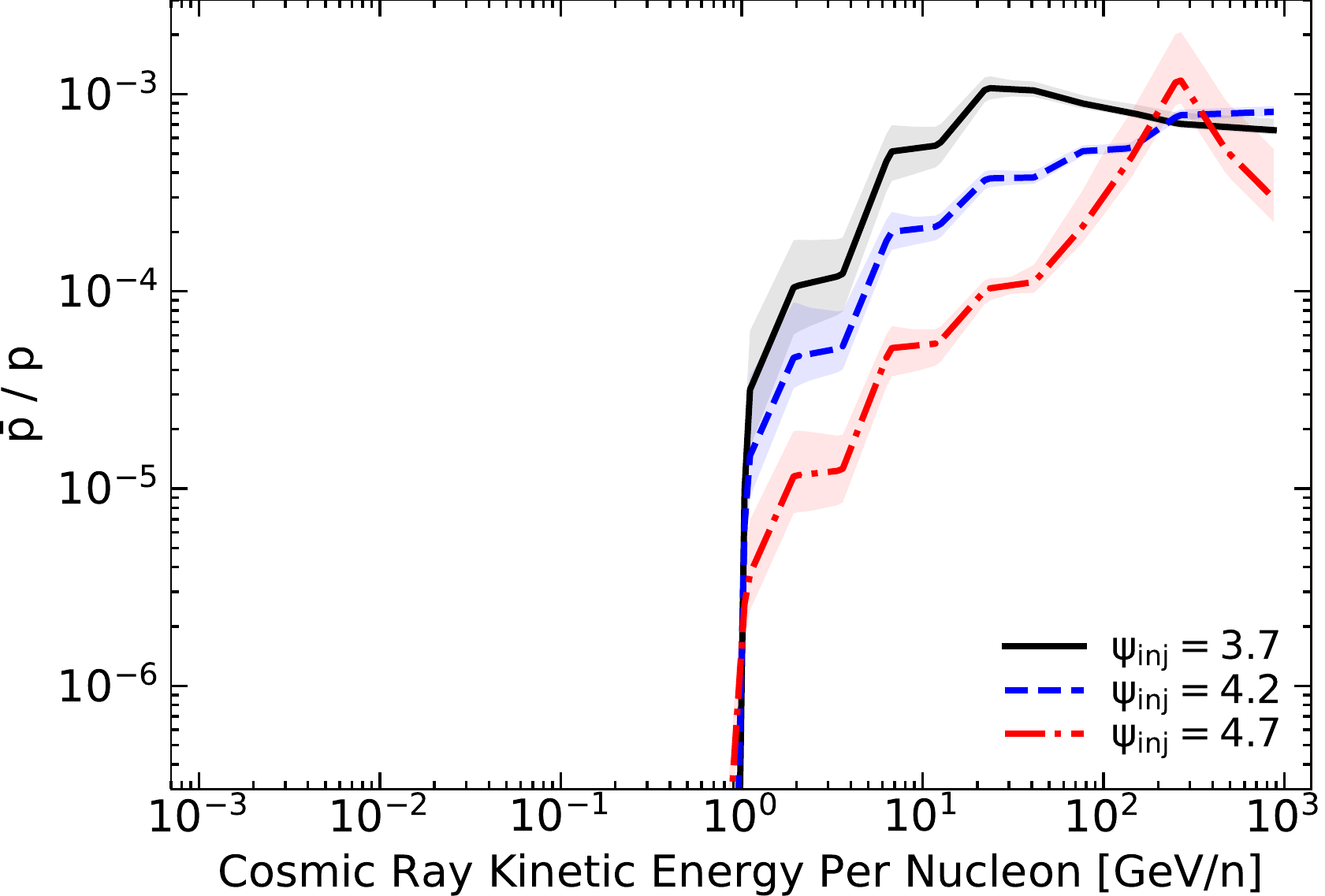}&
	\includegraphics[width=0.33\textwidth]{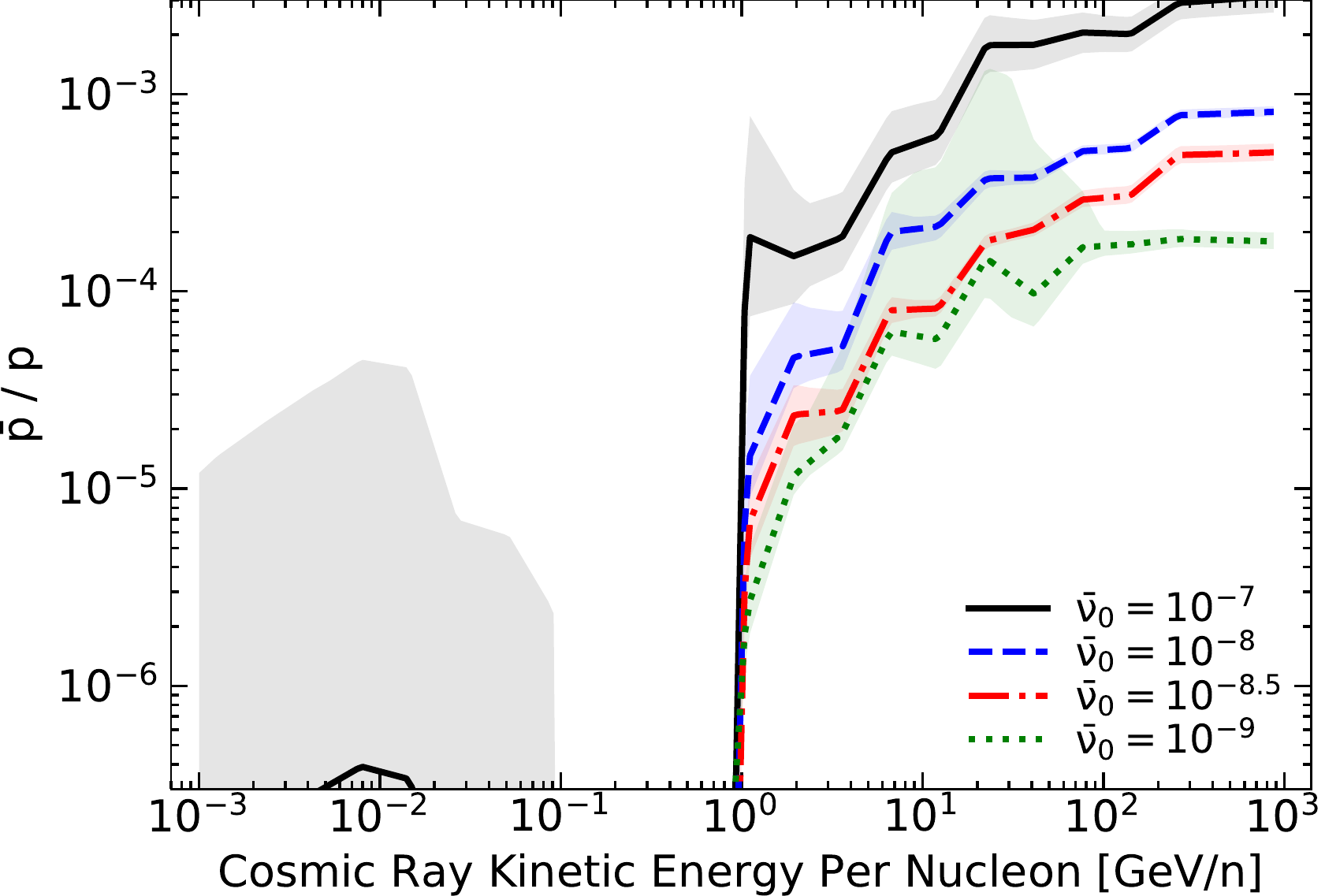}&
	\includegraphics[width=0.33\textwidth]{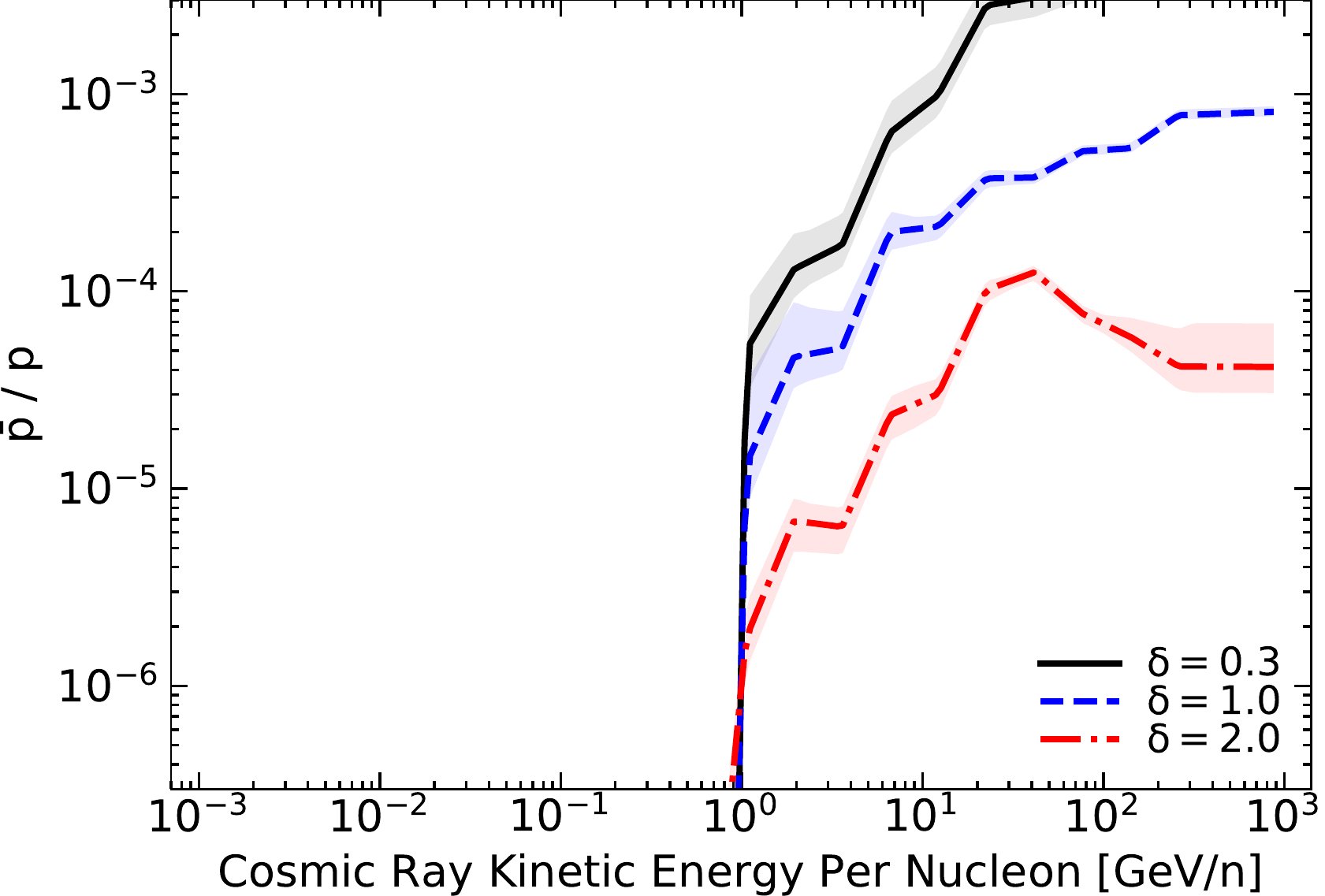} \\
\end{tabular}
	\vspace{-0.3cm}
	\caption{As Fig.~\ref{fig:spec.compare.kappa}, comparing CR spectra in simulations with different injection slopes $\psi_{\rm inj}$, scattering rate normalization $\bar{\nu}_{0}$, and dependence of scattering rates on rigidity $\delta$. The ``reference'' parameters here are different from Fig.~\ref{fig:spec.compare.kappa}: we vary about a ``reference'' model with $\bar{\nu} \sim 10^{-8}\,{\rm s}^{-1}\,\beta\,R_{\rm GV}^{-1}$ -- re-fitting $\delta$ to compensate as best as possible for a higher $\bar{\nu}_{0}$ (lower diffusivity at $\sim 1\,$GV). This model is a notably poorer fit to the observations in Fig.~\ref{fig:demo.cr.spectra.fiducial} compared to that in the main text, but represents a model ``re-tuned'' to at least reasonably fit B/C with different parameters. Systematically varying the parameters about the reference model defaults allows us to see that all the qualitative conclusions from Fig.~\ref{fig:spec.compare.kappa} regarding the systematic effects of these parameter variations are robust to the ``reference'' model or other parameter choices.
	\label{fig:spec.compare.kappa.alt}\vspace{-0.4cm}}
\end{figure*}

\begin{figure*}
\begin{tabular}{r@{\hspace{0pt}}r@{\hspace{0pt}}r}
	\includegraphics[width=0.33\textwidth]{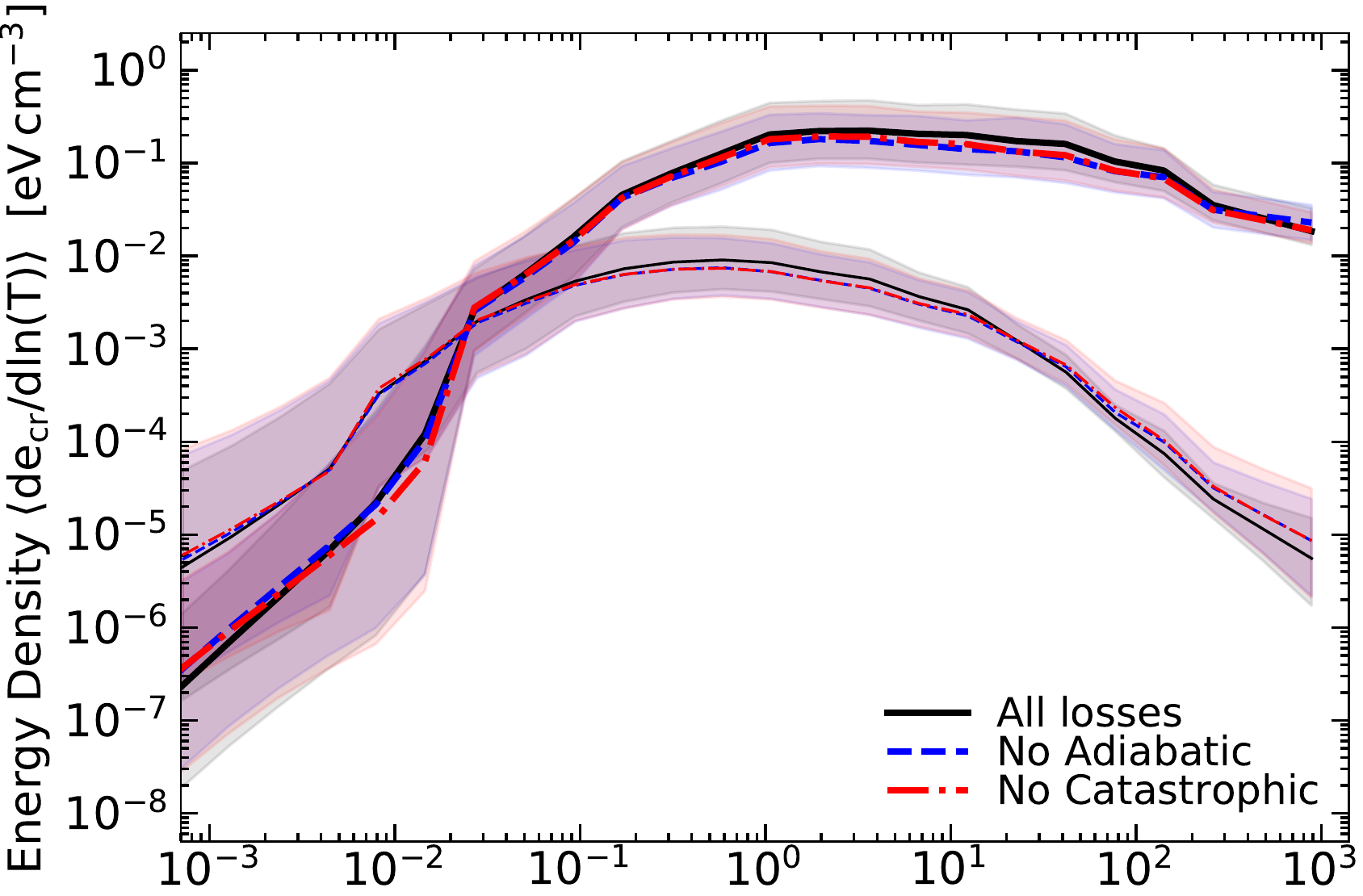}&
	\includegraphics[width=0.32\textwidth]{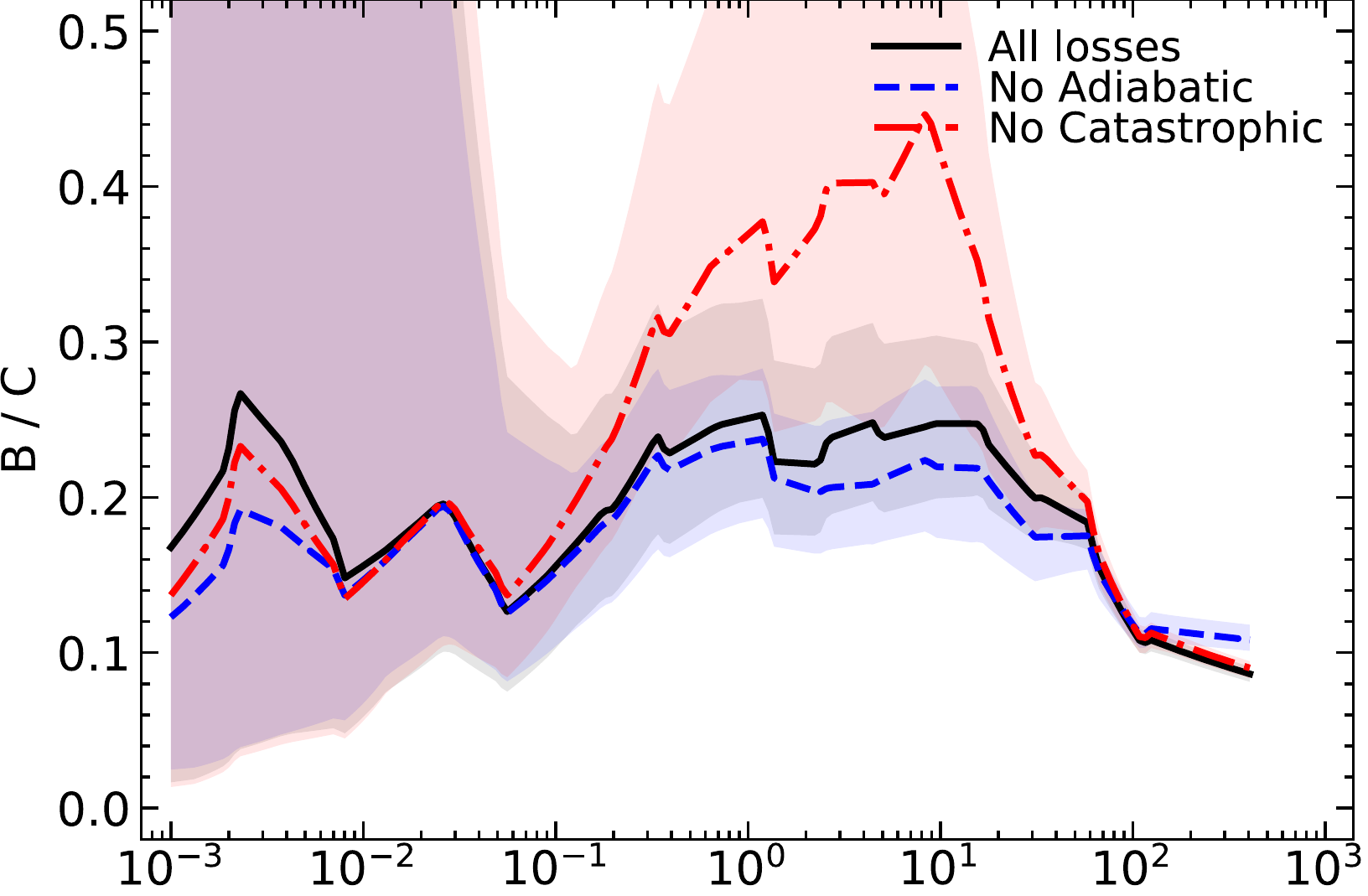}&
	\includegraphics[width=0.33\textwidth]{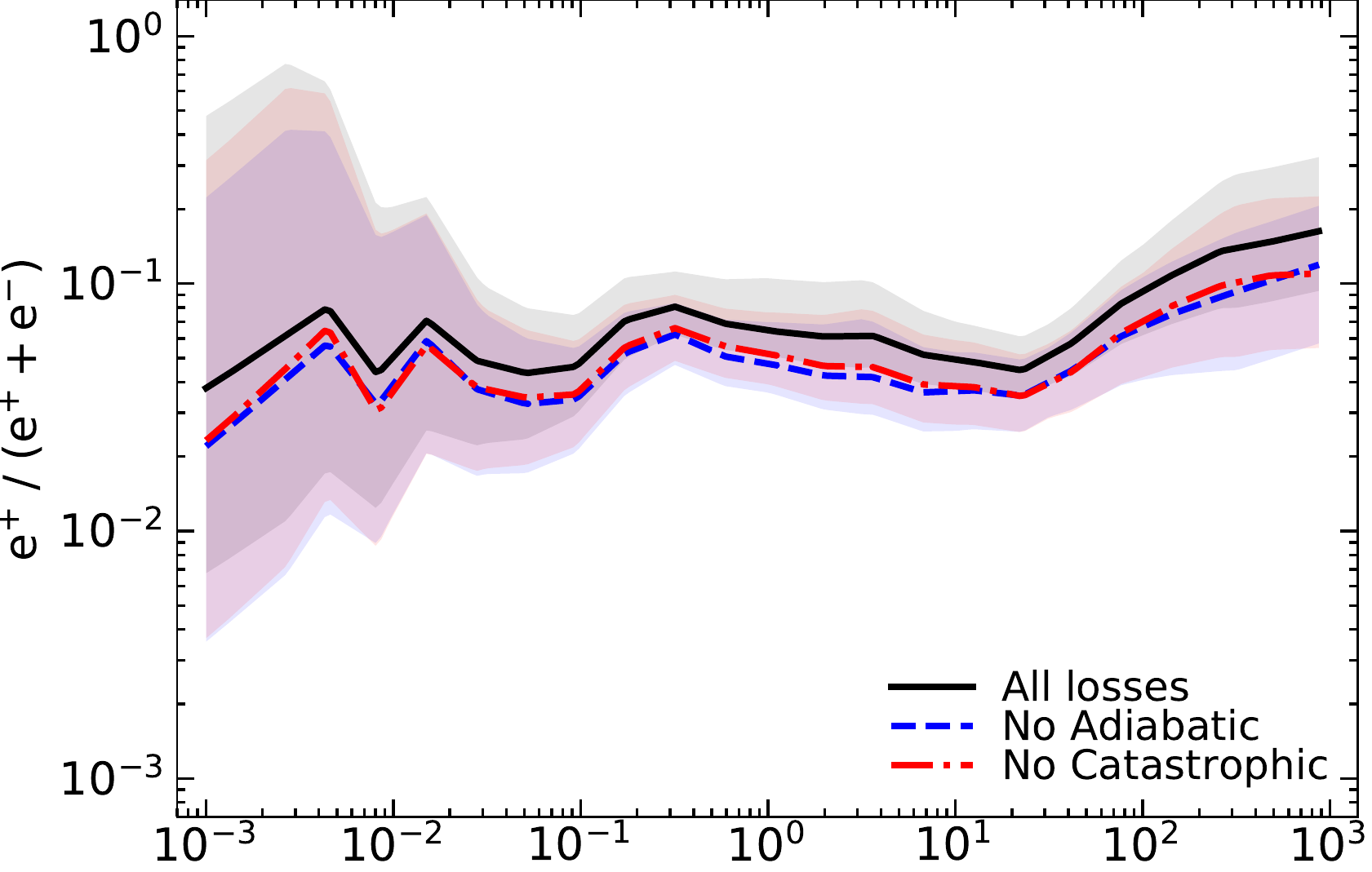}
	\\
	\includegraphics[width=0.33\textwidth]{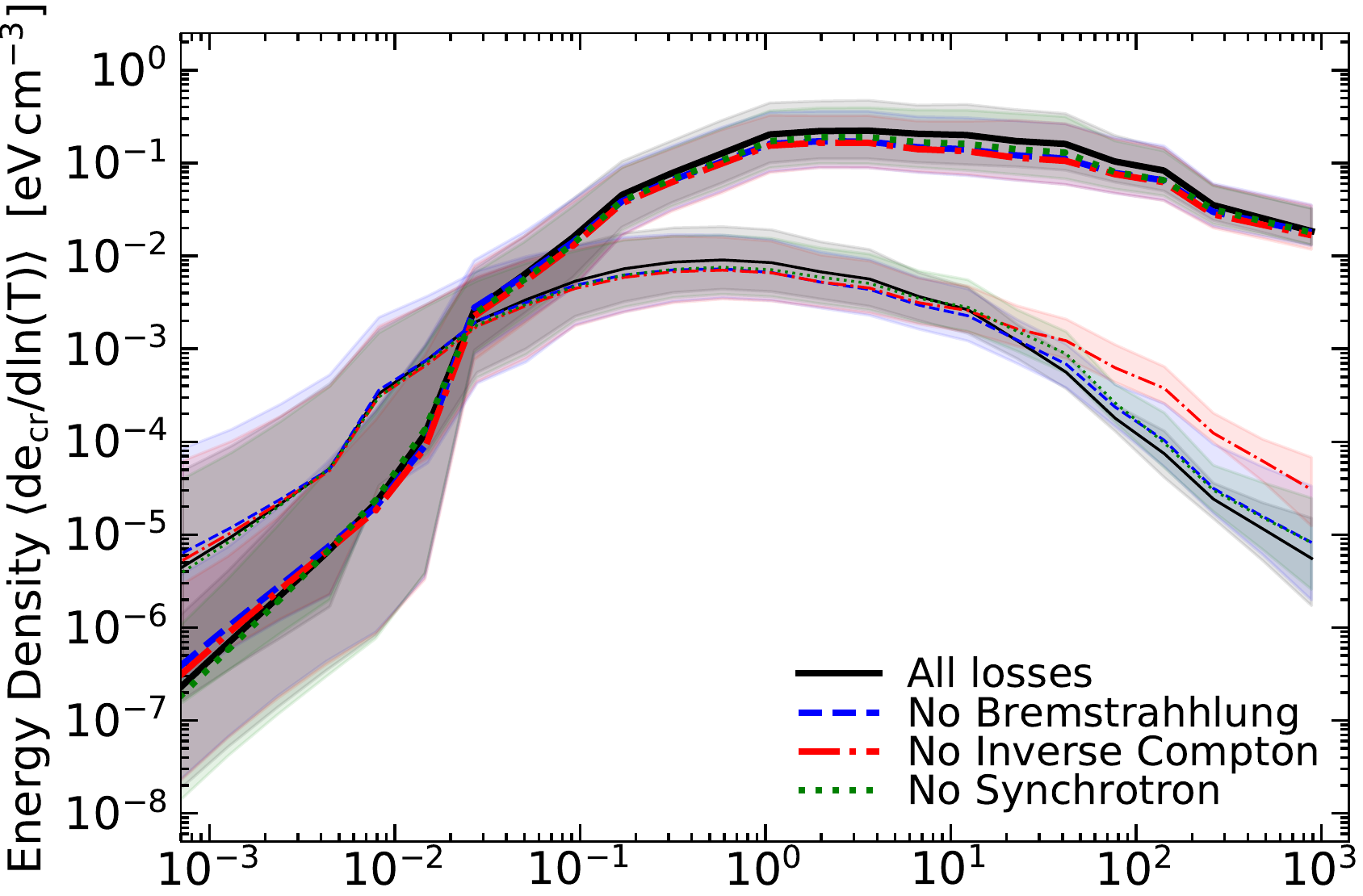}&
	\includegraphics[width=0.32\textwidth]{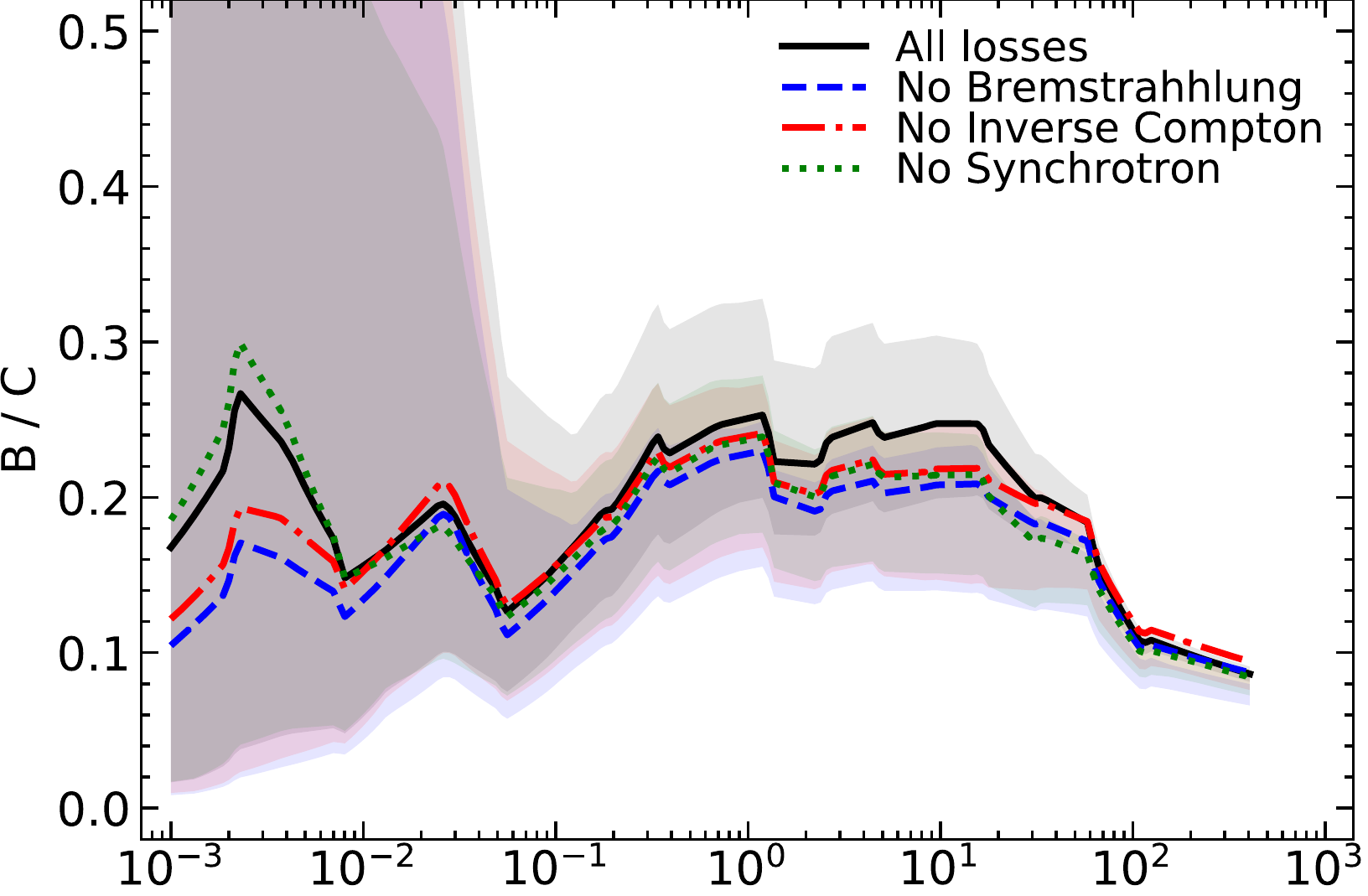}&
	\includegraphics[width=0.33\textwidth]{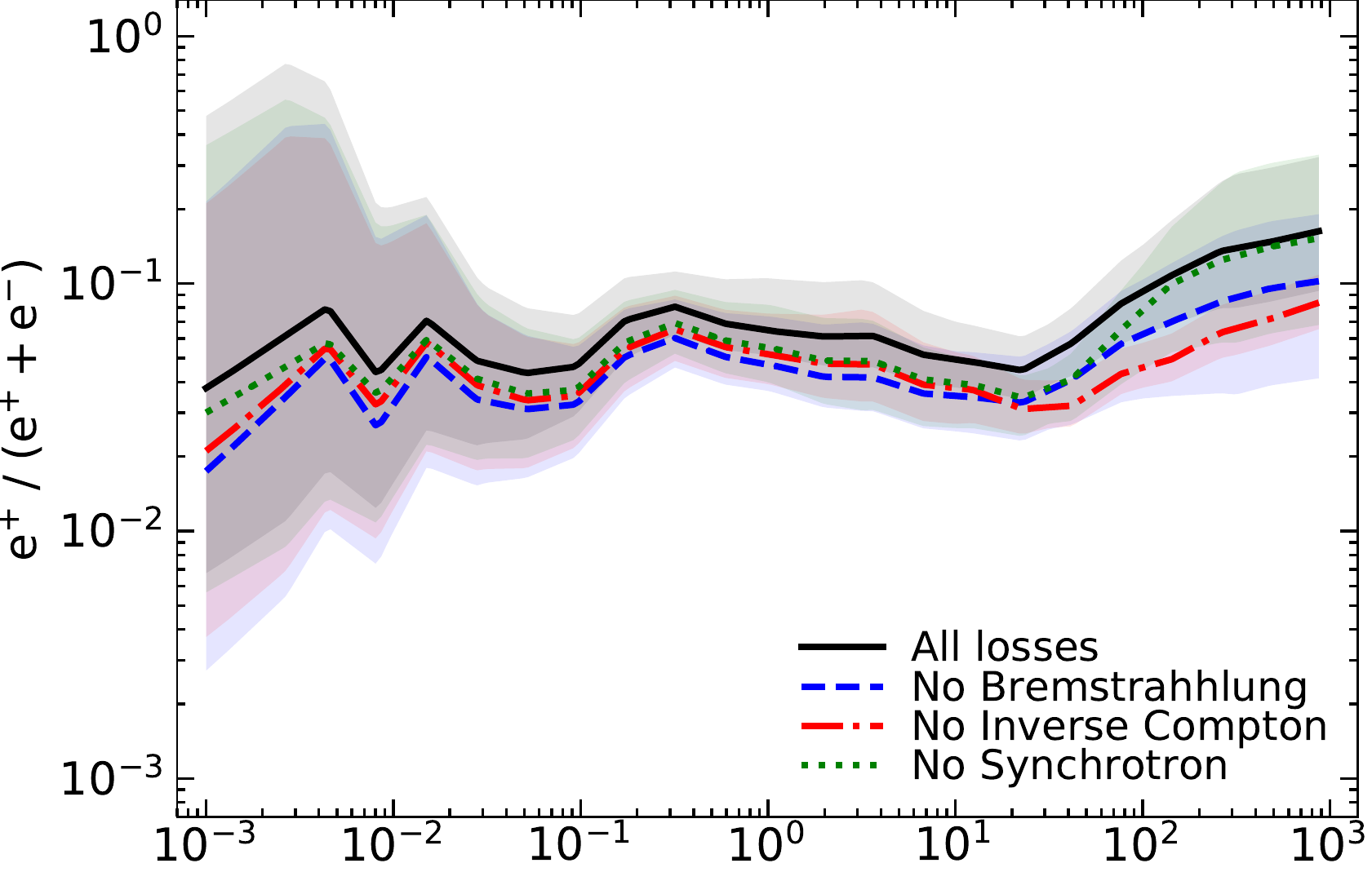}
	\\
	\includegraphics[width=0.33\textwidth]{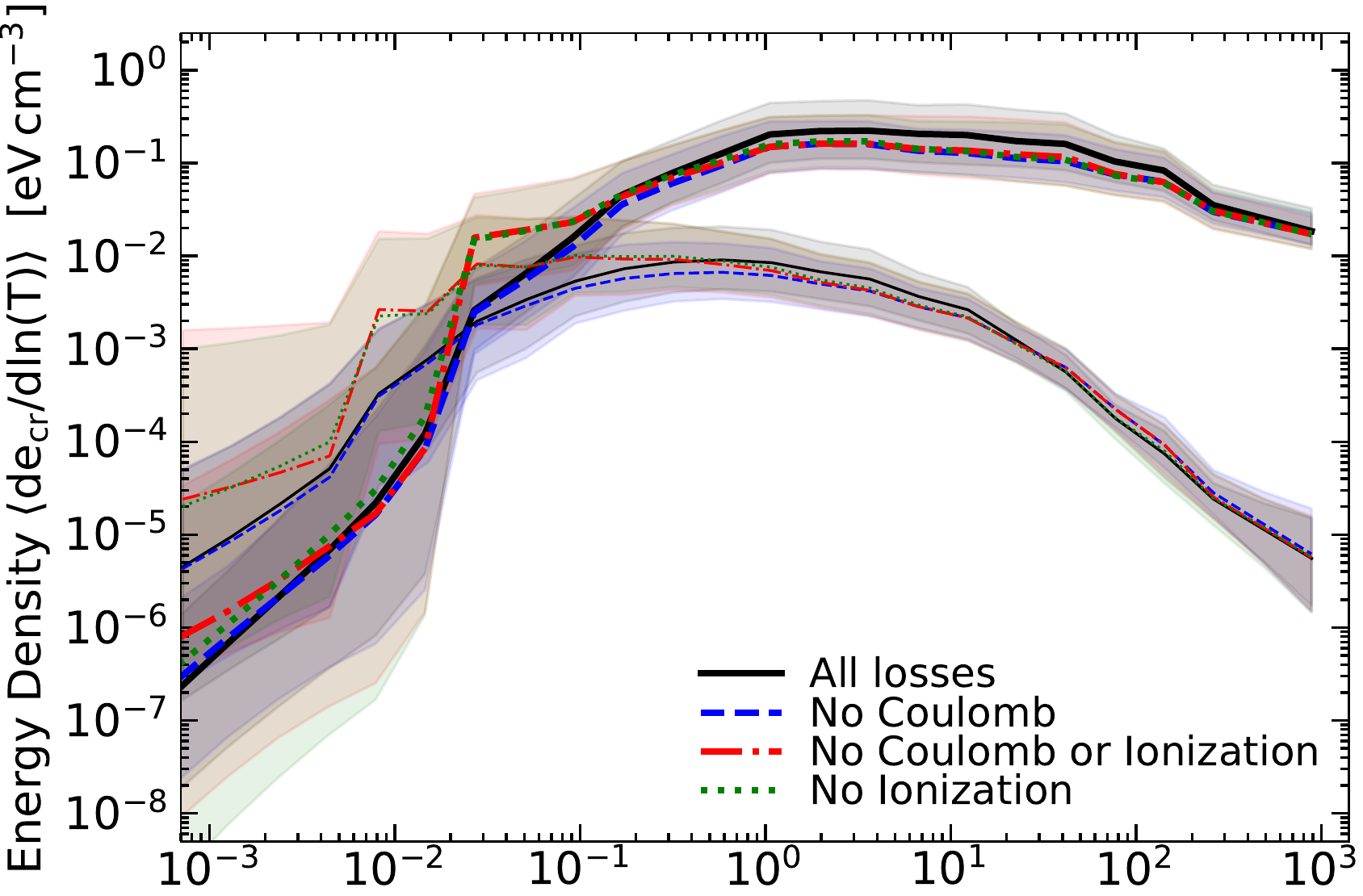} &
	\includegraphics[width=0.32\textwidth]{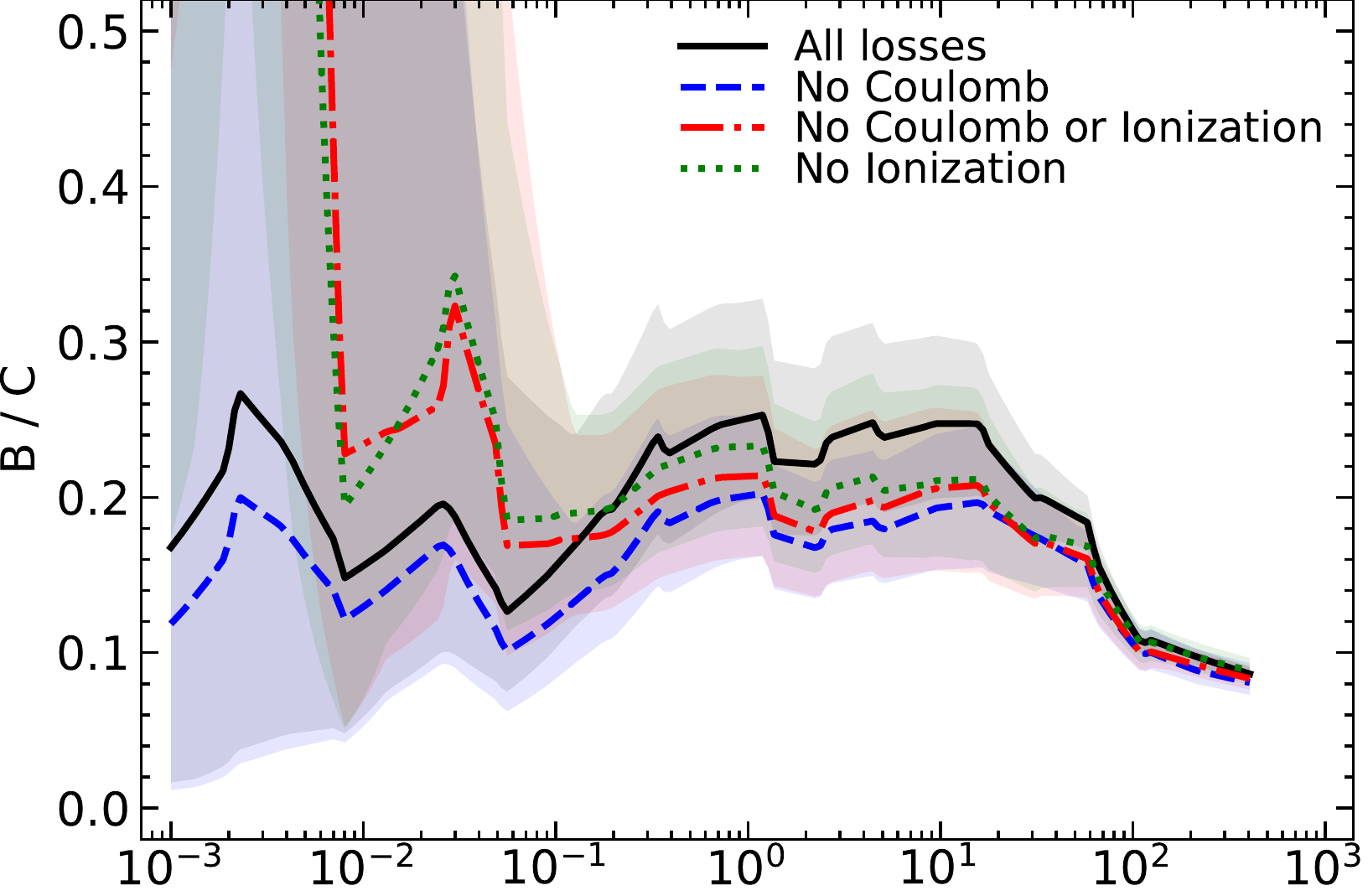} &
	\includegraphics[width=0.33\textwidth]{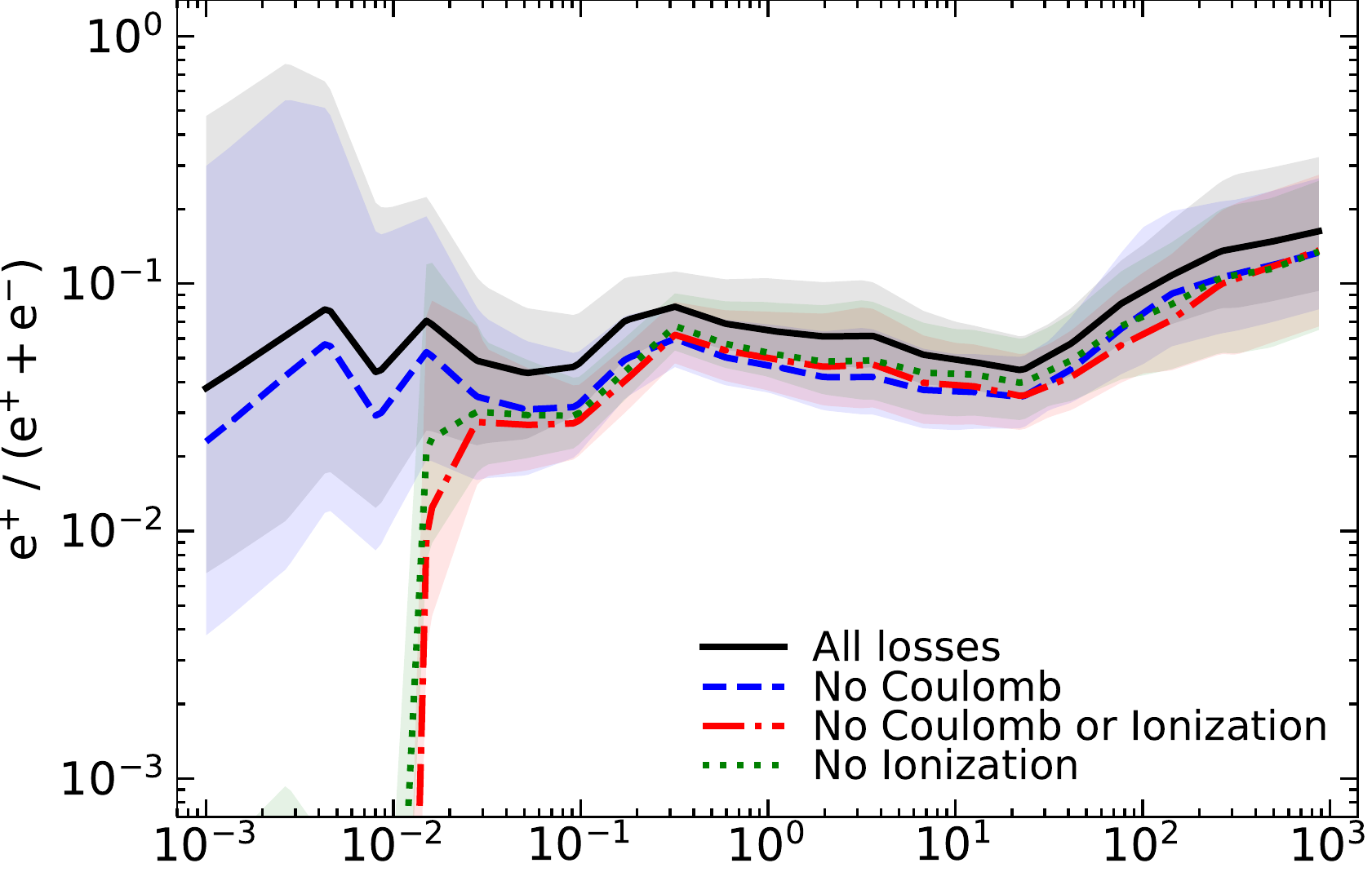} 
	\\
	\includegraphics[width=0.33\textwidth]{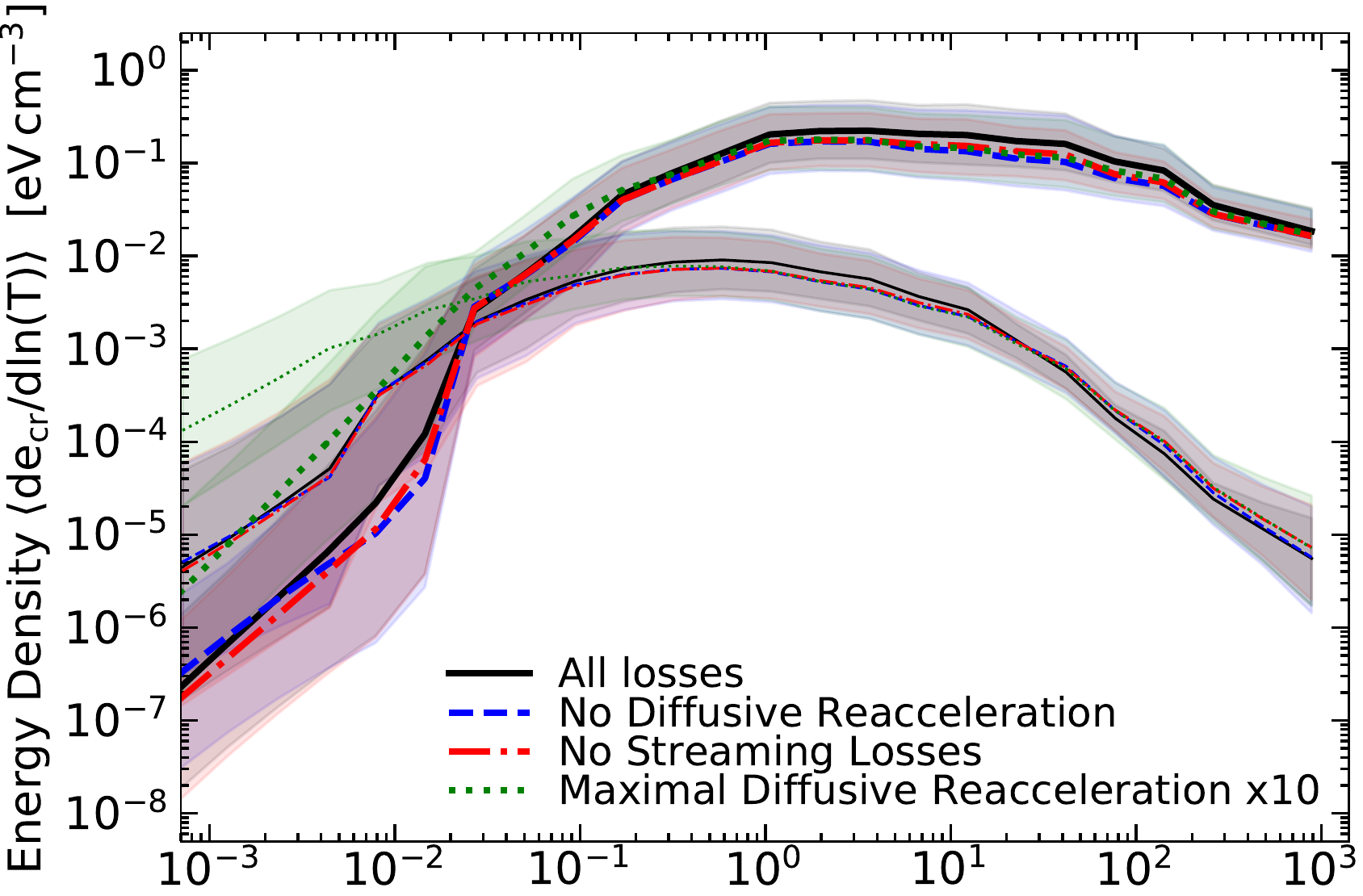}&
	\includegraphics[width=0.32\textwidth]{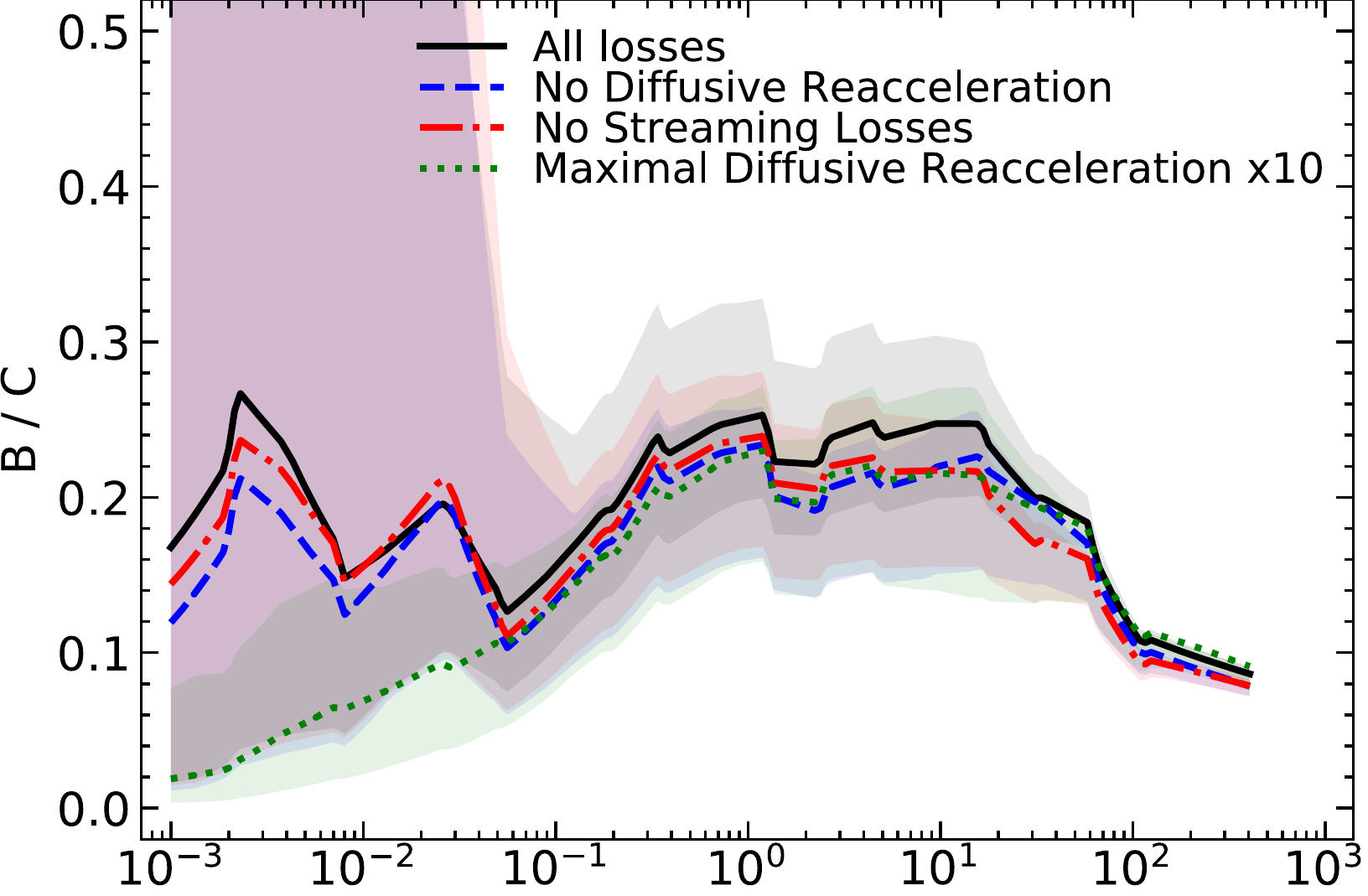}&	
	\includegraphics[width=0.33\textwidth]{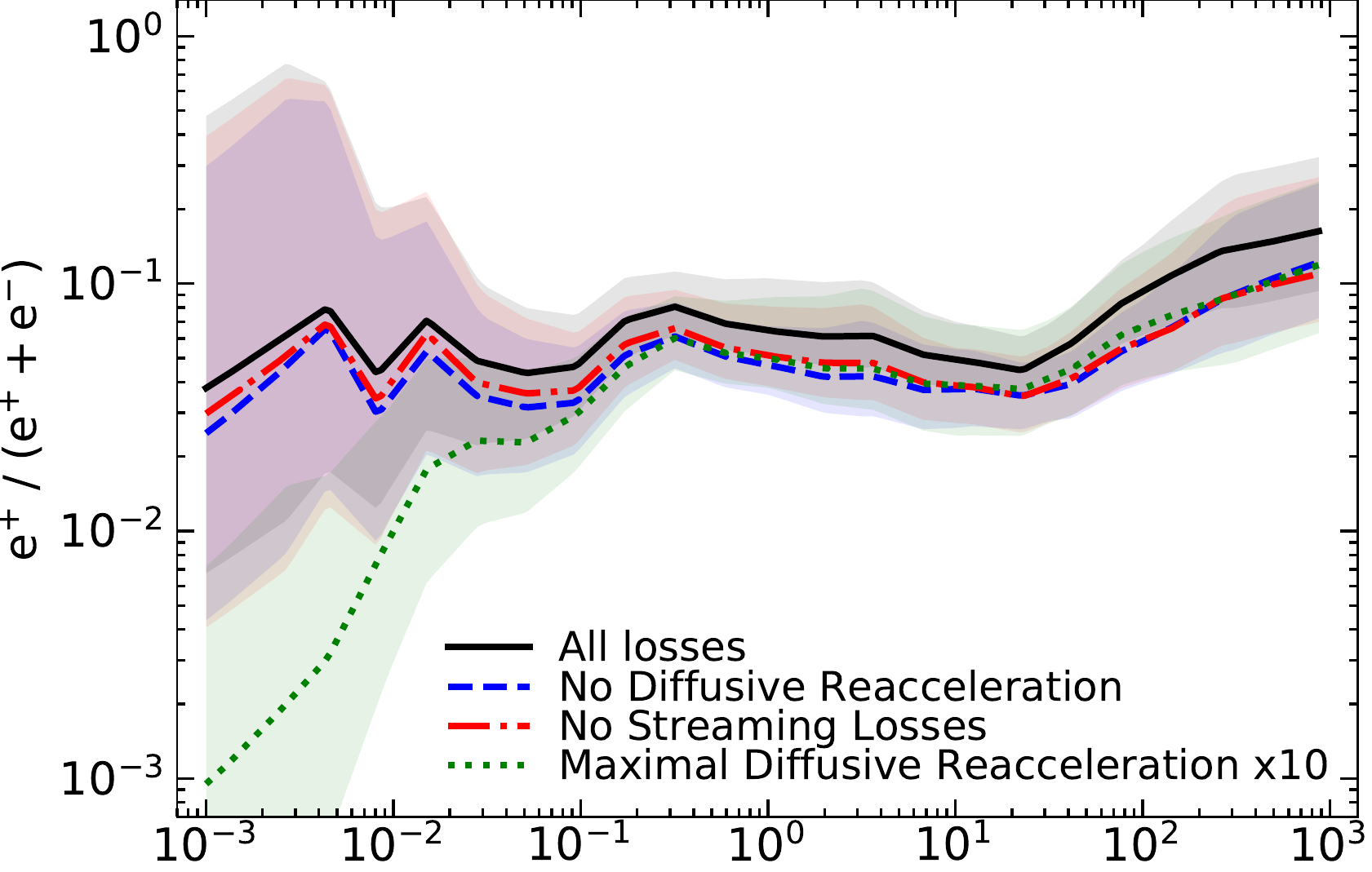}
	\\
	\includegraphics[width=0.33\textwidth]{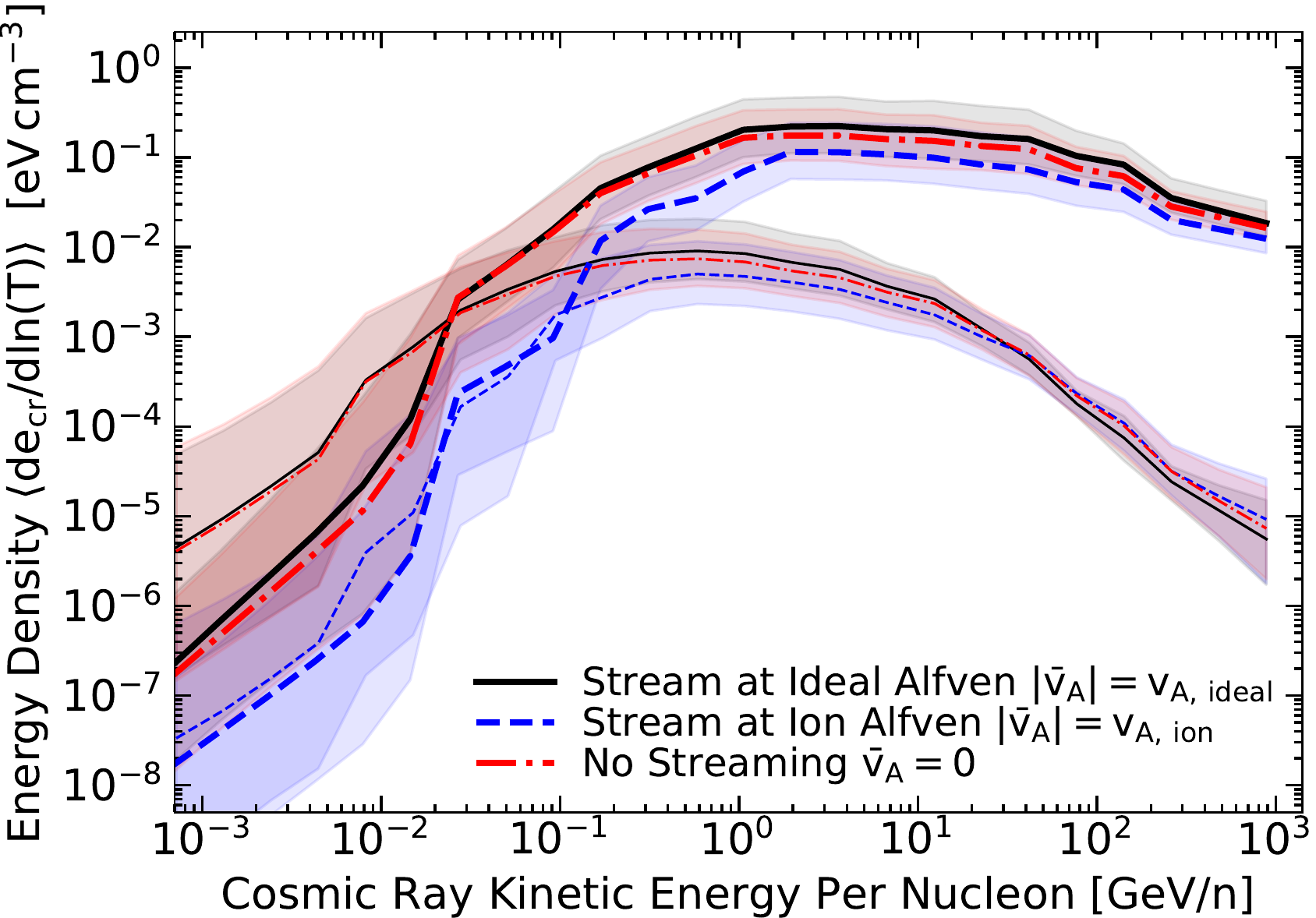}&
	\includegraphics[width=0.32\textwidth]{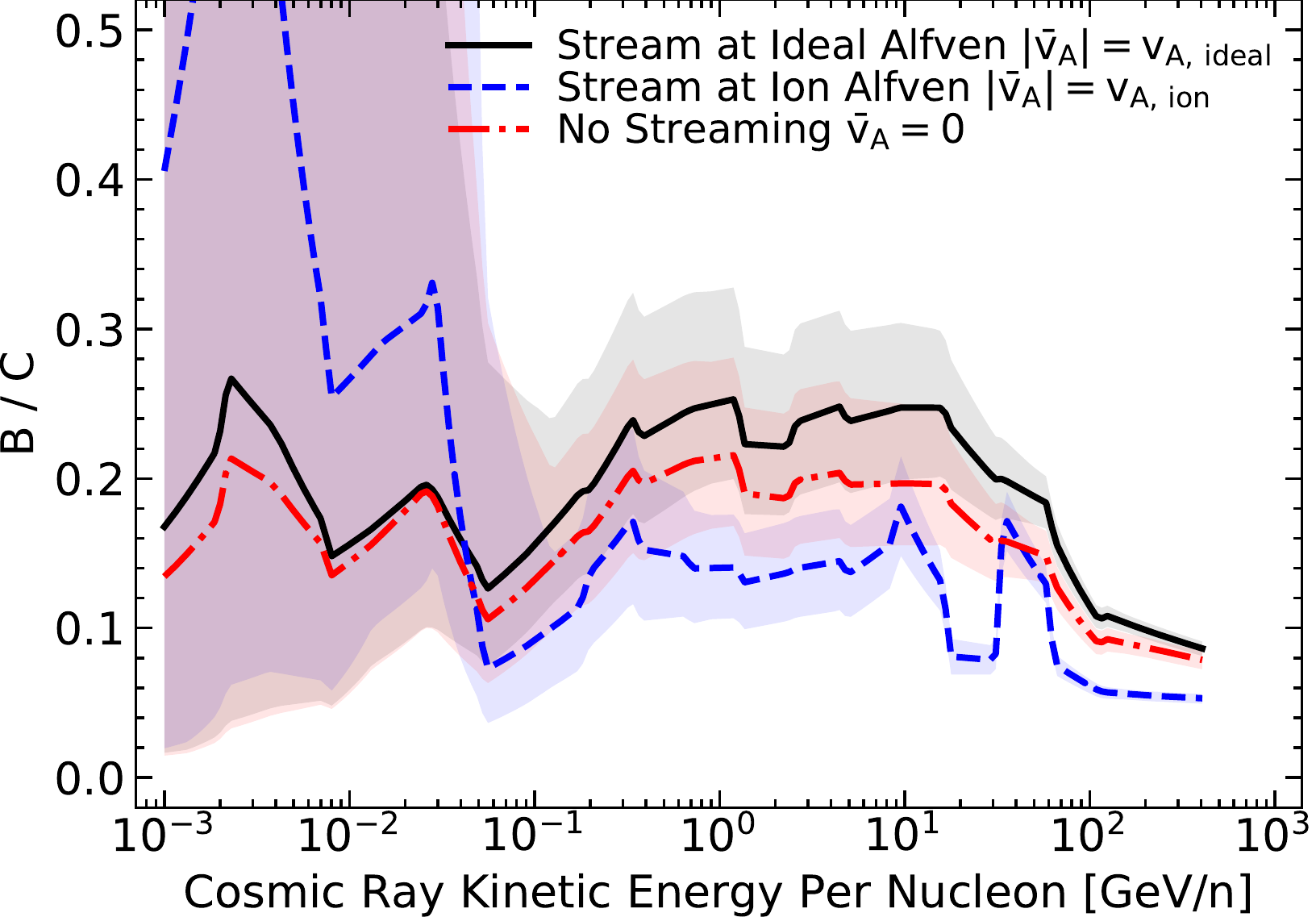}&
	\includegraphics[width=0.33\textwidth]{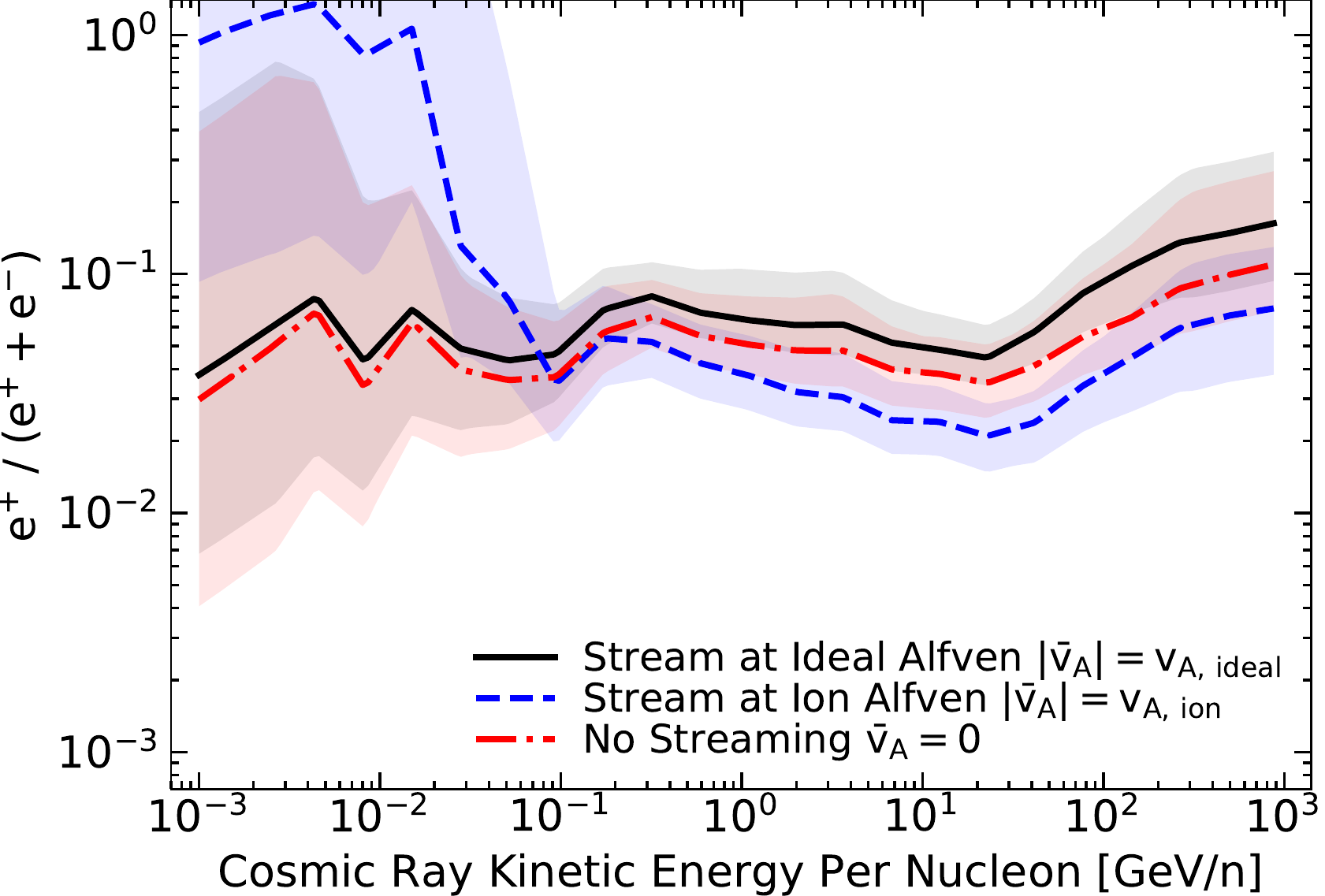}
\end{tabular}
	\vspace{-0.1cm}
	\caption{CR spectra with different loss terms and other physics disabled, as Fig.~\ref{fig:spec.compare.losses}, but with the variations being with respect to the alternative ``reference model'' from Fig.~\ref{fig:spec.compare.kappa.alt}. Again, despite the systematically different reference-model parameters, the systematic effects of these physics variations are consistent with Fig.~\ref{fig:spec.compare.losses}.
	\label{fig:spec.compare.losses.alt}\vspace{-0.4cm}}
\end{figure*}

\begin{figure*}
\begin{tabular}{r@{\hspace{0pt}}r@{\hspace{0pt}}r}
	\includegraphics[width=0.33\textwidth]{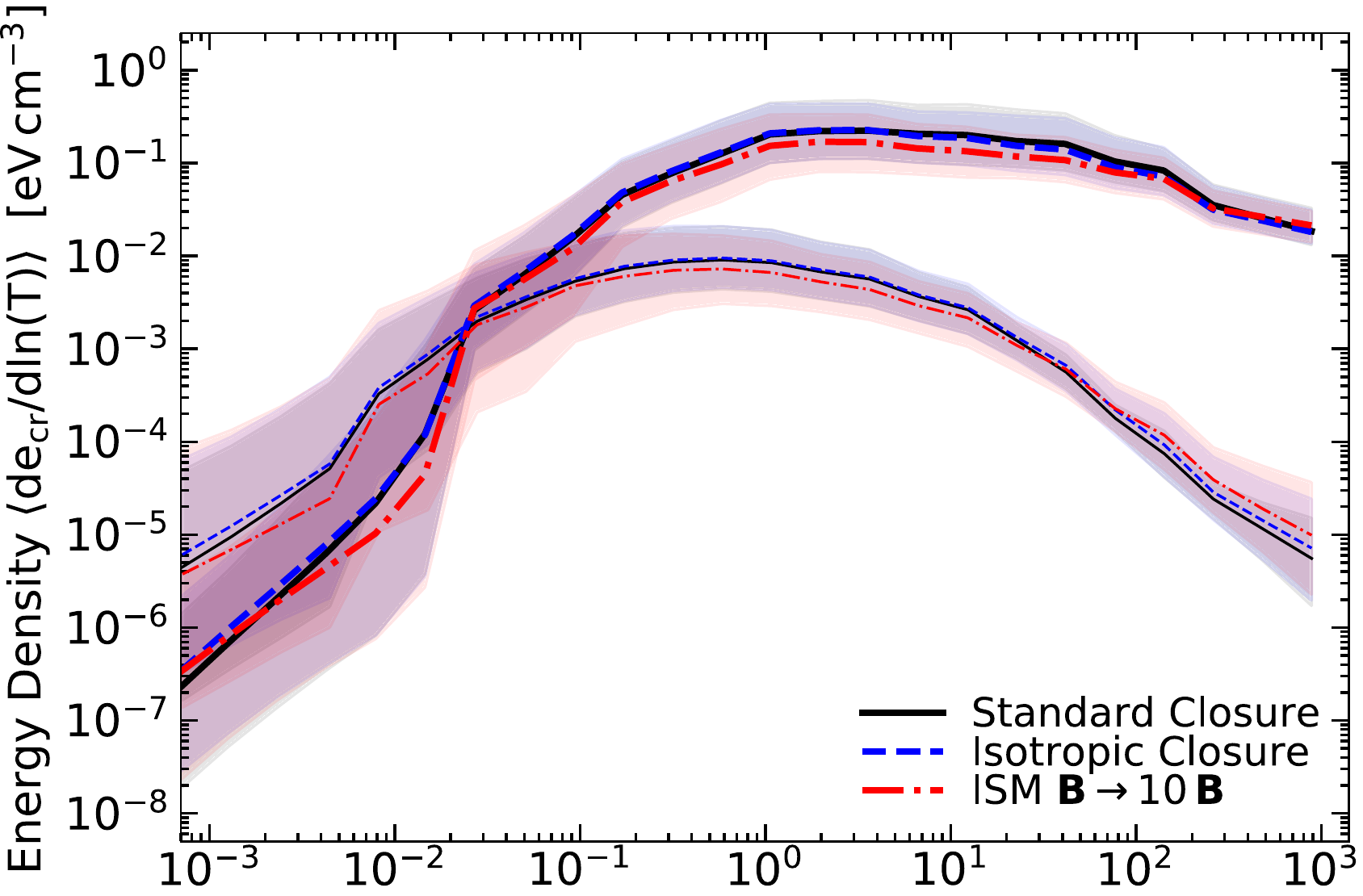}&
	\includegraphics[width=0.32\textwidth]{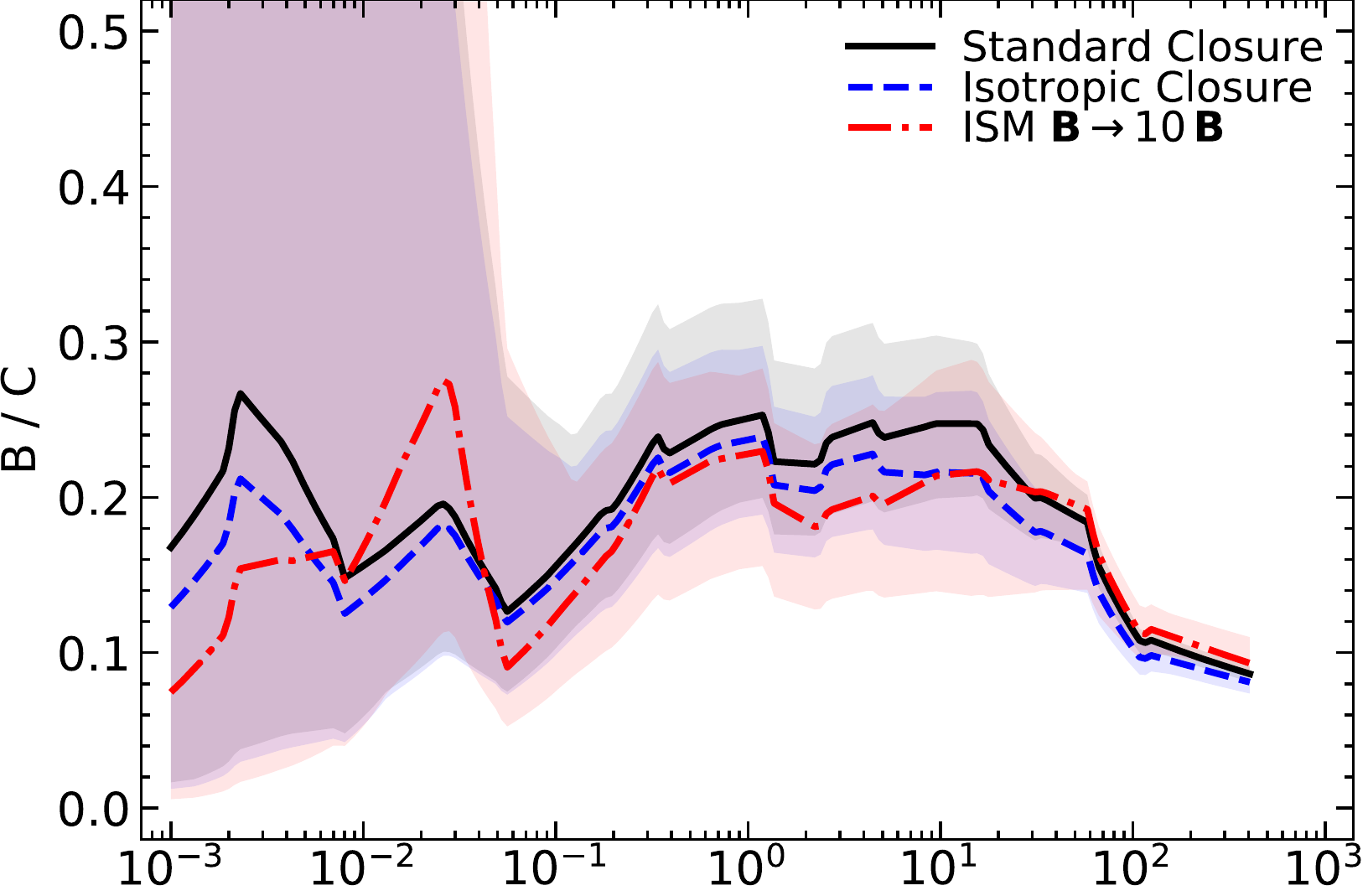}&
	\includegraphics[width=0.33\textwidth]{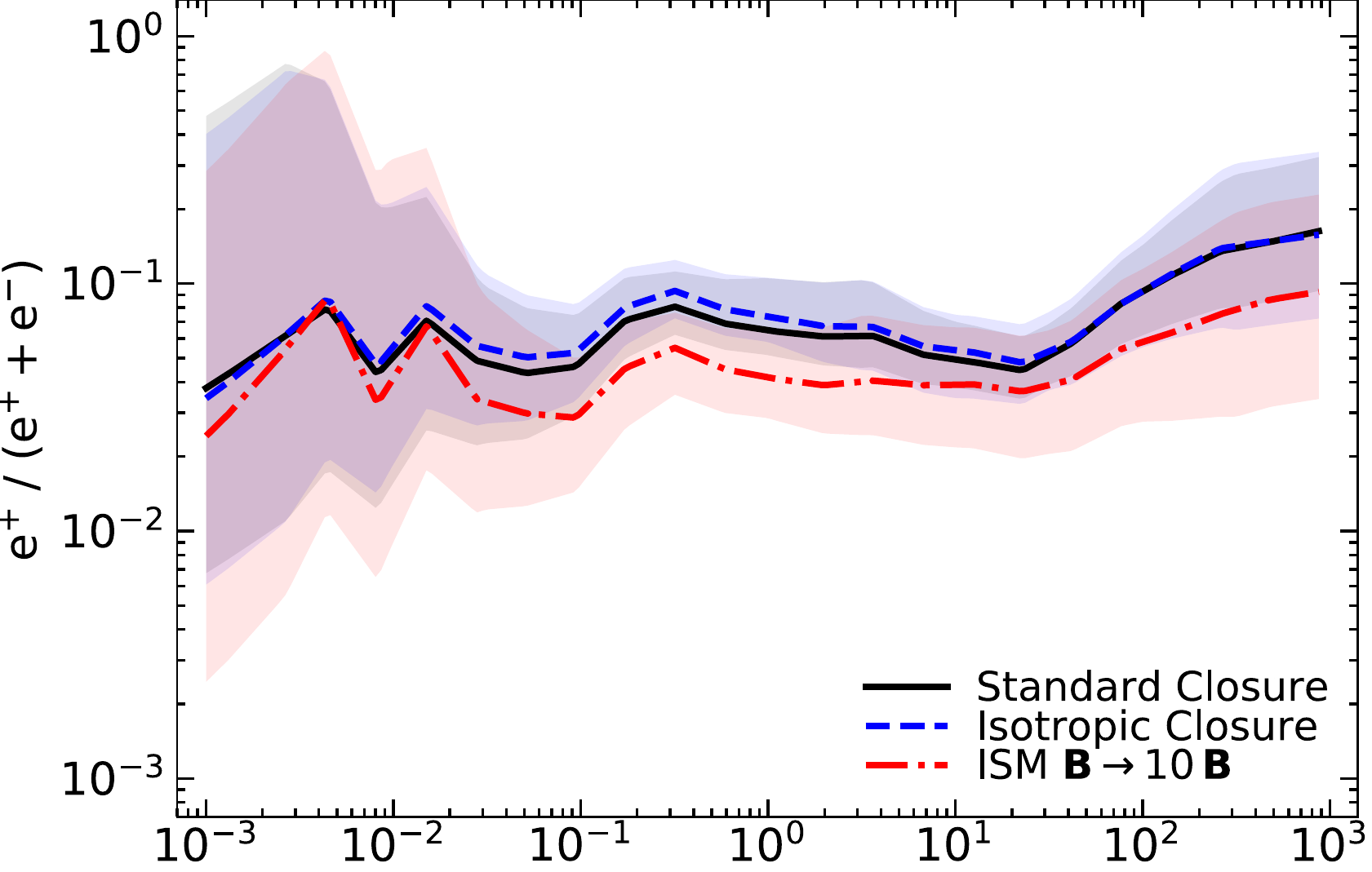}
	\\
	\includegraphics[width=0.33\textwidth]{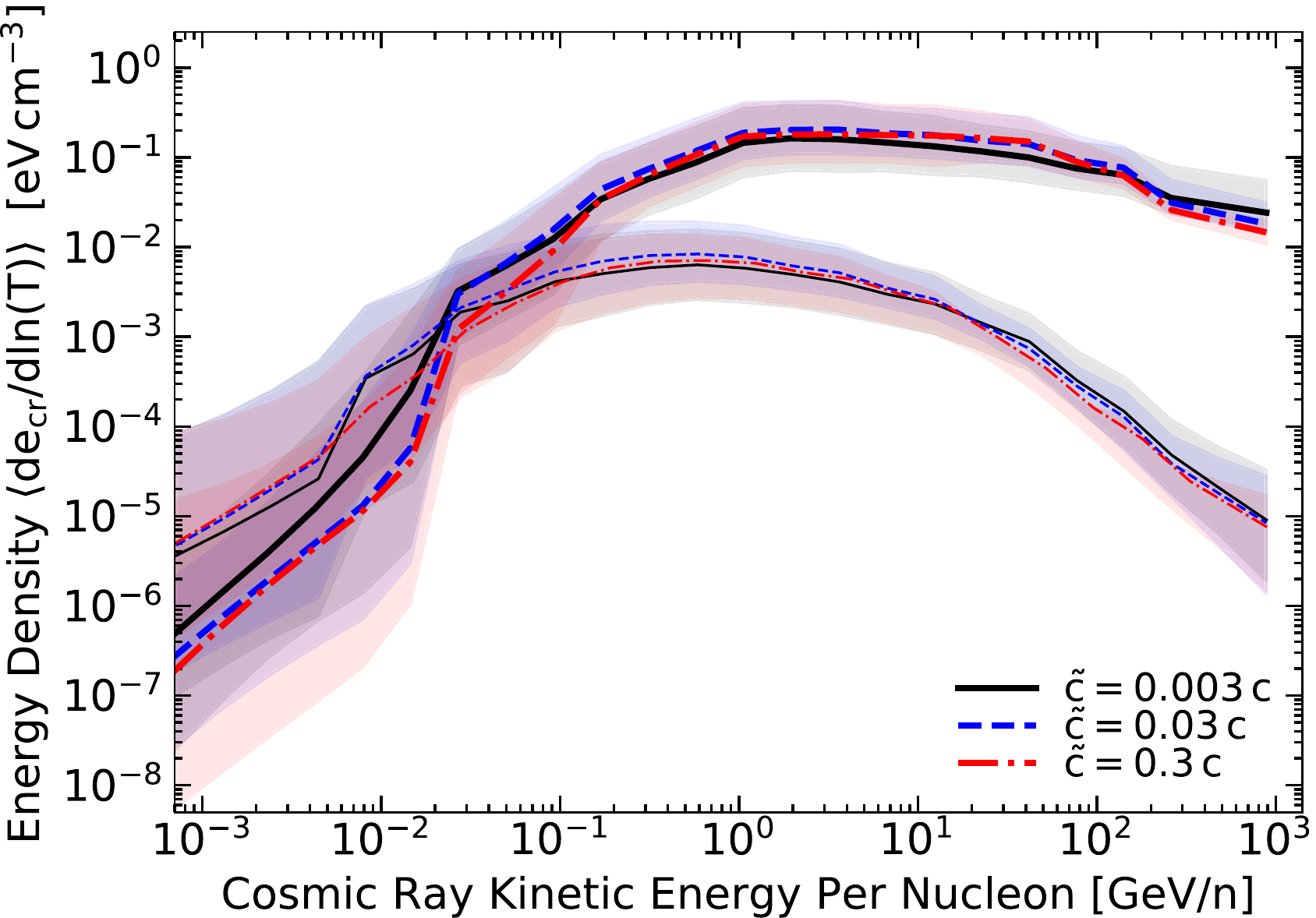} &
	\includegraphics[width=0.32\textwidth]{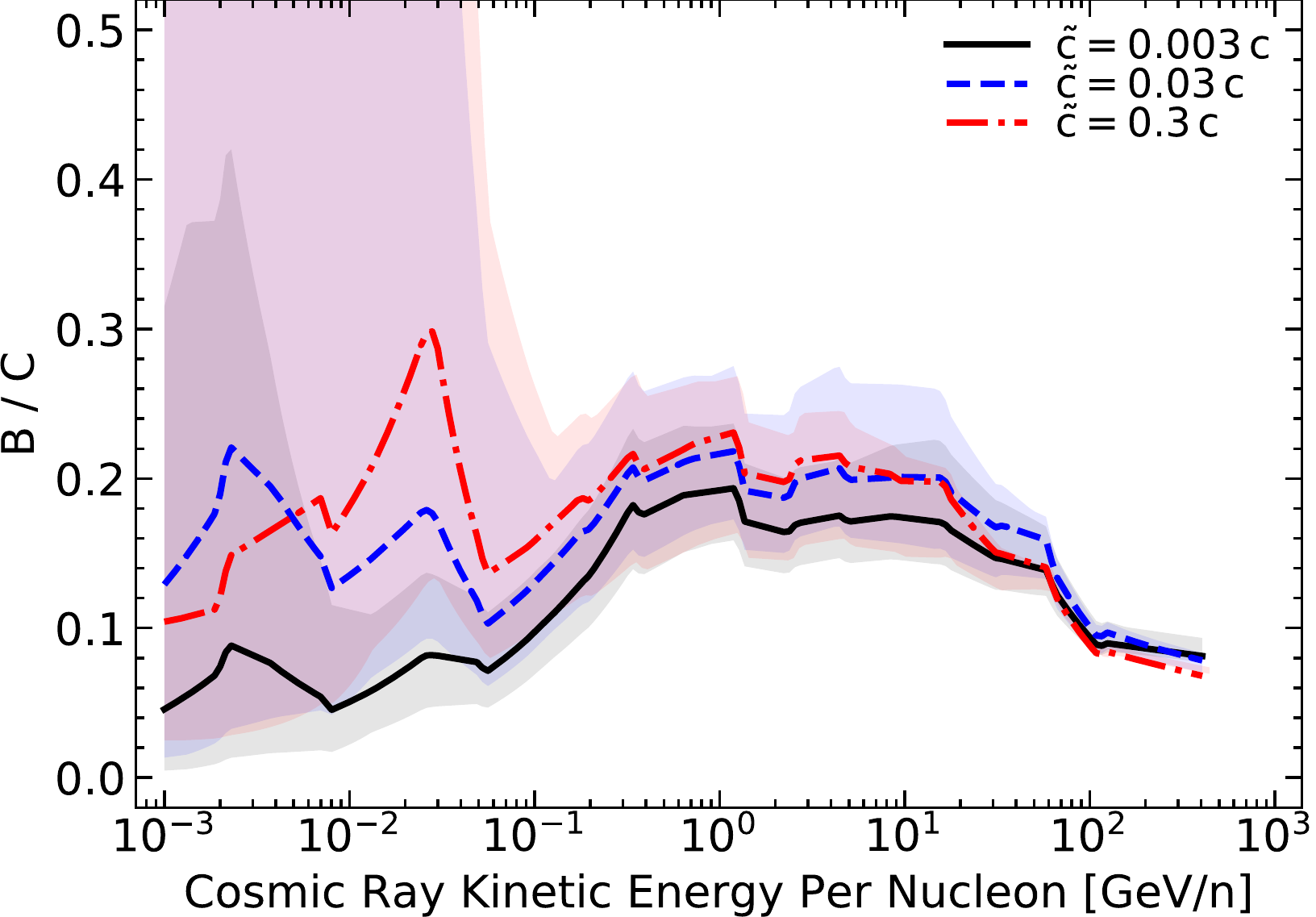} &
	\includegraphics[width=0.33\textwidth]{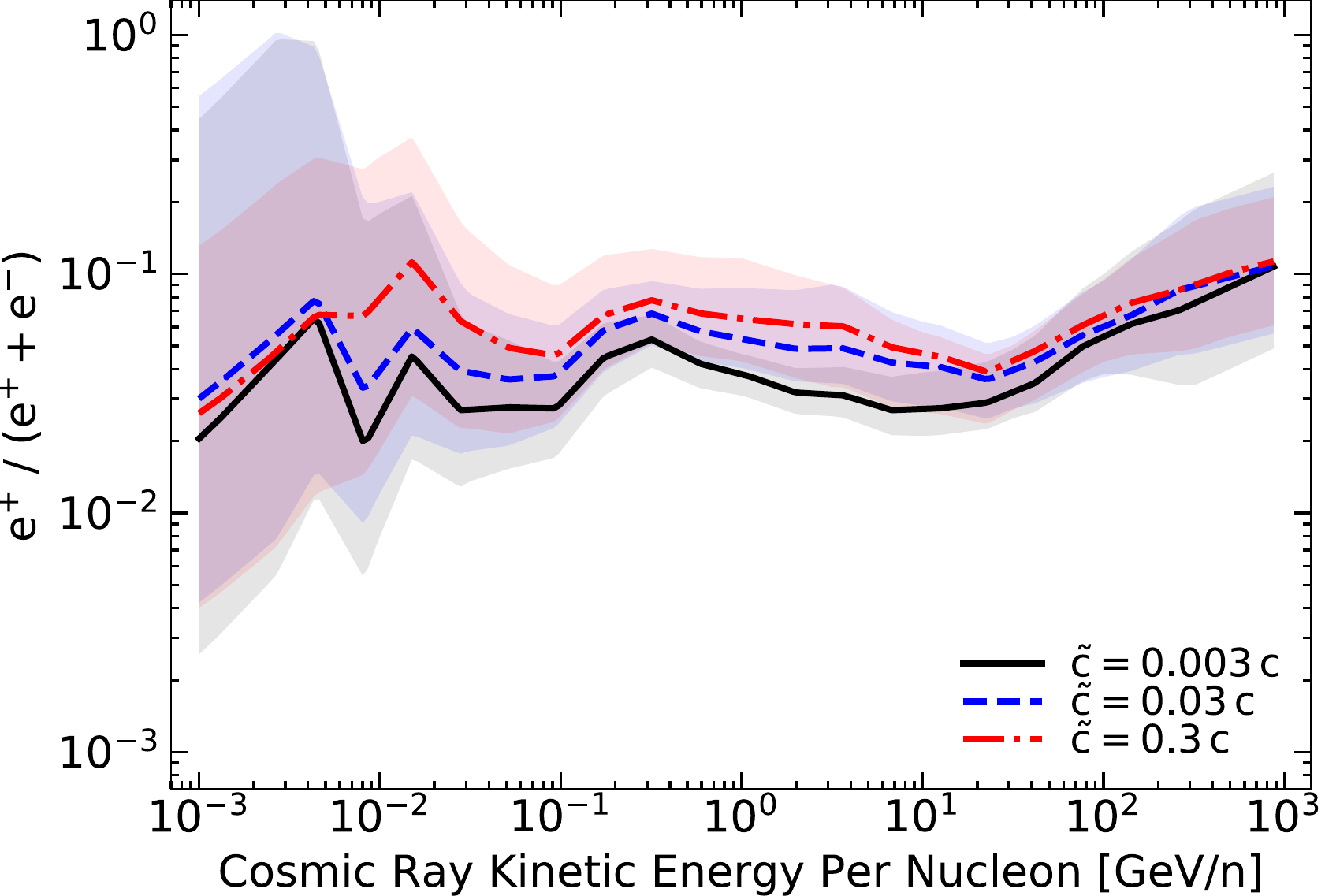} 
\end{tabular}
	\vspace{-0.3cm}
	\caption{CR spectra with different closure assumptions, arbitrarily re-normalized magnetic field strengths, and reduced-speed-of-light, as Fig.~\ref{fig:spec.compare.numerics}, with the variations being with respect to the alternative ``reference model'' from Fig.~\ref{fig:spec.compare.kappa.alt}. As in Fig.~\ref{fig:spec.compare.losses.alt}, the systematic effects of these physics variations are consistent with Fig.~\ref{fig:spec.compare.numerics}.
	\label{fig:spec.compare.numerics.alt}\vspace{-0.4cm}}
\end{figure*}

In \S~\ref{sec:physics}, Figs.~\ref{fig:spec.compare.kappa}, \ref{fig:spec.compare.losses}, \&\ \ref{fig:spec.compare.numerics} compared the effects of different parameter and physics variations on CR observables, by turning on and off different physics or varying different parameters with respect to the ``reference'' or best-fit model in Fig.~\ref{fig:demo.cr.spectra.fiducial}. 

We have explored a number of other variations as well, as described in the main text, in order to identify robust trends and the best-fit model compared to observations. Figs.~\ref{fig:spec.compare.kappa.alt}, \ref{fig:spec.compare.losses.alt}, \&\ \ref{fig:spec.compare.numerics.alt} illustrate some of these. These are identical to Figs.~\ref{fig:spec.compare.kappa}, \ref{fig:spec.compare.losses}, \&\ \ref{fig:spec.compare.numerics}, except that we consider a different ``reference'' model as the baseline about which parameters and physics are varied. Specifically here we take a model with a fixed higher scattering rate (lower diffusivity) normalization $\bar{\nu}_{0}=10^{-8}\,{\rm s^{-1}}$, which is then re-tuned (fitting $\delta$ and $\psi_{\rm inj}$) to try and reproduce the spectra and B/C ratios as best as possible, giving $\bar{\nu} \sim 10^{-8}\,{\rm s}^{-1}\,\beta\,R_{\rm GV}^{-1}$ (i.e.\ $\delta=1$, with slightly-different $\psi_{\rm inj}=4.3$), as compared to main-text default $\bar{\nu}_{0}=10^{-9}\,{\rm s^{-1}}$, $\delta=0.5-0.6$, $\psi_{\rm inj}=4.2$. We stress that directly comparing this reference model to the observations as in Fig.~\ref{fig:demo.cr.spectra.fiducial} shows that even with ``re-fitting'' $\delta$ and $\psi_{\rm inj}$ at this $\bar{\nu}_{0}$, the fit (comparing to Solar circle LISM data) is significantly more poor than our default main-text model: B/C is too flat between $\sim 0.3-100\,$GeV (under-predicting B/C at $<3$\,GeV and over-predicting B/C at $>3\,$GeV), $^{10}$Be/$^{9}$Be is systematically too-high at $\sim 0.03-100\,$GeV, $e^{+}/(e^{+}+e^{-})$ is ``too flat'' (it does not feature the ``curvature'' observed from $\sim 0.5-300\,$GeV), and the spectra are too hard, under (over)-predicting the intensity of $e^{-}$ and $p$ at $<100\,$GeV ($>100\,$GeV). 

Nonetheless, this provides a useful reference case to consider the systematic effects of different physics and parameter variations in Figs.~\ref{fig:spec.compare.kappa}, \ref{fig:spec.compare.losses}, \&\ \ref{fig:spec.compare.numerics}. Because of non-linear interactions between the different physics, as described in the text, it is not totally obvious that changing one of the physics or assumptions would have the same systematic effect if we also change the ``reference'' model. For example, since the diffusivity at low CR energies is much lower here than in our main-text reference model, certain losses in dense ISM environments could be qualitatively more important, and this can order-of-magnitude change the ratio of e.g.\ diffusive reacceleration to streaming loss terms. Nonetheless, Figs.~\ref{fig:spec.compare.kappa.alt}, \ref{fig:spec.compare.losses.alt}, \&\ \ref{fig:spec.compare.numerics.alt} confirm that all of our qualitative conclusions in the text, regarding the systematic effects of these variations as well as their qualitative importance, appear to be robust.

\section{Additional Numerical Details}
\label{sec:numerics}

\begin{footnotesize}
\ctable[caption={{\normalsize CR Spectral Momentum Range/Intervals Followed Explicitly}\label{tbl:intervals}},center,star
]{lcccccccccccc}{
}{
\hline\hline
\multicolumn{13}{l}{Leptons ($e^{\pm}$): Rigidity $R$, Kinetic Energy $T$, Lorentz factor $\gamma$, for each momentum interval boundary $p^{-}_{n,s}$, $p^{+}_{n,s}$, for electrons and positrons} \\ 
\hline
%$10^{-3}$ & $5.62\times 10^{-3}$ & $1.78\times10^{-2}$ & $5.62 \times10^{-2}$ & $1.78 \times10^{-1}$ \\ % & 5.62 \times10^{-1} & 1.78
$R$ (GV) & 0.001 & 0.00562 & 0.0178 & 0.0562 & 0.178 & 0.562 & 1.78 & 5.62 & 17.8 & 56.2 & 178 & 1000 \\
$T$ (GeV) & 0.000612 & 0.00513 & 0.0173 & 0.0570 & 0.177 & 0.561 & 1.78 & 5.62 & 17.8 & 56.2 & 178& 1000 \\
$\gamma$ & 2.2 & 11 & 35 & 110 & 350 & 1100 & 3500 & 11000 & 35000 & 1.1$\times10^{5}$ & 3.5$\times10^{5}$ & 2.0$\times10^{6}$ \\
\hline
\multicolumn{13}{l}{Hadrons: All hadronic species are discretized on the same intervals in $R$; the corresponding $T$, $\gamma$, and $\beta$ are given for protons (H), but differ between species} \\ 
\hline
$R$ (GV) & -- & -- & -- & 0.0316 & 0.178 & 0.562 & 1.78 & 5.62 & 17.8 & 56.2 & 178 & 1000 \\
$T$ (GeV, H) & -- & -- & -- & 0.000533 & 0.0167 & 0.155 & 1.07 & 4.76 & 16.9 & 55.3 & 177 & 1000 \\
$\gamma$ (H) & -- & -- & -- & 1.0006 & 1.018 & 1.17 & 2.1 & 6.1 & 19 & 60 & 190 & 1100 \\
$\beta$ (H) & -- & -- & -- & 0.034 & 0.19 & 0.51 & 0.88 & 0.99 & 0.999 & 0.9999 & 0.99999 & 0.9999996 \\
\hline\hline
}
\end{footnotesize}

Here we outline various technical numerical details of the methods used for CR evolution in our simulations, originally presented in other papers we refer to below. We direct the interested reader to these and other papers cited for additional numerical tests.

To begin, as shown in \citet{hopkins:m1.cr.closure}, Eqs.~\ref{eqn:f0}-\ref{eqn:f1} can be re-written in the convenient form:
\begin{align}
\label{eqn:f0.rev} 
D_{t} \bar{f}_{0,s} 
&= - \nabla \cdot  (v\,\bar{f}_{1,s}\,\bhat) + j_{0,s} +  \frac{1}{p^{2}}\frac{\partial }{\partial p}\left[ Q_{s}\,p^{2}\,\bar{f}_{0,s} \right] \\
\nonumber Q_{s} &\equiv S_{\ell} + p\,\mathbb{D}_{s}:\nabla{\bf u} + \tilde{D}_{p \mu,s}\,\frac{\bar{f}_{1,s}}{\bar{f}_{0,s}} + \frac{\tilde{D}_{p p,s}}{\bar{f}_{0,s}}\, \frac{\partial \bar{f}_{0,s}}{\partial p} 
\end{align}
with
\begin{align}
\label{eqn:f1.rev} 
D_{t} \bar{f}_{1,s} +  
v\,\Delta(\bar{f}_{0,s}) &= - \left[ \tilde{D}_{\mu\mu,s}\,\bar{f}_{1,s} + \tilde{D}_{\mu p,s}\,\frac{\partial \bar{f}_{0,s}}{\partial p} \right]
+ {j_{1,s}} 
\end{align}
This is just a matter of definitions and some algebra to re-arrange terms, but it will be useful below. Note that we explicitly include the subscript $s$ indicating species here. The total distribution function can be reconstructed from 
\begin{align}
\bar{f}_{0} &\equiv \sum_{s} \bar{f}_{0,s}, \\
\nonumber \bar{f}_{1} &\equiv \sum_{s} \bar{f}_{1,s},
\end{align} 
but the important point is that these equations are completely separable in species (there are no ``cross terms'' to be integrated, except for secondary injection which we detail below). This means that we simply repeat the identical numerical exercise for each separate species (calculated as a simple loop of species in every numerical step described below), and all our numerical methods are completely agnostic to the actual species being followed. In principle, one could trivially extend our method to a completely arbitrary list of species, with the only constraint being computational memory limitations and the cost of repeating so many computations in the relevant loops.

As noted in \S~\ref{sec:methods:crs}, we will operator split these equations, according to each of the three terms on the right-hand side of Eq.~\ref{eqn:f0.rev}: (1) the ``spatial'' or coordinate-space integration term $- \nabla \cdot  (v\,\bar{f}_{1,s}\,\bhat)$ (and all of the $D_{t} \bar{f}_{1}$ or ``flux'' equation except for the $j_{1}$ term); (2) the $j_{0}$ and $j_{1}$ terms which describe injection by SNe, catastrophic losses, and secondary production; (3) the terms inside $\partial_{p}[...]$ (i.e.\ in $Q_{s}$), which describe continuous evolution in momentum-space (integrating the CR spectral evolution).

\begin{comment}
\begin{align}
\nonumber
\tilde{D}_{p p,s}  = \chi_{s}\,\frac{p^{2}\,v_{A}^{2}}{v^{2}}\,\bar{\nu}_{s}
\  , \  \
& \tilde{D}_{p \mu,s} = \frac{p\,\bar{v}_{A}}{v}\,\bar{\nu}_{s}
\  , \ \
\tilde{D}_{\mu\mu,s} = \bar{\nu}_{s} 
\  , \  \
\tilde{D}_{\mu p,s} = \chi_{s}\,\frac{p\,\bar{v}_{A}}{v}\,\bar{\nu}_{s}
\end{align}
\end{comment}

Also recall the definition of the conserved quantities we integrate: CR number $N_{j,n,s}$ and kinetic energy $E_{j,n,s}$, integrated over a spatial domain/cell $j$ and momentum interval/bin $n$ for one species $s$. 
\begin{align}
\label{eqn:Ndef} N_{j,\,n,\,s}(t) &\equiv \int _{V_{j}} n_{j,\,n,\,s}\,d^{3}{\bf x} \equiv \int_{V_{j}}\int_{p_{n,\,s}^{-}}^{p_{n,\,s}^{+}} f_{j,\,n,\,s}(...) \, d^{3}{\bf x} \,d^{3}{\bf p}  \\ 
\label{eqn:Edef} E_{j,\,n,\,s}(t) &\equiv \int_{V_{j}} \epsilon_{j,\,n,\,s}\,d^{3}{\bf x} \equiv \int_{V_{j}} \int_{p_{n,\,s}^{-}}^{p_{n,\,s}^{+}} T_{s}(p)\,f_{j,\,n,\,s}(...) \, d^{3}{\bf x} \,d^{3}{\bf p}
\end{align}
These are what we actually evolve (computing fluxes etc.), in order to ensure manifest conservation. But as noted in the main text, for any bin $n,s$, there is a one-to-one correspondence between the values of $N_{j,n,s}$ and $E_{j,n,s}$ and the equivalent power-law form of 
\begin{align}
\bar{f}_{0,j,n,s} &= \bar{f}_{0,j,n,s}[p^{0}_{n,s}]\,(p/p_{n,s}^{0})^{\psi_{j,n,s}}
\end{align} 
(where $p_{n,s}^{0}$ is the ``bin center,'' as defined in the main text, being the geometric mean of the bin boundaries $p^{-}_{n,s}$, $p^{+}_{n,s}$, i.e.\ mean $p_{n,s}$ in log-space). We pre-compute the mapping from ($N_{j,n,s}$, $E_{j,n,s}$) to ($\bar{f}_{0,j,n,s}[p^{0}_{n,s}]$, $\psi_{j,n,s}$) (and vice-versa) in a look-up table for every species and bin at simulation startup.

Recall the bin ``edges'' $p^{\pm}_{n,s}$ for both leptons and hadrons are defined in rigidity, so e.g.\ two different hadronic nuclei species $s$, $s^{\prime}$ with different charge and atomic weights, such as H and Be, will have different ``bin edge'' values of momentum $p^{-}_{n,s}$, $p^{-}_{n,s^{\prime}}$, and of course different boundary values of kinetic energy and $\beta$, even if their bins align exactly in rigidity. These are summarized for reference in Table~\ref{tbl:intervals}.

\subsection{Coordinate-Space Integration}
\label{sec:numerics:spatial}

\subsubsection{Spatial Advection \&\ Flux Equations}

Consider the coordinate-space (advection and flux) terms: 
\begin{align}
\label{eqn:spatial} D_{t} \bar{f}_{0,s} &= -\nabla \cdot (v\,\bar{f}_{1,s}\,\bhat), \\
\nonumber D_{t} \bar{f}_{1,s} &+ v\,\Delta(\bar{f}_{0,s}) = -  \tilde{D}_{\mu\mu,s}\,\bar{f}_{1,s} - \tilde{D}_{\mu p,s}\,{\partial_{p} \bar{f}_{0,s}}.
\end{align} 
From a numerical point of view, this is identical to (and in-code uses the same modular implementation as) the algorithms used previously in {\small GIZMO} for both ``single-bin'' CR transport in previous papers \citep{ji:fire.cr.cgm,ji:20.virial.shocks.suppressed.cr.dominated.halos,chan:2021.cosmic.ray.vertical.balance,su:2018.stellar.fb.fails.to.solve.cooling.flow,su:turb.crs.quench,su:2021.agn.jet.params.vs.quenching}, presented in full detail and extensively numerically tested/validated in e.g.\ \citet{chan:2018.cosmicray.fire.gammaray,hopkins:cr.mhd.fire2,hopkins:cr.transport.constraints.from.galaxies} (see e.g.\ Figs.~A1, B1-B9, and C1 of \citealt{chan:2018.cosmicray.fire.gammaray}, Figs.~17 \&\ A1-A4 in \citealt{hopkins:cr.mhd.fire2}, Appendices D \&\ E in \citealt{hopkins:cr.transport.constraints.from.galaxies}; and see also the extensive implementations of similar algorithms in different codes reviewed in \S~\ref{sec:intro}), as well as existing radiation-hydrodynamics (M1-like) photon transport in {\small GIZMO} \citep[see e.g.][for relevant tests]{lupi:2018.h2.sfr.rhd.gizmo.methods,hopkins:2019.grudic.photon.momentum.rad.pressure.coupling,hopkins:radiation.methods,grudic:starforge.methods}. It is also numerically similar to other anisotropic advection+diffusion operators used for a variety of other transport physics \citep{hopkins:gizmo.diffusion,rennehan:2021.anisotropic.turb.mixing}.

The equivalence is more obvious if we note that (since there are no cross terms) we can (and do) separately solve every CR momentum ``bin'' and species, and volume-integrate over the cell domains to calculate the flux of conserved quantities. 
First consider $N_{j,n,s}$: using the definition of $D_{t}$ (which is defined such that ${\rm d}_{t}\left( \int_{V_{j}}\,d^{3}{\bf x}\,X \right) \equiv \int_{V_{j}}\,d^{3}{\bf x}\,D_{t} X$ for any $X$ in a cell $j$), and $N_{j,n,s}$ (Eq.~\ref{eqn:Ndef} above) and the usual Stokes's law transformation, we immediately have the discrete law: 
\begin{align}
\label{eqn:N.advection} \frac{{\rm d} N_{j,n,s}}{{\rm d} t} &\equiv \int_{V_{j}} d^{3}{\bf x} \, D_{t} \int_{p_{n,\,s}^{-}}^{p_{n,\,s}^{+}} d^{3}{\bf p}\,f_{j,\,n,\,s} \\
\nonumber &= \int_{V_{j}} d^{3}{\bf x}\,\int_{p_{n,\,s}^{-}}^{p_{n,\,s}^{+}} 4\pi\,p^{2}\,dp\, D_{t} \bar{f}_{0,\,j,\,n,\,s} \\
\nonumber &= -\int_{V_{j}}\, d^{3}{\bf x}\, \nabla \cdot {\bf F}_{j,n,s}^{N} = -\sum_{j^{\prime}}\,\tilde{\bf F}_{jj^{\prime},n,s}^{N}\cdot {\bf A}_{jj^{\prime}},
\end{align} 
where 
\begin{align}
{\bf F}_{j,n,s}^{N} &\equiv F_{j,n,s}^{N}\,\bhat = \bhat\,\int_{p_{n,\,s}^{-}}^{p_{n,\,s}^{+}} 4\pi\,p^{2}\,dp\,v\,\bar{f}_{1,\,j,\,n,\,s}
\end{align}
is the flux of $N$, ${\bf A}_{jj^{\prime}}$ is the usual oriented interface area between neighboring cells $j$ and $j^{\prime}$, and $\tilde{\bf F}_{jj^{\prime}}$ is the interface flux. Multiplying by $T(p)$ before integrating we trivially have the analogous energy equation 
\begin{align}
\label{eqn:E.advection} \frac{{\rm d} E_{j,n,s}}{{\rm d}t} &= -\sum_{j^{\prime}}\,\tilde{\bf F}_{jj^{\prime},n,s}^{E}\cdot {\bf A}_{jj^{\prime}}.
\end{align} 
Numerically, these are just advection equations (akin to any other advection term being solved simultaneously in-code, and solved via the same second-order method per \citealt{hopkins:gizmo,hopkins:mhd.gizmo}). 

So we simply need $F^{N,E}_{j,n,s}$, which is explicitly evolved like in any two-moment method and determined by the $D_{t} \bar{f}_{1}$ equation. The challenge here is, as noted by e.g.\ \citet{girichidis:cr.spectral.scheme,ogrodnik:2021.spectral.cr.electron.code} and \citet{hanasz:2021.cr.propagation.sims.review}, that we cannot trivially integrate $D_{t} v\,\bar{f}_{1}$ over $d^{3}{\bf p}$ and arrive at an equation for a ``single'' $F^{N,E}_{j,n,s}$ which can be evolved if the bin has a finite width in momentum-space (we would need infinitesimally small bins). It is significantly less complex, computationally less expensive, more numerically stable, and directly analogous to the previous ``single-bin'' CR studies to calculate instead the ``bin-centered'' fluxes as in both those previous studies -- essentially solving for $F_{j,n,s}^{N,E}$ by taking the values of  $\bar{\nu}_{j,n,s}(p)$, $\langle \mu\rangle$ at the bin center $p=p^{0}_{n,j,s}$. This automatically means quantities like $\chi \rightarrow \chi_{0,\,j,n,s}$ are constant over the bin (within one bin, one species, one cell: they can and do vary between different species, bins, and cells). Then we can take Eq.~\ref{eqn:spatial} for $\bar{f}_{1}$, multiply through by $v$ and insert the definitions of $\tilde{D}_{\mu\mu}$, $\tilde{D}_{\mu p}$, and $\Delta(q)\equiv\bhat\cdot\nabla \cdot (\mathbb{D}\,q)$ from Eq.~\ref{eqn:f1}, and use our local-power-law definition of $\bar{f}_{0}$ to treat the $\partial_{p} \bar{f}_{0}$ terms. Integrating then immediately yields the desired bin-centered equation:
$D_{t} F_{j,n,s}^{N} + c^{2}\,\bhat\cdot\nabla \cdot \mathbb{N}_{j,n,s}  = -\bar{\nu}_{0,j,n,s}\,\left[ F_{j,n,s}^{N} - v_{{\rm st},0,j,n,s}\,n_{j,n,s} \right]$ 
where $n_{j,n,s} \equiv N_{j,n,s}/V_{j}$ is the CR number density, 
$\mathbb{N}_{j,n,s} \equiv \mathbb{D}_{0,j,n,s}\,\beta_{0,j,n,s}^{2}\,n_{j,n,s}$ is a second-moment tensor, and 
$v_{{\rm st},0,j,n,s} \equiv -\chi_{0,j,n,s}\,\psi_{j,n,s}\,\bar{v}_{A,j}$ is an effective ``streaming speed'' (note $\psi_{j,n,s}<0$ at all energies here, so $-\psi_{j,n,s}=|\psi_{j,n,s}|$).
Given this bin-centered approximation (which necessarily means $F_{j,n,s}^{E}$ and $F_{j,n,s}^{N}$ do not simultaneously follow the full integrals of $\bar{f}_{1}$ except for infinitesimal bins), we have implicitly assumed a constant drift velocity $v_{d} = F^{N}/N = F^{E}/E$ over the bin, so consistency requires that 
\begin{align}
\label{eqn:Fe.Fn} F_{j,n,s}^{E} &= \left( \frac{E_{j,n,s}}{N_{j,n,s}} \right)\,F_{j,n,s}^{N}. 
\end{align}
 Alternatively, we could first derive $F_{j,n,s}^{E}$ directly from the $f_{1}$ equation and infer $F_{j,n,s}^{N}$ from the Eq.~\ref{eqn:Fe.Fn} relation, and this would be equally valid/consistent at the level of the bin-centered approximation, giving: 
\begin{align}
\label{eqn:FluxE.eqn} D_{t} F_{j,n,s}^{E} + c^{2}\,\bhat\cdot \nabla \cdot \tilde{\mathbb{P}}_{j,n,s} = -\bar{\nu}_{0,j,n,s}\,\left[ F_{j,n,s}^{E} - v_{{\rm st},0,j,n,s}\,e_{j,n,s} \right]\ ,
\end{align}
where $\tilde{\mathbb{P}}_{j,n,s}\equiv  \mathbb{D}_{j,n,s}\,\beta^{2}_{0,j,n,s}\,e_{j,n,s}$ is akin to the CR pressure tensor.\footnote{We note here that $\tilde{\mathbb{P}} \equiv \mathbb{D}\,\beta^{2}\,e$ refers to either the kinetic or total energy, whichever is evolved by $e$ in the expressions above. But the usually-defined scalar ``CR pressure'' as it appears in e.g.\ the gas momentum equation in the tightly-coupled limit is defined as $\beta^{2}/3$ times the total CR energy density (or $=(1+\gamma^{-1})/3$ times the kinetic); see \citealt{hopkins:m1.cr.closure} for details.}

Cast in this representation (e.g.\ Eqs.~\ref{eqn:E.advection} \&\ \ref{eqn:FluxE.eqn}), we can now see directly that the numerical two-moment equations for the spatial evolution of the CRs, for each individual species and bin, are numerically identical to the equations solved in many of our previous ``single-bin'' CR studies in the references above. We simply repeat the previous ``single bin'' flux and advection computation numerically about $N_{\rm bins} \times N_{\rm species} \sim 70$ times per interface (once for each bin, for each species), and evolve the entire set  of fluxes and time derivatives. With this in mind, the interface flux which appears in the Riemann problem is determined in exactly the same way as in those previous single-bin studies. We compute this interface flux first for $F_{jj^{\prime},n,s}^{E}$, then if the net flux flows from $j\rightarrow j^{\prime}$ (i.e.\ CRs move from cell $j$ to cell $j^{\prime}$), we take $F_{jj^{\prime}}^{N}=(N_{j,n,s}/E_{j,n,s})\,F_{jj^{\prime}}^{E}$ (i.e.\ the CRs carry the same energy as $j$), if it flows from $j^{\prime}\rightarrow j$, $F_{jj^{\prime}}^{N}=(N_{j^{\prime},n,s}/E_{j^{\prime},n,s})\,F_{jj^{\prime}}^{E}$. The updates to conserved quantities are then drifted and kicked using the standard scheme for all conserved advected radiation-MHD quantities (identical to our previous single-bin implementation). Every time $N_{j,n,s}$ or $E_{j,n,s}$ is updated, we immediately recompute the ``primitive variables'' $\bar{f}_{0,j,n,s}[p^{0}_{n,s}]$, $\psi_{j,n,s}$.

Finally, this also gives us everything we need to evaluate the CR force on gas. Per \citet{hopkins:m1.cr.closure}, Eq.~41 therein, we can rewrite the combined  Lorentz+scattering term from Eq.~\ref{eqn:gas.momentum} as:
\begin{align}
\label{eqn:gas.momentum.alt} D_{t}(\rho\,{\bf u}) + ...  + \nabla \cdot \mathbb{P}^{\rm tot} = -\frac{\bhat}{c^{2}} \sum_{s,n}\,D_{t} F_{n,s}^{E_{\rm tot}}
\end{align}
where $\mathbb{P}^{\rm tot} \equiv \sum_{s,n}\,\mathbb{P}_{n,s} = \sum_{s,n}\,\mathbb{D}_{n,s}\,P_{0,n,s}$ is just the sum of the CR pressure over all bins, and likewise for $D_{t} F^{E_{\rm tot}}$ (here in terms of the total energy, so $D_{t} F^{E_{\rm tot}}_{n,s} = D_{t} F^{E_{\rm kin}}_{n,s} + m_{s}\,c^{2}\,D_{t} F^{N_{\rm tot}}$). In-code, we simply add $\mathbb{P}^{\rm tot}$ to the total pressure tensor used in the Riemann problem, then (in each cell immediately following the Riemann problem update to the momentum fluxes) add the $D_{t} F_{n,s}^{E_{\rm tot}}$ as a source term (using the discrete value of $D_{t} F_{n,s}^{E_{\rm tot}}$ we would have at the end of the timestep). The detailed treatment of the latter makes little difference, since we see from this form that it acts as a correction which (a) is suppressed by $\sim 1/c^{2}$ (so is typically suppressed relative to other terms in the gas momentum equation by $\mathcal{O}(n_{\rm cr}/n_{\rm gas})$), and (b) vanishes when the CRs approach flux-steady-state (which generally occurs on the scattering time $\sim \bar{\nu}^{-1}$).

\subsubsection{Timestep Condition}

As discussed in the papers above, this imposes the usual Courant condition on the cell timesteps 
\begin{align}
\Delta t^{\rm cell}_{j} \le C_{\rm cour}\,\Delta x_{j}/\tilde{c}\ , 
\end{align}
where $\tilde{c}$ ($\tilde{c}= 10^{4}\,{\rm km\,s^{-1}}$ is our default, with $C_{\rm cour}=0.25$ to be conservative; for various tests see \citealt{hopkins:gizmo,hopkins:mhd.gizmo,hopkins:cg.mhd.gizmo,hopkins:gizmo.diffusion,hubber:gandalf.gizmo.methods,panuelos:gizmo.kd.methods,deng:2019.mri.turb.sims.gizmo.methods,bonnerot:2021.gizmo.rhd.tde.sims}). As in all other applications in {\small GIZMO} the actual timesteps are determined by the smallest of all possible constraints, including e.g.\ gravity, MHD, and any other constraints \citep[see][]{hopkins:gizmo}, but in these runs, that is almost always set by this condition given the extremely large $\tilde{c}$.

\subsection{Injection, Catastrophic Losses, and Secondary Production}
\label{sec:numerics:injection}

Now consider the $j$ (injection, catastrophic loss, and secondary production) terms:
\begin{align}
\label{eqn:injection} D_{t} \bar{f}_{0,s} &= j_{0,s}\ , \\
\nonumber D_{t} \bar{f}_{1,s} &= j_{1,s}\ .
\end{align}

\subsubsection{Injection}

For injection from discrete (point) sources, i.e.\ SNe and stellar mass-loss, the injection terms $j_{0}$ and $j_{1}$ are handled in the exact same manner as our previous single-bin CR studies \citep{chan:2018.cosmicray.fire.gammaray,hopkins:cr.mhd.fire2,hopkins:cr.transport.constraints.from.galaxies,ji:fire.cr.cgm,ji:20.virial.shocks.suppressed.cr.dominated.halos,chan:2021.cosmic.ray.vertical.balance,su:2018.stellar.fb.fails.to.solve.cooling.flow,su:turb.crs.quench,su:2021.agn.jet.params.vs.quenching} or any other injection of scalar quantities such as mass, metals, passive scalars/tracers, or thermal energy \citep{hopkins:sne.methods}. Briefly, in a timestep for a star particle $\Delta t_{\ast}$, if a SNe occurs some total CR energy $\Delta E_{\rm cr}^{\ast}$ (or $\Delta E_{\rm cr}^{\ast}=\dot{E}_{\rm cr}^{\ast}\,\Delta t_{\ast}$, for continuous sources like O/B mass-loss) will be injected into the surrounding cells each receiving some fraction 
\begin{align}
\Delta E_{{\rm cr},j}^{\ast} = w_{j}\,\Delta E_{{\rm cr}}^{\ast} 
\end{align} 
according to an appropriate weight function (such that $\sum_{j} w_{j} = 1$), and we then immediately calculate exactly the corresponding $\Delta N_{j,n,s}$, $\Delta E_{j,n,s}$ according to the specified injection spectra as defined in the main text (these are equivalent to the integral of $j_{0}$ over each bin). 
\begin{align}
\Delta N_{j,n,s} &\equiv \int_{p^{-}_{n,s}}^{p^{+}_{n,s}}\,4\pi\,p^{2}\,dp\,j^{\ast}_{0,n,s}(p)\,\Delta t_{\ast} \\ 
\Delta E_{j,n,s} &\equiv \int_{p^{-}_{n,s}}^{p^{+}_{n,s}}\,4\pi\,p^{2}\,dp\,T(p)\,j^{\ast}_{0,n,s}(p)\,\Delta t_{\ast} \\ 
\label{eqn:sum} \sum_{n,s}\,\Delta E_{j,n,s}  &\equiv \Delta E_{{\rm cr},j}^{\ast} = w_{j}\,\Delta E_{{\rm cr}}^{\ast} 
\end{align}
Recall, we assume $j^{\ast}_{0} \propto p^{-\psi_{\rm inj}}$ (with $\psi_{\rm inj}=4.2$) is a simple power-law, so these can easily be calculated exactly, with the normalization of $j^{\ast}_{0,n,s}$ for each species given by (1) Eq.~\ref{eqn:sum}, which normalizes the total sum energy deposited to be exactly that desired,\footnote{Because for our injection spectra, the total energy is totally dominated by protons/H, if we instead defined $\Delta E_{{\rm cr},j}^{\ast}$ as specifically that injected into protons our results are nearly indistinguishable.} and (2) the ratios of different species following the ratios specified in \S~\ref{sec:methods:injection} \&\ \ref{sec:methods:params}.\footnote{As noted in the text, the injection abundances of antimatter, B, Be are negligible, so of the species we follow this is only important for $e^{-}$, where $j^{\ast}_{0,n,e^{-}} \equiv 0.02\,j^{\ast}_{0,n,{\rm H}}$ and CNO where the abundance is scaled relative to H (with $dN_{s}(\beta)/d\beta = (N_{s,j}/N_{{\rm H},j})\,dN_{{\rm H}}(\beta)/d\beta$ at each $\beta$) according to the ambient ISM and ejecta abundances $(N_{s,j}/N_{{\rm H},j})$ as given in \S~\ref{sec:methods:injection} \&\ \ref{sec:methods:params}.} These $\Delta N_{j,n,s}$ and $\Delta E_{j,n,s}$ are added to the cell $N_{j,n,s}$ and $E_{j,n,s}$ respectively, and the primitive variables are re-computed.

For consistency, we also update the flux (e.g.\ include the $j_{1} = \langle \mu \rangle_{\rm inj}\,j_{0}$ term), assuming the newly-injected CRs are streaming radially away from the star (so given our bin-centered approximation, $\Delta F_{j,n,s}^{N} = v_{0,j,n,s}\,\Delta N_{j,n,s}\,\hat{\bf r}\cdot\bhat$), but because the CR flux equation evolves towards local equilibrium on the (very rapid) scattering timescale $\sim \bar{\nu}^{-1}$, it makes no detectable difference if we ignore this flux update. 

Extensive tests of the actual numerical injection algorithm for arbitrary scalar fields demonstrating its numerical stability, manifest conservation, numerical isotropy (ability to avoid imprinting preferred directions), and accuracy are given in \citet{hopkins:sne.methods}.

\subsubsection{Catastrophic Losses}

Catastrophic/fragmentation/pionic losses, annihilation, radioactive decay, and secondary production also fall into this term, as they can be described by some $j = \dot{f}$ term effectively ``adding'' or ``removing'' CRs, rather than slowly and continuously increasing/decreasing their individual momenta $p$. So for example as in the main text, for some catastrophic collisional loss process, $\dot{f}_{s} = -\sigma\,v_{s}\,n_{\rm x}\,f_{s}$, where $n_{\rm x}$ is the local ISM density of nucleons or whatever relevant ``target'' species. These are integrated alongside the momentum-space terms, on the same subcycle timestep $\Delta t^{\rm sub}_{j} \le \Delta t^{\rm cell}_{j}$ (where $\Delta t^{\rm cell}_{j}$ is the cell timestep for other operations). Assume on the sub-cycle timestep (always smaller than e.g.\ the CR transport timestep $\sim 0.25\,\Delta x_{j}/\tilde{c}$, which is itself almost always much smaller than any other evolved MHD fluid evolution timescales) that the background (non-CR) fluid state vector\footnote{For the CR momentum-space and catastrophic loss update, we adopt the value of ${\bf U}_{j}$ drifted to the midpoint of the timestep of $j$. However because our CR timesteps are so small compared to macroscopic MHD evolution timescales ($\gtrsim 10^{6}$\,yr), it makes no perceptible difference if we use the value of ${\bf U}_{j}$ at the beginning or end of each step.} ${\bf U}_{j}({\bf x},\,t)$ (e.g.\ magnetic field, density, gas velocity, etc.) is constant over the substep (a good assumption for the reasons above). 

First consider the loss terms, and define the total CR number loss rate as
\begin{align}
 \dot{f}_{{\rm loss},j,s} = \dot{f}_{{\rm catastrophic},j,s} + \dot{f}_{{\rm annihilation},j,s} + \dot{f}_{{\rm decay},j,s}\ ,
 \end{align} 
summing all the relevant expressions for each species as a function of $p$ for each species $s$ (all given explicitly in \S~\ref{sec:collisional.losses}). This can be written 
\begin{align}
\dot{f}_{{\rm loss},j,s}(p) = -\mathcal{R}_{j,s}(p,\,s,\,{\bf U}_{j},\,...)\,f_{j,s} 
\end{align} 
for all terms considered. Then for every cell $j$, for every subcycle timestep $\Delta t^{\rm sub}_{j}$, for every species $s$, for every CR momentum bin $n$, we compute 
\begin{align}
\label{eqn:ndot.catastrophic} \dot{N}_{j,n,s} &\equiv -\langle \mathcal{R}_{j,s} \rangle_{n}^{N}\,N_{j,n,s} \equiv -\int_{V_{j}} \int_{p_{n,\,s}^{-}}^{p_{n,\,s}^{+}} d^{3}{\bf x}\,d^{3}{\bf p}\,\mathcal{R}_{j,s}(...)\,{f}_{j,n,s} \\
\nonumber \dot{E}_{j,n,s} &\equiv -\langle \mathcal{R}_{j,s} \rangle_{n}^{E}\,E_{j,n,s}  \equiv -\int_{V_{j}} \int_{p_{n,\,s}^{-}}^{p_{n,\,s}^{+}} d^{3}{\bf x}\,d^{3}{\bf p}\,\mathcal{R}_{j,s}(...)\,T_{s}(p){f}_{j,n,s}\ . 
\end{align}
For most of the terms in $\dot{f}$ the $\langle \mathcal{R}_{j,s} \rangle_{n}^{N,E}$ terms can be computed analytically but even if not, they can be pre-computed to arbitrary precision again as a function of $\psi_{n,s}$ (and ${\bf U}_{j}$, usually entering just in the normalization of $\langle \mathcal{R}_{j,s} \rangle_{n}^{N,E}$) in a look-up-table. We then integrate the change in $N_{j,n,s}$ and $E_{j,n,s}$ as: 
\begin{align}
\label{eqn:cata.update} N_{j,n,s}(t_{j,0} + \Delta t^{\rm sub}_{j}) &= N_{j,n,s}(t_{j,0})\,\exp{\left[-\langle \mathcal{R}_{j,s} \rangle_{n}^{N}\,\Delta t^{\rm sub}_{j}\right]}\ ,\\
\nonumber 
E_{j,n,s}(t_{j,0} + \Delta t^{\rm sub}_{j}) &= E_{j,n,s}(t_{j,0})\,\exp{\left[-\langle \mathcal{R}_{j,s} \rangle_{n}^{E}\,\Delta t^{\rm sub}_{j}\right]}\ .
\end{align}

While this is  numerically stable for arbitrary timesteps, for accuracy we impose a subcycle timestep restriction (together with the restrictions for continuous losses below) of 
\begin{align}
\Delta t^{\rm sub}_{j} &\le \frac{C_{\rm cour}}{{\rm MAX}\left( \langle R_{j,s} \rangle^{N}_{n}, \, \langle R_{j,s} \rangle^{E}_{n} \right)}
\end{align} 
where the maximum is taken over all $n$ and $s$. 

For consistency with the ``bin-centered'' approximation described in \S~\ref{sec:numerics:spatial}, which effectively takes $\langle \mu \rangle_{j,n,s}(p)=\langle \mu\rangle_{0,\,j,n,s}$, we simply reduce $F_{j,n,s}^{N}$ (and $F_{j,n,s}^{E}$) by the same fractional amount as $N_{j,n,s}$ ($E_{j,n,s}$) after each subcycle (but for the same reasons above regarding the rapid response of $D_{t} F_{j,n,s}^{N,E}$, this update has almost no effect on our results).

\subsubsection{Secondary Production}

Secondary (or tertiary or any other successive) production is then handled immediately following and consistent with these loss terms. Consider a loss event which acts upon a CR with momentum $p=p_{s}$ (in bin $n$), species $s$, within cell $j$, and leaves a product with $p_{s^{\prime}}^{\prime}$ (or $T^{\prime}_{s^{\prime}} \equiv \sqrt{ p^{2}_{s^{\prime}}\,c^{2} + m^{2}_{s^{\prime}}\,c^{4}} - m_{s^{\prime}}\,c^{2}$), $n^{\prime}$, $s^{\prime}$ in $j$. Obviously the species conversion $s\rightarrow s^{\prime}$ is specified by the reaction. As detailed in the main text, we simplify for each secondary-producing reaction by assuming a fixed energy ratio $T_{s^{\prime}}^{\prime} = \alpha_{ss^{\prime}}\,T_{s}$ for that reaction: knowing $s^{\prime}$ and $T^{\prime}$ (hence $p^{\prime}$) we can then determine the bin $n^{\prime}$ into which the secondaries should be deposited.\footnote{For numerical convenience, since the map $T_{s^{\prime}}^{\prime} = \alpha_{ss^{\prime}}\,T_{s}$ and set of secondary-producing processes is fixed at runtime, we pre-compute a lookup table for each secondary-producing process which specifies the corresponding bin(s) $n^{\prime}$, $s^{\prime}$ for each $n$, $s$.}
We then calculate the number and energy of secondaries going into bin $n^{\prime}$, $s^{\prime}$. For an effective production cross-section $\sigma_{s\rightarrow s^{\prime}}$, by definition we have $d\dot{n}_{s^{\prime}}(p^{\prime}) \equiv d^{3}{\bf p}^{\prime}\,\dot{f}_{s^{\prime}}(p^{\prime})  = d^{3}{\bf p}\,\sigma_{s\rightarrow s^{\prime}}\,v_{s}\,n_{\rm n}\,f_{s}(p)$, and $d\dot{e}_{s^{\prime}}(p^{\prime}) = T^{\prime}_{s^{\prime}}(p^{\prime})\,d\dot{n}_{s^{\prime}}(p^{\prime})$. 
Defining\footnote{We also include radioactive decay production, with $\mathcal{R}_{j,s\rightarrow s^{\prime}} \equiv 1/(\gamma\,t_{1/2,s}/\ln{2})$, but this is negligible for the species followed (see \S~\ref{sec:collisional.losses}).} $\mathcal{R}_{j,s\rightarrow s^{\prime}} \equiv  \sigma_{s\rightarrow s^{\prime}}\,v_{s}\,n_{\rm n}$ and integrating, we have:
\begin{align}
\dot{N}_{j,n\rightarrow n^{\prime},s\rightarrow s^{\prime}} &\equiv  \int_{V_{j}} \int_{p_{\rm min}}^{p_{\rm max}} d^{3}{\bf x}\,d^{3}{\bf p}\,\mathcal{R}_{j,s\rightarrow s^{\prime}}(p,\,s,\,...)\,{f}_{j,n,s} \\
\nonumber \dot{E}_{j,n\rightarrow n^{\prime},s\rightarrow s^{\prime}} &\equiv  -\int_{V_{j}} \int_{p_{\rm min}}^{p_{\rm max}} d^{3}{\bf x}\,d^{3}{\bf p}\,\mathcal{R}_{j,s\rightarrow s^{\prime}}(p,\,s,\,...)\,T^{\prime}_{s^{\prime}} {f}_{j,n,s}\  \\ 
\nonumber &= -\alpha_{ss^{\prime}}\,\int_{V_{j}} \int_{p_{\rm min}}^{p_{\rm max}} d^{3}{\bf x}\,d^{3}{\bf p}\,\mathcal{R}_{j,s\rightarrow s^{\prime}}(p,\,s,\,...)\,T_{s}\, {f}_{j,n,s}
\end{align}
where $p_{\rm min} \equiv {\rm MAX}[p_{n,s}^{-},\,p_{s}(p^{\prime,-}_{n^{\prime},s^{\prime}})]$, $p_{\rm max} \equiv {\rm MIN}[p_{n,s}^{+},\,p_{s}(p^{\prime,+}_{n^{\prime},s^{\prime}}) ]$ represent the appropriate minimum/maximum range of either the ``primary'' bin ($n$) itself, or of the primary $p_{s}$ which would producing a secondary $p^{\prime}$ within the bounds of the ``target'' bin ($n^{\prime}$).\footnote{This accounts for the fact that from one primary bin $n$, the products can be split across two secondary bins $n^{\prime}$, depending on the map $p^{\prime}(p)$, which we treat in two successive secondary injection ``steps'' for each bin $n$, $s$.} 
These integrals have the exact same form as the catastrophic loss terms, so we evaluate them identically\footnote{For consistency, we calculate the ratio $\Phi^{N}_{ss^{\prime}} \equiv |\dot{N}_{j,n\rightarrow n^{\prime},s\rightarrow s^{\prime}}| / |\dot{N}_{j,n,s}^{\rm tot}|$ where $\dot{N}_{j,n,s}^{\rm tot}$ is the total catastrophic loss rate from bin $n$ from Eq.~\ref{eqn:ndot.catastrophic}, then take 
$\Delta N_{j,n\rightarrow n^{\prime},s\rightarrow s^{\prime}} = \Phi^{N}_{ss^{\prime}}\,|\Delta N_{j,n,s}|$ where $\Delta N_{j,n,s} = N_{j,n,s}(t_{j,0}) - N_{j,n,s}(t_{j,0} + \Delta t_{j}^{\rm sub})$ from Eq.~\ref{eqn:cata.update}, and do the same for energy $\Delta E_{j,n\rightarrow n^{\prime},s\rightarrow s^{\prime}} = \Phi^{E}_{ss^{\prime}}\,|\Delta E_{j,n,s}|$ (with $\Phi^{E}_{ss^{\prime}} \equiv |\dot{E}_{j,n\rightarrow n^{\prime},s\rightarrow s^{\prime}}| / |\dot{E}_{j,n,s}^{\rm tot}|$). Thus the correct fraction of the total loss is always assigned to $n^{\prime}$, $s^{\prime}$.} to obtain the number $\Delta N_{j,n\rightarrow n^{\prime},s\rightarrow s^{\prime}}$ and energy $\Delta E_{j,n\rightarrow n^{\prime},s\rightarrow s^{\prime}}$ added to bin $n^{\prime}$, $s^{\prime}$. 
We then immediately update the conserved variables for bin $n^{\prime}$ of species $s^{\prime}$: 
$N_{j,n^{\prime},s^{\prime}} \rightarrow N_{j,n^{\prime},s^{\prime}} + \Delta N_{j,n\rightarrow n^{\prime},s\rightarrow s^{\prime}}$, 
$E_{j,n^{\prime},s^{\prime}} \rightarrow E_{j,n^{\prime},s^{\prime}} + \Delta E_{j,n\rightarrow n^{\prime},s\rightarrow s^{\prime}}$, 
and recompute the primitive variables for $n^{\prime}$, $s^{\prime}$. This is looped over each bin $n$ for each primary-producing species $s$ in turn, alongside the catastrophic losses.

Note that for some species (e.g.\ $\bar{p}$), the highest values of the secondary momentum we evolve (e.g.\ $p^{\prime,{\rm max}}_{s^{\prime}} = p^{\prime,+}_{n^{\prime},\,s^{\prime}}$ for the highest-energy bin $n^{\prime}$) correspond to higher values of the primary momentum $p_{s}(p^{\prime}_{s^{\prime}})$ outside of the range we evolve. To account for this, when we consider the highest-energy bin $n$ for each secondary-producing primary species $s$, we calculate $\dot{N}_{j,n\rightarrow n^{\prime},s\rightarrow s^{\prime}}$ up to the maximum $p^{\prime,{\rm max}}_{s^{\prime}}$ by simply extrapolating the primary spectrum in that highest-energy bin to arbitrarily-large  $p$ (i.e.\ assuming $\bar{f}_{0,\,j,n,s}[p_{n,s}^{0}]\,(p/p_{n,s}^{0})^{\psi_{j,n,s}}$ simply continues to $p \gg  p_{n,s}^{+}$). This is consistent with our spectral boundary conditions defined below (\S~\ref{sec:spectral.bcs}).

For some species (e.g.\ $\bar{p}$, B, etc.), it is possible in principle that even-heavier nuclei which we do not follow explicitly could produce some secondaries. We have attempted to assess the importance of this with the following method: first we assumed the spectra of all heavier hadrons to follow the same {relative} normalization to whatever the ``primary'' of interest is (e.g.\ H or CNO), as observed in the fits to the spectra of different species in \citet{cummings:2016.voyager.1.cr.spectra,bisschoff:2019.lism.cr.spectra}. Then, we combined this with the production cross-sections from \citet{moskalenko:2003.galprop.cx,tommassetti:2015.cr.frag.cx.uncertainties,korsmeier:2018.antiproton.cx,evoli:2018.dragon2.cr.cross.sections} to calculate the mean ``additional'' production at each $p^{\prime}$, relative to the channels we follow. Then assume this ratio is constant, so that the number produced $d\dot{n}_{s^{\prime}}(p^{\prime})$ is enhanced (relative to the rate at the channels we follow) by a factor $1 + \epsilon_{s^{\prime}}(p^{\prime})$. For all the secondary species of interest here ($e^{\pm}$, $\bar{p}$, B, Be), this correction is negligible (always $\lesssim 10\%$ and usually $\lesssim 1\%$),  compared to other uncertainties.

\subsection{Momentum-Space Integration}
\label{sec:numerics:momentum}

\subsubsection{Basic Setup}

Now consider the continuous momentum-space terms, 
\begin{align}
\label{eqn:mom.split.continuous} D_{t} \bar{f}_{0,s} = p^{-2}\,\partial_{p}[ Q_{s}\,p^{2}\,\bar{f}_{0,s} ]\ .
\end{align}
which we can write in terms of the one-dimensional distribution function $\bar{f}_{0}^{\rm 1D} \equiv 4\pi\,p^{2}\,\bar{f}_{0}$ as
\begin{align}
D_{t} \bar{f}_{0}^{\rm 1D} = -\partial_{p}[ (-Q_{s})\,\bar{f}_{0}^{\rm 1D}]\ ,
\end{align}
i.e.\ $\bar{f}_{0}^{\rm 1D}$ is simply being translated or advected one-dimensionally in $|{\bf p}|$-space with a flux $(-Q_{s})\,\bar{f}_{0}^{\rm 1D}$ hence a Lagrangian ``velocity'' or translation speed $\dot{p}$\footnote{One can also see this as shown in \citet{hopkins:m1.cr.closure} by inserting the $f$ for a CR ``packet'' with a single value of $p=\langle p\rangle$, and calculating $D_{t} \langle p\rangle$ directly, to obtain Eq.~\ref{eqn:pdot}. Note that technically the statement $\dot{p} = -Q_{s}$ and our treatment of these terms is only valid if $Q_{s}$ can be written in a manner that does not depend explicitly on $f_{0,s}$ itself (but instead on terms which depend on $p$, $s$, $\mu$, and external/background local plasma properties), but this is trivially satisfied for all terms we consider as shown in Eqs.~\ref{eqn:pdot}-\ref{eqn:S.ell.pdot}.} 
\begin{align}
\frac{{\rm d} p}{{\rm d} t} &= \dot{p} = -Q_{s}\ .
\end{align}
Combining this with the definitions of $\tilde{D}_{p\mu}$, $\tilde{D}_{pp}$, $\bar{f}_{1} = (v_{d}/v)\,\bar{f}_{0}$, and using our local power-law representation of $\bar{f}_{0}$, we can simplify to obtain:
\begin{align}
\label{eqn:pdot} \dot{p} = - Q_{s} = -S_{\ell} - p\,\left[ \mathbb{D}:\nabla{\bf u} + \bar{\nu}\,\left\{ \frac{\bar{v}_{A}\,v_{d}}{v^{2}} + \psi\,\chi\,\frac{v_{A}^{2}}{v^{2}}\right\} \right]
\end{align}
Here the advective/turbulent/convective term\footnote{Calculated using our standard second-order gradient estimator for $\nabla{\bf u}_{j}$ for each cell, with the value of $\mathbb{D}_{j,n,s}$ used for the flux update in \S~\ref{sec:numerics:spatial}.} is $\propto \mathbb{D}:\nabla{\bf u}$, the ``streaming'' or gyro-resonant loss term\footnote{Again in-code we take $v_{d} = F_{j,n,s}^{N}/n_{j,n,s} = F_{j,n,s}^{E}/e_{j,n,s}$ defined in the flux update, taken to be constant over the bin and subcycle step per our ``bin-centered'' approximation.} is $\propto \bar{\nu}\,{\bar{v}_{A}\,v_{d}}/{v^{2}}$, and the diffusive re-acceleration term is $\propto \bar{\nu}\,\psi\,\chi\,{v_{A}^{2}}/{v^{2}}$. All other continuous losses (Coulomb, ionization, Bremstrahhlung, inverse Compton, synchrotron) are in 
\begin{align}
\label{eqn:S.ell.pdot} S_{\ell} \equiv -\left[ \dot{p}_{\rm Coulomb} + \dot{p}_{\rm ion} + \dot{p}_{\rm Brems} + \dot{p}_{\rm IC} + \dot{p}_{\rm synch} \right]
\end{align}
with the expressions given for each species in \S~\ref{sec:collisional.losses}.

\subsubsection{Numerical Integration Method}

This can then be immediately integrated using the method presented in \citet{girichidis:cr.spectral.scheme}  (see also \citealt{ogrodnik:2021.spectral.cr.electron.code}) without modification, but we review that here for completeness. Since we have operator-split these terms, there is no communication between cells here: we are effectively updating a ``one-zone'' model in this step independently within each cell $j$. Recall that on the subcycle timestep $\Delta t_{j}^{\rm sub}$, we assume the background MHD plasma state ${\bf U}_{j}$ is fixed, and note that each CR species in this step is strictly independent (there are no cross-terms), so trivially we only need to define the method for a single species $s$ (we will update each species $s$ in serial in turn, on each substep $\Delta t_{j}^{\rm sub}$). Given this, it also immediately follows that over the sub-step for a single species $s$, $\dot{p} = \dot{p}_{j,s}(p_{j,s},\,...)$ is purely a function of $p$ and numerical constants. 

Now consider each ``bin'' in turn. Assume (temporarily) that $\dot{p}>0$: then for convenience we will evaluate the bins in order of increasing $p_{j,n,s}^{0}$. For any initial $p_{0,j,n,s}$ in the bin ($p_{n,s}^{-} < p_{0,j,n,s} < p_{n,s}^{+}$), one can immediately calculate the final momentum 
\begin{align}
p_{f,j,n,s} &= p_{0,j,n,s} + \int_{t_{0,j}}^{t_{0,j} + \Delta  t_{j}} \dot{p}_{j,s}(p_{j,s},\,{\bf U}_{j},\,...)\,dt\ .
\end{align} 
For some $p_{0,j,n,s} = p^{\prime}_{0,j,n,s}$, we will have 
\begin{align}
p^{\prime}_{f,j,n,s}=p_{n,s}^{+}\ .
\end{align} 
Thus all CRs with $p_{0,j,n,s} < p^{\prime}_{0,j,n,s}$ remain ``in the bin'' though their energy may increase: we can compute their final number and energy as 
\begin{align}
N_{f,j,n,s} &= \int_{p_{n,s}^{-}}^{p^{\prime}_{0,j,n,s}}\,dN_{0,j,n,s} \\ 
\nonumber 
E_{f,j,n,s} &= \int_{p_{n,s}^{-}}^{p^{\prime}_{0,j,n,s}}\,T_{s}(p_{f,j,n,s}[p_{0,j,n,s}])\,dN_{0,j,n,s} 
\end{align}
where 
\begin{align}
dN_{0,j,n,s} & \equiv 
\int_{V_{j}}\,d^{3}{\bf x}\,4\pi\,p_{0,s}^{2}\,dp_{0,s}\,\bar{f}_{0,j,s} \\ 
\nonumber & \equiv V_{j}\,\bar{f}_{0,j,n,s}[p^{0}_{n,s}]\,(p_{0,j,n,s}/p_{n,s}^{0})^{\psi_{j,n,s}}\,4\pi\,p_{0,j,n,s}^{2}\,d p_{0,j,n,s}
\end{align} 
is just the number density per unit ``initial'' momentum $d p_{0,j,n,s}$ (given by the primitive variables for the cell, or equivalently by $N_{0,j,n,s}$, $E_{0,j,n,s}$, at the start of this substep). We can likewise compute a number and energy 
\begin{align}
\Delta N_{j,n\rightarrow n^{\prime},s} &= \int^{p_{n,s}^{+}}_{p^{\prime}_{0,j,n,s}}\,dN_{0,j,n,s} \\ 
\nonumber \Delta E_{j,n\rightarrow n^{\prime},s} &= \int^{p_{n,s}^{+}}_{p^{\prime}_{0,j,n,s}}\,T_{s}(p_{f,j,n,s}[p_{0,j,n,s}])\,dN_{0,j,n,s}\ ,
\end{align}
which are the total number and final energy of the CRs which will increase in momentum sufficiently to move from bin $n$ to the ``next'' bin $n^{\prime}$. The relevant integrals here can be computed numerically to arbitrary desired precision.\footnote{We can operator split all loss processes in $\dot{p}$ and evaluate the integrals to construct a table of $p_{f,j,n,s}[p_{0,j,n,s}]$, which then immediately allows us to evaluate the relevant updates to $N_{j,n,s}$ and $E_{j,n,s}$, each independently either analytically (where this is solveable) or pre-computed in a lookup table. But we find identical results using a simple composite trapezoidal rule quadrature to evaluate the integrals numerically for arbitrary $\dot{p}_{j,s}$. For this we impose a fractional error tolerance of better than $1\%$ in $|p_{f,j,n,s}-p_{0,j,n,s}|$, which is almost always easily satisfied for $\sim 10-12$ integration steps, but because this tolerance is so much smaller than our bin sizes, and our subcycle timesteps are small (so the ``integration'' is usually extremely well-approximated by a simple linear expansion in $\Delta t_{j}^{\rm sub}$, which makes the expressions above trivial for any $\dot{p}$) we find we can make the tolerance much larger (up to $\sim50\%$) before we detect measureable differences in any results.}  
Note that our subcycle timestep condition ensures CRs do not cross multiple bins in a single substep. We then immediately update the conserved quantities in both bins ($N_{j,n,s} \rightarrow N_{f,j,n,s}$, $E_{j,n,s} \rightarrow E_{f,j,n,s}$, $N_{j,n^{\prime},s} \rightarrow N_{j,n^{\prime},s} + \Delta N_{j,n\rightarrow n^{\prime},s}$,  $E_{j,n^{\prime},s} \rightarrow E_{j,n^{\prime},s} + \Delta E_{j,n\rightarrow n^{\prime},s}$), and immediately recalculate the corresponding primitive variables. We then repeat this for the next bin $n^{\prime}$, and so on until all bins for species $s$ in cell $j$ are updated for that subcycle step $\Delta t^{\rm sub}_{j}$ (we then repeat for each species, then repeat for the next subcycle timestep, until the full timestep $\Delta t_{j}^{\rm cell}$ is complete). 
Of course, if $\dot{p}_{j,s}<0$, then the procedure above is numerically identical, but we instead evaluate the number and energy of CRs which ``move down'' to a lower-$p_{j,s}$ bin, working in order of $n$ from the highest-$p_{j,s}$ to lowest-$p_{j,s}$ bin.\footnote{It is possible, though extremely rare, for $\dot{p}_{j,s}(p_{j,s},\,...)$ to change sign as a function of $p_{j,s}$ over the range of $p$ (for a given $s$ and $j$) that we evolve in our simulations for that $s$. In these cases, we evaluate bins in the ``order'' (increasing or decreasing $p$) matching the sign of $\dot{p}_{j,s}$ in the majority of bins for species $s$ in cell $j$, but for each bin we use the correct value of $\dot{p}_{j,s}$ to determine if there is flux to higher or lower-$p$ bins (or both, if the sign-change occurs mid-bin).}

Again we stress that this is just a straightforward implementation of the method from \citet{girichidis:cr.spectral.scheme} (itself an extension of the methods proposed and utilized in e.g.\ \citealt{1999ApJ...511..774J,miniati:2001.cr.piecewise.powerlaw,2007JCoPh.227..776M,2001ApJ...562..233M,jones:2005.cr.shock.spectral.sims,yang:2017.fermi.cr.spectra.sims,winner:2019.cr.electron.post-processing.tracers}), and readers interested in detailed numerical validation and tests should see that paper (e.g.\ their Figs.~8-12). Another implementation of the same method for leptons, demonstrating the applicability to different species and flexibility to handle arbitrary cooling functions is given in \citet{ogrodnik:2021.spectral.cr.electron.code}, and other details of the method are discussed in \citet{hanasz:2021.cr.propagation.sims.review}. These papers, as well as our tests below, also motivate our choice of bin sizes: if we restrict to the dynamic range of CR energies we follow here, both suggest $\sim10$ intervals spanning that particular range as ``optimal,'' though \citet{ogrodnik:2021.spectral.cr.electron.code} show that increasing the bin number only modestly increases the accuracy of the results at a level significantly smaller than e.g.\ galactic variations or variations between different diffusion models shown in Figs.~\ref{fig:demo.cr.spectra.fiducial}-\ref{fig:spec.compare.numerics} in the text.

\subsubsection{Subcycle Timestep Condition}

For stability and accuracy we enforce a subcycle timestep 
\begin{align}
\Delta t_{j}^{\rm sub} \le C_{\rm cour}\,{\rm MIN}\left(\delta t_{j,\,n,\,s} \right)\ ,
\end{align} 
where the minimum is over all bins and species, and $\delta t_{j,\,n,\,s}$ is the time required for a CR to cross from one bin boundary $p_{j,n,s}^{\pm}$ to the next closest bin (higher or lower according to the sign of $\dot{p}_{j,s}$ evaluated at $p_{j,n,s}^{\pm}$), or to cool from the lowest-$p$ bin boundary $p_{j,n=0,s}^{-}|_{\rm min}$ for each species $s$ to $p=0$ (if $\dot{p}_{j,s}<0$ at this $p_{s}=p_{j,n=0,s}^{-}|_{\rm min}$). We pre-calculate this for all bin ``edges'' for each species $s$ in each cell $j$, and take the minimum. We then take the minimum of this and the similar timestep restriction from the $j$ terms in \S~\ref{sec:numerics:injection}, and set the subcycle timestep to the minimum of this or the cell timestep $\Delta t_{j}^{\rm cell}$.

\subsubsection{Spectral Boundary Conditions}
\label{sec:spectral.bcs}

For the lowest and highest-$p$ bins for each species $s$, we adopt a simple inflow/outflow boundary condition. If the flux ($\Delta N_{j,n\rightarrow n^{\prime},s}$, $\Delta E_{j,n\rightarrow n^{\prime},s}$) would move out of the spectral domain (e.g.\ if $\dot{p}<0$ at the $p_{n,s}^{-}$ boundary of the lowest-$p_{j,n,s}$ bin, or $\dot{p}>0$ at the $p_{n,s}^{+}$ boundary of the highest-$p_{j,n,s}$ bin) we simply allow it to be lost (outflow). If there should be a flux ``into'' the domain (e.g.\ if $\dot{p}>0$ at the $p_{n,s}^{-}$ boundary of the lowest-$p_{j,n,s}$ bin, or $\dot{p}<0$ at the $p_{n,s}^{+}$ boundary of the highest-$p_{j,n,s}$ bin; rare at the low-$p$-boundary, but not uncommon at the high-$p$-boundary), then we calculate the flux ($\Delta N_{j,n^{\prime}\rightarrow n,s}$, $\Delta E_{j,n^{\prime}\rightarrow n,s}$) which should flow into the bin by temporarily assuming the existence of a ``ghost bin'' which has a continuous power-law distribution function matched to the same slope and normalization at the bin edge (e.g.\ if in the final ``regular'' bin, the distribution function is given by some $\bar{f}_{0,\,j,n,s}[p_{n,s}^{0}]\,(p/p_{n,s}^{0})^{\psi_{j,n,s}}$, we simply assume this power law continues to $p \ll p_{n,s}^{-}$ or $p \gg p_{n,s}^{+}$ for the lower or higher boundaries respectively). Ignoring this ``boundary flux,'' however, has very minimal effects (de-activating it leads to slightly-steeper slopes for leptons in the highest-rigidity bins, but the effect is small).

\subsection{Reduced Speed of Light Implementation}
\label{sec:numerics:rsol}

As described in the text, we implement a reduced speed of light (RSOL) $\tilde{c}<c$ as is standard practice in the field, to allow larger numerical timesteps. The specific implementation is presented and derived exactly from the Vlasov equation for an arbitrary CR population (obtained by multiplying the $D_{t} f$ term in the Vlasov equation by $c/\tilde{c}$,  then re-deriving all equations) in \citet{hopkins:m1.cr.closure} (their Eq.~50). Mathematically, this amounts to\footnote{Trivially, this is also equivalent to taking $\tilde{\bf F}_{jj^{\prime},n,s}^{N,E} \rightarrow (\tilde{c}/c)\,\tilde{\bf F}_{jj^{\prime},n,s}^{N,E}$ in Eqs.~\ref{eqn:N.advection}-\ref{eqn:E.advection} and ${c}^{2}\,\bhat\cdot\nabla \mathbb{P}_{j,n,s} \rightarrow c\,\tilde{c}\, \bhat\cdot\nabla \mathbb{P}_{j,n,s}$, $\bar{\nu}_{0,j,n,s} \rightarrow (\tilde{c}/c) \bar{\nu}_{0,j,n,s}$ in Eq.~\ref{eqn:FluxE.eqn}; $j\rightarrow (\tilde{c}/c)\,j$ in Eq.~\ref{eqn:injection}; and $Q_{s} \rightarrow (\tilde{c}/c)\,Q_{s}$ in Eq.~\ref{eqn:mom.split.continuous}.}  multiplying $D_{t} \bar{f}_{0,s}$ and $D_{t}\bar{f}_{1,s}$ in Eq.~\ref{eqn:f0.rev}-\ref{eqn:f1.rev} by $(c/\tilde{c})$.
Since $\tilde{c}/c$ is a universal constant in the simulation, this is numerically trivial (it has no effect on the numerical methods above) -- in fact it is, by construction, exactly equivalent to a change of units for certain operations \citep{hopkins:m1.cr.closure}. We stress that this method is identical to the standard implementation of an RSOL in radiation-hydrodynamics simulations (see e.g.\ \citealt{2013ApJS..206...21S,rosdahl:2013.m1.ramses}, and references therein), and recently-developed implementations for MHD-PIC simulations \citep{ji:2021.mhd.pic.rsol,ji:2021.cr.mhd.pic.dust.sims} and as such has been tested in hundreds of applications. As shown therein and in \citet{hopkins:m1.cr.closure}, it is trivial to demonstrate that this necessarily gives the exact solutions (for $\tilde{c}=c$) in local steady-state, and even out-of-steady-state, the solutions for e.g.\ any propagating population exactly match the $\tilde{c}=c$ solution at any fixed distance from a source. And we show explicitly in the text (Figs.~\ref{fig:spec.compare.numerics} \&\ \ref{fig:spec.compare.numerics.alt}) that our results are independent of $\tilde{c}$, as they should be.

\subsection{Simple Numerical Tests}
\label{sec:numerical.tests}

\begin{figure*}
	\includegraphics[width=0.34\textwidth]{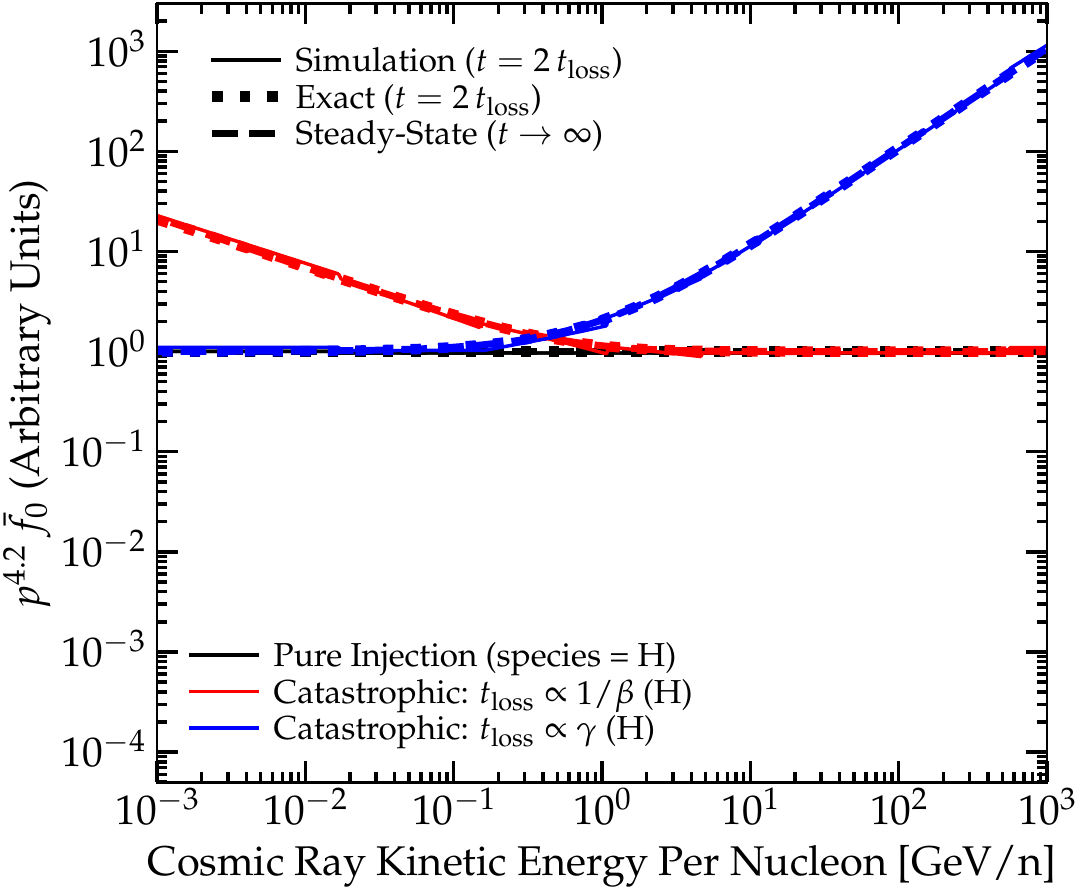} 
	\includegraphics[width=0.32\textwidth]{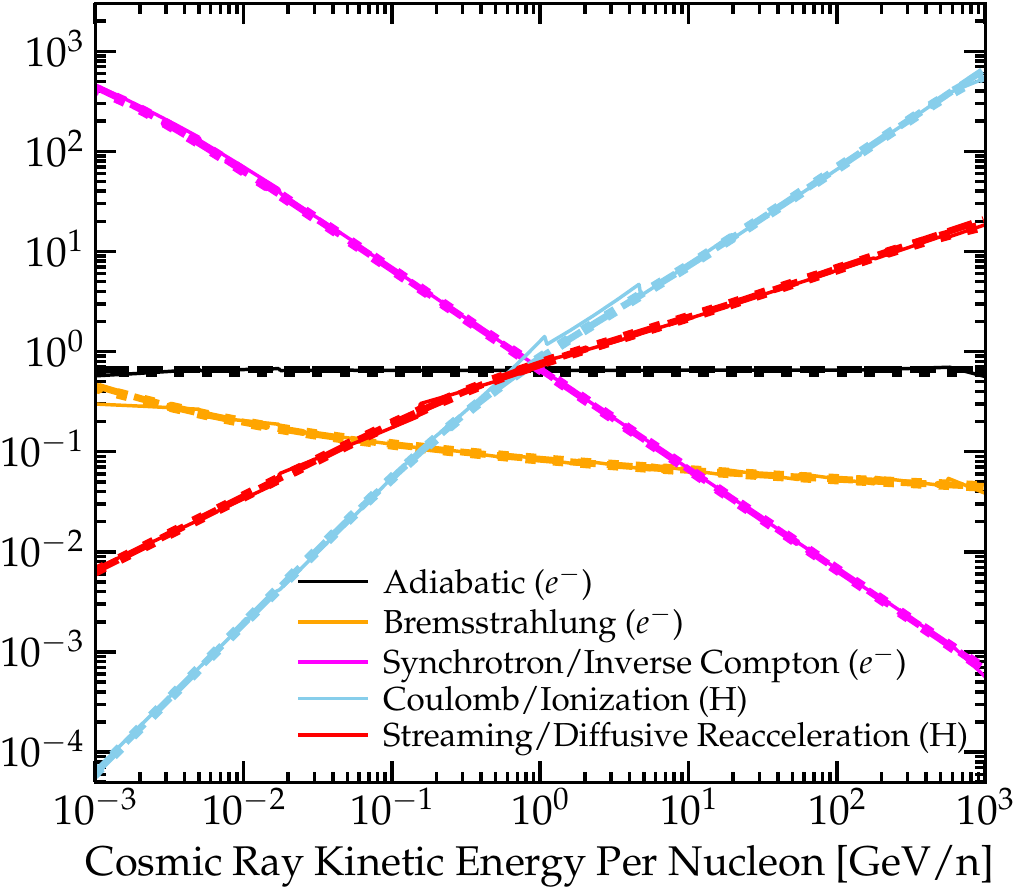} 
	\includegraphics[width=0.32\textwidth]{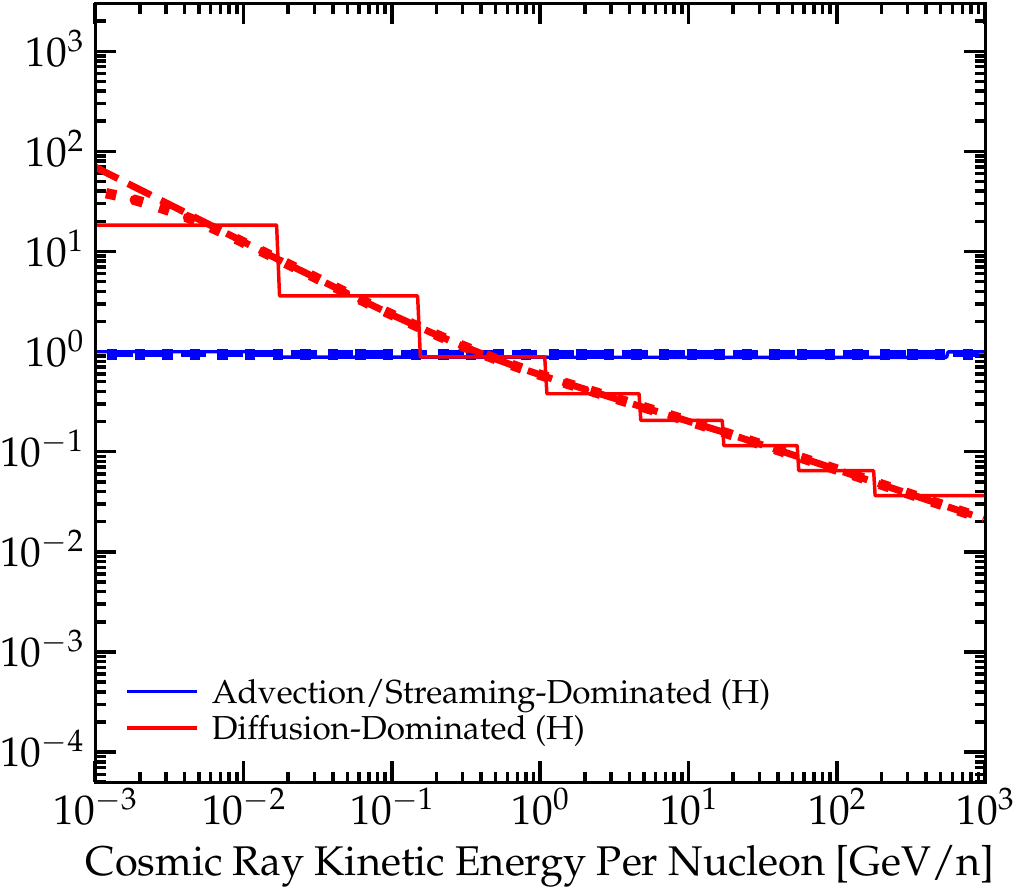} \\ 
	\vspace{-0.2cm}
	\caption{Numerical tests of our implementation of CR spectral evolution, from \S~\ref{sec:numerical.tests}. We consider an idealized homogeneous medium, with $D_{t} f = j_{\rm inj} - \mathcal{L}$ given by continuous injection $j_{\rm inj} \propto p^{-4.2}$ as in the main text and different loss terms ``$ \mathcal{L}$'' considered each in turn, so that we can compare to exact solutions. For each test besides ``pure injection'' we evolve to twice the CR ``loss timescale'' $t_{\rm loss}$. We plot the predicted spectrum $\bar{f}_{0}$ (in arbitrary units, compensated by $p^{4.2}$ for clarity) to see how different terms modify the spectra, and label the assumed scaling of the loss rates. We show either electrons or protons for each (whichever is more relevant). We compare both exact numerical solutions at the same time as the simulation, and steady-state solutions given as $t\rightarrow \infty$. In all cases simulation and exact solutions have converged close to steady-state.
	{\em Left:} Injection and catastrophic losses (\S~\ref{sec:numerical.tests:injection}). For ``injection only'' $\mathcal{L} = 0$. Otherwise we take $\mathcal{L}= f / t_{\rm loss}$ with the arbitrary scaling of $t_{\rm loss}$ shown bracketing the range of collisional and radioactive losses in-text. 
	{\em Center:}  Continuous momentum losses (\S~\ref{sec:numerical.tests:continuous}), $\mathcal{L} = p^{-2}\,\partial_{p}(p^{2}\,\dot{p}\,f)$ with $\dot{p} = -p/t_{\rm loss}$, where $t_{\rm loss}$ scales according to examples of the different continuous loss/gain terms considered in the text.
	{\em Right:}  Spatial advection/streaming/diffusion (\S~\ref{sec:numerical.tests:spatial}), $\mathcal{L} = \nabla \cdot {\bf F} = \nabla \cdot (v\,\bar{f}_{1}\,\bhat)$ where the spatial flux ${\bf F}$ (or $\bar{f}_{1}$) is integrated for a vertically-stratified plane-parallel CR atmosphere. We choose parameters such that the transport is either streaming/advection-dominated (drift speed and $t_{\rm loss}$ independent of CR momentum), or diffusion-dominated (drift speed and $t_{\rm loss}^{-1}\propto \bar{\nu}(p)/v^{2}$). The latter produces the well-known ``step'' artifacts as a result of the ``bin-centered'' diffusion approximation (where the scattering rate is treated as constant across each bin -- varying only bin-to-bin -- for purposes of calculating the spatial flux). 
	\vspace{-0.5cm}
	\label{fig:numerical.tests}}
\end{figure*}

Fig.~\ref{fig:numerical.tests} presents some idealized numerical tests of the methods above. We emphasize that the methods here have all been presented and tested in other papers previously, so these should be regarded as ``validation'' tests our particular implementation, and readers interested in more comprehensive details should see the numerical methods papers referenced above. We consider tests of each operator-split term from \S~\ref{sec:numerics:spatial}-\ref{sec:numerics:momentum} in turn, in our {\small GIZMO} code.

\subsubsection{Injection \&\ Catastrophic Terms}
\label{sec:numerical.tests:injection}

First consider the injection and catastrophic loss processes in $j$ (\S~\ref{sec:numerics:injection}). Without loss of generality, consider a single species $s$, and single cell $j$ (i.e.\ a spatial ``one-zone'' model), in a frame comoving with the cell. To test these terms, assume all other terms are negligible,\footnote{We do this in-code by multiplying all other terms (besides those of interest for our test) in the equations for $D_{t} f$ by some arbitrarily small number.} so the DF evolves according to $D_{t} f \approx j_{\rm inj} - j_{\rm loss}$, with $j_{\rm inj}(p) \propto p^{-4.2}$ (as we assume in the text) and $j_{\rm loss}=-(\sigma\,v\,n)_{\rm loss}\,f = -f/t_{\rm loss}$ (the general form of all loss terms we consider, with $t_{\rm loss}(p)$ some function of $p$), with $j$ and $t_{\rm loss}$ independent of time, and begin at $t=0$ with $f=0$. With these simplifications, the actual units and normalization of $j_{\rm inj}$ and $j_{\rm loss}$ are arbitrary: so we simply work in convenient code units. This has an exact analytic solution $f(p,\,t) =  j_{\rm inj}\,t_{\rm loss}\,(1 - \exp{\{-t/t_{\rm loss} \}})$. We compare the numerical results, using our in-code implementation, for both an early time and/or negligible loss case ($t \ll t_{\rm loss}$, where the spectrum is essentially ``pure injection'' with $f \approx j_{\rm inj}\,t$), and for a time $t \approx 2\,t_{\rm loss}^{\rm max}$ (where $t_{\rm loss}^{\rm max}$ is the maximum $t_{\rm loss}$ over the momentum range we evolve), by which point the spectrum should be close-to-steady-state.

Note that the different catastrophic processes considered in the text mostly have similar dependence on $p$: the different catastrophic hadronic processes generally feature approximate $j_{\rm loss} \propto \beta$ ($t_{\rm loss} \propto 1/\beta$) at high energies (i.e.\ roughly constant in the relativistic limit),\footnote{This also includes positron annihilation: although the Dirac formula features a complicated dependence on $\gamma$, for ultra-relativistic positrons (the case of interest at our energies), this becomes $t_{\rm loss} = {\rm constant} + \mathcal{O}(1/\gamma)$ to leading order. Likewise the residual $T$ dependence for heavier nuclei is generally weak.} with a cutoff at very low energies where $j_{\rm loss}\rightarrow 0$ (which is not interesting for our test). Two exceptions are radioactive decay, where $t_{\rm loss} \propto \gamma$, and $\bar{p}$ annihilation (where $\sigma_{p\bar{p}} \sim $\,constant at high-$p$, but then rises $\propto p^{-1/3}$ at small-$p$, weaker than $\beta^{-1}$ for non-relativistic CRs). So for the sake of completeness we consider both a case with $t_{\rm loss} \propto  1/\beta$ and $\propto \gamma$ at all energies, which bracket the range of different cases for different species.

\subsubsection{Continuous Momentum-Space Terms}
\label{sec:numerical.tests:continuous}

Next consider an analogous experiment for the continuous momentum-space terms (\S~\ref{sec:numerics:momentum}), ignoring all other terms (besides injection) so $D_{t} f = j_{\rm inj} + p^{-2}\,\partial_{p}(p^{2}\,Q_{s}\,f)$. Already analytic solutions become non-trivial here, but the steady-state ($D_{t} \rightarrow 0$) solutions can be easily solved exactly. For the simple (but representative) case of a power-law $Q_{s} = -\dot{p} = p/t_{{\rm loss}}$ with $t_{\rm loss} \equiv t_{{\rm loss,0}}\,(p/p_{0})^{-\psi_{\rm loss}}$ (and $j_{\rm inj}=j_{0}\,(p/p_{0})^{-\psi_{\rm inj}}$), we have $f\rightarrow (j_{0}\,t_{{\rm loss},0}/(\psi_{\rm inj}-3))\,(p/p_{0})^{-(\psi_{\rm loss}+\psi_{\rm inj})} = (\psi_{\rm inj}-3)^{-1}\,j_{\rm inj}\,t_{\rm loss}$. We again begin from $f=0$ and evolve each test until $t\approx 2\,t_{\rm loss}^{\rm max}$ (where the maximum is over $p$ for a given $t_{\rm loss}[p]$ in each test). 

We consider each of the continuous loss processes treated in-code in turn, but since the units/normalization are arbitrary in these tests, we only consider separately those which exhibit a different dependence of $t_{\rm loss}$ on $p$. Thus we have (1) adiabatic ($t_{\rm loss}\sim$\,constant); (2) Bremsstrahlung ($t_{\rm loss} \propto 1/(\ln{(2\gamma)} -1) \propto p^{-(0.1-0.3)}$ for leptons, the case we consider), not very different from adiabatic; (3) inverse Compton \&\ synchrotron ($t_{\rm loss} \propto p/\gamma^{2} \propto p^{-1}$, again for leptons); (4) Coulomb \&\ ionization losses ($t_{\rm loss} \propto p\,\beta\,(1+1/(\gamma\,\beta)^{2})^{-1/2}$, or $\propto p^{3}$ in the non-relativistic limit and $\propto p$ in the ultra-relativistic limit, for hadrons and similarly up to logarithmic corrections for leptons); (5) streaming losses and diffusive reacceleration ($t_{\rm loss} = (v^{2}/\bar{\nu})\,(\bar{v}_{A}\,v_{d} + \psi\,\chi\,v_{A}^{2})^{-1}$ with $v_{d}\equiv F_{n,s}^{N}/n_{j,n,s}$; the pre-factor here depends on the local drift velocity/flux and spectral shape, so for simplicity in this test problem we assume $|\bar{v}_{A}\,v_{d} + \psi\,\chi\,v_{A}^{2}| \sim $\,constant,\footnote{As noted in the text, when the flux equation ($D_{t} \bar{f}_{1}$) is in local steady-state in the near-isotropic limit ($|D_{t} \bar{f}_{1}| \ll |\bar{\nu}\,\bar{f}_{1}|$), which is often a good approximation, then $v_{d}$ takes a value such that the ``streaming + diffusive re-acceleration'' term in Eq.~\ref{eqn:pdot} becomes $\bar{\nu}\,\left\{ ...\right\} \rightarrow (\bar{v}_{A}/3)\,\bar{f}_{0}^{-1}\,\bhat\cdot \nabla \bar{f}_{0} + (\psi\,\bar{\nu}/3)\,[(v_{A}^{2}-\bar{v}_{A}^{2})/v^{2}]$. For our default model assumptions in the main text (allowing \Alf{ic} streaming, $\bar{v}_{A} = \pm v_{A}$) this further simplifies to give $t_{\rm loss} = 3\,\ell_{\nabla}/v_{A}$ where $\ell_{\nabla} \equiv \bar{f}_{0}/|\bhat\cdot\nabla \bar{f}_{0}| \approx n_{n,s}/|\nabla_{\|} n_{n,s}|$. In this limit then, the streaming plus diffusive reacceleration term has $t_{\rm loss} \sim $\,constant, identical to the adiabatic term, and is trivial to accurately integrate. Our test therefore intentionally reflects a strongly ``out-of-equilibrium flux'' or a ``no streaming'' ($\bar{v}_{A}=0$) configuration, which are more challenging to treat accurately. 
% in total generality, $D_{p\mu}\,\bar{f}_{1} + D_{pp}\,\partial_{p} \bar{f}_{0} \rightarrow -p\,\bar{v}_{A}\,\Delta \bar{f}_{0} + \bar{\nu}\,\chi\,(p/v)^{2}\,(v_{A}^{2}-\bar{v}_{A}^{2})\,\partial_{p} \bar{f}_{0}$
} with $\bar{\nu} \propto \beta\,p^{-1/2}$ similar to the observationally-favored values, so $t_{\rm loss} \propto v\,p^{1/2}$).

Since some of these terms depend on $\beta$, $\gamma$ and have different behavior in relativistic and non-relativistic limits, we focus on the most interesting cases by considering Bremsstrahlung and inverse Compton+synchrotron for a leptonic case ($e^{-}$ or $e^{+}$, they are the same here), and Coulomb/ionization and streaming/diffusive reacceleration for a hadronic case so we can see the non-relativistic/relativistic transition (here we take protons [H] as the test case, though the scaling to other hadrons is straightforward). The adiabatic case is entirely independent of species choice in this setup. 

Note that a couple of these ``loss'' terms can have either sign and represent gains (e.g.\ adiabatic), but then (in this highly-simplified test problem) there is no steady-state solution so we only consider the ``loss'' sense. In our full simulations this energy comes from some other term (e.g.\ gas mechanical energy) and other loss/escape terms are always present, so these cannot run away.

\subsubsection{Spatial Flux/Advection Terms}
\label{sec:numerical.tests:spatial}

Next we consider the spatial flux terms (\S~\ref{sec:numerics:spatial}), ignoring all other terms besides injection so $D_{t} \bar{f}_{0} = j_{\rm inj} - \nabla \cdot {\bf F}$ with ${\bf F} = v\,\bar{f}_{1}\,\bhat$. To construct a simple analytically-solveable test problem, consider an infinitely-thin, cylindrically-symmetric source plane in the $xy$ axis with effective ``upward'' $j_{2D} = \delta(z)\,F_{0}$ (i.e.\ some constant injection rate per unit area $F_{0} \equiv dN/dt\,dA\,d^{3}{\bf p}$ in the source plane directed in the $+z$ direction, and no injection elsewhere), and $\bhat = \hat{z}$, with spatially-uniform $\bar{\nu}$ and $\bar{v}_{A}$. This is designed to be directly analogous to classic historical thin disk/leaky-box-type models. Note that (as discussed in \S~\ref{sec:numerics:spatial}) it is trivial to see in this setup that the spatial solutions for each momentum ``bin'' are entirely independent and operator-split here (each repeats the identical numerical procedure). Nonetheless, we show the full CR spectrum to illustrate this independence, and the effect of the ``bin-centered'' approximation discussed below.

We will consider two limits: first, a ``streaming-dominated'' limit, obtained by setting $\bar{\nu}$ to some extremely large numerical value, so that $\bar{f}_{1} \rightarrow (\tilde{D}_{\mu\mu}/\tilde{D}_{\mu p})\,\partial_{p}\,\bar{f}_{0}$ in steady-state. This has steady-state solutions for $\bar{f}_{0}$ with $F=v\,\bar{f}_{1}=\bar{v}_{A}\,\psi_{\rm inj}\,\bar{f}_{0} = F_{0}$, so $\bar{f}_{0} \rightarrow F_{0}/|\psi_{\rm inj}\bar{v}_{A}|$ is just proportional to the injection spectrum ($\psi=-\psi_{\rm inj}$) and the ``loss'' timescale (given by the escape time) to travel some finite distance $\ell_{0}$ is $t_{\rm loss} \sim \ell_{0} / |\psi\,\bar{v}_{A}|$. In other words, the injection spectrum is simply advected. Numerically we treat this with 10 spatial cells (akin to our coarse number of momentum bins for the problems above) along the $z$ direction (the symmetry of the problem means it is one-dimensional), where the lower ($z=0$) boundary cell treats the injection with an inflow boundary of flux $F_{0}$, and the uppermost (defined as $z=\ell_{0}$) uses an outflow boundary. Again the units are arbitrary, so we evolve to $2\,t_{\rm loss} = 2\,\ell_{0}/|\psi_{\rm inj}\,\bar{v}_{A}|$ in code units.

Second, we consider a ``diffusive'' case, obtained by taking $\bar{v}_{A} =0$ with finite $\bar{\nu}$. Now in steady-state, ${\bf F}=v\,\bar{f}_{1}\,\bhat = -(v^{2}/3\,\bar{\nu})\,\partial_{z}\,\bar{f}_{0}\,\hat{z} =  F_{0}\,\hat{z}$, so $\partial_{z}\,\bar{f}_{0} = -(3\,\bar{\nu}/v^{2})\,F_{0}=\,$constant (in space). Here we treat this with the same 10-cell profile with the lower cell having an injection (inflow) boundary and the upper cell enforces $\bar{f}_{0}=0$ at its upper boundary $\ell_{0}$, so $\bar{f}_{0} =  (3\,\bar{\nu}/v^{2})\,F_{0}\,(\ell_{0}-z)$.\footnote{Note this upper boundary is only needed in 1D. For a real 3D problem like our full simulations at distances $r$ far from the source injection, for (spatially) constant diffusivity $\kappa \sim (v^{2}/\bar{\nu})$, the solution is power-law-like, $f \propto F_{0}/\kappa\,r$, so we can have an infinite or open box with $f>0$ everywhere \citep{ji:fire.cr.cgm}.} This gives the usual effective (parallel) diffusivity $\kappa_{\|} = v^{2}/(3\,\bar{\nu})$, and loss/escape time $t_{\rm loss} \sim \ell_{0}^{2}/\kappa_{\rm eff} \sim \ell_{0}^{2}\,(3\,\bar{\nu}/v^{2})$ (and again we evolve to $2\,t_{\rm loss}^{\rm max}$). Here, unlike the streaming case, the transport/drift speed $\sim \kappa_{\|}/\ell_{0} \propto v^{2}/\bar{\nu}$ is not $p$-independent: motivated by the cases in the text we consider protons (so we can see both relativistic and non-relativistic limits for the $v$ terms) and $\bar{\nu} \sim 10^{-9}\,{\rm s^{-1}}\,\beta\,R_{\rm GV}^{-1/2} \propto  v\,p^{-1/2}$ (the actual units are arbitrary and scale out of our simplified test, but the dependence on $p$ we retain), so the effective diffusivity scales as $\kappa_{\|} \propto v\,p^{1/2}$. 

Recall, in our ``bin-centered'' approximation, we take $(v^{2},\,\bar{\nu}) \approx (v_{0}^{2},\,\bar{\nu}_{0})$ in the flux equation (Eq.~\ref{eqn:FluxE.eqn} for $D_{t} \bar{f}_{1}$ or $d_{t}  F^{N,E}$), where $v_{0}$, $\bar{\nu}_{0}$ are the values at the bin centers (geometric mean values of $p^{0}_{n,s} \equiv (p^{-}_{n,s}\,p^{+}_{n,s})^{1/2}$). As a result, numerically, $v^{2}/\bar{\nu} \rightarrow (v_{0}^{2}/\bar{\nu}_{0})$ is effectively constant over each bin, so after injection, that bin is transported from cell-to-cell conserving its slope $\psi$ within the bin. But since e.g. $\nu_{0}=\nu_{0}[p^{0}_{n,s}]$, the transport speed is correct at the center of each bin and each bin diffuses with a different mean speed. As a result, we see that the gross spectrum traces the expected steady-state behavior: if we draw a curve connecting bin centers $p_{n,s}^{0}$, for example, it traces the exact solution. But without accounting for how $\bar{\nu}$ varies {\em within each bin}, the slopes under pure diffusion do not evolve so we produce the ``step'' structure seen. This is evident in the predicted spectra in the main text as well (e.g.\ Figs.~\ref{fig:demo.cr.spectra.fiducial}-\ref{fig:spec.compare.numerics}), in particular where diffusion dominates the residence/escape time (e.g.\ for protons at energies $\gtrsim\,$GeV). 

One can show that the discrepancy here is formally second-order in momentum space, in that the correction terms to the fluxes ($\bar{f}_{1}$ terms) or difference between the bin-centered and exact solutions scales as $\mathcal{O}(|\ln{(p^{+}/p^{-})}|^{2})$. So we could decrease the errors by substantially increasing the number of bins. But this would entail very large computational expense for almost no other gain. Instead, in future work \citep{hopkins:cr.spectra.accurate.integration}, we present a modified version of the method here which incorporates a series of second-order correction terms to modify the evolve the spectrum appropriately ``within each bin'' when momentum-dependent diffusion dominates, in a way that retains manifest conservation of CR number and energy, stability, and the trivially operator-split bin-to-bin behaviors desired. We have run preliminary (low-resolution) experiments with this method, and find that the results are nearly identical to our bin-centered model except the ``step'' structures are smoothed.

\subsubsection{Continuous versus Discrete Injection}
\label{sec:injection}

In the experiments above, we treat the injection as continuous (numerically, we operator split and include an injection step and then subsequent CR operator steps in each cell timestep $\Delta t^{\rm cell}$; see \S~\ref{sec:pseudo.code}), both because this allows comparison with simple steady-state solutions and behaviors discussed in the main text, and because it is a numerically difficult ``stress test.'' It is well-known in many contexts that accurately representing steady-state solutions in the continuous-injection limit is more challenging (compared to e.g.\ free-decay of CR spectra from some initial condition, without injection) for numerical methods like ours which operator-split injection and losses. But in our simulations in the main text, recall that the delay between individual injection events (i.e.\ SNe) in some patch of the ISM is time-resolved,\footnote{At our fiducial resolution, a single star particle representing a young stellar population has a mean time between individual supernova events of $\sim 1$\,Myr, compared to a numerical timestep of $\sim 10^{3}\,$yr.} so achieving  exact balance in continuous injection is not particularly important to our results. We can immediately test the free-decay limit by simply taking any of our tests above and at some time turning off the injection (setting it to zero or some numerically small number). As expected, the solutions in this case are at least as accurate as those in Fig.~\ref{fig:numerical.tests}, or formally better (though the numerical errors are small in any case, except for the ``bin centered'' effects, which are similar in both steady-state and free-escape limits). 

\subsection{Summary}

\subsubsection{List of Operations in Pseudo-Code}
\label{sec:pseudo.code}

To summarize, the CR-specific operations taken on every cell timestep $\Delta t_{j}^{\rm cell}$ are, in-code:

\begin{itemize}

\item{Compute cell timesteps $\Delta t_{j}^{\rm cell}$ and list of active cells.}

\item{Perform first half-step kick for evolved fluxes (CR and MHD), drift and synch cells. Calculate where sources (e.g.\ SNe) will occur.}

\item{Update neighbor lists and re-compute volume decomposition, and cell primitive variables (e.g.\ $V_{j}$).}

\item{Inject CRs from discrete (stellar \&\ black hole) sources, alongside other mechanical feedback (e.g.\ $\Delta N_{j,n,s}^{\rm inj}$, per \S~\ref{sec:numerics:injection}). Update CR variables (conservative and primitive).}

\item{Compute spatial gradients, shielding, and other quantities for inter-cell faces (e.g.\ ${\bf A}_{jj^{\prime}}$) and fluxes.}

\item{Compute inter-cell fluxes (e.g.\ $\tilde{\bf F}_{jj^{\prime}}$) for conserved quantities (e.g.\ $d N_{j,n,s}^{\rm flux}/dt$, per \S~\ref{sec:numerics:spatial}) for all interacting neighbors. MHD and other fluxes also computed.}

\item{Perform second half-step kick for evolved fluxes.}

\item{Compute CR momentum-space update (losses \&\ continuous momentum-space terms; \S~\ref{sec:numerics:injection} \&\ \ref{sec:numerics:momentum}).}

\begin{itemize}

\item{Calculate cell sub-cycle timestep $\Delta t_{j}^{\rm sub}$ from minimum of all constraints.}

\item{Iterate over subcycle timesteps within each cell $j$ until the subcycles reach $\Delta t_{j}^{\rm cell}$.}

\begin{itemize}
\item{Iterate over each species $s$, within the subcycle step.} 
\begin{itemize}
\item{Iterate over each bin $n$, for the species $s$.}
\item{Calculate loss/gain or neighbor-bin flux $n\rightarrow n^{\prime}$ or secondary-bin flux ($n\rightarrow n^{\prime}$, $s\rightarrow s^{\prime}$) for each loss/gain process, from the current cell $n$.}
\item{Update CR conserved and primitive variables for each bin ($j,n,s$, $j,n^{\prime},s^{\prime}$), according to those loss/gain terms.}
\end{itemize}
\end{itemize}

\end{itemize}

\item{Repeat until final simulation time is reached.}

\end{itemize}

\subsubsection{Computational Expense and Typical Timesteps}

In terms of computational expense, the simulations here are typically $\sim 30$ times more expensive than an otherwise identical simulation without any CRs. This difference is almost entirely driven by the Courant condition ($\Delta t_{j}^{\rm cell} \le C_{\rm cour}\,\Delta x_{j}/v_{\rm signal}$), given our very high $\tilde{c} \sim v_{\rm signal}$ adopted, which reduces the timesteps for all cells at all times by a correspondingly large factor. If we compare to, say, a single-bin CR simulation \citep{hopkins:cr.mhd.fire2} or an M1 radiation-hydrodynamics simulation (which solves numerically essentially identical spatial advection/flux equations so imposes the same Courant condition; see \citealt{hopkins:radiation.methods}) with the same $\tilde{c}$, the cost difference is much more modest, a factor $\lesssim 2-3$. Of that added cost, most owes to added communication and associated imbalances, particularly in the gradients and MHD (+CR) flux computation, because the method requires calculating and passing in memory $\sim 100$ times as many CR-specific variables compared to a ``single bin'' CR simulation, as each species and bin requires its own ``set'' of variables (e.g.\ $N_{j,n,s}$, $E_{j,n,s}$, $F_{j,n,s}^{N,E}$, their gradients, etc.) each equivalent to the set we would normally pass for a single-bin CR calculation. 

The momentum-space operations are relatively modest in cost (typically entailing $\sim 10-20\%$ of the total runtime), for three reasons. First, while we invoke subcycling, the cells which dominate the total CPU cost of our simulations (those with the smallest $\Delta t_{j}^{\rm cell}$, generally the most dense, star-forming gas), generally do not require many subcycles because their flux/Courant timesteps are already very small: for e.g.\ a gas cell with $n_{\rm n} \sim 100\,{\rm cm^{-3}}$ at our fiducial mass resolution and reduced speed of light, $\Delta x_{j}/\tilde{c} \sim 10^{3}\,{\rm yr}$ (so the timestep is a couple hundred years). But at this density, the subcycle timestep $\Delta t_{j}^{\rm sub}$ is generally limited by either the ionization loss timescale (the gas being mostly neutral) for hadrons in the lowest-energy bin ($\sim p_{n,s}^{-}/|\dot{p}_{{\rm ion},\,j,n,s}|$ in the lowest-$p$ hadronic bin) which is also $\sim 10^{3}$\,yr, or by the synchrotron/inverse Compton loss timescale in the highest-energy leptonic bin which is $\sim{\rm a\ few}\times10^{3}\,{\rm yr}$ (for magnetic+radiation energies of $\sim 100\,{\rm eV\,cm^{-3}}$ at these gas densities, per Fig.~\ref{fig:egy.corr.density} in the text). So these cells typically only feature a few subcycles at most. Second, the momentum-space operations are embarrassingly parallel at the cell $j$ and  species $s$ level (involving no communication between threads). And third, the complicated sub-operations (e.g.\ numerical integration and evaluation of the relevant cooling/loss functions) are almost entirely floating-point operations that are very high-efficiency. 

There are some cases in e.g.\ very low-density  cells where  $\Delta t_{j}^{\rm cell}$ is larger but the radiation energy density is still relatively high (as it cannot fall much below the ISRF in the ISM or CMB in the CGM; see Fig.~\ref{fig:egy.corr.density}), so more subcycles are required, reaching $\gtrsim 100-1000$ in rare cases. But since the overall timesteps are so much larger to begin with and this is a small fraction of all cells, it has little effect on the total CPU cost.

\subsubsection{Timescales for Convergence}
\label{sec:numerics:timescales}

Briefly, we review here the timescales over which we expect the simulations to converge to local-steady-state behavior for the CR spectra. In steady state in the disk we have injection $\dot{f} \sim \dot{j}_{\rm inj}$ balanced by some ``effective loss'' rate $\dot{f} \sim -f/t_{\rm loss}$, where ``loss timescale'' $t_{\rm loss}$ is given by whatever process dominates the loss rate (i.e.\ the fastest of all loss processes), including escape (i.e.\ diffusive/streaming/advective losses from the disk to the halo/CGM/IGM). Focusing on just the terms which dominate at most of the energies and most of the species of interest, and using the expressions in the text or the values from the simulations shown in \S~\ref{sec:loss.timescales}, the Coulomb/ionization loss timescale is given roughly by: 
$t_{\rm loss,\,ion} \sim 0.1\,{\rm Myr}\,( {\rm cm^{-3}}/{n_{\rm n}} )\,( {T}/{\rm MeV} )$;
the synchrotron-plus-inverse Compton loss timescale (for leptons) is given by 
$t_{\rm loss,\,synch/IC} \sim 0.3\,{\rm Myr}\,({\rm eV\,cm^{-3}}/{(u_{\rm rad}+u_{\rm B})})\,({R}/{\rm TV})^{-1}$;
and the escape loss timescale (for the best-fit scattering rates, ignoring advection and streaming, for a simple smooth vertical profile and tangled fields) is approximately: 
$t_{\rm loss,\,escape} \sim {\rm Myr}\,({\ell_{\rm cr}}/{\rm kpc})^{2}\,({R}/{\rm GV})^{-1/2}$,
where $\ell_{\rm cr}(R) \sim \bar{f}_{0,s}(R)/|\nabla \bar{f}_{0,s}(R)|$ is the CR gradient scale length at some rigidity. 

First, note that these are extremely ``well-resolved'' timescales: our typical numerical timestep in our fiducial simulations for gas at densities $\sim 1\,{\rm cm^{-3}}$ is $\sim 10^{3}\,$yr, so we do not have to worry about e.g.\ the limit where the CRs converge to equilibrium on much faster timescales than are numerically tractable (where implicit solvers may be useful, compared to our explicit spectral integration). 

Second, recall that the typical timescale for the CR properties to converge to steady-state at some $R$ is given by the relevant (shortest) loss timescale at that $R$:\footnote{Formally, if $D_{t} f = j - f/t_{\rm loss}$ with $j$ and $t_{\rm loss}$ constant, then any deviations in the initial condition from the equilibrium solution are damped exponentially with a decay time $=t_{\rm loss}$.} $\sim 0.1$\,Myr at the lowest and highest CR energies we evolve, and $\sim$\,a few Myr at the peak of the spectrum ($\sim$\,GeV) where the {\em net} loss+escape timescale is maximized (this, of course, is in part the reason why the spectum peaks at these energies). So we expect to converge to numerical equilibrium in $\lesssim 10\,$Myr at {\em all} evolved CR energies in the Galactic disk. But as discussed in the text, to be conservative we should account for the reduced-speed-of-light slowing down CR escape. As shown in \citet{hopkins:m1.cr.closure}, for our numerical formulation the steady-state solutions are guaranteed to be invariant to the choice of $\tilde{c}$, and we confirm this explicitly in e.g.\ Figs.~\ref{fig:spec.compare.numerics} \&\ \ref{fig:spec.compare.numerics.alt}, but in the worst-case scenario, the simulation time it takes for the simulation to accurately converge to steady-state can be increased systematically by a factor $=c/\tilde{c}$. But for our highest-$\tilde{c}$ tests in Figs.~\ref{fig:spec.compare.numerics} \&\ \ref{fig:spec.compare.numerics.alt}, this is just a factor $\sim 3$, so we would expect worst-case convergence times, at the slowest-converging energies, still $\ll 100\,$Myr, i.e.\ shorter than one Galactic dynamical time at the Solar circle (and far shorter than the $\sim 500\,$Myr for which we typically run our simulations). Of course, as discussed in the text (\S~\ref{sec:variation:env}, \ref{sec:variation:time}, \ref{sec:gamma.rays}), we have also confirmed directly that the simulations reached steady-state (as expected), by verifying that the results are statistically time-independent (e.g.\ independent of snapshot number, up to small stochastic and, as we noted, effectively ergodic variations at a given location corresponding to phenomena such as individual SNe bubbles) for the last several hundred Myr over which we evolve them. 

This does, however, introduce the caveat (discussed in the main text) that in very low-density gas, at distances far from the Galactic center ($\gg 10\,$kpc), e.g.\ in the CGM, where one might have very low $n_{\rm n}$ and $u_{\rm rad}+u_{\rm B}$ and very large $\ell_{\rm cr} \gg {\rm kpc}$, these timescales could become $\gtrsim$\,Gyr, in which case we would not expect the local CR spectra to have converged to equilibrium. Indeed, if the loss/escape times exceed $\sim$\,Gyr in these regions, then it is not clear if the CGM  or IGM can converge to true steady state at all in a Hubble time. In this regime (the distant CGM), fully-cosmological simulations are required to capture both the relevant timescales and non-equilibrium accretion/outflow/galaxy formation effects.

\subsubsection{Approximate ``Error Budget''}

For highly non-linear, chaotic, multi-physics simulations such as those in this paper, it is impossible to rigorously define a theoretical error or systematic uncertainty ``breakdown'' uniquely assigned to different terms. But we can make some estimates from our analytic derivations, idealized tests, and full-physics tests in the paper. 

On the purely numerical side, quantities such as errors from our assumed closure relation for the Vlasov hierarchy of the CR moments equations, or reduced-speed-of-light assumption, or finite time needed for the CR equations to converge to local steady-state, or numerical integration, are all formally demonstrably small, and we confirm this directly in our tests (both idealized but also full-physics, where we vary the closure, $\tilde{c}$, run-time or snapshots analyzed, resolution, etc.). From examination of our numerical derivations or full-physics results in Figs.~\ref{fig:demo.cr.spectra.fiducial}-\ref{fig:spec.compare.numerics} \&\ \ref{fig:spec.compare.kappa.alt}-\ref{fig:spec.compare.numerics}, the most significant numerical error is likely the ``bin-centered'' approximation for the spatial fluxes (\S~\ref{sec:numerics:spatial}). As is well-known (see \S~\ref{sec:numerical.tests} above and e.g.\ Fig.~12 in \citealt{girichidis:cr.spectral.scheme} and Fig.~7 in \citealt{ogrodnik:2021.spectral.cr.electron.code}), this produces the ``step'' structure between bins of a given $s$ (where the spectra are not perfectly smooth between bin edges), which in turn produces the more noticeable step features when taking ratios of e.g.\ B/C (where the bin edges do not exactly align in units of energy-per-nucleon).
We could reduce this error by increasing the number of bins by a large factor, but that is highly inefficient. 
A more efficient approach would be to evolve the fluxes without making such an approximation; but this introduces both conceptual difficulties (e.g.\ one must invoke some ``closure'' or model assumption for how $\langle \mu \rangle$ varies across each bin) and practical numerical challenges (it is difficult to construct a numerically-stable reconstruction and integration; see e.g.\ \citealt{girichidis:cr.spectral.scheme}).
So this is outside the scope of our study here.

In any case, Figs.~\ref{fig:demo.cr.spectra.fiducial}-\ref{fig:spec.compare.galaxies} \&\ \ref{fig:spec.compare.kappa.alt}-\ref{fig:spec.compare.numerics} show that physical uncertainties are generally much larger than these pure-numerical uncertainties. These are investigated in detail in the main text, but we briefly summarize here. One uncertainty is whether our galaxies are ``realistic'' in properties like the phase structure (which influence loss rates), source (SNe/massive-star) distribution, magnetic field structure, etc. For this reason we consider a wide range of models where we vary different loss terms, make different assumptions about streaming and diffusive reacceleration, change the magnetic fields by an order of magnitude, compare different local regions of the galaxy and different times/snapshots (separated by e.g.\ several galaxy dynamical times, over which the phase structure and local field geometry, etc, will vary), as well as entirely different galaxies. In general, these differences are larger than the numerical errors from e.g.\ the bin-centered flux approximation, but still small compared to the differences that arise from modest changes to the assumed scaling of the CR scattering rates (e.g.\ Figs.~\ref{fig:spec.compare.kappa} \&\ \ref{fig:spec.compare.kappa.alt}), and small compared to what would be needed to change our key qualitative conclusions (as demonstrated, for example, by the fact that our conclusions and favored $\bar{\nu}(R)$ are very similar to what is inferred in classic models which assume a static, much more highly-simplified analytic model for the Galaxy). So the dominant physical uncertainty, as we discuss in the text, is almost certainly our assumption that the CR scattering rates can be approximated as a universal-in-time-and-space function of rigidity. That is not the prediction of any physically-motivated model (in either extrinsic turbulence or self-confinement motivated scenarios), but is a common phenomenological approximation. In future work \citep{hopkins:2021.sc.et.models.incompatible.obs}, we will lift this assumption and explore different physically-motivated models.

\section{Magnetic Field Comparisons}
\label{sec:bfields}

\begin{figure}
	\includegraphics[width=0.98\columnwidth]{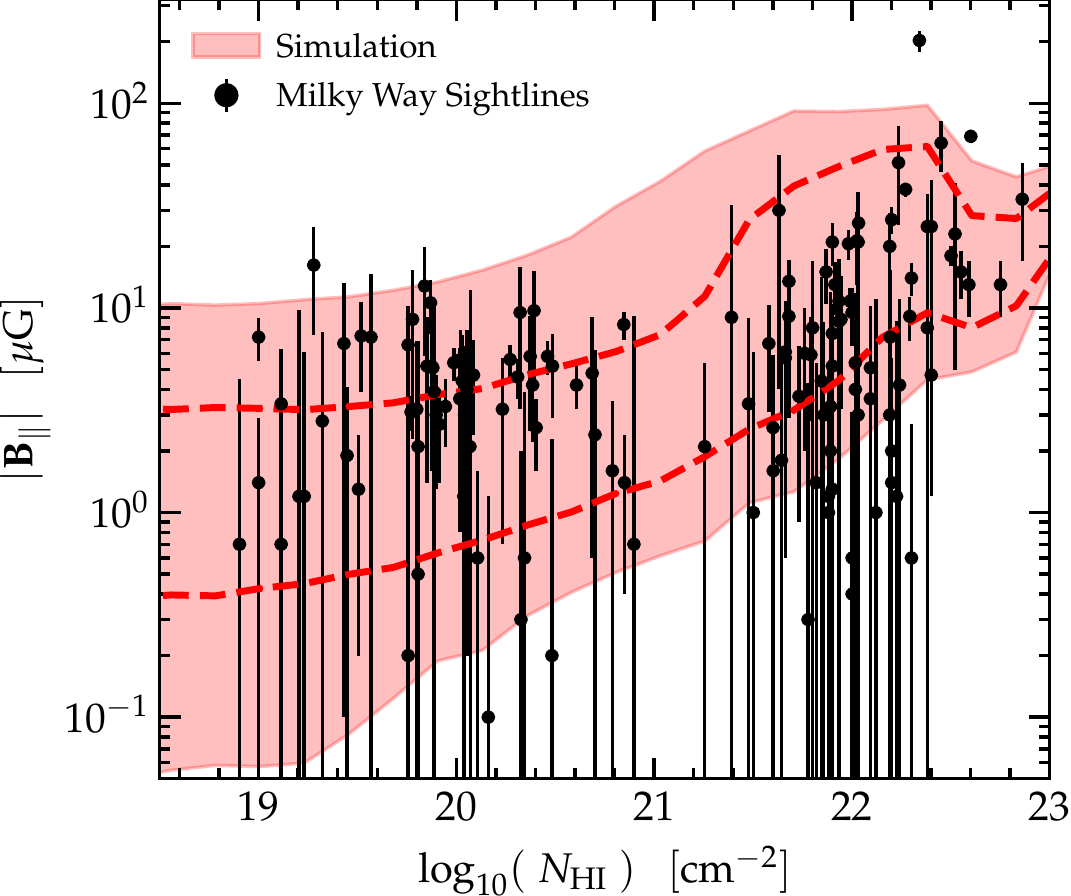} \\ 
	\includegraphics[width=0.98\columnwidth]{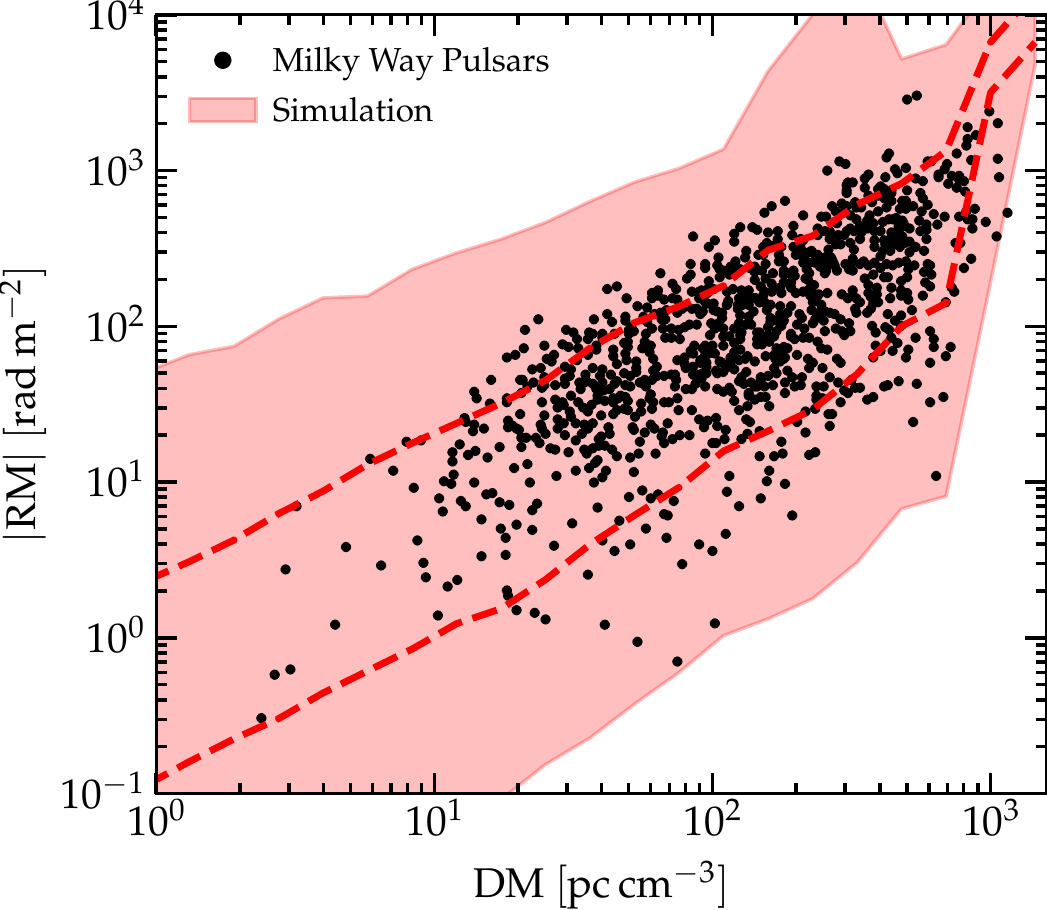} \\ 
	\vspace{-0.4cm}
	\caption{Comparison of mock-observational diagnostics of ISM magnetic fields to observations in our fiducial simulation (same as Fig.~\ref{fig:egy.corr.density}), from Ponnada et al.\ (in prep.). 
	{\em Top:} Simulation sightlines (dashed lines show $\pm 1\sigma$ range, shaded range shows $\pm 2\,\sigma$ or $5-95\%$ range, at each $N_{\rm HI}$) compared to Galactic Zeeman absorption observations from \citet{crutcher:2012.zeeman}.
	{\em Bottom:} Simulation sightlines (same style) compared to RM \&\ DM measurements of Milky Way pulsars from the ATNF catalogue.
	We construct mock sightlines through the simulation ISM (restricting to gas within $\pm1\,$kpc of the midplane and $<10\,$kpc from the galactic center), and calculate column densities of atomic HI ($N_{\rm HI}\equiv \int n_{\rm HI}\,d\ell$) and free electrons (${\rm DM}\equiv \int n_{e}\,d\ell$), as well as appropriately-weighted integrals along the line of sight to obtain ${\rm RM}\equiv (e^{3}/2\pi\,m_{e}^{2}\,c^{4})\,\int n_{e}\,{\bf B}_{\|}\,d\ell$ ({\em bottom}) and the cool HI-column-weighted $|{\bf B}_{\|}|$ ({\em top}).
	\label{fig:compare.bfields}}
\end{figure}

One can immediately read off the median magnetic field strength $|{\bf B}|$ as a function of ISM gas density $n$ in the Solar neighborhood from Fig.~\ref{fig:egy.corr.density} in the main text: fitting a power law we have roughly $|{\bf B}| \sim 6.3^{+3.5}_{-2.3}\,{\rm \mu G}\,(n/{\rm cm^{-3}})^{0.4}$. This is consistent with (perhaps slightly higher than, at the tens of percents level) typical observational estimates at gas densities $\sim 0.01-100\,{\rm cm^{-3}}$ \citep[e.g.][and references therein]{2009ASTRA...5...43B,sun:2010.galactic.bfield.halo.strong.upper.limits.microGauss,crutcher:cloud.b.fields,2015ASSL..407..483H,beck:2015.b.field.review,beck:2016.galactic.random.bfields.strong,ordog:2017.b.field.mw.disk}. 

More detailed comparison of the simulated magnetic fields to observational diagnostics for the ``single-bin'' FIRE+CR simulations and FIRE simulations without CRs can be found in \citet{su:2016.weak.mhd.cond.visc.turbdiff.fx,su:fire.feedback.alters.magnetic.amplification.morphology,guszejnov:fire.gmc.props.vs.z,chan:2021.cosmic.ray.vertical.balance}. In an independent study, Ponnada et al.\ (in prep) extend these by considering mock observations of Zeeman absorption (from the compilation in \citealt{crutcher:2012.zeeman}), Galactic (pulsar) RMs and DMs\footnote{The pulsar RM+DM points are taken from the Australia Telescope National Facility (ATNF) pulsar catalogue \citep{manchester:atnf.pulsar.catalogue}, using version 1.63 of the updated catalogue available at \href{http://www.atnf.csiro.au/research/pulsar/psrcat}{\url{http://www.atnf.csiro.au/research/pulsar/psrcat}}, as compiled in \citet{seta:2021.turb.sims.vs.rm.dm.obs}.}, as well as extragalactic RM constraints from indirect synchrotron modeling \citep{fletcher:RM.maps.from.synchrotron} and upper limits in the CGM inferred from e.g.\ fast radio bursts \citep[FRBs;][]{prochaska:2019.weak.magnetization.low.Bfield.rm.massive.gal.frb,lan:2020.cgm.b.fields.rm}. Ponnada et al.\ focuses on the simulations run fully-cosmologically from $z=100$ to $z=0$, i.e.\ our prior cosmological simulations with ``single-bin'' CR treatments (but otherwise identical physics and numerics to the simulations here), and therefore does not include the specific simulations (the controlled restarts) studied here. We therefore include here in Fig.~\ref{fig:compare.bfields} a direct comparison of the simulation magnetic field constraints from Galactic Zeeman and RM+DM observations, using the identical methodology applied to our ``default'' or fiducial simulation in this paper (from Fig.~\ref{fig:demo.cr.spectra.fiducial}) at $z=0$ (the end of the simulation). We refer to Ponnada et al.\ for details of the mock observational methodology. Together, the range of free electron column (DM) and atomic HI column ($N_{\rm HI}$) correspond in the simulations to a range of total (ionized+atomic+molecular) column densities from $\sim 10^{18.5} - 10^{24.3}\,{\rm cm^{-2}}$, almost six orders of magnitude. 

In brief, the agreement is good at a given DM or $N_{\rm HI}$ (and the simulations lie below the upper limits measured for the CGM from FRBs). Nonetheless, as emphasized in the text we still consider systematic variations of a factor of $\sim10$ in $|{\bf B}|$ to larger or smaller values -- well outside the range allowed by observations, but instructive as a counterfactual test case. As shown clearly in Figs.~\ref{fig:spec.compare.numerics} \&\ \ref{fig:spec.compare.numerics.alt}, even order-of-magnitude changes in $|{\bf B}|$ have little effect on our results (much smaller than modest changes to the assumed scattering rates).

\end{appendix}

\end{document}